\documentclass[final,3p,times]{elsarticle}

\usepackage{amsmath}
\usepackage{graphicx,bm}
\usepackage{ulem}
\usepackage{amssymb}
\usepackage{color}
\usepackage{version}

\usepackage[english,francais]{babel}

\usepackage{array} 
\usepackage{hyperref}
\hypersetup{breaklinks=true}
\hypersetup{colorlinks=true}
\hypersetup{urlcolor=blue}
\hypersetup{citecolor=black}
\hypersetup{linkcolor=black}
\usepackage{cleveref}
\usepackage{tikz}
\usepackage{float}

\usetikzlibrary{calc,patterns,decorations.pathmorphing,decorations.markings}

\newcommand{\be}{\begin{equation}}
\newcommand{\ee}{\end{equation}}
\newcommand{\bmul}{\begin{multline}}
\newcommand{\emul}{\end{multline}}
\newcommand{\bea}{\begin{eqnarray}}
\newcommand{\eea}{\end{eqnarray}}
\newcommand{\rr}{\mathbf{r}}
\newcommand{\kk}{\mathbf{k}}
\newcommand{\qq}{\mathbf{q}}
\newcommand{\vv}{\mathbf{v}}

\newcommand{\zero}{\mathbf{0}}
\newcommand{\bra}[1]{\langle #1|}
\newcommand{\ket}[1]{|#1\rangle}

\newcommand{\norm}{\mathcal{N}}

\newcommand{\meanv}[1]{\langle #1 \rangle}

\newcommand{\g}[1]{``#1''}

\newcommand{\bb}[1]{\left( #1 \right)}
\newcommand{\bbr}[1]{\left. #1 \right)}
\newcommand{\bbl}[1]{\left( #1 \right.}
\newcommand{\bbcro}[1]{\left[ #1 \right]}
\newcommand{\bbcror}[1]{\left. #1 \right]}
\newcommand{\bbcrol}[1]{\left[ #1 \right.}
\newcommand{\bbaco}[1]{\left\{ #1 \right\}}

\newcommand{\ii}{\textrm{i}}
\newcommand{\eee}{\textrm{e}}
\newcommand{\dd}{\mathrm{d}}

\newcolumntype{R}[1]{>{\raggedright\arraybackslash}m{#1}}
\newcolumntype{L}[1]{>{\raggedleft\arraybackslash}m{#1}}
\newcolumntype{V}[1]{>{\centering\arraybackslash}m{#1}}
\newcolumntype{C}[1]{>{\centering}m{#1}}

\def\bleu{\color{black}}
\def\rouge{\color{black}}
\def\vers{\color{black}}
\def\violet{\color{black}}
\def\rosso{\color{black}}

\DeclareMathOperator\im{Im}
\DeclareMathOperator\re{Re}

\newcounter{noterenormalisation}
\newcounter{notepasundemi}

\begin{document}
\selectlanguage{english}
\begin{frontmatter}

\title{Three-phonon and four-phonon interaction processes in a pair-condensed Fermi gas}
\author{H. Kurkjian}
\author{Y. Castin}
\author{A. Sinatra}
\address{Laboratoire Kastler Brossel, ENS-PSL, CNRS, UPMC-Sorbonne Universit\'es and Coll\`ege de France, Paris, France}

\begin{abstract}
We study the interactions among phonons and the phonon lifetime in a pair-condensed Fermi gas in the BEC-BCS crossover {\bleu in the collisionless regime}.
To compute the phonon-phonon coupling amplitudes we use a microscopic model based on a generalized BCS  Ansatz including moving pairs, which allows for a systematic expansion around the mean field BCS approximation of the ground state. 
We show that the quantum hydrodynamic expression of the amplitudes obtained by Landau and Khalatnikov apply only on the energy shell, that is
for resonant processes that conserve energy.
The microscopic model yields the same excitation spectrum as the 
Random Phase Approximation, with a linear (phononic) start and a concavity at low wave number that changes from upwards to downwards in the BEC-BCS crossover. 
When the concavity of the dispersion relation is upwards at low wave number, the leading damping mechanism at low temperature is the Beliaev-Landau process 2 phonons $\leftrightarrow$ 1 phonon while, when the concavity is downwards, it is the Landau-Khalatnikov process 2 phonons $\leftrightarrow$ 2 phonons. In both cases, by rescaling the wave vectors to absorb the dependence on the interaction strength, we obtain a universal formula for the damping rate. This universal formula corrects and extends the original analytic results of Landau and Khalatnikov [ZhETF {\bf 19}, 637 (1949)] for the $2\leftrightarrow2$ processes in the downward concavity case. In the upward concavity case, for the Beliaev 1$\leftrightarrow$ 2 process for the unitary gas at zero temperature, we calculate the damping rate of an excitation with wave number $q$ including the first correction proportional to $q^7$ to the $q^5$ hydrodynamic prediction, which was never done before in a systematic way.
\noindent
\vskip 0.5\baselineskip
\end{abstract}
\end{frontmatter}

\section{Introduction}

In several many-body systems the low-lying collective excitations are phonons.
At low temperature, interactions among phonons determine their lifetime, correlation time and mean-free path. Therefore, they play a central role in transport phenomena, such as thermal conduction in dielectric solids, in hydrodynamic properties, such as temperature dependent viscosity and attenuation of sound in liquid helium \cite{Khalatnikov1949,Khalatnikov1966,PhysRevLett.25.220,Sherlock1975,Smith2005,1367-2630-9-3-052}, and in macroscopic coherence properties of degenerate gases, as they determine the intrinsic coherence time of the condensate both in bosonic and fermionic gases \cite{CastinSinatra2009,KCS2015}.   
The dominant phonon decay channel differs among physical systems, and depends in particular on the curvature of the phonon excitation branch \cite{Wyatt1992,Wyatt2009,KCS2016}. For a convex dispersion relation at low wave-number, Beliaev-Landau $2 \leftrightarrow 1$ processes  involving three {\bleu quasiparticles} dominate \cite{Beliaev1958,Salasnich2015}, while for a concave dispersion relation at low wave-number, Beliaev-Landau interactions are not resonant, and the Landau-Khalatnikov $2 \leftrightarrow 2$ processes involving four quasiparticles dominate the dynamics. 

Compared to other many-body systems, cold atomic gases offer the unique possibility to control and tune some microscopic parameters, in particular the interaction strength. For a cold atomic Fermi gas in two spin states $\uparrow$ and $\downarrow$, interactions occur only in the $s$-wave between fermions of opposite spins. {\violet We restrict ourselves to the universal regime of interactions of negligible range $b$,} fully characterized by a single parameter, the $s$-wave scattering length $a$, {\violet $|a|\gg b$}. {\violet Using an external magnetic field, one can adjust this parameter of the interaction strength} close to a so-called Feshbach resonance. This feature has allowed cold Fermi gases experiments to study the crossover between the Bardeen-Cooper-Schrieffer (BCS) regime $1/a\to -\infty$, where the superfluid pairs 
{\violet $(\kk \uparrow,-\kk\downarrow)$} are localized in Fourier space close to the Fermi surface, and the Bose-Einstein Condensate (BEC) regime $1/a\to +\infty$, of dilute tightly bound dimers that behave as bosonic particles
\cite{Thomas2002,Salomon2003,Grimm2004,Grimm2004b,Ketterle2004,Ketterle2005,Salomon2010,Zwierlein2012,Grimm2013}. 
For the unpolarized gas of spin-$1/2$ fermions, the excitation spectrum consists of two branches: a gapped fermionic branch describing the excitation of the internal degrees of freedom of the $\uparrow\downarrow$ pairs, and a bosonic branch describing the collective motion of the pair center of mass, which has a phononic behavior at small wave vectors \cite{Anderson1958,Strinati1998,CKS2006,Tempere2011,Randeria2014,KCS2016}. If the density of excitations is sufficiently small, the elementary excitations are long-lived weakly interacting quasiparticles.

In cold gases of spin-$1/2$ fermions, the dispersion relation of phonons changes 
from convex in the BEC regime {\vers \cite{CKS2006}} to concave in the BCS regime {\vers \cite{Strinati1998}}; 
{\vers it is generally expected at unitarity to be convex at low wave number \cite{Salasnich2008,Manes2009,Rupak2009,Salasnich2011,Tempere2011,KCS2016}, although the predicted
values of the curvature parameter significantly differ and there is no experimental verdict yet.}
A complete experimental investigation of the interactions between phonons, in both the concave and convex cases, seems {\vers in any case}
possible within this same physical system. In particular, experiments should observe a sharp increase in the phonon lifetime by switching from the convex to the concave case through a tuning of the scattering length. Experimental studies of collective excitations in cold gases have been performed both in the spectral domain with Bragg scattering \cite{Davidson2002,Davidson2002b} and in the time domain \cite{Dalibard2002,Kinast2005}. 
Those studies combined with the possibility to realize homogeneous systems in flat-bottom traps \cite{Hadzibabic2013} 
{\bleu make} the measurement of the decay caused by either the $2 \leftrightarrow1$ or the $2 \leftrightarrow 2$ processes a concrete prospect.

To date, an exhaustive theoretical study of phonon interactions and a general expression of the phonon damping rates at low temperature for any interaction strength is missing.
The most natural example of a convex dispersion relation is the Bogoliubov excitation branch of the weakly interacting Bose gas. In this case, Beliaev obtained the coupling amplitudes in a microscopic framework and, from those, the damping rate associated to the $1\to2$ processes \cite{Beliaev1958}. The contribution of the $2\to1$ Landau processes, which exist only at nonzero temperature, was obtained later in \cite{Martin1965,Kondor1974,Stringari1997}. In superfluid spin-$1/2$ Fermi gases, reference~\cite{Salasnich2015} calculated the zero-temperature Beliaev $1\to2$ damping rate by a phenomenological low-energy effective theory going one step beyond quantum hydrodynamics in order to include a curvature in the phonon dispersion relation. However, the authors omitted to include corrections to the phonon coupling amplitude, which makes their treatment inconsistent.
In the bosonic case of phase II of $^4{\rm He}$, whose dispersion relation was originally believed to be concave,
Landau and Khalatnikov obtained the coupling amplitude for the $2\leftrightarrow2$ processes, including the contribution of virtual off-resonant $3$-phonon processes and introducing by hand corrections to hydrodynamics in the form of a cubic term in the excitation spectrum. However, the fact that they computed the damping rates only in the low and high wave-number limits makes this study incomplete and motivates us to revisit it. Studies of low-energy $2\leftrightarrow2$ processes in $^4{\rm He}$ were put on hold after 1970, when it was discovered that the dispersion relation of $^4{\rm He}$ is in fact convex at low wave-number
\cite{PhysRevLett.25.220,Sherlock1975} {\vers at least at not too high pressure \cite{Sherlock1975,Smith2005}}. Studies on processes in the wave-number ranges where the dispersion relation becomes concave again, particularly challenging due to the existence of a small convex region, are still an active research topic \cite{Wyatt2009}. To our knowledge, the study of the $2\leftrightarrow2$ processes in cold Fermi gases is mainly at the unitary limit and in the collisional hydrodynamic regime where the relevant measure for dissipation is the shear viscosity of the gas 
\cite{Schafer2007,Zwerger2011,Manuel2013}.
In this article, {\violet we restrict to temperatures low enough to be in the collisionless regime. There,} we  present a complete study of the interaction processes between phonons in cold Fermi gases,
for any interaction strength {\violet in the BEC-BCS crossover}, therefore in both the concave and convex cases. 

We describe the unpolarized pair-condensed Fermi gas using a microscopic semiclassical model, based on a variational state including moving pairs. This model allows for a systematic expansion of the Hamiltonian in terms of two canonically conjugate fields $\beta$ and $\beta^\ast$ that are weak if the density of excitations in the gas is low. 
The microscopic model and the principle of the method are exposed in section \ref{sec:methode}. 

In section \ref{sec:quadratique} we consider the expansion of the Hamiltonian to the quadratic order in the fields $\beta$ and $\beta^\ast$. By diagonalizing the quadratic problem, we find the spectrum of the excitations \cite{Anderson1958,Strinati1998,CKS2006,Tempere2011,Randeria2014,KCS2016}.
Beyond the purely spectral results known in the literature, we derive here explicitly the {\bleu quasiparticle} modes and the annihilation/creation amplitudes, which we then quantize. This is rather elegantly done thanks to the Hamiltonian character of the equations in the semiclassical model. 

Section \ref{sec:interactions} is devoted to the calculation of the coupling amplitudes for three-phonon and four-phonon interaction processes.
We give the quantum hydrodynamics \cite{Khalatnikov1949} predictions for the amplitudes of the $2 \leftrightarrow 1$ and the $2 \leftrightarrow 2$ processes, and compare them on the energy shell to a microscopic derivation, based for $2 \leftrightarrow 1$ on the fermionic model of section \ref{sec:quadratique}, and for $2 \leftrightarrow 2$ on a model of bosons with finite range interactions designed to have a concave dispersion relation at low wave numbers. Our microscopic test of the hydrodynamic predictions is particularly valuable for the $2 \leftrightarrow 2$ processes, as $(i)$ these processes involve nonresonant $2 \leftrightarrow 1$ and $3 \leftrightarrow 0$ processes, whose amplitudes are not correctly given by hydrodynamics, $(ii)$ genuine hydrodynamics leads to a divergence in the $2 \leftrightarrow 2$ amplitude that Landau and Khalatnikov had to regularize ``by hand'' by introducing a curvature in the phonon dispersion relation.

In section \ref{sec:amortissement} we present a direct application of the previous results by calculating the damping rate of phonons in a Fermi gas in the whole BEC-BCS crossover {\bleu in the collisionless regime}. To this aim we use the effective on-shell coupling amplitudes for $2 \leftrightarrow 1$ and $2 \leftrightarrow 2$ processes calculated in section \ref{sec:interactions} within a Master equation approach.
The {\violet three} main results of section \ref{sec:amortissement} are that $(i)$ both for a convex and concave dispersion, by introducing properly normalized dimensionless quantities, we obtain a universal curve, giving the damping rate as a function of the wave vector, for any interaction strength in the BEC-BCS crossover, $(ii)$ we give all the analytic asymptotic behaviors of the phonon damping rates, for $\hbar cq \ll k_{\rm B}T$ and $\hbar cq \gg k_{\rm B}T$ where $c$ is the speed of sound and $T$ is the temperature, correcting in particular the result of Landau and Khalatnikov \cite{Khalatnikov1949}
in the concave case, {\violet and} $(iii)$ for the unitary gas and at zero temperature only, we calculate the first correction to the hydrodynamic prediction for the phonon damping {\violet rate}, using the effective field theory of Son and Wingate \cite{SonWingate2006,Manuel2010}, and improving the result of {\violet reference} \cite{Salasnich2015}. 



\section{The microscopic approach}
\label{sec:methode}

\subsection{Interaction Hamiltonian in the $s$-wave}


We consider a gas of fermions in two internal states $\uparrow$ and $\downarrow$ on a cubic lattice of step $l$ with periodic boundary conditions in the finite-size volume $[0,L]^3$. Opposite spin fermions have an on-site interaction characterized by a coupling constant $g_0$
\be
V(\rr,\rr')=g_0\frac{\delta_{\rr,\rr'}}{l^3}
\label{eq:potentiel}
\ee
{\vers The discretisation of space automatically provides a momentum cut-off, restricting the single-particle wave vectors to the first
Brillouin zone $\mathrm{FBZ}=[-\pi/l,\pi/l[^3$ of the lattice, while allowing one to keep the simplicity of contact interactions.}
The Hamiltonian of the system in the grand canonical ensemble of chemical potential $\mu$ is given by:
\be
\hat{H}= l^3  \sum_{\rr, \sigma=\uparrow/\downarrow} \hat{\psi}_\sigma^\dagger(\rr) \left( - \frac{\hbar^2}{2m}  \Delta_{\rr} -\mu\right)  \hat{\psi}_\sigma(\rr) \\
+ g_0 l^3 \sum_{\rr} \hat{\psi}_\uparrow^\dagger(\rr) \hat{\psi}_\downarrow^\dagger(\rr)\hat{\psi}_\downarrow(\rr){\hat{\psi}_\uparrow}(\rr)
\label{eq:hamiltonien}
\ee
where the discrete Laplacian operator $\Delta_\rr$ has the eigenfunctions $\eee^{\ii\kk \cdot \rr}$ 
with eigenvalues $-k^2$ 
and the field operator of the fermions has the discrete anticommutation relations $\{ \hat{\psi}_\sigma(\rr) , \hat{\psi}_{\sigma'}^\dagger({\rr'}) \} = \delta_{\sigma \sigma'} \delta_{\rr \rr'}/l^3$, 
with $\sigma, \sigma'=\uparrow {\rm or} \downarrow$. 
{\rouge The desired regime of zero-range interactions is obtained by taking the continuous space limit $l\to 0$ for a scattering length $a$
fixed to the experimental value. In practice, one eliminates the bare coupling constant $g_0$}
using the following {\rouge relation}, derived from the {\vers open space {\violet ($L=\infty$)}} scattering theory applied to the potential \eqref{eq:potentiel} 
\cite{CastinLesHouches2004,Varenna}:
\begin{equation}
\label{eq:g0}
\frac{1}{g_0} = \frac{m}{4 \pi \hbar^2 a} - \int_{\rm FBZ} \frac{\dd^3k}{(2\pi)^3} \frac{m}{\hbar^2 k^2} 
\end{equation}

\subsection{Ground state in the BCS approximation}

The idea behind the theory of Bardeen-Cooper-Schrieffer (BCS) \cite{BCS1957} is to search for an approximation of the ground state of the Hamiltonian \eqref{eq:hamiltonien} within the family of states
\begin{equation}
\ket{\psi^{\rm BCS}}=  \prod_{\kk\in\mathcal{D}}  \bb{ U_\kk - V_\kk \hat{a}_{\kk \uparrow}^\dagger \hat{a}_{-\kk \downarrow}^\dagger} \ket{0}
\label{eq:ansatzBCS2}
\end{equation}
where $\mathcal{D}=\frac{2\pi}{L}\mathbb{Z}^{3}\cap[-\pi/{l},\pi/{l}[^3$ is the ensemble of wavevectors of the first Brillouin zone satisfying the periodic boundary conditions, the operator $\hat{a}_{\kk \sigma}$ is a Fourier component of the field operator $\hat{\psi}_\sigma(\rr)$ and annihilates a fermion of wavevector $\kk$ and spin $\sigma$, $V_\kk$ is the probability amplitude of finding a $\kk\sigma$ fermion in $\ket{\psi^{\rm BCS}}$ 
and $U_\kk=\sqrt{1-|V_\kk|^2}$. The locus of the minimizers of the classical energy functional $\bra{\psi^{\rm BCS}} \hat{H} \ket{\psi^{\rm BCS}}$ is the circle
\be
U_\kk=U_\kk^0 \qquad V_\kk=V_\kk^0 \eee^{\ii \phi}
\label{eq:paramBCS}
\ee
where $\phi\in[0,2\pi[$ and the values of reference $U_\kk^0$ {\vers and} $V_\kk^0$ are chosen real.

To study the low-energy behavior of the system, we select on this circle the state of phase $\phi=0$ 
\be
\ket{\psi^{\rm BCS}_0}=  \prod_\kk  \bb{ U_\kk^0 - V_\kk^0 \hat{a}_{\kk \uparrow}^\dagger \hat{a}_{-\kk \downarrow}^\dagger} \ket{0}
\label{eq:psiBCS0}
\ee
as the origin of our expansion.
In this symmetry-breaking state, we define the \textit{order parameter} of the BCS theory
\be
\Delta\equiv g_0 \bra{\psi^{\rm BCS}_0}{\hat{\psi}_\downarrow \hat{\psi}_\uparrow}\ket{\psi^{\rm BCS}_0} = - \frac{g_0}{L^3} \sum_{\kk\in\mathcal{D}} U_\kk^0 V_\kk^0,
\label{eq:defDelta}
\ee
This quantity replaces $g_0$ or $a$ as the most natural parameter characterizing the interaction strength in the BCS theory. It can be used to express the coefficients $U_\kk^0,V_\kk^0$ in the form
\be
V_\kk^0=\sqrt{\frac{1}{2}\bb{1-\frac{\xi_\kk}{\epsilon_\kk}}} \quad \mbox{and} \quad U_\kk^0=\sqrt{\frac{1}{2}\bb{1+\frac{\xi_\kk}{\epsilon_\kk}}}
\label{eq:Vk0}
\ee
with the energies
\bea
\xi_\kk &=& \frac{\hbar^2k^2}{2m}-\mu+ \frac{g_0{\rho}}{2} \label{eq:defxik} \\
\epsilon_\kk &=& \sqrt{\Delta^2+\xi_\kk^2} \label{eq:defepsilonk}
\eea
With the {\rouge useful} relation $U_\kk^0 V_\kk^0= \Delta/2\epsilon_\kk$, the gap equation \eqref{eq:defDelta} takes its more usual form:
\be
\frac{1}{g_0} = - \frac{1}{L^3}\sum_{\kk\in\mathcal{D}} \frac{1}{2\epsilon_\kk}
\label{eq:gap}
\ee
In the state \eqref{eq:psiBCS0}, the average total density ${\rho}$, or average total number of particles $\meanv{\hat{N}}$ per unit volume, is given by
\be
{\rho}\equiv\frac{\meanv{\hat{N}}}{L^3}\equiv\frac{k_{\rm F}^3}{3\pi^2}=\frac{2}{L^3}\sum_{\kk\in\mathcal{D}} (V_\kk^0)^2
\label{eq:rhobar}
\ee
where $k_{\rm F}$ is the Fermi wave number of the ideal gas of density $\rho$.
Combined with equation~\eqref{eq:Vk0} this leads to the BCS equation of state which relates ${\rho}$ to $\mu$ and $\Delta$.

Performing a Bogoliubov rotation on the particle creation and annihilation operators, one finally defines the creation and annihilation operators of fermionic excitations:
\bea
\hat{{\vers\gamma}}_{\kk\uparrow}     &=& U_\kk^0 \hat{a}_{\kk\uparrow} + V_\kk^0 \hat{a}_{-\kk\downarrow}^\dagger \label{eq:bup} \\
\hat{{\vers\gamma}}_{-\kk\downarrow} &=& -V_\kk^0 \hat{a}_{\kk\uparrow}^\dagger + U_\kk^0 \hat{a}_{-\kk\downarrow}  \label{eq:bdown}
\eea
The BCS ground state \eqref{eq:psiBCS0} is the vacuum of these operators, that {\bleu annihilate} $\kk\uparrow$ and $-\kk\downarrow$ quasiparticles of energy $\epsilon_\kk$.
{\rosso For completeness we give the corresponding modal expansion of the fermionic fields:
\be
\begin{pmatrix}\psi_\uparrow(\rr) \\ \psi_\downarrow^\dagger(\rr) \end{pmatrix}=
\frac{1}{L^{3/2}} \sum_{\kk\in\mathcal{D}} \hat{\gamma}_{\kk\uparrow} \begin{pmatrix} U_\kk^0 \\ V_\kk^0\end{pmatrix}
\eee^{\ii\kk\cdot\rr} + \hat{\gamma}_{\kk\downarrow}^\dagger \begin{pmatrix} -V_{\kk}^0 \\ U_\kk^0 \end{pmatrix}
\eee^{-\ii\kk\cdot\rr}
\label{eq:devmodal}
\ee
}


\subsection{Ansatz of moving pairs}


{\vers To study the fluctuations of the system, we view the symmetry-breaking BCS ground state as the {\violet quasiparticle  vacuum}
and we construct a coherent state of pairs of quasiparticles, or squeezed vacuum, from it:}
\be
\ket{\psi{\rosso(t)}}=\mathcal{N}(t) \exp\bb{\sum_{\rosso\kk,\qq} z_{\kk+\qq,\kk}{\bleu(t)} \hat{\vers\gamma}_{\kk+\qq\uparrow}^\dagger 
\hat{\vers\gamma}_{-\kk\downarrow}^\dagger } \ket{\psi_{\rm BCS}^0}
\label{eq:Ansatz}
\ee
{\rouge where the independent variational parameters $z_{\kk+\qq,\kk}$ are the coherent state complex amplitudes and 
$\mathcal{N}(t)$ is the normalisation factor. This Ansatz is inspired from section 9.9b of reference \cite{Ripka1985}.}
Contrarily to the BCS ground state \eqref{eq:psiBCS0} in which all the pairs are at rest {\rouge it} includes pairs of quasiparticles with nonzero center-of-mass wavevector $\qq$. {\rouge This will allow us to study the gapless, phononlike, collective modes of the system and their interactions.
We do not include in the Ansatz elementary fermionic excitations obtained by the direct action of $\hat{\vers\gamma}_{\kk\sigma}^\dagger$ on the BCS ground state
$\ket{\psi_{\rm BCS}^0}$ since they are gapped, and we are interested here in excitation energies much smaller than the gap.
Note that the quasiparticle {\bleu annihilation and creation} operators $\hat{\vers\gamma}_{\kk\sigma}$ and $\hat{\vers\gamma}_{\kk\sigma}^\dagger$ are linear combinations
of the particle {\bleu annihilation and creation} operators $\hat{a}_{\kk\sigma}$ and $\hat{a}_{\kk\sigma}^\dagger$, see equations~(\ref{eq:bup},\ref{eq:bdown}),
so that considering a coherent state of quasiparticle pairs (\ref{eq:Ansatz}) is equivalent to considering a coherent state of pairs of particles,
that is {\sl in fine} a general time-dependent BCS Ansatz. In the end, all these Ansatz are Gaussian in the fermionic field operators, but
the form (\ref{eq:Ansatz}) leads to much simpler calculations in the weakly excited regime as it immediately identifies
the $z_{\kk+\qq,\kk}$ as small parameters of the expansion.}

References \cite{Ripka1985} and \cite{TheseHK} explain how to apply the variational principle to the state \eqref{eq:Ansatz}. One cleverly 
introduces the variables
\be
\beta_{\kk,\kk'}=-\bb{z(1+z^\dagger z)^{-1/2}}_{\kk,\kk'}
\label{eq:defbeta}
\ee
with $z$ the matrix $(z_{\kk,\kk'})_{\kk,\kk'\in\mathcal{D}}$, so that the equations of motion take a Hamiltonian form
\bea
\ii\hbar\frac{\dd\beta_{\kk',\kk}}{\dd t} &=& \frac{\partial E}{ \partial \beta_{\kk',\kk}^* } \label{eq:hamiltonienne1}\\
-\ii\hbar\frac{\dd\beta_{\kk',\kk}^*}{\dd t} &=& \frac{\partial E}{\partial \beta_{\kk',\kk}}  \label{eq:hamiltonienne2}
\eea
One evaluates the associated classical Hamiltonian
\be
E\equiv\bra{\psi}\hat{H}\ket{\psi}
\label{eq:defE}
\ee
through Wick's theorem, {\rosso that relates $E$ to averages of operators that are bilinear in $\hat{\vers\gamma}_{\kk\sigma}$}. 
Finally one obtains an elegant expression for {\rosso these} averages 
by performing a Schmidt decomposition of the matrix $z$ and expressing the result in terms of the field $\beta$, 
{\violet in matrix notation:}
{\rosso
\bea
\meanv{\hat{\rosso\gamma}_{-\kk\downarrow} \hat{\rosso\gamma}_{\kk'\uparrow}} &=& - \bb{\beta(1-\beta^\dagger\beta)^{1/2}}_{\kk',\kk} \label{eq:kappa} \\
\meanv{\hat{\rosso\gamma}_{\kk\uparrow}^\dagger \hat{\rosso\gamma}_{-\kk'\downarrow}^\dagger} &=& - \bb{\beta(1-\beta^\dagger\beta)^{1/2}}_{\kk,\kk'}^* \label{eq:kappa_star} \\
\meanv{\hat{\rosso\gamma}_{\kk'\uparrow}^\dagger \hat{\rosso\gamma}_{\kk\uparrow}} &=& \bb{\beta\beta^\dagger}_{\kk,\kk'} \label{eq:rho_up} \\
\meanv{\hat{\rosso\gamma}_{-\kk'\downarrow}^\dagger \hat{\rosso\gamma}_{-\kk\downarrow}} &=&  \bb{\beta^\dagger\beta}_{\kk',\kk}.\label{eq:rho_down}
\eea
}


\section{Quadratic order of the classical Hamiltonian: normal variables of the bosonic branch and phonon operators}
\label{sec:quadratique}


\subsection{Linearized equations of motion}


To linearize the equations of motion \eqref{eq:hamiltonienne1} and \eqref{eq:hamiltonienne2}, we assume that the state \eqref{eq:Ansatz} differs only slightly from the BCS ground state, so that
\be
\forall \kk,\kk',\,|z_{\kk,\kk'}|\ll 1 \quad \mbox{and} \quad |\beta_{\kk,\kk'}|\ll 1, \label{eq:petitesamplitudes}
\ee
We expand the energy functional in powers of the field $\beta$
\be
E=E_0+E_2+O(\beta^3)
\label{eq:devE}
\ee
where $E_0=\bra{\psi^{\rm BCS}_0}\hat{H}\ket{\psi^{\rm BCS}_0}$ is a constant {\violet and} $E_2$ is bilinear in $\beta$. There are no linear terms in this expansion since the BCS ground state (corresponding to $\beta=0$) is a minimizer of $E$. As in the Random Phase Approximation (RPA) \cite{Anderson1958}, the linearized equations of motion, which follow from \eqref{eq:hamiltonienne1} {\vers and} \eqref{eq:hamiltonienne2} after replacement of $E$ by $E_2$, are decoupled according to the total wavevector $\qq$. For this reason, we rewrite the coordinates of the field {\violet $\beta_{\kk_1,\kk_2}$}
in the Anderson fashion, with the relative wavevector {\violet $(\kk_1+\kk_2)/2$} in the index and the center-of-mass one 
{\violet $\kk_1-\kk_2$} in the exponent, {\bleu keeping in mind that $\beta_{\kk_1,\kk_2}$ actually involves 
{\violet the physical wave vectors $\kk_1$ and $-\kk_2$, see equation (\ref{eq:kappa}), so that:}}
\be
\beta_{\kk}^\qq\equiv\beta_{\kk+\qq/2,\kk-\qq/2}
\label{eq:notationsRPA}
\ee
In terms of the vectors $\beta^\qq=(\beta_\kk^\qq)_{\kk\in\mathcal{D}}$ 
and $\bar{\beta}^\qq=((\beta_{\kk}^{-\qq})^*)_{\kk\in\mathcal{D}}$,
the equations of motion take the matrix form
\be
\ii\hbar \frac{\dd}{\dd t} \begin{pmatrix}
\beta^\qq \\
\bar{\beta}^\qq
\end{pmatrix}=
\mathcal{L}^\qq
 \begin{pmatrix}
\beta^\qq \\
\bar{\beta}^\qq
\end{pmatrix},
\label{eq:systvaria}
\ee
with an evolution operator $\mathcal{L}^\qq$ that is symplectic
\be
\sigma_z \mathcal{L}^\qq \sigma_z = (\mathcal{L}^\qq)\dagger
\label{eq:symplectique}
\ee
and particle-hole symmetric
\be
\sigma_x \mathcal{L}^\qq \sigma_x = -(\mathcal{L}^\qq)^*
\label{eq:renversementtemps}
\ee
We have introduced $\sigma_x=\begin{pmatrix}0&1\\1&0\end{pmatrix}$ and $\sigma_z=\begin{pmatrix}1&0\\0&-1\end{pmatrix}$ in block notations,
{\vers and we termed (\ref{eq:renversementtemps}) a particle-hole symmetry because $\sigma_x$ exchanges the field $\beta$ with
the complex-conjugate field $\bar{\beta}$, {\violet which is reminiscent of the fact that, in the quantum world,} the hole field operator is the hermitian conjugate of the particle field operator.}

We give an explicit expression of the evolution operator $ \mathcal{L}^\qq$ in the sum-and-difference basis
\bea
y_{\kk}^\qq=\beta_{\kk}^\qq - \bar{\beta}_{\kk}^{\qq} \\
s_{\kk}^\qq=\beta_{\kk}^\qq + \bar{\beta}_{\kk}^{\qq}
\eea
We have
\be
\ii\hbar\frac{\dd y_{\kk}^\qq }{\dd t}  =  \epsilon_{\kk\qq} s_{\kk}^\qq    +  \frac{g_0}{L^3} \sum_{\kk'\in\mathcal{D}} \bb{W_{\kk\qq}^-  W_{\kk'\qq}^- + w_{\kk\qq}^+  w_{\kk'\qq}^+} {s}_{\kk'}^\qq \label{eq:systvaria1} 
\ee
\be
\ii\hbar\frac{\dd{s_{\kk}^\qq }}{\dd t}  =  \epsilon_{\kk\qq} y_{\kk}^\qq    +   \frac{g_0}{L^3} \sum_{\kk'\in\mathcal{D}} \bb{W_{\kk\qq}^+ W_{\kk'\qq}^+ - w_{\kk\qq}^- w_{\kk'\qq}^-} y_{\kk'}^\qq 
\label{eq:systvaria2}
\ee
The coefficients $W_{\kk,\qq}^\pm$ and $w_{\kk,\qq}^\pm$ are {\vers cleverly chosen} combinations of the coefficients $U_\kk^0$ and $V_\kk^0$
\bea
W_{\kk\qq}^\pm=U_{\kk+\qq/2}^0U_{\kk-\qq/2}^0\pm V_{\kk+\qq/2}^0V_{\kk-\qq/2}^0 \label{eq:defW} \\
w_{\kk\qq}^\pm=U_{\kk+\qq/2}^0V_{\kk-\qq/2}^0\pm U_{\kk-\qq/2}^0V_{\kk+\qq/2}^0 \label{eq:defw} 
\eea
and the energies $\epsilon_{\kk\qq}$ are those of the continuum of two fermionic quasiparticles,
\be
\epsilon_{\kk\qq}=\epsilon_{\kk+\qq/2}+\epsilon_{\kk-\qq/2}
\ee
The equations of motion (\ref{eq:systvaria1},\ref{eq:systvaria2}) contain two terms: first, an individual part that couples the amplitudes of same relative and center-of-mass wavevectors $\kk$ and $\qq$ and contains the trivial evolution of the operators $\hat{\vers\gamma}_{\kk\sigma}$ under the BCS Hamiltonian $\hat{H}_{\rm BCS}=E_0+\sum_{\kk\sigma} \epsilon_\kk \hat{\vers\gamma}_{\kk\sigma}^\dagger \hat{\vers\gamma}_{\kk\sigma} $, and second, a collective part that couples the normal amplitudes $ y_{\kk}^\qq$ and $ s_{\kk}^\qq$ to collective amplitudes of same total wavevector $\qq$.
Our semi-classical equations of motion coincide with the quantum average in state \eqref{eq:Ansatz} of the RPA equations \cite{Anderson1958}. This can be seen by using the definitions \eqref{eq:bup} and \eqref{eq:bdown} to express equations~(78a--d) of reference~\cite{Anderson1958} in terms of the quasiparticle operators  and by neglecting the expectation values of the $\hat{\vers\gamma}_{\kk\sigma}^\dagger \hat{\vers\gamma}_{\kk'\sigma}$ operators, which, by virtue of equations~\eqref{eq:rho_up} and \eqref{eq:rho_down}, is justified at the linear order of the variational theory.


\subsection{Branch of collective excitations}


We now look for the eigenmodes of the linear system (\ref{eq:systvaria1},\ref{eq:systvaria2}) with a positive energy $\hbar\omega_\qq$ below the continuum $k\mapsto\epsilon_{\kk+\qq/2}+\epsilon_{\kk-\qq/2}$ of two fermionic excitations, which is the spectrum we would obtain by restricting ourselves to the individual part of the system (\ref{eq:systvaria1},\ref{eq:systvaria2})
\be
0<\hbar\omega_\qq<\inf_{\kk}  (\epsilon_{\kk+\qq/2}+\epsilon_{\kk-\qq/2})
\ee
The eigenvalue problem associated to (\ref{eq:systvaria1},\ref{eq:systvaria2}) is solved in full generality in reference~\cite{TheseHK}. 
We give the main steps of the solution {\vers in the simplifying continuous space limit $l\to 0$ in 
\ref{appen:RPA}.}
The implicit equation on the mode eigenfrequency reads
\be
I_{++}(\omega_\qq,q) I_{--}(\omega_\qq,q) = {\hbar^2\omega^2_\qq} \left[I_{+-}(\omega_\qq,q)\right]^2
\label{eq:dedispersion}
\ee
{\vers where we introduced the integrals}
\bea
I_{++}(\omega,q) \!\!&=&\!\!\int_{\mathbb{R}^3} \!\!\dd^3k\left[\frac{\epsilon_{\kk\qq}(W_{\kk\qq}^+)^2}{(\hbar\omega)^2
-(\epsilon_{\kk\qq})^2}+\frac{1}{2\epsilon_\kk}\right]  \label{eq:intpp}\\
I_{--}(\omega,q) \!\!&=&\!\! \int_{\mathbb{R}^3} \!\!\dd^3k\left[\frac{\epsilon_{\kk\qq}(W_{\kk\qq}^-)^2}
{(\hbar\omega)^2-(\epsilon_{\kk\qq})^2}+\frac{1}{2\epsilon_\kk}\right] \label{eq:intmm} \\
\!\!\!\! I_{+-}(\omega,q) \!\!&=&\!\! \int_{\mathbb{R}^3} \!\!\dd^3k\frac{W_{\kk\qq}^+ W_{\kk\qq}^-}{(\hbar\omega)^2-(\epsilon_{\kk\qq})^2}
\label{eq:intpm}
\eea
{\vers We note that the same equation \eqref{eq:dedispersion} can be obtained by the RPA; this was done by Anderson \cite{Anderson1958} in the weak coupling limit and by reference~\cite{TheseHK} in the general case. Although the linear system of the RPA contains extra terms proportional to the operators $\hat{\vers\gamma}_{\kk+\qq\uparrow}^\dagger \hat{\vers\gamma}_{\kk\uparrow}$ and $\hat{\vers\gamma}_{-\kk\downarrow}^\dagger \hat{\vers\gamma}_{-\kk-\qq\downarrow}$, which are absent in 
(\ref{eq:systvaria1},\ref{eq:systvaria2}) because their expectation values in the Ansatz \eqref{eq:Ansatz} are quadratic in the field $\beta$,
these extra terms can be treated as source terms and do not affect the spectrum. Equation~(\ref{eq:dedispersion}) can also be obtained by} a Gaussian approximation of the action in a path integral framework \cite{Strinati1998,Randeria2014} and a Green’s functions approach associated with a diagrammatic approximation 
\cite{CKS2006}.
The conditions on $q$ for the existence of a solution $\omega_\qq$ are discussed in reference~\cite{CKS2006}, while the concavity of the spectrum is studied in reference~\cite{KCS2016}. Beyond those previous works, we construct here {in the next section} the quantum operators associated to the collective modes.


\subsection{Construction of the normal variables of the collective branch}
\label{sec:variablesnormales}


\subsubsection{General case}


Using the symmetries \eqref{eq:symplectique} and \eqref{eq:renversementtemps} of the evolution operator, we obtain the normal amplitudes exactly as in the bosonic case with the Bogoliubov theory \cite{CastinDum1998}.
The first step is to derive the eigenvector $\vec{e}_{+}$ of energy $\hbar\omega_\qq>0$
\be
\vec{{e}}_{+}(\qq) =
\begin{pmatrix} 
M_{\kk}^{\qq} \vspace{2mm}\\
N_{\kk}^{\qq}
\end{pmatrix}_{\kk\in\mathcal{D}}
\ee
To obtain the analytic expression of the coefficients $M$ {\vers and} $N$ in the continuous limit $l\to0$, we solve the system (\ref{eq:systvaria1},\ref{eq:systvaria2}) setting $\dd/\dd t \to - \ii \omega_\qq$ and we use equations \eqref{eq:yqsq} and \eqref{eq:proprecollective} to eliminate the collective amplitudes. We have
\bea
\!\!\!\!\!\!M_{\kk}^{\qq}-N_{\kk}^{\qq} &=& \frac{2\Delta\bb{ \epsilon_{\kk\qq} W_{\kk\qq}^+  - W_{\kk\qq}^- \frac{I_{++}(\omega_\qq,q)}{I_{+-}(\omega_\qq,q)}   }
}{ \sqrt{\norm_\qq}({\rosso \epsilon_{\kk\qq}^2 - \hbar^2\omega_\qq^2} )} \label{eq:Ykq} \\
\!\!\!\!\!\!M_{\kk}^{\qq}+N_{\kk}^{\qq} &=&  \frac{2\Delta \bb{ \hbar^2\omega_\qq^2 W_{\kk\qq}^+  -  \epsilon_{\kk\qq} W_{\kk\qq}^- \frac{I_{++}(\omega_\qq,q)}{I_{+-}(\omega_\qq,q)}   }}{ \hbar\omega_\qq \sqrt{\norm_\qq}({\rosso \epsilon_{\kk\qq}^2 - \hbar^2\omega_\qq^2})} 
\label{eq:Skq}
\eea
where the dimensionless normalization constant $\norm_\qq$ will be determined by equation~\eqref{eq:normalisation}. Important properties of $M$ and $N$ are their invariance by internal ($\kk\to-\kk$) and external ($\qq\to-\qq$) parity, where internal and external refers to the structure of the pairs:
\bea
M_{\kk}^{\qq} &=& M_{-\kk}^{\qq}   = M_{\kk}^{-\qq} \label{eq:invartranspo1}\\
N_{\kk}^{\qq} &=& N_{-\kk}^{\qq}   = N_{\kk}^{-\qq} \label{eq:invartranspo2}
\eea
This is a consequence of the parity invariance of the problem and of the $s$-wave pairing.
Due to the {\violet particle-hole} symmetry \eqref{eq:renversementtemps},
one can associate to the eigenvector $\vec{e}_{+}$ another eigenvector $\vec{e}_{-}$ of energy $-\hbar\omega_\qq$ by a multiplication by $\sigma_x$:
\be
\vec{{e}}_{-}(\qq) =
\sigma_x
\begin{pmatrix} 
M_{\kk}^{\qq} \vspace{2mm}\\
N_{\kk}^{\qq}
\end{pmatrix}_{\kk\in\mathcal{D}}
=
\begin{pmatrix} 
N_{\kk}^{\qq} \vspace{2mm}\\
M_{\kk}^{\qq}
\end{pmatrix}_{\kk\in\mathcal{D}}
\ee
We now derive the dual vectors with which we will project the field $\beta$ on the eigenmodes $\vec{e}_{+}$ and $\vec{e}_{-}$. The symplectic symmetry \eqref{eq:symplectique} ensures that they are obtained by a mere multiplication by $\sigma_z$
\bea
\vec{{d}}_{+}(\qq)= \sigma_z
\begin{pmatrix} 
M_{\kk}^{\qq} \vspace{2mm}\\
N_{\kk}^{\qq}
\end{pmatrix}_{\kk\in\mathcal{D}} &=& 
 \begin{pmatrix} 
M_{\kk}^{\qq} \vspace{2mm}\\
-N_{\kk}^{\qq}
\end{pmatrix}_{\kk\in\mathcal{D}} \label{eq:dual+}
\\
\vec{{d}}_{-}(\qq)= - \sigma_z
\begin{pmatrix} 
N_{\kk}^{\qq} \vspace{2mm}\\
M_{\kk}^{\qq}
\end{pmatrix}_{\kk\in\mathcal{D}} &=&
 \begin{pmatrix} 
-N_{\kk}^{\qq} \vspace{2mm}\\
M_{\kk}^{\qq} 
\end{pmatrix}_{\kk\in\mathcal{D}} \label{eq:dual-}
\eea
Finally the eigenvectors and their dual vectors are normalized by choosing $\mathcal{N}_\qq$ so that
\be
\bb{\vec{{d}}_{\pm }}^* \cdot \vec{e}_{\pm } =  \sum_{\kk\in\mathcal{D}}{\rosso[}(M_\kk^\qq)^2-(N_\kk^\qq)^2{\rosso]}=1
\label{eq:normalisation}
\ee

To obtain the amplitudes $b_\qq$ of the collective modes, we project the field $\beta$:
\bea
\!\!\!\!\!\!b_\qq=\vec{{d}}_{+}(\qq) \cdot \begin{pmatrix}  \beta^{\qq} \\ \bar{\beta}^{\qq}\end{pmatrix} \!\!\!&=&\!\!\! 
\sum_{\kk\in\mathcal{D}} {\rosso[}M_\kk^\qq \beta_\kk^\qq - N_{\kk}^\qq ({\beta}_\kk^{-\qq})^*{\rosso]} \label{eq:BSC1}\\
\!\!\!\!\!\!b_{-\qq}^*=\vec{{d}}_{-}(\qq) \cdot \begin{pmatrix}  \beta^{\qq} \\ \bar{\beta}^{\qq}\end{pmatrix} \!\!\!&=&\!\!\! 
\sum_{\kk\in\mathcal{D}} {\rosso[}-N_\kk^\qq \beta_\kk^\qq +M_{\kk}^\qq ({\beta}_\kk^{-\qq})^*{\rosso]}
\label{eq:BSC2}
\eea
This is the first central result of this paper.
The equality $b_{\qq}^*=(b_{\qq})^*$ suggested by our notation is a consequence of the invariance by external parity (\ref{eq:invartranspo1}--\ref{eq:invartranspo2}).

Conversely, to express the classical field in terms of the phonon amplitudes, we expand it on the eigenvectors:
\be
\begin{pmatrix} 
\beta^{\qq} \\
\bar{\beta}^{\qq}
\end{pmatrix} = 
b_\qq \vec{e}_{+}(\qq) + b^*_{-\qq} \vec{e}_{-}(\qq) +\ldots
\label{eq:champsurmodes}
\ee
where in the ellipsis $\ldots$ we omitted the component of the field on the other eigenmodes of total wavevector $\qq$. One can show \cite{TheseHK} that in the continuous limit those omitted modes are nothing else than the continuum $\kk\mapsto\epsilon_{\kk+\qq/2}+\epsilon_{\kk-\qq/2}$ of {\violet fermionic-quasiparticle} biexcitations with center-of-mass wavevector $\qq$.
Projecting the vectorial equation \eqref{eq:champsurmodes}, we obtain:
\bea
\label{eq:devmodbetajoli1}
\beta_{\kk+\qq/2,\kk-\qq/2} &=& M_{\kk}^{\qq}   b_\qq + N_{\kk}^{\qq} b^*_{-\qq}   + \ldots  \\
{\beta}_{\kk-\qq/2,\kk+\qq/2}^* &=& N_{\kk}^{\qq}   b_\qq + M_{\kk}^{\qq} b^*_{-\qq}   + \ldots
\label{eq:devmodbetajoli2}
\eea
After quantization (see Sec.\ref{sec:phonons_quantification} below), $M$ and $N$ 
appear in these expressions as the coefficients of a new Bogoliubov rotation rearranging the fermionic bilinear operators 
{\rouge appearing in equations}~(\ref{eq:kappa},\ref{eq:kappa_star}) into bosonic quasiparticle operators. It comes on top of the rotation induced by the $U_\kk^0$ and $V_\kk^0$ coefficients rearranging the particle operators into fermionic quasiparticle operators. This new rotation acts on the pairs of fermions, hence the two indices of $M$ and $N$.


\subsubsection{Long-wave limit}
 

In the long-wave limit $q\to 0$, we recall the expansion of the energy $\hbar\omega_\qq$ obtained by reference~\cite{KCS2016}:
\be
\hbar \omega_\qq \underset{q\to 0}{=} \hbar c q \left[1+\frac{\gamma}{8} \bb{\frac{\hbar q}{mc}}^2+ O\bb{\frac{\hbar q}{mc}}^4
\right]. \label{eq:fleur}
\ee
In this expression, $c$ is the speed of sound, derived, as in any superfluid, from the equation of state via the hydrodynamic formula
\be
mc^2=\rho \frac{\dd\mu}{\dd\rho},\label{eq:vitesseson}
\ee
where the derivative with respect to $\rho$ is taken for a fixed scattering length $a$. The coefficient  $\gamma$ of the cubic order of the 
expansion (\ref{eq:fleur}) is a rational fraction {\cite{KCS2016}} of the variables
\be
x = \frac{\Delta}{\mu} \ \ \ \ \mbox{and}\ \ \ \ 
y = \frac{\dd\Delta}{\dd\mu}=\frac{\int_{\mathbb{R}^3}\dd^3k \frac{\xi_\kk}{\epsilon_\kk^3}}{\int_{\mathbb{R}^3}\dd^3k \frac{\Delta}{\epsilon_\kk^3}}
\label{eq:xety}
\ee
Note that $y$ can be related to $x$ by the use of the BCS equation of state \eqref{eq:rhobar} in the last equality of \eqref{eq:xety}.

The {\bleu $q\to 0$} expansion of the sum {\bleu \eqref{eq:Ykq}} and difference \eqref{eq:Skq} of $M_\kk^\qq$ and $N_\kk^\qq$ then read:
\bea
\sqrt{\norm_\qq} (M_{\kk}^{\qq}-N_{\kk}^{\qq}) &=&  \frac{\Delta}{\epsilon_\kk} + O(q^2) \label{eq:devYkq} \\
\sqrt{\norm_\qq} (M_{\kk}^{\qq}+N_{\kk}^{\qq}) &=&  {\rosso-} \frac{\hbar c q \epsilon_\kk}{2\Delta} \frac{\dd W_{\kk\zero}^-}{\dd \mu}  +O(q^3) \label{eq:devSkq} \\
\norm_\qq/L^3 &=&  \frac{\hbar c q}{2}  \frac{\dd{\rho}}{\dd\mu}  +O(q^3) \label{eq:devnorm}
\eea
Note that to lowest order these expressions coincide with the coefficients of the zero-energy mode $\vec{e}_n$ and of the anomalous mode $\vec{e}_a$ 
of the zero-momentum subspace evolution operator $\mathcal{L}^\zero$ obtained in reference~\cite{KCS2013}.

{\rosso As an application of the modal expansion (\ref{eq:devmodbetajoli1},\ref{eq:devmodbetajoli2})
and of the above low-$q$ limit (\ref{eq:devYkq},\ref{eq:devSkq}) of the corresponding amplitudes, we derive a modal
expansion for the density 
\be
\rho(\rr,t)=\sum_\sigma \langle \hat{\psi}_\sigma^\dagger(\rr) \hat{\psi}_\sigma(\rr)\rangle
\label{eq:defrhomicro}
\ee
and for the phase $\theta(\rr,t)$ of the order parameter, such that
\be
g_0\langle\hat{\psi}_\downarrow(\rr)\hat{\psi}_\uparrow(\rr)\rangle =
|g_0\langle\hat{\psi}_\downarrow(\rr)\hat{\psi}_\uparrow(\rr)\rangle|
\eee^{\ii \theta(\rr,t)}, 
\label{eq:defthetamicro}
\ee
where the expectation value is taken in the time-dependent variational Ansatz (\ref{eq:Ansatz}). Expanding the Ansatz to first order
in the amplitudes $z_{\kk+\qq,\kk}(t)\simeq -\beta_{\kk+\qq,\kk}(t)$, 
using Wick's theorem to calculate the expectation value in the BCS ground state and replacing the fermionic fields
$\hat{\psi}_\sigma(\rr)$ by their modal expansion from equation (\ref{eq:devmodal}), we obtain to leading order
the deviations from the BCS ground state values:
\bea
\delta\rho(\rr,t) &\!\!\!=\!\!\!& \frac{1}{L^3}\sum_{\kk,\qq} w_{\kk\qq}^+ [\beta_{\kk+\frac{\qq}{2},\kk-\frac{\qq}{2}}(t)\eee^{\ii\qq\cdot\rr}+\mbox{c.c.}]+O(\beta^2) \\
\delta\theta(\rr,t)&\!\!\!=\!\!\!&\frac{-g_0}{2\ii\Delta L^3} \sum_{\kk,\qq} W_{\kk\qq}^+ [\beta_{\kk+\frac{\qq}{2},\kk-\frac{\qq}{2}}(t)\eee^{\ii\qq\cdot\rr}
-\mbox{c.c.}]+O(\beta^2) \nonumber \\
&&
\eea
where the coefficients $w_{\kk\qq}^+$ and $W_{\kk\qq}^+$ are defined by equations (\ref{eq:defW},\ref{eq:defw}).
Next, we replace the field $\beta$ by its expansion (\ref{eq:devmodbetajoli1},\ref{eq:devmodbetajoli2}) on the collective modes 
to obtain, using the symmetry property (\ref{eq:invartranspo2}),
\bea
\label{eq:devmodrho}
\delta\rho(\rr,t)   &=& \frac{1}{L^{3/2}}     \sum_\qq \mathcal{B}_\qq (b_\qq+b_{-\qq}^*) \eee^{\ii\qq\cdot\rr}+O(\beta^2)\\
\delta\theta(\rr,t) &=& \frac{-\ii}{L^{3/2}} \sum_\qq \mathcal{C}_\qq (b_\qq-b_{-\qq}^*) \eee^{\ii\qq\cdot\rr}+O(\beta^2)
\label{eq:devmodtheta}
\eea
with the coefficients
\bea
\mathcal{B}_\qq &=& \frac{1}{L^{3/2}} \sum_\kk w_{\kk\qq}^+ (M_\kk^\qq+N_\kk^\qq) \\
\mathcal{C}_\qq &=& \frac{g_0}{2\Delta L^{3/2}} \sum_\kk W_{\kk\qq}^+ (N_\kk^\qq-M_\kk^\qq)
\eea
In the low-$q$ limit, also using the density (\ref{eq:rhobar}), the gap equation (\ref{eq:gap}) 
and the thermodynamic expression (\ref{eq:vitesseson}) of the sound velocity, we obtain the simple scaling laws
\bea
\mathcal{B}_\qq &=& \left(\frac{\hbar q\rho}{2mc}\right)^{1/2}[1 + O(q^2)] \\
\mathcal{C}_\qq &=&  \left(\frac{2mc}{\hbar q\rho}\right)^{1/2}[1 + O(q^2)]
\eea
This result will be compared to the quantum hydrodynamic theory in section \ref{subsubsec:approchehydro}.
}


\subsection{Quantization of the normal variables}
\label{sec:phonons_quantification}

To quantize the amplitudes of the bosonic modes obtained in our semi-classical approach, we remember that,
{\rosso at the linear order of the small amplitude approximation \eqref{eq:petitesamplitudes},
the field $\beta_\kk^\qq$ reduces to $-\meanv{\hat{\gamma}_{-\kk+\qq/2\downarrow} \hat{\gamma}_{\kk+\qq/2 \uparrow}}$,
that is to the expectation value of a quasiparticle pair operator, see equation (\ref{eq:kappa}).}
We therefore perform the substitution
\be
\beta_{\kk}^\qq=\beta_{\kk+\qq/2,\kk-\qq/2}\rightarrow - {\hat{\vers\gamma}_{-\kk+\qq/2\downarrow} \hat{\vers\gamma}_{\kk+\qq/2 \uparrow}} 
\label{eq:substitution}
\ee
which transforms the amplitude $b_\qq$ in a quantum operator:
\be
{
\hat{b}_\qq= - \sum_{\kk\in\mathcal{D}} \bb{M_{\kk}^{\qq} \hat{\vers\gamma}_{-\kk+\qq/2\downarrow} \hat{\vers\gamma}_{\kk+\qq/2 \uparrow} -  N_{\kk}^{\qq} {\hat{\vers\gamma}_{\kk-\qq/2 \uparrow}^\dagger \hat{\vers\gamma}_{-\kk-\qq/2\downarrow}^\dagger } } \label{eq:BqQuantique}}
\ee
This intuitive action can in fact be justified by several arguments. First, the quantum operator \eqref{eq:BqQuantique} can also be obtained in the framework of the RPA by diagonalizing the homogeneous system for the quasiparticle pair creation and annihilation operators $\hat{\vers\gamma}_{\kk\uparrow} \hat{\vers\gamma}_{\kk'\downarrow}$ and $\hat{\vers\gamma}_{\kk'\downarrow}^\dagger \hat{\vers\gamma}_{\kk\uparrow}^\dagger$, and treating the operators $\hat{\vers\gamma}_{\kk\sigma}^\dagger\hat{\vers\gamma}_{\kk'\sigma}$, whose dynamics is trivial in the RPA and whose average value is sub-leading in the variational theory, as source terms.
Second, one can apply the quantization procedure {\bleu based on bosonic images as} described in chapter 11 of reference~\cite{Ripka1985} where the {\rouge quantum version of the} field $\beta$ is {\bleu taken to be} a bosonic field operator $\mathbb{B}$ (not to be confused with the bosonic excitation operators $\hat{b}_\qq$). The expression of the {\it bosonic image} of a two-body fermionic operator such as $\hat{\vers\gamma}_{-\kk+\qq/2\downarrow} \hat{\vers\gamma}_{\kk+\qq/2 \uparrow}$ in terms of the field $\mathbb{B}$, which would provide an exact version of the substitution \eqref{eq:substitution}, is not simple in the general case since it involves an infinite series expansion in powers of $\mathbb{B}$. Fortunately, in the limit of a weakly excited gas, one can neglect the operator population of the bosonic images $\mathbb{B}\mathbb{B}^\dagger$,
and thus justify the substitution \eqref{eq:substitution}, {\bleu once} we assimilate the two-body fermionic operators and their bosonic images.
Last, we highlight the bosonic nature of the operator $\hat{b}_\qq$ \eqref{eq:BqQuantique} when the gas is weakly excited. The commutator of $\hat{b}_\qq$ and $\hat{b}^\dagger_\qq$ reads:
\be
\bbcro { \hat{b}_\qq,\hat{b}_\qq^\dagger   }-1= \\ \sum_{\kk\in\mathcal{D}} \bbcrol{ \bb{{N}_\kk^\qq}^2 
\bb{  \hat{\vers\gamma}_{\kk-\qq/2\uparrow}^\dagger \hat{\vers\gamma}_{\kk-\qq/2\uparrow} + \hat{\vers\gamma}_{-\kk-\qq/2\downarrow}^\dagger \hat{\vers\gamma}_{-\kk-\qq/2\downarrow}  } } \\ \bbcror{
- \bb{{M}_\kk^\qq}^2 \bb{  \hat{\vers\gamma}_{\kk+\qq/2\uparrow}^\dagger \hat{\vers\gamma}_{\kk+\qq/2\uparrow} + \hat{\vers\gamma}_{-\kk+\qq/2\downarrow}^\dagger \hat{\vers\gamma}_{-\kk+\qq/2\downarrow}  } }
\label{eq:commbq}
\ee
It differs from unity by fermionic quasiparticle population operators, that
are exactly zero in the BCS ground state and of second order in the field $z=O(\beta)$ in a quasiparticle coherent state such as \eqref{eq:Ansatz}.


\section{Beyond the quadratic order: Interactions between phonons and comparison to hydrodynamics}
\label{sec:interactions}

{\bleu The central actor in this work is the phonon-phonon interaction, that eventually causes phonon damping. 
It is derived in this section, in the form of three-phonon or four-phonon coupling amplitudes in phonon-interaction Hamiltonians
that are cubic or quartic in the phonon field. Although the results involve sometimes lengthy formulas, the calculations
are conceptually very simple: starting on one side from the microscopic energy functional (\ref{eq:defE}) or the easier-to-handle
bosonic model (\ref{eq:hamiltonienB}) and on the other side from the quantum hydrodynamic
Hamiltonian (\ref{eq:hamiltonienhydro}), one performs an expansion up to order three or four in the fields, and one inserts in the 
cubic or quartic terms the expansion of the fields on the phonon normal modes involving as linear coefficients the phonon creation and
annihilation operators. That is all for the three-phonon processes in subsection \ref{sec:3phonons}. In the four-phonon
processes of subsection \ref{sec:4phonons}, there is the additional subtlety that the three-phonon processes
treated to second order have to be added to the direct four-phonon ones to get the correct effective four-phonon coupling
as already understood in reference \cite{Khalatnikov1949}.
An important conclusion of this section will be that the three-phonon and four-phonon coupling amplitudes of the microscopic
model and {\bleu of} quantum hydrodynamics coincide in the low wave number limit on the energy shell (that is for processes conserving the 
unperturbed phonon energy), although they strongly differ {\violet outside} the energy shell.
}


\subsection{Three-phonon processes}
\label{sec:3phonons}
In this subsection, we study the processes involving three bosonic quasiparticles: both the $2\leftrightarrow1$ Beliaev-Landau processes and the off-resonant $3\leftrightarrow0$ processes. Our goal is to identify {\bleu these} processes in the expansion of the Hamiltonian in powers of the excitation field, and to extract the associated matrix elements. We shall use two independent theories and compare them. First, we will use our 
microscopic variational theory and expand the energy functional $E$ \eqref{eq:defE} up to order three in $\beta$,
\be
E=E_0+E_2+E_3+O(\beta^4)
\label{eq:devE3}
\ee
and we will insert the expansion \eqref{eq:champsurmodes} of the field on the collective eigenmodes in the trilinear term $E_3$. We shall perform this
microscopic calculation in the case $2\leftrightarrow1$ only. Second, we will use the quantum hydrodynamics of Landau and Khalatnikov \cite{Khalatnikov1949}. This mesoscopic theory treats the pairs of fermions at large spatial scale as bosonic 
{\violet fields}. It has the advantage of relying on the exact equation of state. It will be applied to the $2\leftrightarrow1$ and
$3\leftrightarrow0$ processes. The comparison of the two theories
will allow us to discuss the validity of quantum hydrodynamics.

\subsubsection{Microscopic approach}


The idea behind the microscopic calculation of the three-phonon coupling amplitudes is simple: we express the classical Hamiltonian \eqref{eq:defE} in terms of the amplitudes $b_\qq$ of the collective modes (see Sec.\ref{sec:variablesnormales}), isolate the terms containing the creation and annihilation amplitudes $b_\qq^*$ and $b_\qq$ corresponding to the processes we study, and extract their coefficient. For the three-phonon processes, we focus on the cubic terms gathered in $E_3$.


\paragraph{General case}


The cubic part of $E$ can be written as
\be
E_3=\frac{g_0}{L^3}\!\!\sum_{\kk,\kk',\qq\in\mathcal{D}}\!\!\mathcal{T}_{\kk,\kk'}^{\qq} 
 \bbcrol{ \bbl{\beta_{\kk'+\qq/2,\kk'-\qq/2} 
{\rosso \meanv{\hat{\gamma}_{\kk+\qq/2\uparrow}^\dagger\hat{\gamma}_{\kk-\qq/2\uparrow}}}
 }}
\bbcror{\bbr{+ {\beta}_{\kk'-\qq/2,\kk'+\qq/2} 
{\rosso \meanv{\hat{\rosso\gamma}_{-\kk-\qq/2\downarrow}^\dagger \hat{\rosso\gamma}_{-\kk+\qq/2\downarrow}} }}
+ {\rm c.c.}} \label{eq:E3}
\ee
where we introduced the tensor
\be
\mathcal{T}_{\kk,\kk'}^\qq = \frac{\rosso W_{\kk\qq}^-w_{\kk'\qq}^+ - w_{\kk\qq}^+W_{\kk'\qq}^- - w_{\kk\qq}^-W_{\kk'\qq}^+ - W_{\kk\qq}^+w_{\kk'\qq}^-}{2}.
\label{eq:TkkPq}
\ee 
We {\rosso use the relations (\ref{eq:rho_up},\ref{eq:rho_down})} and we insert the expansion \eqref{eq:champsurmodes} in the expression \eqref{eq:E3} {\rosso to} obtain the following result \footnote{Everywhere in this article the coupling amplitudes $\mathcal{A}^{m\leftrightarrow n}$ are symmetric functions of their $m$ first arguments, the wavevectors $\qq_1,\qq_2,\ldots,\qq_m$ of the incoming phonons, as well as of their $n$ last arguments, the wavevectors $\qq_{m+1},\qq_{m+2},\ldots,\qq_{m+n}$ of the outgoing phonons. With this property the amplitudes are uniquely determined.}:
\be
E_{\rm 3}= \frac{m c^2} {(\rho L^3)^{1/2}} \sum_{\qq_1,\qq_2,\qq_3\in\mathcal{D}} \delta_{\qq_1+\qq_2,\qq_3} \bb{\mathcal{A}_{\rm micro}^{2\leftrightarrow1} (\qq_1,\qq_2;\qq_3) b^*_{\qq_1} b_{\qq_2}^* b_{\qq_3} + {\rm c.c.}} + \ldots
\label{eq:E3Beliaev}
\ee
The vectors $\qq_1$, $\qq_2$ and $\qq_3$ are the three wavevectors involved in the Beliaev-Landau process, 
with $\qq_3$ the wavevector of the phonon which decays into phonons of wavevectors $\qq_1$ and $\qq_2$ or which originates from the merging of 
two such phonons.
In the ellipsis $\ldots$, we omit the three-body processes involving non-bosonic excitations and the terms proportional to $b_{\qq_1} b_{\qq_2} b_{\qq_3}$ or $b_{\qq_1}^* b_{\qq_2}^* b_{\qq_3}^*$ describing the off-resonant $3\leftrightarrow0$ processes which we do not study with the microscopic theory. 
We have factorized the quantity $m c^2/(\rho L^3)^{1/2}$ and defined {\violet the} dimensionless coupling amplitude 
{\violet $\mathcal{A}_{\rm micro}^{2\leftrightarrow1} (\qq_1,\qq_2;\qq_3)$} of the process 
{\violet $b^*_{\qq_1} b^*_{\qq_2} b_{\qq_3}$,} {\violet such that it} is finite and nonzero in the thermodynamic limit:
\begin{multline}
\frac{mc^2}{(\rho L^3)^{1/2}}  \mathcal{A}_{\rm micro}^{2\leftrightarrow1} (\qq_1,\qq_2;\qq_3) = \frac{g_0}{L^3} \sum_{\kk,\kk'\in\mathcal{D}} 
\left[\mathcal{T}_{\kk,\kk'}^{\qq_1} {M}_{\kk'}^{\qq_1} \bb{ {M}_{\kk-\qq_2/2}^{\qq_3}  {M}_{\kk-\qq_3/2}^{\qq_2} + {N}_{\kk+\qq_3/2}^{\qq_2}  {N}_{\kk+\qq_2/2}^{\qq_3} } \right.\\ 
\left. +\mathcal{T}_{\kk,\kk'}^{\qq_1} {N}_{\kk'}^{\qq_1} \bb{ {M}_{\kk+\qq_3/2}^{\qq_2}  {M}_{\kk+\qq_2/2}^{\qq_3} + {N}_{\kk-\qq_2/2}^{\qq_3}  {N}_{\kk-\qq_3/2}^{\qq_2} }
+\mathcal{T}_{\kk,\kk'}^{\qq_3} \bb{{N}_{\kk'}^{\qq_3} {N}_{\kk+\qq_1/2}^{\qq_2} {M}_{\kk-\qq_2/2}^{\qq_1} + {M}_{\kk'}^{\qq_3} {M}_{\kk+\qq_1/2}^{\qq_2} {N}_{\kk-\qq_2/2}^{\qq_1}} 
+ \qq_1 \leftrightarrow \qq_2 \vphantom{\mathcal{T}_{\kk,\kk'}^{\qq_1} } \right]
\label{eq:amplitudemicro}
\end{multline}
where the notation $+\qq_1 \leftrightarrow \qq_2$ means that one must add to the terms already present
in \eqref{eq:amplitudemicro} those obtained by exchanging $\qq_1$ and $\qq_2$ while leaving $\qq_3$ unchanged.
{\violet Although the principle of our microscopic approach to calculate the $2\leftrightarrow1$ coupling amplitude is fairly simple, we obtain a relatively cumbersome expression (\ref{eq:amplitudemicro}). This is due to the resummation over the internal wave vectors $\kk$ and $\kk'$ of the colliding pairs. A much more tractable expression is obtained in the long-wave limit in the next subsection.}


\paragraph{Long-wave limit}


At low temperature, the low-wavevector phonons dominate the kinetics of the gas. This motivates a specific study of $ \mathcal{A}^{2\leftrightarrow1} $ in the limit $q_1,q_2,q_3\to0$, where a comparison to hydrodynamics is meaningful. We perform the expansion of the expression \eqref{eq:amplitudemicro} in this limit and for a continuous space, $l\to 0$, which allows us to integrate over the internal degrees of freedom $\kk$ and $\kk'$ of the pairs and to obtain an expression of the coupling amplitude that depends only of the external wavevectors $\qq_1$, $\qq_2$ and $\qq_3$. The microscopic calculation is done in detail in reference~\cite{TheseHK}; the underlying fermionic nature of the problem makes it rather tedious. 
We give here only the final result, expressed in terms of the energies $\hbar\omega_\qq$ instead of the wavevectors,
with the shorthand notation ${\omega}_i \equiv \omega_{\qq_i}, \ i=1,2,3$. The two sets of variables are connected by the dispersion relation
\eqref{eq:fleur} and momentum conservation relations. In the limit $\omega_i\to 0$, we obtain:
\begin{align}
&\mathcal{A}_{\rm micro}^{2\leftrightarrow1} (\qq_1,\qq_2;\qq_3) = \bb{\frac{\hbar}{mc^2}}^{-1/2} \frac{1}{2^{3/2} \sqrt{{\omega}_1 {\omega}_2 {\omega}_3} }  \Bigg[   2J(x,y)    \bb{{\omega}_1 +{\omega}_2 - {\omega}_3}   \notag
  \\    
   &+\bb{\frac{\hbar}{mc^2}}^{2} \bbaco{A(x,y) {({\omega}_1 +{\omega}_2-{\omega}_3)({\omega}_1^2 + {\omega}_2^2+{\omega}_3^2) }   + B(x,y) \bb{{\omega}_3^3-{\omega}_1^3-{\omega}_2^3  }  
  + C(x,y) {\omega}_1 {\omega}_2 {\omega}_3} + O(\omega^5)   \Bigg]  \label{eq:finaleA}
\end{align}
where the rational fractions $A$, $B$, $C$ and $J$ of the variables $x$ and $y$ defined in \eqref{eq:xety} are given in \ref{ann:fractionsrationnelles}. The denominator $({\omega}_1 {\omega}_2 {\omega}_3)^{1/2}$ leads to a divergence of the coupling amplitude in the long-wave limit whenever the process is not on the energy shell, that is when it does not obey the energy conservation relation
\be
\omega_{3}=\omega_{1}+\omega_{2}.
\label{eq:cons_energie}
\ee
One can check that the Bose-Einstein Condensate (BEC) limit ($a\to0^+$) of our result \eqref{eq:finaleA} coincides with the prediction of Bogoliubov theory \cite{Giorgini1998,CastinSinatra2009} for a weakly interacting gas of bosonic dimers
\footnote{The BEC limit corresponds to $x\to0^+$ and $y\sim-4/x$ \cite{KCS2016}; 
in \ref{ann:fractionsrationnelles} we give the expressions \eqref{eq:limBEC} of the rational fractions in this limit. It is then straightforward to check that our expression of $\mathcal{A}_{\rm micro}^{2\leftrightarrow1}$ in the BEC limit coincides with equation~(16) of reference~\cite{CastinSinatra2009} expanded to third order in $q$ and expressed in terms of the Bogoliubov excitation energies of a gas of $\meanv{\hat{N}_{\rm B}}=\meanv{\hat{N}}/2$ bosonic dimers of mass $m_{\rm B}=2m$, and speed of sound $c$.}.


\paragraph{Resonant processes}

\begin{figure}[t] 
\begin{center}
\includegraphics[width=0.52\textwidth,clip=]{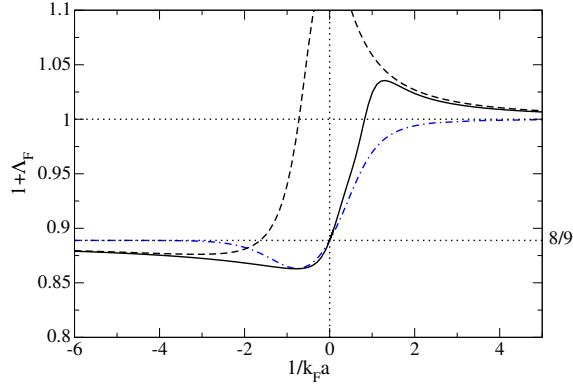} 
\end{center}
\caption{ \label{fig:lambdaF} 
The factor $1+\Lambda_{\rm F}$, which is the Fermi gas equation-of-state dependent part in the $2\leftrightarrow1$ phonon coupling amplitude (see equation \eqref{eq:lambdaF}), is plotted as a function of the interaction strength $1/k_{\rm F} a$. The black solid line is obtained
from the equation of state measured in \cite{Salomon2010}. In the BEC ($k_{\rm F} a \to 0^+$) and BCS ($k_{\rm F} a \to 0^-$) limits the dashed
black lines are the asymptotic behaviors given by the Lee-Huang-Yang corrections \cite{Yang1957,Khalatnikov1957,Galitskii1958} (that include
the first correction to the mean field energy, scaling as $(k_{\rm F} a)^2$ on the BCS side and as $\rho a\sqrt{\rho a^3}$ on the BEC side).
For comparison the prediction of the BCS equation of state \eqref{eq:rhobar} is plotted as a dash-dotted blue line. Note that $1+\Lambda_{\rm F}$
tends to $1$ in the BEC limit, as for a weakly interacting Bose gas. It is equal to $8/9$ in the unitary limit,
and tends to the same value in the BCS limit, since in both cases $\mu\propto{\rho}^{2/3}$.}
\end{figure}

To conclude this microscopic study, we evaluate the coupling amplitude on the energy shell, that is for processes satisfying the energy conservation \eqref{eq:cons_energie}. Such processes are allowed in the limit $\omega_3\to0$ for a positive $\gamma$ parameter only, that is when the dispersion relation $q\mapsto\omega_\qq$ is convex at low $q$. We collect the rational fractions $B$ and $C$ to form the thermodynamic quantity
\be
2B(x,y)+\frac{2}{3}C(x,y) =  1+\frac{{\rho}}{3} {\frac{\dd^2\mu}{\dd{\rho}^2}} \bb{\frac{\dd\mu}{\dd{\rho}}}^{-1} \equiv 1+\Lambda_{\rm F}  \label{eq:lambdaF}
\ee
Its expression as a function of $x$ and $y$ is given in \ref{ann:fractionsrationnelles} and can also be obtained by differentiating
twice the BCS equation of state \eqref{eq:rhobar} with respect to $\mu$. This leads to an elegant expression of the on-shell coupling amplitude:
\be
\mathcal{A}_{\rm {OnS}}^{2\leftrightarrow1}(\qq_1,\qq_2;\qq_3) = 3 \bb{1+\Lambda_{\rm F}} \sqrt{\frac{\hbar^3 {q}_1{q}_2 {q}_3}{32m^3c^3}}   + O\bb{q_3^{7/2}} \label{eq:ACM}
\ee
The thermodynamic quantity $1+\Lambda_{\rm F}$ {\rosso is related to the so-called Gr\"uneisen parameter $u=\frac{\partial \ln c}{\partial\ln\rho}$ by
$1+\Lambda_{\rm F}=\frac{2}{3} (1+u)$. Its occurrence as a factor in equation (\ref{eq:ACM}) }
is the only difference between this fermionic formula and its equivalent for a gas of weakly interacting bosons obtained by the Bogoliubov theory (see equations (D8) and (D9) of reference~\cite{CastinSinatra2009}). It is plotted as a function of the interaction strength
in the BEC-BCS crossover in figure~\ref{fig:lambdaF}.  It tends to $1$ in the BEC limit ($x\to0^+$, $y\sim-4/x$) like in a gas of weakly interacting bosons where $\mu\propto{\rho}$, and it is equal to $8/9$ at unitarity ($x=y$) and in the BCS limit ($x\to0$, $y\to0$) since in both cases $\mu\propto{\rho}^{2/3}$. 

\subsubsection{Hydrodynamic approach}
\label{subsubsec:approchehydro}


We now compare our microscopic result \eqref{eq:finaleA} to the irrotational quantum hydrodynamics of Landau and Khalatnikov. This theory performs
a large-scale description of the gas in terms of two hermitian quantum fields $\hat{\rho}(\rr,t)$ and $\hat{\vv}(\rr,t)$,
directly neglecting the exponentially {\violet weak} density of fermionic quasiparticles at low $T$. The velocity field $\hat{\vv}(\rr,t)$ 
is {\violet assumed} to be irrotational and is written as the (discrete) gradient of the phase field operator $\hat{\phi}(\rr,t)$,
\be
\hat{\vv}(\rr,t)=\frac{\hbar}{m}\nabla \hat{\phi}(\rr,t),
\ee
canonically conjugated to the density field operator $\hat{\rho}(\rr,t)$:
\be
[\hat{\rho}(\rr,t),\hat{\phi}(\rr',t)]=\ii\frac{\delta_{\rr,\rr'}}{{l}^3}.
\label{eq:rhophi}
\ee
The dynamics of these fields is governed by the hydrodynamic Hamiltonian
\begin{equation}
\hat{H}_{\rm hydro} = l^3\sum_{\rr} \left[\frac{\hbar^2}{2m} {\vers (}\nabla \hat{\phi}{\vers)}\cdot \hat{\rho}\ 
{\vers(}\nabla \hat{\phi}{\vers)}
+ e_{0,0}(\hat{\rho})\right],
\label{eq:hamiltonienhydro}
\end{equation}
where $e_{0,0}$ is the bare energy density, that shall be renormalized by the eigenmodes zero-point energy as described in \cite{KCS2015}
to give rise to the true energy density $e_0$  in the ground state, related to the zero-temperature chemical potential by $\mu=\dd e_0/\dd \rho$.

The procedure to follow is standard and similar to the microscopic approach. We linearize the equations of motion for weak spatial fluctuations of
the density and phase fields:
\bea
\hat{\rho}(\rr,t) & =&  \hat{\rho}_0 +\delta{\hat{\rho}}(\rr,t)  \label{eq:fluc1} \\ 
\hat{\phi}(\rr,t) & =&  \hat{\phi}_0(t) +\delta{\hat{\phi}}(\rr,t) \label{eq:fluc2}
\eea
where $\hat{\rho}_0$ {will be replaced} by the mean density $\rho$. We then expand the fields on the eigenmodes of the linearized dynamics:
\begin{eqnarray}
\label{eq:devmod1}
\!\!\!\!\!\!\delta{\hat{\rho}}(\rr,t)&=&\frac{{\rho}^{1/2}}{L^{3/2}}  \!\!
\sum_{\qq \in \mathcal{D}^*} \!\!
\left(\frac{\hbar q}{2 m {c}}\right)^{1/2} \!\! (\hat{b}_\qq+\hat{b}_{-\qq}^\dagger) \,\mathrm{e}^{\ii \qq\cdot \rr} \\
\label{eq:devmod2}
\!\!\!\!\!\!\delta{\hat{\phi}}(\rr,t)&=&\frac{-\ii}{{\rho}^{1/2} L^{3/2}} \!\!
\sum_{\qq \in \mathcal{D}^*} \!\!
\left(\frac{m {c}}{2\hbar q}\right)^{1/2} \!\! (\hat{b}_\qq-\hat{b}_{-\qq}^\dagger) \,\mathrm{e}^{\ii \qq\cdot \rr}
\end{eqnarray}
where the bosonic operators $\hat{b}_\qq$ are the hydrodynamic counterpart of those in equation \eqref{eq:BqQuantique}.
The corresponding bosonic excitations have a purely linear spectrum $\hbar \omega^{\rm hydro}_\qq = \hbar c q$ where the sound velocity
$c$ at density $\rho$ is still given by the hydrodynamic expression \eqref{eq:vitesseson}.
Next, we insert in the cubic part of the Hamiltonian \eqref{eq:hamiltonienhydro}
\begin{equation}
\hat{H}_{\rm hydro}^{(3)} = l^3\sum_{\rr} \left[ \frac{\hbar^2}{2m} \nabla\delta\hat{\phi}\cdot \delta\hat{\rho} \nabla\delta\hat{\phi}
+ \frac{1}{6} \frac{\dd^2\mu}{\dd\rho^2} (\delta\hat{\rho})^3\right],
\label{eq:hamiltonienhydro3}
\end{equation}
the modal expansions (\ref{eq:devmod1}) and (\ref{eq:devmod2}) to obtain
\begin{multline}
\hat{H}_{\rm hydro}^{(3)} = \frac{m{c}^2}{\bb{{\rho}L^3}^{1/2}} 
\sum_{\qq_1,\qq_2,\qq_3\in\mathcal{D}^*}  \bbcrol{ \delta_{\qq_1 + \qq_2,\qq_3} \mathcal{A}^{2\leftrightarrow 1}_{\rm hydro}({\qq_1,\qq_2;\qq_3}) 
\left(\hat{b}_{\qq_1}^\dagger \hat{b}_{\qq_2}^\dagger \hat{b}_{\qq_3} +\mbox{h.c.}\right) }  \\
+\bbcror{  \delta_{\qq_1 +\qq_2 + \qq_3,\zero}\mathcal{A}^{3\leftrightarrow 0}_{\rm hydro} ({\qq_1,\qq_2,\qq_3})   \left(\hat{b}^\dagger_{\qq_1}\hat{b}^\dagger_{\qq_2}\hat{b}_{\qq_3}^\dagger +\mbox{h.c.}\right) }
\label{eq:H3hydro}
\end{multline}
where the coupling amplitudes of the $2\leftrightarrow1$ and $3\leftrightarrow0$ processes are given by 
\be
\mathcal{A}^{2\leftrightarrow 1}_{\rm hydro}({\qq_1,\qq_2;\qq_3})  \!=\!  \sqrt{\frac{\hbar^3 {q}_1{q}_2 {q}_3}{32m^3c^3}} \bb{3\Lambda_{\rm F}\!+\!u_{12}\!+\!u_{13}\!+\!u_{23} } \label{eq:Ahydro21} 
\ee
\be
\mathcal{A}^{3\leftrightarrow0}_{\rm hydro}({\qq_1,\qq_2,\qq_3})  \!=\!  \frac{1}{3} \sqrt{\frac{\hbar^3 {q}_1{q}_2 {q}_3}{32m^3c^3}} \bb{3\Lambda_{\rm F}  \!+\!u_{12}\!+\!u_{13}\!+\!u_{23} }
\label{eq:Ahydro30}
\ee
in terms of the parameter $\Lambda_{\rm F}$ defined in equation \eqref{eq:lambdaF} and of the cosine of the angle between the wave vectors $\qq_i$ and $\qq_j$,
\be
u_{ij}=\frac{\qq_i \cdot \qq_j}{q_i q_j}
\ee
The $2\leftrightarrow1$ amplitude clearly differs from the amplitude \eqref{eq:finaleA} obtained by the microscopic approach (in particular
it does not diverge in the large wavelength limit). 
{\rosso Physically, this may be surprising at first sight. The collective-mode component of the field in the microscopic theory
is indeed expected to correspond to the field of the quantum hydrodynamic theory: for example, the modal expansions of the
density and phase fields (\ref{eq:devmodrho},\ref{eq:devmodtheta}) 
in the microscopic theory seem to match, after quantization of the normal modes $b_\qq$, the
modal expansions (\ref{eq:devmod1},\ref{eq:devmod2}) of quantum hydrodynamics 
(apart from an overall factor $2$ in equation (\ref{eq:devmodtheta})
due to the fact that the phase $\theta(\rr,t)$ of the pairing field is conjugated to the pair density, 
rather than to the particle density),
so that the operators $\hat{b}_\qq$ in equation (\ref{eq:BqQuantique}) indeed seem to correspond to those in equations 
(\ref{eq:devmod1},\ref{eq:devmod2}). 

This discrepancy between the three-phonon coupling amplitudes can be attributed to two reasons: $(i)$ the most obvious
one is the difference of the Hamiltonians of the microscopic and hydrodynamic models, $(ii)$ the somehow more hidden one 
is the fact that the modal expansions (\ref{eq:devmodrho},\ref{eq:devmodtheta}) in the microscopic model 
are valid only to first order in $\beta$, whereas the modal expansions (\ref{eq:devmod1},\ref{eq:devmod2}) in quantum hydrodynamics 
are exact, with the consequence that the operators $\hat{b}_\qq$ coincide in the two points of view only
to first order in $\beta$. 
The first effect is due to the fact that quantum hydrodynamics is an effective field theory only valid at small wave numbers.
The second effect is due to the fact that quantum hydrodynamics directly use the density field $\hat{\rho}(\rr,t)$
and the phase field $\hat{\phi}(\rr,t)$ as the canonically conjugated variables of the theory (this corresponds to a modulus-phase
point of view), whereas the microscopy theory uses the pair fields $\beta_{\kk,\kk'}(t)$ and $\beta^*_{\kk,\kk'}(t)$ as conjugate
variables, in which case the field density $\rho(\rr,t)$ and the phase field $\theta(\rr,t)$ are only specific subfields
that are nonlinear functions of $\beta$ and $\beta^*$, see equations (\ref{eq:defrhomicro},\ref{eq:defthetamicro}).
To qualitatively assess the relative importance of the two effects in the coupling amplitude discrepancies, 
the simplest way is to study the weakly interacting Bose gas case, which
leads to less tedious calculations and reproduces the fermionic case in the BEC limit $k_\mathrm{F} a\to 0^+$.
This study is performed in \ref{appen:diffmicrohydro}.
We then conclude that the leading effect is effect $(ii)$, that is the slight difference between the operators
$\hat{b}_\qq$ in the microscopic and hydrodynamic theories, as it explains the out-of-the-energy-shell
divergence of the microscopic model coupling amplitude at vanishing wave numbers $q_i$. On the contrary, effect $(i)$ only gives
a negligible correction $O(q_i^{7/2})$ to the $\approx q_i^{3/2}$ hydrodynamic coupling amplitude.}

One can check however that {\rosso the microscopic-theory and the hydrodynamic results for the $2\leftrightarrow 1$ coupling amplitude}
agree when the energy is conserved, see equation \eqref{eq:ACM} and equation \eqref{eq:Ahydro21} {\rosso written on the energy shell},
{\rosso 
\be
\mathcal{A}^{2\leftrightarrow 1}_{\rm hydroOnS}({\qq_1,\qq_2;\qq_3})  \!=\!  
3(1+\Lambda_{\rm F}) \sqrt{\frac{\hbar^3 {q}_1{q}_2 {q}_3}{32m^3c^3}}
\label{eq:Ahydro21surlacouche} 
\ee
}
In the hydrodynamic theory on the energy shell
the three wave vectors $\qq_1$, $\qq_2$ and $\qq_3$ must indeed be colinear in the same direction {\violet in order 
to satisfy the equality $|\qq_1+\qq_2|=q_1+q_2$ imposed by energy conservation for a linear spectrum (this is the 
well-known equality case in the triangular inequality). {\rosso This is a satisfactory and important conclusion as, we shall see it,
the Beliaev-Landau phonon damping rates $\Gamma_\qq$ of Sec.\ref{sec:ccbld} only involve the on-shell coupling amplitudes. 
In this respect, the microscopic and
hydrodynamic theories lead as expected to the same low-energy physics.}
}



\subsection{Four-phonon processes}
\label{sec:4phonons}


We now consider the $2\leftrightarrow2$ four-phonon process. {W}hen the excitation branch $q\mapsto\omega_\qq$ is concave at low wave numbers,
this is the resonant process involving the minimal number of phonons because the $1\leftrightarrow 2$ and $1\leftrightarrow 3$ processes
are now forbidden by energy conservation. Also, this process is more intriguing on a theoretical point of view since it involves virtual nonresonant
$1\leftrightarrow 2$ or $3\leftrightarrow0$ intermediate processes. In this case, the equivalence of hydrodynamics with the microscopic approach
is not obvious, as hydrodynamics does not correctly describe the $1\leftrightarrow 2$ processes off-shell.
In this section, we give the hydrodynamic prediction for the $2\leftrightarrow2$ effective coupling amplitude that includes the virtual
processes, then we validate the result with a microscopic model. Since the fermionic microscopic model would be quite cumbersome,
we use a boson model with finite range interactions, such that the excitation branch is concave at low $q$.


\subsubsection{Transition amplitude}


We need to calculate the transition amplitude between an initial state of energy $E_i$, which is an arbitrary Fock state of bosonic quasiparticles,
\be
\ket{i}=\ket{(n_\qq)_{\qq\in \mathcal{D}}}
\ee
and a final state of energy $E_f$ where two phonons of wave vectors $\qq_1$ and $\qq_2$ were annihilated and replaced by phonons of
wavevectors $\qq_3$ and $\qq_4$:
\be
\ket{f}= \frac{\hat{b}_{\qq_3}^\dagger \hat{b}_{\qq_4}^\dagger \hat{b}_{\qq_1} \hat{b}_{\qq_2}}{\sqrt{n_{\qq_1} n_{\qq_2} (1+n_{\qq_3})(1+n_{\qq_4})}} \ket{i}
\ee
Whatever the specific model, the Hamiltonian can be expanded as
\be
\hat{H}=E_0+\hat{H}_2+\hat{H}_3+\hat{H}_4+\ldots
\label{eq:hamiltoniendev}
\ee
where $E_0$ is a constant, $\hat{H}_2$ is the {\bleu free} quasiparticle Hamiltonian and $\hat{H}_3$, $\hat{H}_4$ are the third order and fourth order terms.
$\hat{H}_3$ cannot directly couple $\ket{i}$ to $\ket{f}$, so  we calculate the coupling to second order in perturbation theory,
which amounts to treating $\hat{H}_4$ to first order and $\hat{H}_3$ to second order to construct an effective Hamiltonian \cite{Cohen}:
\be
\bra{f} \hat{H}^{2\leftrightarrow2,\rm eff} \ket{i} = \bra{f} \hat{H}_4 \ket{i}+\sum_\lambda \frac{\bra{f} \hat{H}_3 \ket{\lambda} \bra{\lambda} \hat{H}_3 \ket{i}}{E_i-E_\lambda}\equiv\mathcal{A}_{i\to f}
\label{eq:Aif}
\ee
There exist 6 intermediate states $\ket{\lambda}$, labeled from I to VI, that can be accessed at zero temperature that is when all the
modes $\qq\neq\qq_1,\qq_2$ are initially empty. These states correspond to the creation and reabsorption of a virtual phonon 
by a three-phonon nonresonant process. They are represented by the diagrams on the left part of figure \ref{fig:diagrammes}, with
the virtual intermediate phonon plotted as a dashed line. 
To these six intermediate states $\lambda$=I--VI correspond six other diagrams $\lambda$=I'--VI', shown on the right part of figure \ref{fig:diagrammes},
where the intermediate phonon has the same wave vector but is annihilated and recreated rather than being created then annihilated.
These intermediate states exist only at nonzero temperature since the intermediate phonon must preexist in state $\ket{i}$.


\subsubsection{Effective $2\leftrightarrow2$ coupling amplitude}


The effective coupling amplitude $\mathcal{A}^{2\leftrightarrow2,{\rm eff}}$ of the $2\leftrightarrow2$ process is defined by the
following writing of the effective Hamiltonian:
\be
\hat{H}^{2\leftrightarrow2,{\rm eff}} = \frac{mc^2}{\rho L^3} \sum_{\qq_1,\qq_2,\qq_3,\qq_4\in\mathcal{D}}  \delta_{\qq_1+\qq_2,\qq_3+\qq_4}\mathcal{A}^{2\leftrightarrow2,{\rm eff}}(\qq_1,\qq_2;\qq_3,\qq_4)   \hat{b}_{\qq_3}^\dagger \hat{b}_{\qq_4}^\dagger \hat{b}_{\qq_1} \hat{b}_{\qq_2}
\label{eq:H2donne2eff}
\ee
By construction, {the} matrix element {of $\hat{H}^{2\leftrightarrow2,{\rm eff}}$} between $\ket{i}$ and $\ket{f}$ is the transition amplitude $\mathcal{A}_{i\to f}$. We thus {have the relation}
\be
\mathcal{A}_{i\to f} = \sqrt{n_{\qq_1} n_{\qq_2} (1+n_{\qq_3})(1+n_{\qq_4})} \frac{4mc^2}{\rho L^3} \mathcal{A}^{2\leftrightarrow2,{\rm eff}}
\ee
where the factor $4$ is a counting factor.

From now one we restrict to on-shell processes, such that
\be
\omega_1+\omega_2=\omega_3+\omega_4,
\ee
with the shorthand notation $\omega_i=\omega_{\qq_i}$, $i=1,2,3,4$.
In this case, a simplification occurs between each of the $\lambda$=I,II,III,IV,V,VI diagrams and its $\lambda'$=I',II',III',IV',V',VI' counterpart
on the right column of figure \ref{fig:diagrammes}, which formally reduces the problem to zero temperature. Each diagram and its counterpart
have indeed opposite energy denominators, with numerators that differ only through the factor involving the occupation number
of the intermediate phonon $\qq$, $(1+n_\qq)$ in the left column (where the intermediate phonon is first created) and
$n_\qq$ in the right column (where the intermediate phonon is first annihilated).
Taking I and I' as an example, one has $E_i-E_{\rm I}=\omega_1+\omega_2-\omega_{\qq_1+\qq_2}$ and $E_i-E_{\rm I'}=\omega_{\qq_3+\qq_4}-\omega_3-\omega_4=-(E_i-E_{\rm I})$; in the matrix element $\bra{f} \hat{H}_3 \ket{{\rm I}} \bra{{\rm I}} \hat{H}_3 \ket{i}$ one has the factor $(1+n_{\qq_1+\qq_2})$
and in $\bra{f} \hat{H}_3 \ket{{\rm I'}} \bra{{\rm I'}} \hat{H}_3 \ket{i}$ one has the factor $n_{\qq_1+\qq_2}$. Collecting the diagrams
by pairs, we obtain an effective coupling amplitude identical to the zero-temperature one, that is with $n_\qq=0$ for all $\qq\neq\qq_1,\qq_2$.
{In terms of}
the direct on-shell $2\leftrightarrow2$ coupling amplitude $\mathcal{A}^{2\leftrightarrow2,{\rm dir}}_{\rm {OnS}}$
({that is} related to $\hat{H}_4$ in the same way as $\mathcal{A}^{2\leftrightarrow2,{\rm eff}}$ is related to $\hat{H}^{2\leftrightarrow2,{\rm eff}}$)
and of the amplitudes $\mathcal{A}^{2\leftrightarrow1}$ and $\mathcal{A}^{3\leftrightarrow0}$ introduced in section \ref{sec:3phonons}, {the effective on-shell amplitude can finally be written as follows}:
\begin{multline}
\!\!\!\!\mathcal{A}^{2\leftrightarrow2,{\rm eff}}_{\rm {OnS}}(\qq_{1},\qq_{2},\qq_{3},\qq_{4}) = \mathcal{A}^{2\leftrightarrow2,{\rm dir}}_{\rm {OnS}} (\qq_{1},\qq_{2},\qq_{3},\qq_{4}) \\
+ \frac{\mathcal{A}^{2\leftrightarrow1}(\qq_1,\qq_2;\qq_{1}+\qq_{2}) \mathcal{A}^{2\leftrightarrow1}(\qq_3,\qq_4;\qq_{1}+\qq_{2}) }{\check{\omega}_1+\check{\omega}_2-\check{\omega}  _{1+2}} 
+ \frac{9\mathcal{A}^{3\leftrightarrow0}(\qq_3,\qq_4,-\qq_{1}-\qq_{2})  \mathcal{A}^{3\leftrightarrow0}(\qq_1,\qq_2,-\qq_{1}-\qq_{2}) }{-(\check{\omega}_1+\check{\omega}_2+\check{\omega}_{1+2})} 
\\  +\frac{\mathcal{A}^{2\leftrightarrow1}(\qq_3,\qq_{1}-\qq_3;\qq_{1}) \mathcal{A}^{2\leftrightarrow1}(\qq_{1}-\qq_3,\qq_2;\qq_{4}) }{\check{\omega}_1-\check{\omega}_3-\check{\omega}_{1-3}}
+ \frac{\mathcal{A}^{2\leftrightarrow1}(\qq_4,\qq_{3}-\qq_1;\qq_{2}) \mathcal{A}^{2\leftrightarrow1}(\qq_{3}-\qq_1,\qq_1;\qq_{3}) }{\check{\omega}_3-\check{\omega}_1-\check{\omega}_{3-1}}\\
+ \frac{\mathcal{A}^{2\leftrightarrow1}(\qq_4,\qq_{1}-\qq_4;\qq_{1}) \mathcal{A}^{2\leftrightarrow1}(\qq_{1}-\qq_4,\qq_2;\qq_{3}) }{\check{\omega}_1-\check{\omega}_4-\check{\omega}_{1-4}}
+ \frac{\mathcal{A}^{2\leftrightarrow1}(\qq_3,\qq_{4}-\qq_1;\qq_{2}) \mathcal{A}^{2\leftrightarrow1}(\qq_{4}-\qq_1,\qq_1;\qq_{4}) }{\check{\omega}_4-\check{\omega}_1-\check{\omega}_{4-1}}
\label{eq:A22eff}
\end{multline}
where the energies were rescaled by $mc^2$,
\be
\check{\omega}\equiv\frac{\hbar\omega}{mc^2} \quad \mbox{with the notation} \quad \omega_{i\pm j}\equiv\omega_{\qq_i\pm\qq_j}.
\ee
Note that the coupling due to diagram VI (whose contribution is written here immediately after the one of diagram I)
has a counting factor equal to 9.

\tikzset{->-/.style={decoration={
  markings,
  mark=at position .5 with {\arrow{>}}},postaction={decorate}}}
\tikzset{-<-/.style={decoration={
  markings,
  mark=at position .5 with {\arrow{<}}},postaction={decorate}}}

\setlength{\tabcolsep}{8pt}
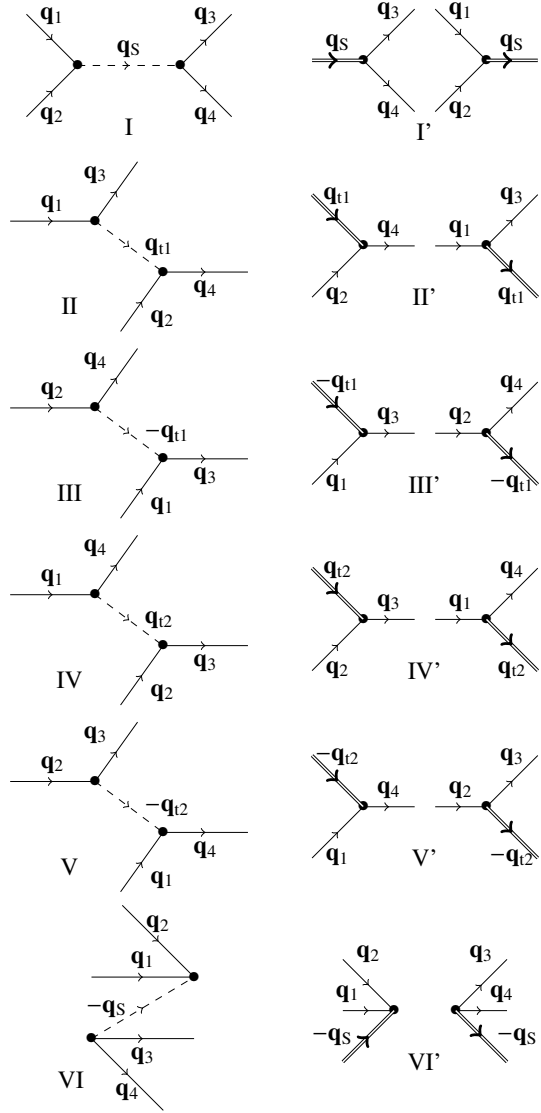
\begin{figure}[p]
\begin{center}
\begin{tabular}{V{0.22\linewidth}V{0.22\linewidth}}
{
\begin{tikzpicture}[scale=1.45]
\coordinate (A) at (-0.5,0);
\coordinate (B) at (0.5,0);
\def\xA{-0.5}
\def\yA{0}
\def\xB{0.5}
\def\yB{0}
\draw (A) node{\textbullet};
\draw (B) node{\textbullet};
\draw[->-,dashed] (A) -- node[above]{$\qq_{\rm S}$} (B);
\draw[->-] (\xA-0.5,\yA+0.5) -- node[above=0.1cm]{$\qq_{1}$} (A);
\draw[->-] (\xA-0.5,\yA-0.5) -- node[below=0.1cm]{$\qq_{2}$} (A);
\draw[->-] (B) -- node[above=0.1cm]{$\qq_{3}$} (\xB+0.5,\yB+0.5);
\draw[->-] (B) -- node[below=0.1cm]{$\qq_{4}$} (\xB+0.5,\yB-0.5);
\draw (\xA*0.5+\xB*0.5,\yA*0.5+\yB*0.5-0.6) node{I};
\end{tikzpicture}
} &
{\centering \begin{tikzpicture}[scale=1.45]
\coordinate (A) at (-0.6,0);
\coordinate (B) at (0.6,0);
\def\xA{-0.6}
\def\yA{0}
\def\xB{0.6}
\def\yB{0}
\draw (A) node{\textbullet};
\draw (B) node{\textbullet};
\draw[->-] (\xB-0.5,\yB+0.5) -- node[above=0.1cm]{$\qq_{1}$} (B);
\draw[->-] (\xB-0.5,\yB-0.5) -- node[below=0.1cm]{$\qq_{2}$} (B);
\draw[->-,double] (B)  -- node[above]{$\qq_{\rm S}$} (\xB+0.5,\yB);
\draw[->-,double] (\xA-0.5,\yA) -- node[above]{$\qq_{\rm S}$} (A);
\draw[->-] (A)  -- node[above=0.1cm]{$\qq_{3}$} (\xA+0.5,\yA+0.5) ;
\draw[->-] (A)  -- node[below=0.1cm]{$\qq_{4}$} (\xA+0.5,\yA-0.5);
\draw (\xA*0.5+\xB*0.5,\yA*0.5+\yB*0.5-0.7) node{I'};
\end{tikzpicture}} \tabularnewline
{
\begin{tikzpicture}[scale=1.2]
\coordinate (A) at (-0.4,0.3);
\coordinate (B) at (0.4,-0.3);
\def\xA{-0.4}
\def\yA{0.3}
\def\xB{0.4}
\def\yB{-0.3}
\draw (A) node{\textbullet};
\draw (B) node{\textbullet};
\draw[->-,dashed] (A) -- node[right=0.1cm]{$\qq_{\rm t1}$} (B);
\draw[->-] (\xA-1,\yA) -- node[above]{$\qq_{1}$} (A);
\draw[->-]  (A) -- node[above left]{$\qq_{3}$} (\xA+0.5,\yA+0.7);
\draw[->-] (\xB-0.5,\yB-0.7) -- node[below right]{$\qq_{2}$} (B);
\draw[->-] (B) -- node[below]{$\qq_{4}$} (\xB+1,\yB);
\draw (\xA*0.5+\xB*0.5-0.7,\yA*0.5+\yB*0.5-0.7) node{II};
\end{tikzpicture}
} &
{\begin{tikzpicture}[scale=1.45]
\coordinate (A) at (0.6,0);
\coordinate (B) at (-0.6,0);
\def\xA{0.6}
\def\yA{0}
\def\xB{-0.6}
\def\yB{0}
\draw (A) node{\textbullet};
\draw (B) node{\textbullet};
\draw[->-,double] (\xB-0.5,\yB+0.5) -- node[above=0.1cm]{$\qq_{\rm t1}$} (B);
\draw[->-] (\xB-0.5,\yB-0.5) -- node[below=0.1cm]{$\qq_{2}$} (B);
\draw[->-] (B)  -- node[above]{$\qq_{\rm 4}$} (\xB+0.5,\yB);
\draw[->-,] (\xA-0.5,\yA) -- node[above]{$\qq_{1}$} (A);
\draw[->-] (A)  -- node[above=0.1cm]{$\qq_{3}$} (\xA+0.5,\yA+0.5) ;
\draw[->-,double] (A)  -- node[below=0.1cm]{$\qq_{\rm t1}$} (\xA+0.5,\yA-0.5);
\draw (\xA*0.5+\xB*0.5,\yA*0.5+\yB*0.5-0.5) node{II'};
\end{tikzpicture}} \tabularnewline
{
\begin{tikzpicture}[scale=1.2]
\coordinate (A) at (-0.4,0.3);
\coordinate (B) at (0.4,-0.3);
\def\xA{-0.4}
\def\yA{0.3}
\def\xB{0.4}
\def\yB{-0.3}
\draw (A) node{\textbullet};
\draw (B) node{\textbullet};
\draw[->-,dashed] (A) -- node[right=0.1cm]{$-\qq_{\rm t1}$} (B);
\draw[->-] (\xA-1,\yA) -- node[above]{$\qq_{2}$} (A);
\draw[->-]  (A) -- node[above left]{$\qq_{4}$} (\xA+0.5,\yA+0.7);
\draw[->-] (\xB-0.5,\yB-0.7) -- node[below right]{$\qq_{1}$} (B);
\draw[->-] (B) -- node[below]{$\qq_{3}$} (\xB+1,\yB);
\draw (\xA*0.5+\xB*0.5-0.7,\yA*0.5+\yB*0.5-0.7) node{III};
\end{tikzpicture}
} &
{\begin{tikzpicture}[scale=1.45]
\coordinate (A) at (0.6,0);
\coordinate (B) at (-0.6,0);
\def\xA{0.6}
\def\yA{0}
\def\xB{-0.6}
\def\yB{0}
\draw (A) node{\textbullet};
\draw (B) node{\textbullet};
\draw[->-,double] (\xB-0.5,\yB+0.5) -- node[above=0.1cm]{$-\qq_{\rm t1}$} (B);
\draw[->-] (\xB-0.5,\yB-0.5) -- node[below=0.1cm]{$\qq_{1}$} (B);
\draw[->-] (B)  -- node[above]{$\qq_{\rm 3}$} (\xB+0.5,\yB);
\draw[->-,] (\xA-0.5,\yA) -- node[above]{$\qq_{2}$} (A);
\draw[->-] (A)  -- node[above=0.1cm]{$\qq_{4}$} (\xA+0.5,\yA+0.5) ;
\draw[->-,double] (A)  -- node[below=0.1cm]{$-\qq_{\rm t1}$} (\xA+0.5,\yA-0.5);
\draw (\xA*0.5+\xB*0.5,\yA*0.5+\yB*0.5-0.5) node{III'};
\end{tikzpicture}} \tabularnewline
{\begin{tikzpicture}[scale=1.2]
\coordinate (A) at (-0.4,0.3);
\coordinate (B) at (0.4,-0.3);
\def\xA{-0.4}
\def\yA{0.3}
\def\xB{0.4}
\def\yB{-0.3}
\tikzset{->-/.style={decoration={
  markings,
  mark=at position .5 with {\arrow{>}}},postaction={decorate}}}
\tikzset{-<-/.style={decoration={
  markings,
  mark=at position .5 with {\arrow{<}}},postaction={decorate}}}
\draw (A) node{\textbullet};
\draw (B) node{\textbullet};
\draw[->-,dashed] (A) -- node[right=0.1cm]{$\qq_{\rm t2}$} (B);
\draw[->-] (\xA-1,\yA) -- node[above]{$\qq_{1}$} (A);
\draw[->-]  (A) -- node[above left]{$\qq_{4}$} (\xA+0.5,\yA+0.7);
\draw[->-] (\xB-0.5,\yB-0.7) -- node[below right]{$\qq_{2}$} (B);
\draw[->-] (B) -- node[below]{$\qq_{3}$} (\xB+1,\yB);
\draw (\xA*0.5+\xB*0.5-0.7,\yA*0.5+\yB*0.5-0.7) node{IV};
\end{tikzpicture}} &
{\begin{tikzpicture}[scale=1.45]
\coordinate (A) at (0.6,0);
\coordinate (B) at (-0.6,0);
\def\xA{0.6}
\def\yA{0}
\def\xB{-0.6}
\def\yB{0}
\draw (A) node{\textbullet};
\draw (B) node{\textbullet};
\draw[->-,double] (\xB-0.5,\yB+0.5) -- node[above=0.1cm]{$\qq_{\rm t2}$} (B);
\draw[->-] (\xB-0.5,\yB-0.5) -- node[below=0.1cm]{$\qq_{2}$} (B);
\draw[->-] (B)  -- node[above]{$\qq_{\rm 3}$} (\xB+0.5,\yB);
\draw[->-] (\xA-0.5,\yA) -- node[above]{$\qq_{1}$} (A);
\draw[->-] (A)  -- node[above=0.1cm]{$\qq_{4}$} (\xA+0.5,\yA+0.5) ;
\draw[->-,double] (A)  -- node[below=0.1cm]{$\qq_{\rm t2}$} (\xA+0.5,\yA-0.5);
\draw (\xA*0.5+\xB*0.5,\yA*0.5+\yB*0.5-0.5) node{IV'};
\end{tikzpicture}} \tabularnewline
{\begin{tikzpicture}[scale=1.2]
\coordinate (A) at (-0.4,0.3);
\coordinate (B) at (0.4,-0.3);
\def\xA{-0.4}
\def\yA{0.3}
\def\xB{0.4}
\def\yB{-0.3}
\tikzset{->-/.style={decoration={
  markings,
  mark=at position .5 with {\arrow{>}}},postaction={decorate}}}
\tikzset{-<-/.style={decoration={
  markings,
  mark=at position .5 with {\arrow{<}}},postaction={decorate}}}
\draw (A) node{\textbullet};
\draw (B) node{\textbullet};
\draw[->-,dashed] (A) -- node[right=0.1cm]{$-\qq_{\rm t2}$} (B);
\draw[->-] (\xA-1,\yA) -- node[above]{$\qq_{2}$} (A);
\draw[->-]  (A) -- node[above left]{$\qq_{3}$} (\xA+0.5,\yA+0.7);
\draw[->-] (\xB-0.5,\yB-0.7) -- node[below right]{$\qq_{1}$} (B);
\draw[->-] (B) -- node[below]{$\qq_{4}$} (\xB+1,\yB);
\draw (\xA*0.5+\xB*0.5-0.7,\yA*0.5+\yB*0.5-0.7) node{V};
\end{tikzpicture}} &
{\begin{tikzpicture}[scale=1.45]
\coordinate (A) at (0.6,0);
\coordinate (B) at (-0.6,0);
\def\xA{0.6}
\def\yA{0}
\def\xB{-0.6}
\def\yB{0}
\draw (A) node{\textbullet};
\draw (B) node{\textbullet};
\draw[->-,double] (\xB-0.5,\yB+0.5) -- node[above=0.1cm]{$-\qq_{\rm t2}$} (B);
\draw[->-] (\xB-0.5,\yB-0.5) -- node[below=0.1cm]{$\qq_{1}$} (B);
\draw[->-] (B)  -- node[above]{$\qq_{\rm 4}$} (\xB+0.5,\yB);
\draw[->-] (\xA-0.5,\yA) -- node[above]{$\qq_{2}$} (A);
\draw[->-] (A)  -- node[above=0.1cm]{$\qq_{3}$} (\xA+0.5,\yA+0.5) ;
\draw[->-,double] (A)  -- node[below=0.1cm]{$-\qq_{\rm t2}$} (\xA+0.5,\yA-0.5);
\draw (\xA*0.5+\xB*0.5,\yA*0.5+\yB*0.5-0.5) node{V'};
\end{tikzpicture}} \tabularnewline
{\begin{tikzpicture}[scale=1.45]
\coordinate (A) at (-0.5,-0.3);
\coordinate (B) at (0.5,0.3);
\def\xA{-0.5}
\def\yA{-0.3}
\def\xB{0.5}
\def\yB{0.3}
\draw (A) node{\textbullet};
\draw (B) node{\textbullet};
\draw[->-,dashed] (A) -- node[left=0.1cm]{$-\qq_{\rm S}$} (B);
\draw[->-] (A) -- node[below]{$\qq_{3}$} (\xA+1,\yA);
\draw[->-] (A) -- node[below]{$\qq_{4}$} (\xA+0.7,\yA-0.7);
\draw[->-] (\xB-1,\yB) -- node[above]{$\qq_{1}$} (B);
\draw[->-] (\xB-0.7,\yB+0.7) -- node[above]{$\qq_{2}$} (B);
\draw (\xA*0.5+\xB*0.5-0.7,\yA*0.5+\yB*0.5-0.7) node{VI};
\end{tikzpicture}} &
{\begin{tikzpicture}[scale=1.45]
\coordinate (A) at (-0.3,0);
\coordinate (B) at (0.3,0);
\def\xA{-0.3}
\def\yA{0}
\def\xB{0.3}
\def\yB{0}
\draw (A) node{\textbullet};
\draw (B) node{\textbullet};
\draw[->-] (\xA-0.5,\yA) -- node[above left]{$\qq_{1}$} (A);
\draw[->-] (\xA-0.5,\yA+0.5) -- node[above=0.2cm]{$\qq_{2}$} (A);
\draw[->-,double] (\xA-0.5,\yA-0.5) -- node[left=0.1cm]{$-\qq_{\rm S}$} (A);
\draw[->-] (B) -- node[above right]{$\qq_{4}$} (\xB+0.5,\yB);
\draw[->-] (B) -- node[above=0.2cm]{$\qq_{3}$} (\xB+0.5,\yB+0.5);
\draw[->-,double] (B) -- node[right=0.1cm]{$-\qq_{\rm S}$} (\xB+0.5,\yB-0.5);
\draw (\xA*0.5+\xB*0.5,\yA*0.5+\yB*0.5-0.5) node{VI'};
\end{tikzpicture}} \tabularnewline
\end{tabular}
\end{center}
\caption{Second order diagrams for the $4$-phonon process $(\qq_1,\qq_2)\to(\qq_3,\qq_4)$ with $\qq_1+\qq_2=\qq_3+\qq_4$, considered as
two successive three-phonon processes. The incoming wave vectors $\qq_1$ and $\qq_2$ and the emergent ones $\qq_3$ and $\qq_4$ are plotted
as a solid line with an arrow. On the left column the diagrams include a virtual intermediate phonon, plotted as a dashed line with an arrow.
On the right column they include a real intermediate phonon, plotted as a double line with an arrow. On any given row, the two diagrams 
have the same intermediate phonon: in I and I' $\qq_{\rm S}=\qq_1+\qq_2=\qq_3+\qq_4$, in II and II' $\qq_{\rm t1}=\qq_1-\qq_3=\qq_4-\qq_2$, 
in III and III' $-\qq_{\rm t1}$, in IV and IV' $\qq_{\rm t2}=\qq_1-\qq_4=\qq_3-\qq_2$, in V and V' $-\qq_{\rm t2}$, in VI and VI' $-\qq_{\rm S}$.
\label{fig:diagrammes}}
\end{figure}

\subsubsection{Effective amplitude in hydrodynamics}
\label{sec:A22effhydro}

At low temperature, we need the large wavelength limit of {\bleu the on-shell} expression \eqref{eq:A22eff}. 
If one wants to get it from quantum hydrodynamics, 
one must introduce in quantum hydrodynamics a correction grasping some element of microscopic physics. 
{\violet Indeed, the hydrodynamic excitation spectrum is purely linear, which causes the energy
denominators in \eqref{eq:A22eff} to vanish, and the effective coupling amplitude to diverge,} 
when the wave vectors are colinear. This is an artefact of hydrodynamics.
In reality the spectrum has a nonzero $\gamma$ curvature parameter, here $\gamma<0$, so that the energy denominators do not vanish.
For almost colinear wave vectors, it is then natural to regularize the coupling amplitude by replacing the hydrodynamic dispersion relation
$\hbar\omega_\qq^{\rm hydro}=\hbar c q$ with the expansion \eqref{eq:fleur} as done by Landau and Khalatnikov \cite{Khalatnikov1949}.
We shall bring a microscopic justification to this: for colinear wave vectors, our bosonic microscopic model (see subsection \ref{sec:modelebosons}) 
gives a large wavelength equivalent of formula \eqref{eq:A22eff} that indeed agrees with the Landau-Khalatnikov modified hydrodynamics.

First we determine the direct $2\leftrightarrow2$ process amplitude from the quartic terms of the Hamiltonian:
\begin{equation}
\hat{H}_{\rm hydro}^{(4)} = \frac{1}{24} \frac{\dd^3\mu}{\dd\rho^3} l^3\sum_{\rr}  \delta\hat{\rho}^4
\label{eq:H4hydro}
\end{equation}
As we did previously, we insert the expansions (\ref{eq:devmod1}) and (\ref{eq:devmod2}) into $\hat{H}_{\rm hydro}^{(4)}$ and
we keep only the $2\leftrightarrow2$ terms $\hat{b}_{\qq_3}^\dagger  \hat{b}_{\qq_4}^\dagger \hat{b}_{\qq_1} \hat{b}_{\qq_2}$:
\be
\hat{H}_{\rm hydro}^{2\leftrightarrow2,{\rm dir}} = \frac{mc^2}{\rho L^3} \sum_{\qq_1,\qq_2,\qq_3,\qq_4\in\mathcal{D}}  \delta_{\qq_1+\qq_2,\qq_3+\qq_4}
\mathcal{A}^{2\leftrightarrow2,{\rm dir}}_{\rm hydro}(\qq_1,\qq_2;\qq_3,\qq_4)   \hat{b}_{\qq_3}^\dagger \hat{b}_{\qq_4}^\dagger \hat{b}_{\qq_1} \hat{b}_{\qq_2}
\label{eq:H2donne2hydro}
\ee
to obtain the rescaled direct $2\leftrightarrow2$ coupling amplitude
\be
\mathcal{A}^{2\leftrightarrow2,{\rm dir}}_{\rm hydro}(\qq_1,\qq_2;\qq_3,\qq_4)  = \frac{{\rm \Sigma_F}}{16} \sqrt{\frac{\hbar^4 q_1 q_2 q_3 q_4}{m^4c^4}}
\label{eq:A22dirhydro}
\ee
where we introduced
\be
{\rm \Sigma_F} \equiv \frac{\rho^3}{mc^2}\frac{\dd^3\mu}{\dd\rho^3}
\ee

Second we combine the amplitude \eqref{eq:A22dirhydro} with our previous expressions for the $2\leftrightarrow1$ \eqref{eq:Ahydro21}
and $3\leftrightarrow0$ \eqref{eq:Ahydro30} amplitudes to obtain
\begin{multline}
\mathcal{A}^{2\leftrightarrow2,{\rm eff}}_{\rm hydro\ corr, {OnS}}(\qq_1,\qq_2;\qq_3,\qq_4)=
\frac{1}{16}\sqrt{\frac{\hbar^4\omega_1\omega_2\omega_3\omega_4}{m^4c^8}} \bbl{\vphantom{\frac{1}{2}}{\rm \Sigma_F} } 
+\frac{(\omega_1+\omega_2)^2 A_{1234}+\omega_{1+2}^2 B_{1234}}{(\omega_1+\omega_2)^2-\omega_{1+2}^2} \\
+\frac{(\omega_1-\omega_3)^2 A_{1324}+\omega_{1-3}^2 B_{1324}}{(\omega_1-\omega_3)^2-\omega_{1-3}^2} 
+\bbr{\frac{(\omega_1-\omega_4)^2 A_{1423}+\omega_{1-4}^2 B_{1423}}{(\omega_1-\omega_4)^2-\omega_{1-4}^2}} \label{eq:A2donne2effhydro}
\end{multline}
where the index \g{corr} means that one goes beyond the hydrodynamic approximation for the dispersion relation in the denominators of expression \eqref{eq:A2donne2effhydro} by using the cubic approximation \eqref{eq:fleur}, {and} we introduced the coefficients
\bea
A_{ijkl}&=&(3\Lambda_{\rm F}+u_{ij})(1+u_{kl})+(3\Lambda_{\rm F}+u_{kl})(1+u_{ij})+(1+u_{ij})(1+u_{kl}) \\
B_{ijkl}&=&(3\Lambda_{\rm F}+u_{ij}) (3\Lambda_{\rm F}+u_{kl})
\eea


\subsubsection{Effective amplitude in a weakly interacting Bose gas with nonzero range interactions}

\label{sec:modelebosons}

To study the $2\leftrightarrow2$ process with a microscopic approach and understand how the divergence in hydrodynamics must be regularised,
we shall not use the fermionic variational theory of section \ref{sec:methode}, which would be very heavy to manipulate due to the internal
degrees of freedom of the pairs. We rather use a bosonic model with a large enough interaction range $b$ so that
the Bogoliubov excitation branch is concave at low $q$. 
{\vers {\violet In \ref{appen:modelebosons}, we describe this model in detail and we calculate}
the $2\leftrightarrow 2$ effective coupling amplitude in the weakly interacting limit from Bogoliubov theory.
{\violet On shell,} in the limit of small wave vectors,} we then recover the prediction
\eqref{eq:A2donne2effhydro} {\violet of corrected quantum hydrodynamics} 
specialised to the bosonic equation of state \eqref{eq:etatB}, {\violet that is} with $\Lambda_{\rm F}={\rm \Sigma_F}=0$
and $(m,c,\rho)\to (m_{\rm B},c_{\rm B},\rho_{\rm B})$, where $m_{\rm B}$ is the boson mass, $c_{\rm B}$ the 
sound velocity in the Bose gas and $\rho_{\rm B}$ the density of the bosons.
{\violet The curvature of the spectrum, which eliminates the divergence of the $2\leftrightarrow 2$ effective coupling amplitude, is present from the start in the microscopic model and does not need to be added by hand.}
{\violet This provides a microscopic justification} to the prescription of Landau and Khalatnikov \cite{Khalatnikov1949} described in the beginning of subsection \ref{sec:A22effhydro}. 
{The result} was not obvious {from the start}, and {it} is due to a subtle cancellation between the large wavelength divergences of
the direct coupling term \eqref{eq:A2donne2Bdir} and of the second-order-perturbation-theory virtual coupling term.
The details of the microscopic calculations are thus very different from the hydrodynamic ones, where the direct coupling term
is zero for the considered equation of state \eqref{eq:etatB}.


\section{Application: phonon damping in the BEC-BCS crossover}
\label{sec:amortissement}


\subsection{A general master-equation expression of the damping rates}

To calculate the damping rate, we view the phonon mode of wave vector $\qq$ as a harmonic oscillator coupled to the reservoir formed by the other 
quasiparticle modes \cite{CastinSinatra2007}, assumed to be at thermal equilibrium at temperature $T$.
We thus split the low-energy effective Hamiltonian as
\be
\hat{H} = \hbar\omega_\qq \hat{b}_\qq^\dagger \hat{b}_\qq + \bb{\sum_{\qq'\neq\qq} \hbar\omega_{\qq'} \hat{b}_{\qq'}^\dagger \hat{b}_{\qq'}}+(\hat{R}^\dagger \hat{b}_\qq +  \hat{b}_\qq^\dagger \hat{R})+\ldots
\ee
The first and the second terms, originating from $\hat{H}_2$, describe the free evolution of the mode $\qq$  and of the reservoir, respectively.
The third term, originating from the part of $\hat{H}_3$ or $\hat{H}^{2\leftrightarrow2,\rm eff}$ involving $\hat{b}_\qq$ or $\hat{b}_\qq^\dagger$,
gives the coupling between the reservoir and the mode  $\qq$. The ellipsis $\ldots$ includes 
{\violet $(i)$} higher order nonlinear processes,
{\violet $(ii)$} processes that do not involve the mode $\qq$, {\violet and $(iii)$}
terms of the form $\hat{R}' \hat{b}_\qq^\dagger\hat{b}_\qq$, where $\hat{R}'$ is an operator of the reservoir,
{\violet that} shift the energy of the mode $\qq$. In the Born-Markov approximation, 
{\bleu valid in the weak coupling and collisionless limit of low temperature $T$ and {\violet low} wave number $q$ at fixed nonzero $\hbar c q/k_B T$,}
\footnote{{\bleu 
When $T$ and $q$ tend to zero, the wave numbers $q_i$ of the intermediate phonons contributing to $\Gamma_\qq$ tend to zero, and
so do in general the phonon coupling amplitudes appearing in the operator $\hat{R}$. Still a validity condition of the result
(\ref{eq:Gammaq_epil}) is that $\omega_\qq \gg \Gamma_{\qq_{\rm th}}$, with $\hbar c q_{\rm th}=k_B T$, that is the angular frequency
$\omega_\qq$ of the considered mode must greatly exceed the typical relaxation rate of the thermal phonons, which defines
the so-called collisionless regime (as opposed to the hydrodynamic regime).}}
one gets for the equation of motion of the mean number of excitations $\meanv{\hat{n}_\qq}$ in the mode $\qq$ \cite{Cohen}:
\be
\frac{\dd}{\dd t}\meanv{\hat{n}_\qq}=-\Gamma_\qq (\meanv{\hat{n}_\qq}-\bar{n}_\qq)
\ee
where $\bar{n}_{\qq}$ is the thermal equilibrium population of {the} mode:
\be
\bar{n}_{\qq} = \frac{1}{\exp\bb{\frac{\hbar\omega_{\qq}}{k_{\rm B} T}}-1} \label{eq:moyennecanB}
\ee
The damping rate $\Gamma_\qq$ is given by
\be
\Gamma_\qq = \int_{-\infty}^{+\infty} \frac{\dd t}{\hbar^2} \eee^{-\ii\omega_\qq t}\mbox{Tr}_{\rm R}\bb{[\hat{R},\hat{R}^\dagger(t)]\sigma_{\rm R}^{\rm eq}(t)}
\label{eq:Gammaq_epil}
\ee
where $\mbox{Tr}_{\rm R}$ is the trace over the states of the reservoir, $\sigma_{\rm R}^{\rm eq}$ is the thermal equilibrium
density operator of the reservoir, $[\hat{A},\hat{B}]$ is the commutator of operators $\hat{A}$ and $\hat{B}$, and the time evolution
of the operator $\hat{R}^\dagger(t)$ of the reservoir is calculated in the interaction picture with the Hamiltonian 
$\sum_{\qq'\neq\qq} \hbar\omega_{\qq'} \hat{b}_{\qq'}^\dagger \hat{b}_{\qq'}$.

\subsection{Convex case: Beliaev-Landau damping}
\label{sec:ccbld}

\begin{figure}[t] 
\begin{center}
\includegraphics[width=0.49\textwidth,clip=]{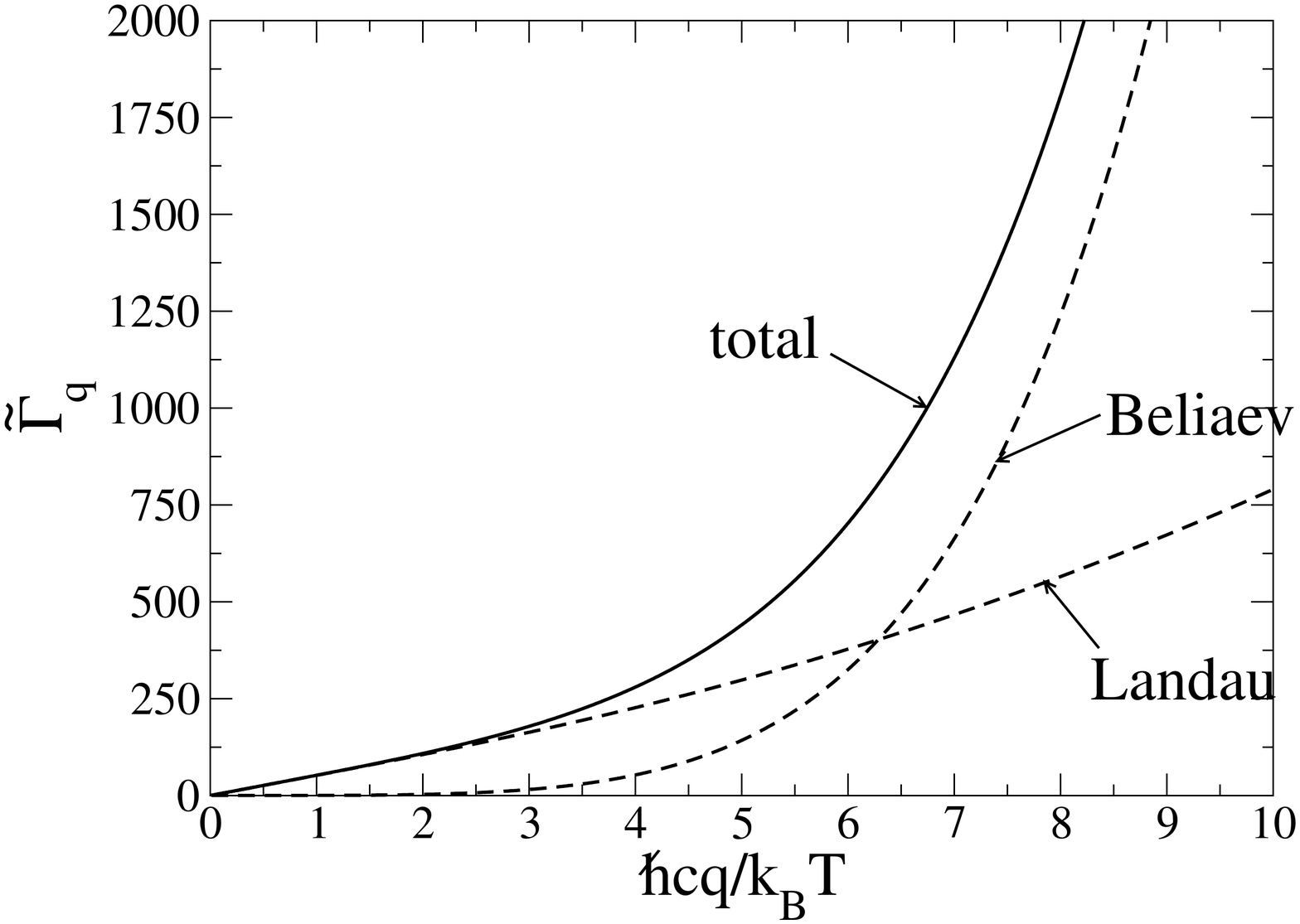}
\end{center}
\caption{\label{fig:GammaBelLan} Low-temperature leading behavior of the total phonon damping rate {\violet (solid line)}
as a function of the rescaled wave number $\tilde{q}=\hbar c q /k_{\rm B} T$ for a convex dispersion relation. 
This is the sum of the Beliaev \eqref{eq:GammaBel} and Landau \eqref{eq:GammaLan} rates (dashed lines). 
The rates are rescaled as in equations \eqref{eq:GammaBeladim} and \eqref{eq:GammaLanadim}.}
\end{figure}

For a convex dispersion relation, the bilinear terms in $\hat{R}$ lead to resonant processes, so they give the leading contribution to $\Gamma_\qq$
at low temperature. {\rouge We thus insert in (\ref{eq:Gammaq_epil}) the following expression for $\hat{R}^\dagger$:
\be
\hat{R}^\dagger=\frac{mc^2}{(\rho L^3)^{1/2}} \left[\sum_{\qq_2,\qq_3}\mathcal{A}^{2\leftrightarrow 1}(\qq_2,\qq_3;\qq) 
\hat{b}_{\qq_2}^\dagger \hat{b}_{\qq_3}^\dagger \delta_{\qq,\qq_2+\qq_3} \right.\\
\left. +\sum_{\qq_1,\qq_2} 2\mathcal{A}^{2\leftrightarrow 1}(\qq_2,\qq;\qq_1) \hat{b}_{\qq_1}^\dagger \hat{b}_{\qq_2} \delta_{\qq_2+\qq,\qq_1}\right]
\label{eq:Rbellan}
\ee
}
We split the contribution of the Beliaev process ${\qq \leftrightarrow (\qq',(\qq-\qq'))}$ {\rouge (first contribution in (\ref{eq:Rbellan}))}
from the one of the Landau process $(\qq,\qq')\leftrightarrow\qq+\qq'$ {\rouge (second contribution in (\ref{eq:Rbellan}))}:
\bea
\Gamma_q^{\rm Bel}&=&\frac{(mc^2)^2}{2\pi^2\hbar^2 \rho} \int_{\mathbb{R}^3} \dd^3 q' |\mathcal{A}^{2\leftrightarrow1}_{\rm {OnS}}(\qq',\qq-\qq';\qq)|^2
\delta(\omega_{\qq'}+\omega_{\qq-\qq'}-\omega_{\qq})(1+\bar{n}_{\qq-\qq'}+\bar{n}_{\qq'}) \label{eq:GamqBel}\\
\Gamma_q^{\rm Lan}&=&\frac{(mc^2)^2}{\pi^2\hbar^2 \rho} \int_{\mathbb{R}^3} \dd^3 q'  |\mathcal{A}^{2\leftrightarrow1}_{\rm {OnS}}(\qq',\qq;\qq'+\qq)|^2 
\delta(\omega_{\qq+\qq'}-\omega_{\qq'}-\omega_{\qq})(\bar{n}_{\qq'}-\bar{n}_{\qq'+\qq})
\label{eq:GamqLan}
\eea
and {\rosso in the low temperature limit} we use the {\rosso leading order}
expression \eqref{eq:ACM} of the on-shell coupling amplitude $2\leftrightarrow1$, {\rosso or equivalently the hydrodynamic on-shell
expression (\ref{eq:Ahydro21surlacouche}).}
We integrate over the wave vector $\qq'$ in spherical coordinates of polar axis $\qq$ to obtain the low temperature equivalents
\footnote{{\rosso If one directly uses
the quantum hydrodynamic linear dispersion relation $\omega_\qq=c q$ in the argument of the Dirac distribution in
equations (\ref{eq:GamqBel},\ref{eq:GamqLan}),
we face the polar integral $\int_{-1}^1 \dd u \delta(u-1)$ 
($u$ is the cosine of the angle between $\qq$ and $\qq'$ and the energy is conserved for $u=1$ only), which seems to
be equal to $1/2$. If one correctly takes into account the strict convexity of the dispersion relation, through
the curvature parameter $\gamma>0$, we obtain the correct value
$\int_{-1}^1 \dd u \delta(u-u_0)=1$ with the energy conserving root $u_0\in]-1,1[$ arbitrarily close to $1$
in the low temperature limit. 
Obviously this value does not depend on $\gamma$ because the Dirac distribution always has a unit mass, 
which explains why the results (\ref{eq:GammaBeladim},\ref{eq:GammaLanadim}) are $\gamma$-independent,
but in a subtle way it utilises the sign of $\gamma$: if $\gamma$ was negative, there would be no energy conserving root $u_0$ 
in the interval of integration, and the resulting integral would
vanish, indicating the absence of Beliaev-Landau damping for a concave dispersion relation.}}
\setcounter{notepasundemi}{\thefootnote} 
\bea
\!\!\!\!\!\!\!\!\!\!\!\!\Gamma_q^{\rm Bel} \!\!\!&\underset{T\to0}{\sim}& \!\!\! \frac{9(1+\Lambda_{\rm F})^2}{32\pi} \frac{mc^2}{\hbar\rho} \bb{\frac{mc}{\hbar}}^3 \bb{\frac{k_{\rm B}T}{mc^2}}^5 \tilde{\Gamma}^{\rm Bel} (\tilde{q}) \label{eq:GammaBeladim}\\
\!\!\!\!\!\!\!\!\!\!\!\!\Gamma_q^{\rm Lan} \!\!\!&\underset{T\to0}{\sim}&\!\!\! \frac{9(1+\Lambda_{\rm F})^2}{32\pi} \frac{mc^2}{\hbar\rho} \bb{\frac{mc}{\hbar}}^3 \bb{\frac{k_{\rm B}T}{mc^2}}^5 \tilde{\Gamma}^{\rm Lan} (\tilde{q}) \label{eq:GammaLanadim}
\eea
The wave numbers are here rescaled by the typical thermal wave number as follows:
\be
\tilde{q}=\frac{\hbar c q}{k_{\rm B}T}
\label{eq:tildeq}
\ee
and the $T\to0$ limit is taken at fixed $\tilde{q}$.
The functions $\tilde{\Gamma}^{\rm Bel}$ and $\tilde{\Gamma}^{\rm Lan}$ are universal functions of $\tilde{q}$, {that can be expressed} in terms of the Bose 
functions $g_\alpha(z)=\sum_{n=1}^{+\infty} z^n/n^\alpha$, also called polylogarithms ${\rm Li}_\alpha(z)$, and of the Riemann $\zeta$
function, $\zeta(\alpha)=g_\alpha(1)$, \footnote{One just has to series expand $1/(\eee^{\tilde{\qq}'}-1)$ and $1/(\eee^{\tilde{\qq}'+\tilde{\qq}}-1)$
in powers of the variable $\eee^{-\tilde{\qq}'}$ and to exchange summation and integration. We recall that $\zeta(4)=\pi^4/90$.}
\bea
\tilde{\Gamma}^{\rm Bel} (\tilde{q})&=& \frac{\tilde{q}^5}{30}- \frac{4\pi^4}{15} \tilde{q}  +48\bbcro{\zeta(5)-g_5(\eee^{-\tilde{q}})}
-24\tilde{q}g_4(\eee^{-\tilde{q}}) + 4\tilde{q}^2[\zeta(3)-g_3(\eee^{-\tilde{q}})] \label{eq:GammaBel}\\
\tilde{\Gamma}^{\rm Lan}(\tilde{q}) &=& \tilde{\Gamma}^{\rm Bel}(\tilde{q}) - \frac{\tilde{q}^5}{30} + \frac{8\pi^4}{15} \tilde{q} \label{eq:GammaLan}
\eea
This leads to the following limiting behaviors:
\bea
\tilde{\Gamma}^{\rm Bel} \!\!\!&\underset{\tilde{q}\to0}{=}& \!\!\!\frac{\tilde{q}^4}{6} + \frac{\tilde{q}^6}{360} + O(\tilde{q}^8) \\ 
\tilde{\Gamma}^{\rm Bel} \!\!\!&\underset{\tilde{q}\to+\infty}{=}& \!\!\!\frac{\tilde{q}^5}{30} \!+\! 4\zeta(3)\tilde{q}^2\!-\!\frac{4\pi^4}{15}\tilde{q}+48\zeta(5) \!+\! O(\tilde{q}^2\eee^{-\tilde{q}}) \label{eq:GamqBel_lim}
\eea
The corresponding ones for $\tilde{\Gamma}^{\rm Lan}$ can be trivially deduced from equation \eqref{eq:GammaLan}.
The two damping rates and their sum are plotted as functions of $q$ in figure \ref{fig:GammaBelLan}.

\subsection{Concave case: Landau-Khalatnikov damping}
\label{subsec:ccadlk}

For a concave dispersion relation, the mode $\qq$ is damped at sufficiently low temperature by the $2\leftrightarrow2$ process only;
in equation \eqref{eq:Gammaq_epil},
we thus keep in {\rouge $\hat{R}^\dagger$} the terms $\hat{b}_{\qq_3}^\dagger \hat{b}_{\qq_4}^\dagger \hat{b}_{\qq_2}$, that originate from
the quartic Hamiltonian \eqref{eq:H2donne2eff}. {\rouge More explicitly, we replace $\hat{R}^\dagger$ in equation
(\ref{eq:Gammaq_epil}) by the following expression:
\be
\hat{R}^\dagger = \frac{mc^2}{\rho L^3}\!\!\!\sum_{\qq_2,\qq_3,\qq_4}\!\!\! 2\mathcal{A}_{\rm eff}^{2\leftrightarrow 2}(\qq,\qq_2;\qq_3,\qq_4) 
\hat{b}_{\qq_3}^\dagger \hat{b}_{\qq_4}^\dagger \hat{b}_{\qq_2} \delta_{\qq+\qq_2,\qq_3+\qq_4}
\ee
}
This leads to
\be
\Gamma_\qq^{2\leftrightarrow2}=\frac{(mc^2)^2}{4\pi^5\hbar^2\rho^2}  \int_{\mathbb{R}^6} \dd^3q_2 \dd^3q_3 |\mathcal{A}_{\rm {OnS}}^{2\leftrightarrow2,{\rm eff}}(\qq,\qq_2;\qq_3,\qq_4)|^2  \delta(\omega_{\qq_3}+\omega_{\qq_4}-\omega_{\qq_2}-\omega_{\qq})\bbcro{\bar{n}_{\qq_2}(1+\bar{n}_{\qq_3})(1+\bar{n}_{\qq_4})-(1+\bar{n}_{\qq_2})\bar{n}_{\qq_3} \bar{n}_{\qq_4}}
\label{eq:Gamma2d2_epil}
\ee
where the vector $\qq_4$ is expressed in terms of the other wave vectors through momentum conservation:
\be
\qq_4=\qq+\qq_2-\qq_3
\label{eq:impulsion}
\ee 
In what follows, we explain how to obtain a low-temperature equivalent for the rate $\Gamma_\qq^{2\leftrightarrow2}$ at fixed $\tilde{q}=\hbar c q/k_B T$.
We take as the polar $z$ axis of the spherical coordinates the direction of $\qq$; the vectors $\qq_i$ then have coordinates $(q_i,\theta_i,\phi_i)$.
We consider a temperature $T$, controlled by the small parameter
\be
\epsilon\equiv\frac{k_{\rm B} T}{mc^2} \ll 1,
\ee
so low that the typical wave numbers are much smaller than $mc/\hbar$ and the bosonic branch is populated in its quasi-linear part only.
In this case, the coupling amplitude $\mathcal{A}_{\rm {OnS}}^{2\leftrightarrow2}$, that would be divergent at vanishing angles for a linear
dispersion relation as already pointed out in section \ref{sec:A22effhydro}, is extremely peaked around $\theta_2=\theta_3=0$
with a width of order $\epsilon$ in $\theta_2$ and $\theta_3$,
\footnote{The peak width is of order $\epsilon$ because the intermediate-virtual-phonon energy change when the $\theta_i$ vary from $0$ to $\epsilon$
is of the same order as the cubic correction to the hydrodynamic dispersion relation, knowing that the typical wave numbers
are $\approx k_B T/\hbar c$. Taking as an example the intermediate phonon $\qq_{\rm S}=\qq+\qq_2$, one finds that $\hbar c[|\qq+\qq_2|-(q+q_2)]
\sim -\frac{\hbar c q q_2}{q+q_2} \theta_2^2 \approx k_B T \theta_2^2$; this deviation is of the same order as the cubic term 
$\approx k_B T |\gamma| \epsilon^2$ in the equation (\ref{eq:fleur}) when $\theta_2^2 \approx |\gamma| \epsilon^2$, hence the change of variable (\ref{eq:tildetheta}).}
and a height $1/\epsilon^2$ times larger than the typical
amplitude at nonzero angles, as we shall see. Using conservation of energy and momentum \eqref{eq:impulsion}, one also finds that $\theta_4=O(\epsilon)$
over the width of the peak. We thus rescale the wave numbers as in equation \eqref{eq:tildeq} and the polar angles as
\be
\tilde{\theta}_i = \frac{\theta_i}{\epsilon |\gamma|^{1/2}}
\label{eq:tildetheta}
\ee
with $\gamma<0$ the curvature parameter \eqref{eq:fleur}, then we perform a Taylor expansion of the coupling amplitude \eqref{eq:A2donne2effhydro}
for $\epsilon\to0$ at fixed rescaled quantities:
\be
\mathcal{A}_{\rm {OnS}}^{2\leftrightarrow2}(\qq,\qq_2;\qq_3,\qq_4) \underset{\epsilon\to0}{=} \bb{\frac{3(1+\Lambda_{\rm F})}{4}}^2 \frac{(\tilde{q} \tilde{q}_2 \tilde{q}_3 \tilde{q}_4)^{1/2}}{|\gamma|}
\mathcal{A}_{\rm red}^{2\leftrightarrow2}(\tilde{q},\tilde{q}_2,\tilde{q}_3,\tilde{\theta}_2,\tilde{\theta}_3) + O(\epsilon^2)
\label{eq:Adejasimple}
\ee
The $\Sigma_{\rm F}$-dependent {term} of the $\mathcal{A}^{2\leftrightarrow2,{\rm dir}}$ direct amplitude does not contribute at this order.
This property, combined with the clever rescaling \eqref{eq:tildetheta} of the polar angles, allowed us in equation (\ref{eq:Adejasimple})
to pull out the factors $1+\Lambda_{\rm F}$
and $\gamma$ that depend on the interaction strength. This leads to a universal $\mathcal{A}_{\rm red}^{2\leftrightarrow2}$ 
reduced amplitude:
\be
\mathcal{A}_{\rm red}^{2\leftrightarrow2}(\tilde{q},\tilde{q}_2,\tilde{q}_3,\tilde{\theta}_2,\tilde{\theta}_3) = 
\frac{1}{\tilde{q} \tilde{q}_2\bb{\frac{\tilde{\theta}_2^2}{(\tilde{q} +\tilde{q}_2)^2}+\frac{3}{4}}} - \frac{1}{\tilde{q} \tilde{q}_3\bb{\frac{\tilde{\theta}_3^2}{(\tilde{q} -\tilde{q}_3)^2}+\frac{3}{4}}}  - \frac{1}{\tilde{q} (\tilde{q}+\tilde{q}_2-\tilde{q}_3) \bb{\frac{\tilde{\theta}_4^2}{(\tilde{q}_3 -\tilde{q}_2)^2}+\frac{3}{4}}}
\label{eq:Ared}
\ee
The first, second and third terms in \eqref{eq:Ared} originate from the second, third and fourth terms in \eqref{eq:A2donne2effhydro}.
In the last two terms, we carefully distinguished the cases $q>q_3$ and $q<q_3$, $q>q_4$ and $q<q_4$ {\violet before taking the $\epsilon\to 0$ limit.}

Next, the implicit relation issued from energy conservation,
\be
\tilde{q}_4 = \tilde{q}+ \tilde{q}_2 - \tilde{q}_3 - \frac{\epsilon^2 |\gamma|}{8}\bb{\tilde{q}^3+\tilde{q}_2^3-\tilde{q}_3^3-\tilde{q}_4^3}+O(\epsilon^4)
\label{eq:tildeq4}
\ee
is iterated once and combined with a spherical geometry calculus projecting relation \eqref{eq:impulsion} over $\qq$. This gives the following
expression for $\tilde{\theta}_4$:
\be
\tilde{\theta}_4^2= \frac{\tilde{q}_2 \tilde{\theta}_2^2- \tilde{q}_3 \tilde{\theta}_3^2-\frac{1}{4}\bbcro{\tilde{q}^3+\tilde{q}_2^3-\tilde{q}_3^3-(\tilde{q}+\tilde{q}_2-\tilde{q}_3)^3}}{\tilde{q}+ \tilde{q}_2 - \tilde{q}_3} \\ + O(\epsilon^2)
\ee
Note that $\tilde{q}_3<\tilde{q}+\tilde{q}_2$ according to \eqref{eq:tildeq4}. Also, due to the rotational invariance around $\qq$,
the integrand of \eqref{eq:Gamma2d2_epil} depends on the azimuthal angles only through their difference $\phi\equiv\phi_2-\phi_3$.

\begin{figure}[t] 
\begin{center}
\includegraphics[width=0.49\textwidth,clip=]{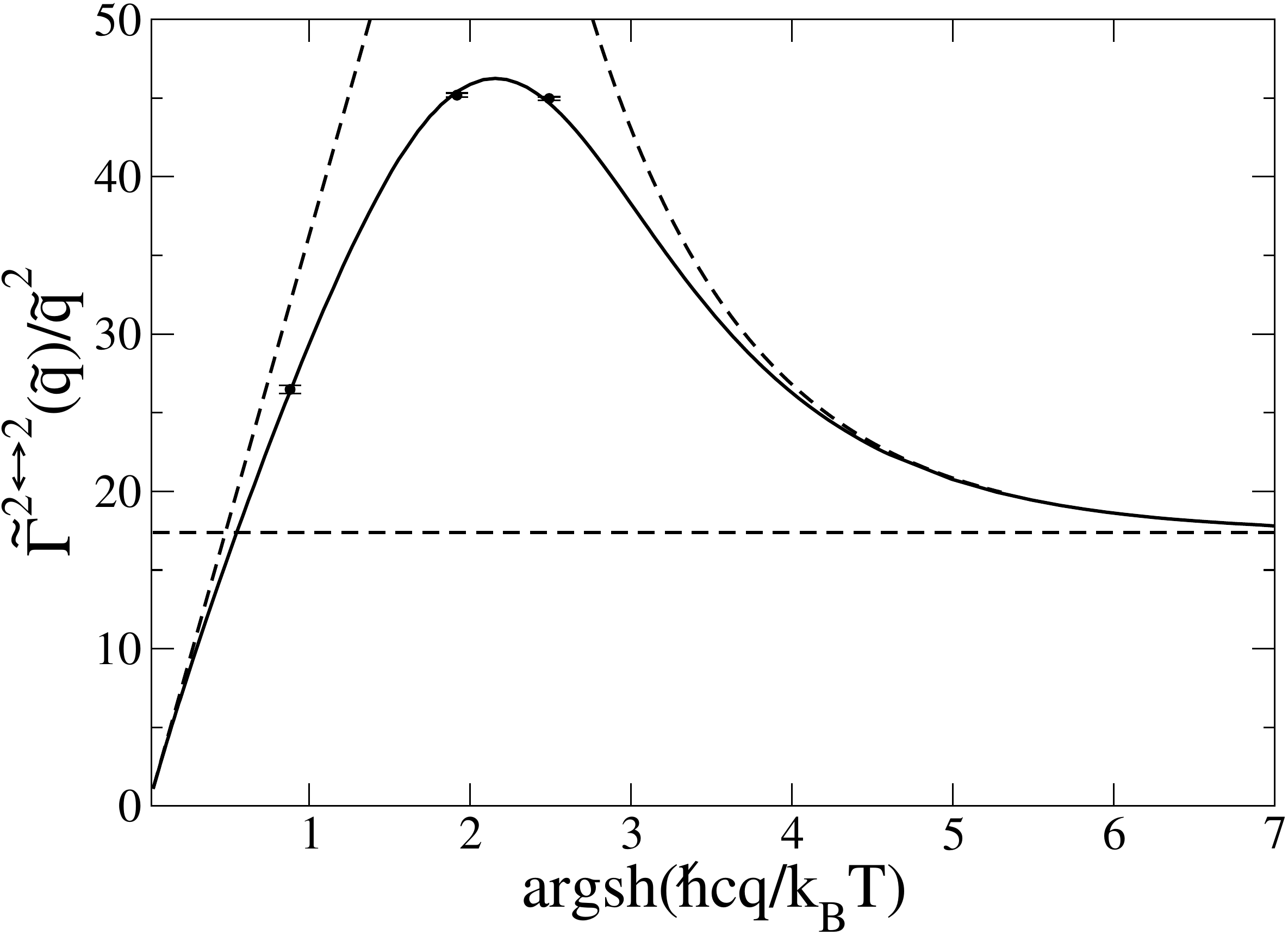}
\end{center}
\caption{\label{fig:F} $2\leftrightarrow2$ damping rate as a function of the rescaled wave number $\tilde{q}{\bleu=\hbar c q/k_B T}$. In order to make the limiting
behaviors more apparent, the rate is divided by $\tilde{q}^2$ and plotted as a function of 
$\mbox{argsh}(\tilde{q}){= \ln(\tilde{q}+\sqrt{1+\tilde{q}^2})}$. The oblique tangent 
at $\tilde{q}=0$ corresponds to the low-$\tilde{q}$ limiting behavior (see equation \eqref{eq:petitsq}). The horizontal straight line 
is the $\tilde{q}\to\infty$ limit of $\tilde{\Gamma}^{2\leftrightarrow2}/\tilde{q}^2$ (see equation \eqref{eq:grandsq})
and the asymptotic curve at large $\tilde{q}$ is a fit of $\tilde{\Gamma}^{2\leftrightarrow2}/\tilde{q}^2$ by an affine function of $1/\tilde{q}$.
{The black symbols are numerical calculations of the integral (\ref{eq:Gamma2d2_epil}). 
{\bleu In the numerics, we} used a dispersion relation of the Bogoliubov form (\ref{eq:Boglike}) restricted to an interval $[0,q_{\rm max}]$ over which its first order derivative is positive. The used dispersion relation coincides with (\ref{eq:fleur}) up to the third order in $q$. The discs are obtained by a linear extrapolation of the value of the integral for $\epsilon=0$, and the error bars halfwidth is given by the difference with a quadratic extrapolation.}
}
\end{figure}

The last step is to integrate the Dirac distribution ensuring energy conservation. 
This is conveniently done in a polar representation of the rescaled angles:
\be
\tilde{\theta}_2 = R \cos\alpha  \qquad \qquad \tilde{\theta}_3 = R \sin\alpha
\label{eq:reppolth}
\ee
We also write the energy difference between the initial state and the final state as
\be
\omega_{\qq_3}+\omega_{\qq_4}-\omega_{\qq_2}-\omega_{\qq} = \frac{mc^2}{\hbar}  \frac{\epsilon^3|\gamma|}{2} \bb{u R^2 + v} +O(\epsilon^5) \label{eq:dansDirac}
\ee
with 
\begin{align}
&u=\frac{\tilde{q}(\tilde{q}_3 \sin^2\alpha - \tilde{q}_2 \cos^2\alpha)+\tilde{q}_2 \tilde{q}_3 (1-\sin2\alpha\cos\phi)}{\tilde{q}+ \tilde{q}_2 - \tilde{q}_3} \label{eq:defualign} \\
&v= \frac{1}{4} \bbcro{\tilde{q}^3+\tilde{q}_2^3-\tilde{q}_3^3-(\tilde{q}+\tilde{q}_2-\tilde{q}_3)^3} \label{eq:defvalign}
\end{align}
In the form \eqref{eq:dansDirac}, the Dirac distribution is readily integrated over $R$. We can finally express the $2\leftrightarrow2$ damping
rate in terms of a universal function $\tilde{\Gamma}^{2\leftrightarrow2}$ of the rescaled wave number $\tilde{q}$:
\be
\frac{\hbar \Gamma_{\qq}^{2\leftrightarrow2}}{mc^2}  \!\!\underset{\epsilon\to0}{\sim} \!\! \frac{81(1+\Lambda_{\rm F})^4}{256\pi^4|\gamma|} \bb{\frac{k_{\rm B} T}{mc^2}}^7 \bb{\frac{mc}{\hbar\rho^{1/3}}}^6  \tilde{\Gamma}^{2\leftrightarrow2}(\tilde{q})
\label{eq:Gamma2d2}
\ee
This is one of the central results of this paper. The function $\tilde{\Gamma}^{2\leftrightarrow2}(\tilde{q})$ is given by a quadruple
integral \footnote{Let us give the explicit intervals of integration $[\alpha_{\rm min},\alpha_{\rm max}]$ and $[\phi_{\rm min},\phi_{\rm max}]$
imposed by the Heaviside function. We set $\chi=\frac{{q}_2 {q}_3+{q} ({q}_3\sin^2\alpha-{q}_2\cos^2\alpha) }{{q}_2 {q}_3\sin2\alpha}$, $A=\frac{{q}}{2{q}_3} +\frac{{q}}{2{q}_2}$, $C=1-\frac{{q}}{2{q}_3}+\frac{{q}}{2{q}_2}$, $A'=\frac{A}{(1+A^2)^{1/2}}$ and $C'=\frac{C}{(1+A^2)^{1/2}}$.
{$(i)$ If $v>0$ and $q<q_3$}, $2\alpha_{\rm min}=\mbox{acos}\, A' - \mbox{acos}\, C' $ and $2\alpha_{\rm max}=\mbox{acos}\, A'  + \mbox{acos}\, C' $.
{$(ii)$ If $v>0$ and $q>q_3$}, $\alpha_{\rm min}=0$ and $2\alpha_{\rm max}=\mbox{acos}\, A'  + \mbox{acos}\, C'$.
{$(iii)$ If $v<0$ and $q<q_3$}, $\alpha_{\rm min}=0$ and $\alpha_{\rm max}=\pi/2$.
{$(iv)$ If $v<0$ and $q>q_3$}, $2\alpha_{\rm min}=\mbox{acos}\, C' - \mbox{acos}\, A' $ and $\alpha_{\rm max}=\pi/2$.
In the cases $(i)$ and $(ii)$ ($v>0$), $\phi_{\rm min}=0$, $\phi_{\rm max}=\pi$ if $\chi<-1$ and $\phi_{\rm max}=\mbox{acos}\, \chi$ otherwise.
In the cases $(iii)$ and $(iv)$ ($v<0$), $\phi_{\rm max}=\pi$, $\phi_{\rm min}=0$ if $\chi>1$ and $\phi_{\rm min}=\mbox{acos}\, \chi$ otherwise.
The integral over $\phi$ can be calculated analytically. One faces $\int_0^{\Phi} \dd\phi\bb{\sum_i \frac{b_i}{a_i+\cos\phi}}^2$
that can be expressed in terms of the primitive $F(\Phi)=\int_0^{\Phi} \dd\phi\frac{1}{a+\cos\phi}$ and of its derivative with respect to $a$.
If $a\in]-1,1[$, $F(\Phi)=\frac{2}{(1-a^2)^{1/2}} \mbox{argth}\,\bbcro{\bb{\frac{1-a}{1+a}}^{1/2} \tan\frac{\Phi}{2}}$. 
If $|a|>1$, $F(\Phi)=\frac{2}{(a^2-1)^{1/2}} \mbox{atan}\,\bbcro{\frac{a-1}{(a^2-1)^{1/2}} \tan\frac{\Phi}{2}}$. 
If $a=1$, $F(\Phi)=\tan\frac{\Phi}{2}$.}
\begin{multline}
\tilde{\Gamma}^{2\leftrightarrow2}(\tilde{q}) =  \int_0^\infty \dd\tilde{q}_2  \int_0^{\tilde{q} + \tilde{q}_2} \dd\tilde{q}_3 \frac{\tilde{q} \tilde{q}_2^3\tilde{q}_3^3(\tilde{q}+\tilde{q}_2-\tilde{q}_3)}{|v|} \frac{[1+f(\tilde{q}_2)] f(\tilde{q}_3) f(\tilde{q}+\tilde{q}_2-\tilde{q}_3)}{f(\tilde{q})}    \\
\times \int_{0}^{\pi/2} \dd\alpha  \int_{0}^\pi \dd\phi \sin\alpha\cos\alpha \, {\vers \Theta}\bb{-\frac{v}{u}} 
 \left\vert \frac{v}{u}\mathcal{A}_{\rm red}^{2\leftrightarrow2}\bb{\tilde{q},\tilde{q}_2,\tilde{q}_3,\left\vert\frac{v}{u}\right\vert^{1/2}\!\!\cos\alpha,\left\vert\frac{v}{u}\right\vert^{1/2}\!\!\sin\alpha} \right\vert^2
\end{multline}
where we introduced the Heaviside function ${\vers \Theta}(x\geq0)=1,\ {\vers \Theta}(x<0)=0$ 
and the reduced Bose function $f(x)=1/(\eee^x-1)$. The occupation numbers in 
\eqref{eq:Gamma2d2_epil} were rewritten using the property $(1+\bar{n}_{\qq_i})/\bar{n}_{\qq_i}=\eee^{\hbar\omega_{\qq_i}/k_{\rm B}T}$
and the conservation of energy. The function $\tilde{\Gamma}^{2\leftrightarrow2}$ is plotted in figure \ref{fig:F}. Its low- and high-$\tilde{q}$
behaviors can be obtained analytically:
\bea
\tilde{\Gamma}^{2\leftrightarrow2}(\tilde{q}) &\underset{\tilde{q}\to0}{=}& \frac{16\pi^5}{135} \tilde{q}^3 +O(\tilde{q}^4) \label{eq:petitsq}\\
\tilde{\Gamma}^{2\leftrightarrow2}(\tilde{q}) &\underset{\tilde{q}\to\infty}{=}& \frac{16\pi\zeta(5)}{3} \tilde{q}^2 + O(\tilde{q}) \label{eq:grandsq}
\eea
These limiting behaviors disagree with the results of Landau and Khalatnikov in reference \cite{Khalatnikov1949} (see their
equations (7.6) {\vers and} (7.12) in the version \cite{LandauCollectedPapers}), even for the order in $\tilde{q}$ of the leading terms.
This discrepancy is due to the fact that these authors have neglected in the coupling amplitude at low $\tilde{q}$ and at high $\tilde{q}$ the
contribution of the diagrams II to V of figure \ref{fig:diagrammes} {\bleu (the diagram VI is
nonresonant)}. This is not justified as already {\rouge pointed out} in  {\rouge reference}
\cite{Wyatt1992}, {\rouge where it was observed numerically that the neglected diagrams are comparable in magnitude with the kept one}. 
We find that at leading order,
the neglected diagrams actually interfere destructively with the diagram I, which makes the exact results (\ref{eq:petitsq},\ref{eq:grandsq})
sub-leading with respect to the Landau and Khalatnikov predictions by two orders in $\tilde{q}$.

\subsection{Phonon damping beyond hydrodynamics}

In this section we concentrate on the zero temperature {\violet unitary} Fermi gas. Our goal is to calculate the phonon damping rates
beyond the hydrodynamic result (\ref{eq:GamqBel_lim}), which we rewrite in the form
\begin{equation}
\left(\Gamma_q^{\rm Bel}\right)_{\rm hydro}^{\rm unitary \, gas} = \left( \frac{2}{9\pi}\right) 
\left( \frac{mc^2}{\hbar}\right) \left( \frac{mc}{\hbar \rho^{1/3}}\right)^3 \frac{\check{q}^5}{30} 
\label{eq:GamqBelhydro}
\end{equation}
where we introduced the notations
\be
\check{q} \equiv \frac{\hbar q}{mc} \quad \quad {\rm and} \quad \quad \check{\omega}_q \equiv \frac{\hbar \omega_\qq}{mc^2}
\ee
To this aim, we first calculate the amplitude ${\cal A}^{2 \leftrightarrow 1}_{\rm {OnS}}(\qq_1,\qq_2;\qq_3)$ 
to first order in the curvature parameter $\gamma$ of the excitation spectrum,
{\rosso that is we go in expression (\ref{eq:ACM}) one step beyond the hydrodynamic order in the low wave number limit.} 
Then we calculate the Beliaev rate from (\ref{eq:GamqBel}) including 
first order corrections in $\gamma$ also in the excitation spectrum. 
{\bleu Last, we question
the accuracy of the Fermi golden rule 
delta distribution in (\ref{eq:GamqBel}) and we investigate processes of
higher order than Beliaev.}


\subsubsection{Amplitude of a Beliaev process $2 \leftrightarrow 1$ for the unitary gas}

In this subsection we calculate the amplitude ${\cal A}^{2 \leftrightarrow 1}_{\rm {OnS}}(\qq_1,\qq_2;\qq_3)$ including the first correction in $\gamma$ from the results of Son and Wingate using conformal invariance in an effective field theory \cite{SonWingate2006}.
Son and Wingate {\violet constructed} a Lagrangian that includes the first correction to hydrodynamics, expressed in terms of a phase field
$\phi$. {\rouge For details on the formalism, we refer to their work \cite{SonWingate2006}, and to reference \cite{Manuel2010} where it was used to calculate
nonzero-temperature corrections to the sound velocity.}

{\vers In short, we first quadratize the Lagrangian to determine the phonon excitation spectrum $\omega_\qq$ and the corresponding eigenmode amplitudes.
The spectrum is of the cubic form (\ref{eq:fleur}) with the curvature parameter $\gamma=-\frac{64}{45 c_0}\left(c_1+\frac{3}{2}c_2\right)$,
where the dimensionless constant $c_0$ appears in the leading (hydrodynamic) order of the Lagrangian, and
the dimensionless constants $c_1$ and $c_2$ appear in the next-to-leading order and parameterize the first corrections beyond hydrodynamics.}
The constant $c_0$ is linked to the already measured \cite{Zwierlein2012} Bertsch parameter $\xi_B$ 
relating the chemical potential $\mu$ to the Fermi wave number $k_{\rm F}$ or the total density $\rho$:
\be
\mu=\xi_B \frac{\hbar^2k_{\rm F}^2}{2m} \quad \mbox{and} \quad \rho=\frac{k_{\rm F}^3}{3\pi^2}
\label{eq:Bertsch}
\ee
The constants $c_1$ and $c_2$, on the contrary, have not been measured yet;
{\rouge they have not been calculated either with good precision from a microscopic theory, only estimates are available, see \cite{Manuel2013} and
references therein}.
{\vers Second, we use the cubic terms in the Lagrangian to obtain the coupling between the phonon modes.  This leads to the on-shell Beliaev
coupling amplitude to first order beyond hydrodynamics:}
\be
{\cal A}^{2 \leftrightarrow 1}_{\rm {OnS}}(\kk,\kk';\qq) =-\frac{\sqrt{2}}{3}(\check{\omega}_q\check{\omega}_k\check{\omega}_{k'})^{1/2} 
\bbcrol{1-\frac{7\gamma}{32}(\check{\omega}_q^2+\check{\omega}_k^2+\check{\omega}_{k'}^2) } 
\bbcror{\vphantom{\frac{1}{1}}+ o(\check{\omega}_q^2)}
\label{eq:couplA}
\ee
{\rouge Note that} the same linear combination $\gamma=-\frac{64}{45 c_0}\left(c_1+\frac{3}{2}c_2\right)$ as in the spectrum appears in this result.
{\vers Details of the calculation are given in \ref{appen:SonetWingate}.}


\subsubsection{Phonon damping rate for the $T=0$ unitary gas}
\label{subsec:tadpdlguat0}

Up to cubic order in $q$, we can rewrite the excitation spectrum in a Bogoliubov form
\begin{equation}
 \check{\omega}_q=\check{q}\left(1+\frac{\gamma}{4}\check{q}^2\right)^{1/2}
 \label{eq:Boglike}
\end{equation}
This allows us to recycle the result (A14) in Appendix~A of reference \cite{CastinSinatra2009} to perform the angular integration in (\ref{eq:GamqBel}), 
\be
\left(\Gamma_q^{\rm Bel}\right)^{\rm unitary \, gas}= 
\left( \frac{mc^2}{\pi \hbar }\right) \left( \frac{mc}{\hbar \rho^{1/3}}\right)^3 \\
 \int_0^{\check{q}} \dd \check{k} 
|{\cal A}^{2 \leftrightarrow 1}_{\rm {OnS}}(k,k';q)|^2 \frac{\check{k}}{\check{q}} 
\frac{\check{\omega}_q-\check{\omega}_k}{\left[ 1 + \gamma (\check{\omega}_q-\check{\omega}_k)^2 \right]^{1/2}}
\ee
where we acknowledge {\violet in the notation} the fact that the coupling amplitude (\ref{eq:couplA}), where $\check{\omega}_{k'}=\check{\omega}_q-\check{\omega}_k$, depends on the moduli of the wavevectors only.
By performing the change of variable $\kappa\equiv \check{k}/\check{q}$, we express $[\left(\Gamma_q^{\rm Bel}\right)^{\rm unitary \, gas} \gamma^2 /\check{q}]$ as an integral over $\kappa$ between 0 and 1 of a function  depending only on $\kappa$ and on the small parameter $\gamma \check{q}^2$.
By expanding this function to the sub-leading order, that is to third order in $\gamma \check{q}^2$, and performing the integral over $\kappa$, we {obtain the provisional result}
\be
\left(\Gamma_q^{\rm Bel}\right)^{\rm unitary \, gas}_{\rm prov} \underset{q\to0}{=} \left( \frac{2}{9\pi}\right) 
\left( \frac{mc^2}{\hbar}\right) \left( \frac{mc}{\hbar \rho^{1/3}}\right)^3 
\frac{\check{q}^5}{30}\left[{1 - \frac{25}{112}\gamma \check{q}^2+ o(\check{q}^2)} \right] 
\label{eq:Bel_result_prov} 
\ee
The sub-leading term in our {provisional} result (\ref{eq:Bel_result_prov}) is qualitatively different from the one in \cite{Salasnich2015}, in particular {it predicts} a reduction of the damping rate with respect to the hydrodynamic prediction at low wavevectors {rather than an increase}. 
{The disagreement is unexpected as the calculation of reference \cite{Salasnich2015} is performed with the same methodology as ours and in the same spirit. We think it is due to fact that the dependence of
${\cal A}^{2 \leftrightarrow 1}$ on $\gamma$ was finally} neglected in \cite{Salasnich2015} while it gives a contribution of the same order as the dependence on $\gamma$ in the spectrum.

{Before accepting the result (\ref{eq:Bel_result_prov}), we should ask ourselves which kind of correction originates from the fact that the considered one-phonon state is unstable, of width $\Gamma_\qq/2$, and as a consequence, that the {\bleu free phonon} energy is not exactly conserved in the one-to-two phonon decay process, contrarily to the constraint imposed by the Dirac delta function in equation (\ref{eq:GamqBel}). In order to estimate the order of magnitude of this effect, one can replace the Dirac delta function by a Lorentzian of half-width $\Gamma/2$,
\be
\delta(\omega_{\qq'}+\omega_{\qq-\qq'}-\omega_{\qq}) \to \frac{\Gamma_\qq/2}{(\omega_{\qq'}+\omega_{\qq-\qq'}-\omega_{\qq})^2+\Gamma_\qq^2/4}
\label{eq:heuristique}
\ee
where $\Gamma_\qq$ can be identified to its leading term in $q^5$. One finds then that this effect introduces a correction to $\Gamma_\qq$ that is of order $q^7$ in the limit $q\to0$. \footnote{Using the note \thenotepasundemi, we consider 
the polar integral $J=\int_{-1}^1 \dd u \frac{\eta'/\pi}{\eta'^2+(u-u_0)^2}$ where $\eta'\approx \Gamma_\qq/\check{q}\approx \check{q}^4$ and $1-u_0=1-\cos\theta_0 \approx \check{q}^2$
as in (\ref{eq:thetaitildeq}). Then $J-1\approx \check{q}^2$ gives a correction $O(\check{q}^7)$ to $\Gamma_\qq^{\rm Bel}$ coming from the finite energy width of the initial state.} {\rouge To get the exact contribution of this broadening effect to $\Gamma_\qq$, we of course
do not use the heuristic prescription (\ref{eq:heuristique}) but we do a rigorous calculation} in \ref{appen:1d3}. {\rouge In short,} using the resolvent of the Hamiltonian, we write at the Beliaev order a self consistent equation for the complex energy $z_\qq=\hbar \omega_\qq -\ii \hbar \Gamma_\qq/2$ of the $\qq$ phonon. If one replaces $z_\qq$ by its zeroth order approximation $\hbar \omega_\qq+\ii\eta$ ($\eta\to 0^+$) in the implicit part of the equation, one recovers exactly (\ref{eq:Bel_result_prov}). If one performs one self consistence iteration, that is if one replaces $-2\im z_\qq/\hbar$ by its usual hydrodynamic approximation (\ref{eq:GamqBelhydro}), one obtains the final result at the Beliaev order:
\be
\left(\Gamma^{\rm Bel}_\qq\right)^{\rm unitary \, gas} = \left(\frac{2}{9\pi}\right) \left(\frac{mc^2}{\hbar}\right)
\left(\frac{mc}{\hbar\rho^{1/3}}\right)^3 \frac{\check{q}^5}{30}
\left[1-\frac{25}{112}\gamma\check{q}^2 + \frac{4\sqrt{3}\xi_B^{3/2}}{243\gamma} \check{q}^2+o(\check{q}^2)\right]
\label{eq:Bel_result}
\ee
{\vers where $\xi_B$ is the Bertsch parameter (\ref{eq:Bertsch}) of the unitary gas.}

In order to conclude, and to make the result (\ref{eq:Bel_result}) rigorous, we have to verify that no other process of higher order
than the Beliaev $1\to2$ process, gives a contribution of order $\check{q}^7$.
A natural candidate is the cascade process represented in figure \ref{fig:1d3}, which combines two $1\to 2$ processes induced by the cubic Hamiltonian $\hat{H}_3$, giving rise to a an effective coupling amplitude $1\to 3$, which is of second order with a virtual phonon \footnote{The attentive reader has probably noticed that the quartic Hamiltonian $\hat{H}_4$ provides a
direct $1\to 3$ coupling to first order. The corresponding amplitude is in $q^2$. This direct coupling contributes to $\Gamma_\qq$ at the order  $q^9$, that is here negligible even by taking into account the effect of a small denominator in $q^3$. For this reason we do not mention it in the main text.}. We estimate the corresponding change in the complex energy of the phonon $\qq$ by treating the effective $1\to 3$ coupling to second order in perturbation theory.

At first sight, the result is $O(\check{q}^9)$ hence negligible. One has to integrate over two independent emitted phonon wave vectors, for example $\qq_2$ and $\qq_3$, the third one $\qq_1$ being imposed by momentum conservation. As the wave numbers 
$q_i$ are of order $q$, this gives a factor $q^6$. It comes then the product of four matrix elements of $\hat{H}_3$, as the effective $1\to 3$ coupling contains two matrix elements and it is treated to the second order, which adds a global factor $(q^{3/2})^4=q^6$. Finally, there are three energy denominators, one from second order perturbation theory and the other two from the effective $1\to 3$ coupling. As the phonon energies are of order $\hbar c q$, this provides a factor $q^3$ in the denominator. The whole thing is $O(\check{q}^9)$ as announced.

\begin{figure}[t] 
\begin{center}
\includegraphics[width=0.3\textwidth,clip=]{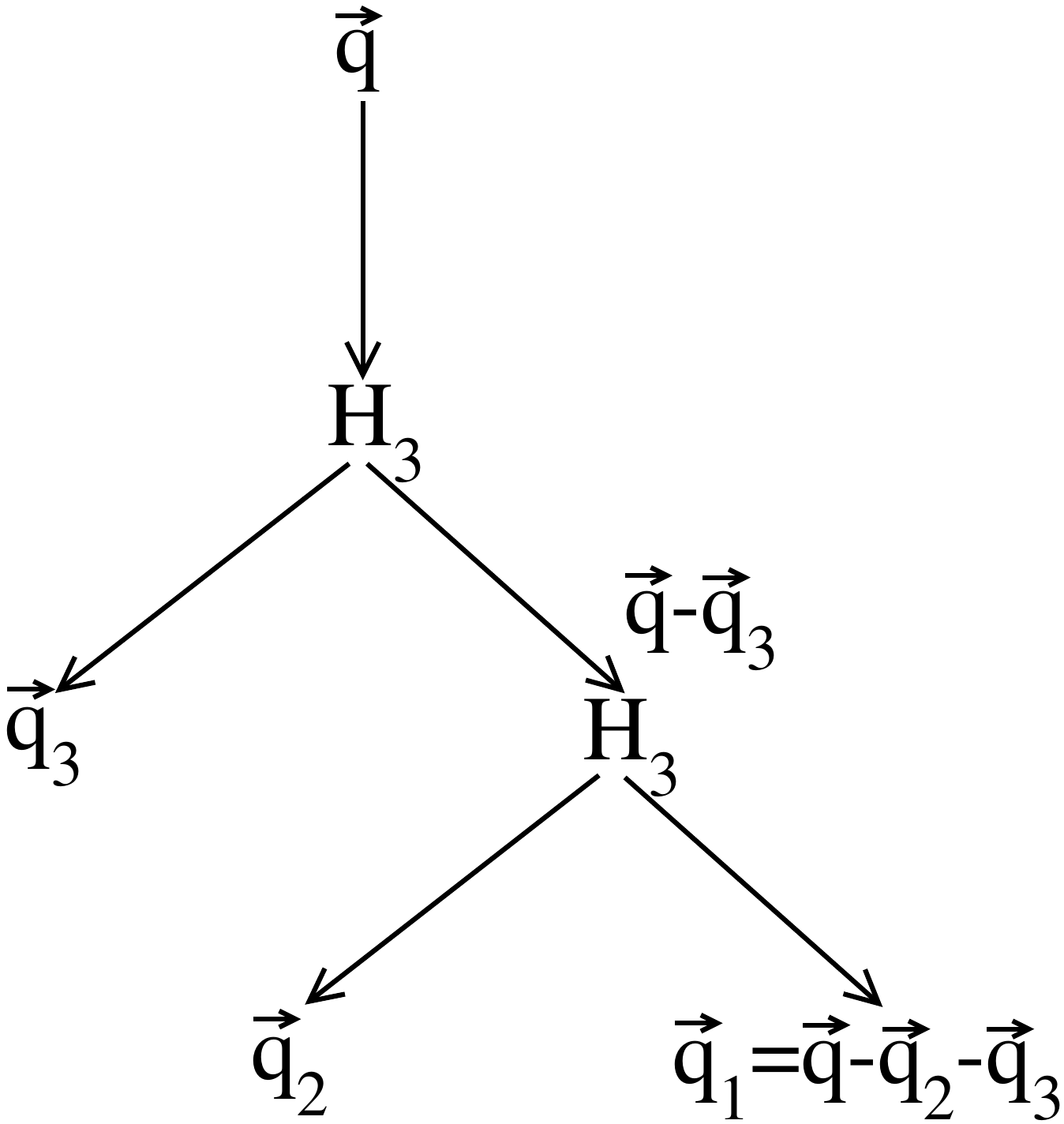}
\end{center}
\caption{\label{fig:1d3} 
{Effective coupling process, of second order in $\hat{H}_3$ involving the emission of a virtual phonon.
As one can deduce from simple power counting (see the text),
in the case of a convex dispersion relation and when the phonons $\qq_1$, $\qq_2$ and $\qq_3$ are
emitted within a small angle $O(\hbar q/mc)$ with respect to $\qq$, this process contributes to the order $(\hbar q/mc)^7$
to the $T=0$ damping rate of a phonon of small wave vector $\qq$. Its contribution has then to be added to the result
(\ref{eq:Bel_result}) as we verify in \ref{appen:1d3} by an explicit calculation.}}
\end{figure} 

The previous reasoning however neglects the enhancement effect of the small denominators of order $q^3$, that occurs when the wave vectors $\qq_i$ are emitted forward, with small angles $\theta_i$ with respect to $\qq$, which already played a crucial role in section \ref{subsec:ccadlk}.
In the limiting case where $\qq$ and the $\qq_i$ are all aligned in the same direction, momentum conservation imposes
\be
q=q_1+q_2+q_3
\label{eq:qq1q2q3}
\ee
so that the energy difference
$\hbar\omega_\qq-(\hbar\omega_{\qq_1}+\hbar\omega_{\qq_2}+\hbar\omega_{\qq_3})$ is not of order $q$ but rather of order $q^3$, taking into account the cubic term in the dispersion relation  (\ref{eq:fleur}). This conclusion extends to all the energy denominators as far as the emission angles are $O(\check{q})$. 
One can check indeed that $q_1=|\qq-\qq_2-\qq_3|$ depends in relative value to second order in the emission angles 
$\theta_2$ and $\theta_3$ with coefficients of order $\check{q}^0$, in the same way as the true dispersion relation deviates in relative value from that of hydrodynamics to second order in $\check{q}$.

Let us then refine the naive estimation $O(\check{q}^9)$ of the previous paragraph, by considering the integration over 
$\qq_2$ and $\qq_3$ within cones of angular aperture $O(\check{q})$ around $\qq$. Each cone occupies a solid angle 
$O(\check{q}^2)$ so that we lose a factor $q^4$ in the integration over polar angles. On the other hand we gain a factor 
$q^{-2}$ for each energy denominator, that is a global factor $q^{-6}$. We then predict a change in the complex energy 
of order $\check{q}^7$, that is the same order as the correction to hydrodynamics appearing  in equation (\ref{eq:Bel_result}). 

In \ref{appen:1d3}, we then explicitly calculate the contribution to the damping rate $\Gamma_\qq$ that comes from the effective {\violet coupling} $1\to 3$ treated to second order. We find that its expression, a rather tedious quintuple integral, 
leads to 
\be
\left(\Gamma^{1\to3}_\qq\right)^{\rm unitary \, gas} = \left(\frac{2}{9\pi}\right) \left(\frac{mc^2}{\hbar}\right)
\left(\frac{mc}{\hbar\rho^{1/3}}\right)^3 \frac{\check{q}^5}{30}
\left[- \frac{2\sqrt{3}\xi_B^{3/2}}{567\gamma} \check{q}^2+o(\check{q}^2)\right]
\label{eq:Gamma1d3}
\ee
In the same appendix, we verify that no other process, of arbitrarily high order in $\hat{H}_3$, $\hat{H}_4$, etc., 
contributes to $\Gamma_\qq$ to the order $\check{q}^7$, even accounting for the enhancement due to small denominators in $q^3$.
We then can sum the contributions (\ref{eq:Bel_result}) and (\ref{eq:Gamma1d3}) to obtain
\be
\left(\Gamma_\qq\right)^{\rm unitary \, gas} = \left(\frac{2}{9\pi}\right) \left(\frac{mc^2}{\hbar}\right)
\left(\frac{mc}{\hbar\rho^{1/3}}\right)^3 \frac{\check{q}^5}{30}
\left[1 - \frac{25}{112}\gamma \check{q}^2 + \frac{22\sqrt{3}\xi_B^{3/2}}{1701\gamma} \check{q}^2+o(\check{q}^2)\right]
\label{eq:amorunit}
\ee
an exact {\rouge expansion}, {\bleu plotted in figure~\ref{fig:BeyondHD},}
to be counted among the successes of this paper. 

}

\begin{figure}[t] 
\begin{center}
\includegraphics[width=0.49\textwidth,clip=]{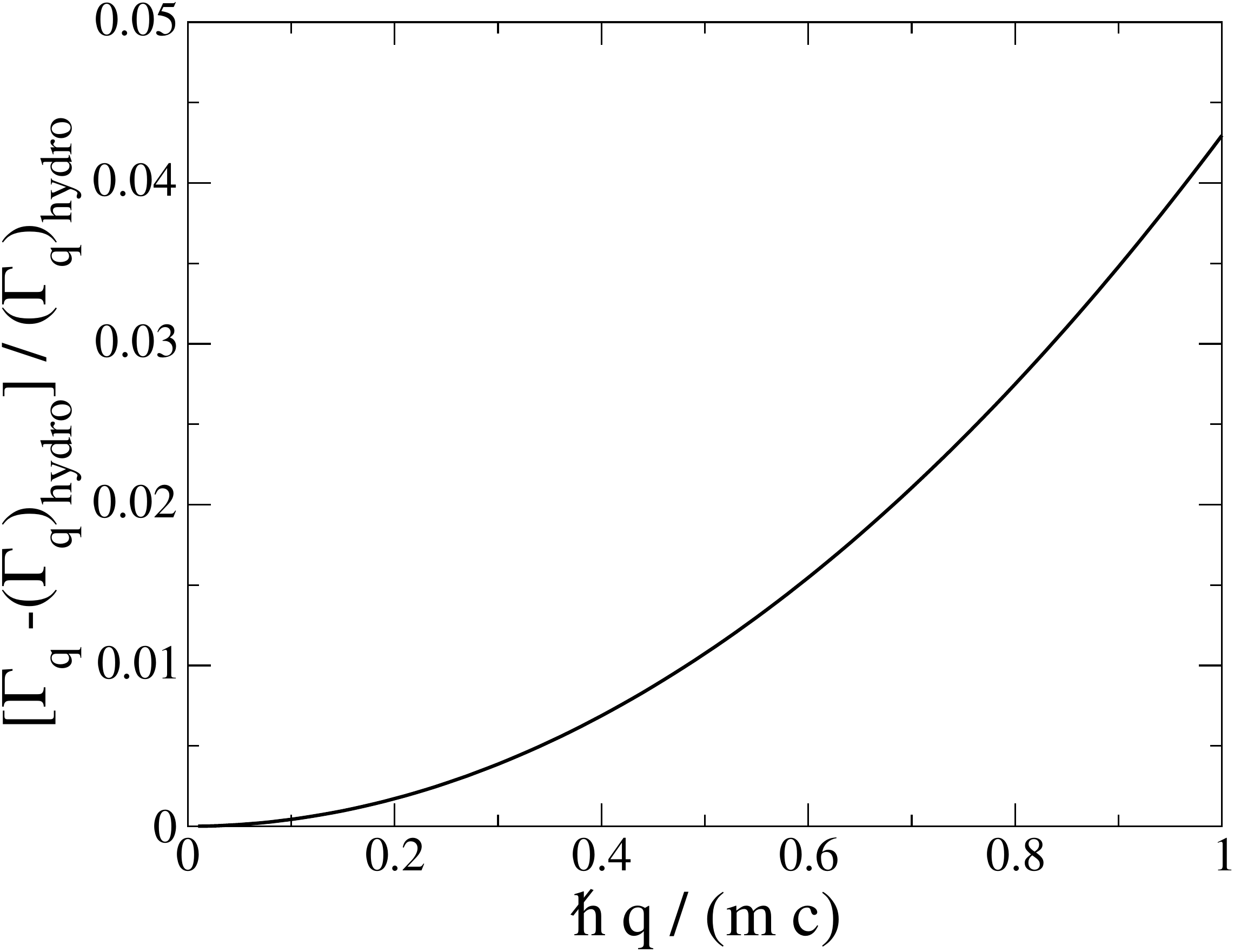}
\end{center}
\caption{\label{fig:BeyondHD} 
We show in relative value the first correction (\ref{eq:amorunit}) to the hydrodynamic prediction for the phonon damping rate of the unitary gas at zero temperature. We took for the curvature parameter the RPA value {$\gamma\simeq 0.083769$ and the Bertsch parameter $\xi_B=0.376$ experimentally measured \cite{Zwierlein2012}.}}
\end{figure} 

\section{Conclusion}


We presented a complete study of interaction processes between phonons in cold Fermi gases at low temperature, for any {\bleu zero-range} interaction strength between fermions, therefore in both the concave and convex cases for the phonon dispersion relation at low wave number $q$.
We clarified the conditions of validity of {\bleu a} low-energy effective {\bleu theory} such as hydrodynamics by comparing {\bleu it} to a microscopic approach which takes into account the internal degrees of freedom of the pairs. Those effective theories correctly predict the phonon
coupling amplitudes only on the energy shell.

One of the main contributions of this study is the microscopic derivation of the $2 \leftrightarrow 2$ coupling amplitude. Indeed,
its expression in second order perturbation theory includes nonresonant $2 \leftrightarrow 1$ and $3 \leftrightarrow 0$ processes. Since these processes can be resonant in the {\bleu quantum} hydrodynamic treatment, where the excitation spectrum is linear, Landau and Khalatnikov had to introduce ``by hand'' a curvature to the dispersion relation to avoid the divergence of the coupling amplitude. At first sight this procedure is risky since, as we just said, quantum hydrodynamics does not predict correctly the coupling amplitudes for nonresonant (off-shell) processes. To provide a microscopic test of this procedure would be particularly cumbersome within our fermionic microscopic approach, so we rather considered a model of weakly interacting bosons with finite range interactions designed to have a concave dispersion relation. We found that although the microscopic expressions of the coupling amplitudes differ from that of hydrodynamics, when one sums up all the nonresonant three-phonon processes to second order and the direct $2 \leftrightarrow 2$ processes to first order, and one restricts on-shell with respect to the effective $2 \leftrightarrow 2$ interaction, the Landau-Khalatnikov prescription and the microscopic result for the effective $2 \leftrightarrow 2$ coupling amplitude agree. 

{As a second result of this paper}, we gave universal formulas for the damping rates of both the $2\leftrightarrow1$ and $2\leftrightarrow2$ processes at low temperature, {\rouge universal in the sense that the introduced reduced-rate functions $\tilde{q}\mapsto \tilde{\Gamma}(\tilde{q})$ 
do not depend on the atomic species, on the interaction strength, or even on the temperature}. The most interesting result is the analytic derivation, as a function of $q$, of the phonon damping rate in the concave case, given to leading order in temperature by the $2\leftrightarrow2$ processes, see \eqref{eq:Gamma2d2} and figure~\ref{fig:F}, that {\rouge is} the subject of a Letter \cite{pubamor}. In the limiting cases $\hbar c q \ll k_{\rm B} T$ and $\hbar c q \gg k_{\rm B} T$, our result disagrees with the one of reference~\cite{Khalatnikov1949} and is sub-leading by two orders in $\hbar c q /k_{\rm B} T$. This is due to the failed assumption in \cite{Khalatnikov1949} that some interaction diagrams are negligible while in reality they destructively interfere with the supposedly leading diagram. 

{Finally we also calculated, at zero temperature and in the unitary limit, the first correction $\propto q^7$ to the hydrodynamic prediction  $\propto q^5$ for the single phonon decay rate. Our calculation allows to refine the prediction of \cite{Salasnich2015} by $(i)$ the actual inclusion of the beyond-hydrodynamics expression of the coupling amplitude \cite{SonWingate2006}, $(ii)$ the inclusion of a finite width $\hbar \Gamma_\qq$ in energy conservation in the Fermi golden rule, that is of a purely  imaginary term of order $q^5$ in the energy denominator of perturbation theory and $(iii)$ the inclusion of other processes, of higher order than the Beliaev process, in particular of the effective coupling $1\to3$ treated to second order.}

All our predictions can be tested in state-of-art experiments on cold fermionic gases. In particular, a discussion of the observability of the Landau-Khalatnikov $2\to2$ damping, using cold atoms trapped in a flat-bottom potential, is done in \cite{pubamor}, {\vers with a proposed experimental protocol for
the selective excitation of a phonon standing wave in the box in a thermodynamic-limit regime, that is
at a wave number $q$ much larger than the inverse of the system size but still in the linear part of the excitation spectrum.}

\appendix



\section{{\vers Implicit equation on the collective excitation spectrum}}
\label{appen:RPA}

{\vers We explain here how to obtain from the linear system (\ref{eq:systvaria1},\ref{eq:systvaria2}) 
the implicit equation (\ref{eq:dedispersion}) on the collective mode angular frequency
$\omega_\qq$ at wave vector $\qq$, in the zero lattice spacing limit $l\to 0$.}
We introduce the collective amplitudes
\bea
Y_\qq = \frac{g_0}{L^3} \sum_{\kk\in\mathcal{D}} W^+_{\kk\qq} y_\kk^\qq &\quad& S_\qq = \frac{g_0}{L^3} \sum_{\kk\in\mathcal{D}} W^-_{\kk\qq} s_\kk^\qq \label{eq:defcoll1}\\
y_\qq = \frac{g_0}{L^3} \sum_{\kk\in\mathcal{D}} w^-_{\kk\qq} y_\kk^\qq   &\quad& s_\qq = \frac{g_0}{L^3} \sum_{\kk\in\mathcal{D}} w^+_{\kk\qq} s_\kk^\qq \label{eq:defcoll2}
\eea
and we solve the $2\times2$ linear system  (\ref{eq:systvaria1},\ref{eq:systvaria2}) to express the unknowns $y_\kk^\qq$ {\vers and} $s_\kk^\qq$ in terms of the collective amplitudes. Next, we replace the result in (\ref{eq:defcoll1},\ref{eq:defcoll2}) and come up with the homogeneous system
\be
\!\!\!\!\!\!\!\!\! \left( \begin{array}{llll}
\Sigma_{W^+ W^+}^\epsilon - 1 & \Sigma_{W^+ W^-}^\omega & -\Sigma_{W^+ w^-}^\epsilon & \Sigma_{W^+ w^+}^\omega \\
\Sigma_{W^- W^+}^\omega  & \Sigma_{W^- W^-}^\epsilon - 1 & -\Sigma_{W^- w^-}^\omega & \Sigma_{W^- w^+}^\epsilon \\
\Sigma_{w^- W^+}^\epsilon  & \Sigma_{w^- W^-}^\omega & -\Sigma_{w^- w^-}^\epsilon -1 & \Sigma_{w^- w^+}^\omega \\
\Sigma_{w^+ W^+}^\omega  & \Sigma_{w^+ W^-}^\epsilon & -\Sigma_{w^+ w^-}^\omega & \Sigma_{w^+ w^+}^\epsilon -1 \\
\end{array} \right)
\begin{pmatrix}
Y_\qq \\
S_\qq \\
y_\qq \\
s_\qq
\end{pmatrix}
=0
\label{eq:auxvalpropres}
\ee
{ where we introduced the notations} 
\bea
\Sigma_{ab}^\epsilon &=& \frac{g_0}{L^3} \sum_{\kk\in\mathcal{D}} \frac{\epsilon_{\kk\qq} a_{\kk\qq} b_{\kk\qq}}{(\hbar\omega_\qq)^2-(\epsilon_{\kk\qq})^2} \\
\Sigma_{ab}^\omega &=& \frac{g_0}{L^3} \sum_{\kk\in\mathcal{D}} \frac{\hbar\omega_{\qq} a_{\kk\qq} b_{\kk\qq}}{(\hbar\omega_\qq)^2-(\epsilon_{\kk\qq})^2} 
\eea
{with} $a$ and $b$ that can be any of $W^+,W^-,w^+,w^-$. 
The system \eqref{eq:auxvalpropres} simplifies in the continuous limit $l\to0$. Since $g_0\to0$ and $w_{\kk\qq}^\pm\underset{k\to+\infty}{=}O(1/k^2)$, all the $\Sigma$ tend to $0$ in the third and fourth lines, and therefore we must have
\be
y_\qq=s_\qq=0
\label{eq:yqsq}
\ee
Next we divide the first two lines of \eqref{eq:auxvalpropres} by $g_0$. The {gap} equation \eqref{eq:gap} ensures that all the divided matrix elements have a finite nonzero limit. The system therefore reduces to its $2\times2$ upper left block, which, in the thermodynamic limit, we write as:
\be
\begin{pmatrix}
I_{++}(\omega_\qq,q) & \hbar\omega_\qq I_{+-}(\omega_\qq,q) \\
\hbar\omega_\qq I_{+-}(\omega_\qq,q) & I_{--}(\omega_\qq,q)
\end{pmatrix}
\begin{pmatrix}
Y_\qq \\
S_\qq 
\end{pmatrix}
=0
\label{eq:proprecollective}
\ee
where the integrals $I$ depend on the eigenfrequency $\omega_\qq$ and on the wavevector $q$, {\vers see equations (\ref{eq:intpp},\ref{eq:intmm},\ref{eq:intpm}).}
The {\vers sought} implicit equation on the eigenfrequency {\vers (\ref{eq:dedispersion}) is the condition that the determinant of the $2\times 2$ matrix
in equation (\ref{eq:proprecollective}) is zero.}

\section{$x$ and $y$ rational fractions in the three-phonon process amplitudes}
\label{ann:fractionsrationnelles}
The rational fractions of the variables $x$ and $y$ that appear in the expression \eqref{eq:finaleA} of the microscopic 
{\bleu phonon coupling} amplitude are given by
\bea
J(x,y) \!\!\!&=&\!\!\! \frac{3 x y}{2 x y+2}   \label{eq:polydexy} \\
A(x,y) \!\!\!&=&\!\!\! \frac{\sum_{i=0}^3 A_i(y) x^i}{360 \left(x^3+x\right) \left(y^2+1\right)^3}   \\
B(x,y) \!\!\!&=&\!\!\! \frac{\sum_{i=0}^3 B_i(y) x^i}{18 \left(x^3+x\right) \left(y^2+1\right)^2}   \\
C(x,y) \!\!\!&=&\!\!\! \frac{y (x y+1)}{6 \left(x^3+x\right) \left(y^2+1\right)^2} \left[x^2 \left(5 y^2+3\right)-x y \left(y^2-3\right)+2\left(y^2+2\right)\right] \notag
\label{eq:Axy}
\eea
with
\bea
A_0(y) &=& -20 y \left(5 y^4+5 y^2+2\right)  \label{eq:polydey} \\
A_1(y) &=& 2 \left(10 y^6+109 y^4+63 y^2+24\right)  \\
A_2(y) &=& -y \left(53 y^4+186 y^2+13\right)  \\
A_3(y) &=& 25 y^6+116 y^4+167 y^2+36  \\
B_0(y) &=& -2 y \left(3 y^2+2\right)   \\
B_1(y) &=& 3 y^4+17 y^2+8  \\
B_2(y) &=& - 2 y \left(y^2+4\right)  \\
B_3(y) &=& 4 y^4+11 y^2+9  
\eea
Their values in the BEC limit $x\to0,y\sim-4/x$ are useful to connect the result \eqref{eq:finaleA} to its equivalent in the Bogoliubov theory:
\bea
A(x,y) &\underset{\substack{x\to0\\y\sim-4/x}}{\to} \frac{1}{8}  \label{eq:limBEC} \qquad
B(x,y) &\underset{\substack{x\to0\\y\sim-4/x}}{\to} \frac{1}{4}  \\
J(x,y) &\underset{\substack{x\to0 \\ y\sim-4/x}}{\to} 2  \qquad
C(x,y) &\underset{\substack{x\to0\\y\sim-4/x}}{\to} \frac{3}{4} \notag 
\eea
For resonant processes, these rational fractions add up to give the thermodynamic coefficient
\be
1+\frac{\rho}{3}{\frac{\dd^2\mu}{\dd{\rho}^2}}\bb{\frac{\dd\mu}{\dd{\rho}}}^{-1}  
   = \frac{\sum_{i=0}^3 D_i(y) x^i}{9(x^3+x)(y^2+1)^2} \label{eq:unpluslambdaF}
\ee
with
\bea
D_0(y) &=& -4 y^3 \\
D_1(y) &=& 4 \left(y^4+6 y^2+2\right) \\
D_2(y) &=& -y \left(y^4-6 y^2+5\right) \\
D_3(y) &=& 9 y^4+14 y^2+9 
\eea

{\rosso

\section{Low wave number out-of-the-energy-shell discrepancy between the microscopic and quantum hydrodynamic 
Beliaev-Landau coupling amplitudes}
\label{appen:diffmicrohydro}

In this appendix, as announced in the penultimate paragraph of section \ref{sec:3phonons}, 
we clarify the physical origin of the discrepancy of the Beliaev-Landau 2 phonons $\leftrightarrow$  1 phonon coupling amplitudes
of the microscopic model and of quantum hydrodynamics in the low wave number limit
out of the energy shell: the former indeed diverges 
whereas the latter tends to zero. To avoid tedious calculations in the fermionic microscopic model, we rather use a bosonic
microscopic model, the weakly interacting spatially homogeneous Bose gas, of grand canonical Hamiltonian $\hat{H}$
at chemical potential $\mu$, with
\be
\hat{H}=l^3\sum_\rr \hat{\psi}^\dagger \left(-\frac{\hbar^2}{2m}\Delta_\rr -\mu\right)\hat{\psi}
+\frac{g_0}{2} \sum_\rr l^3 \hat{\psi}^\dagger\hat{\psi}^\dagger\hat{\psi}\hat{\psi}
\label{eq:hamilmicrobosons}
\ee
We then easily make the link between this microscopic theory and quantum hydrodynamics through the modulus-phase representation
of the bosonic field operator \cite{CastinMora2003}, 
\be
\hat{\psi}(\rr)=\eee^{\ii\hat{\theta}(\rr)} \hat{\rho}^{1/2}(\rr)
\ee
where the phase field $\hat{\theta}(\rr)$ and the density field $\hat{\rho}(\rr)$ are canonically conjugated. 

In the Bogoliubov $U(1)$-symmetry breaking formalism,
one splits the bosonic field as $\hat{\psi}(\rr)=\psi_0+\delta\hat{\psi}(\rr)$, where $\psi_0>0$ is the classical field 
energy minimiser, and one expands the Hamiltonian in powers
of $\delta\hat{\psi}$, $\hat{H}=H_0+\hat{H}_2+\hat{H}_3+\ldots$. The field $\delta\hat{\psi}(\rr)$ is then expanded over the normal
modes of $\hat{H}_2$ (Bogoliubov expansion):
\be
\delta\hat{\psi}(\rr)=\frac{1}{L^{3/2}}\sum_{\qq\ne\mathbf{0}} U_\qq \eee^{\ii\qq\cdot\rr} \hat{b}^{\delta\psi}_\qq + 
V_\qq \eee^{-\ii\qq\cdot\rr} (\hat{b}^{\delta\psi}_\qq)^\dagger
\label{eq:devBogdpsi}
\ee
where we omitted for simplicity the contribution of the anomalous modes associated to symmetry breaking and we explicitly indicated
that the quasi-particle creation and annihilation operators are relative to a $\delta\hat{\psi}$-expansion of $\hat{H}$.
Insertion of the expansion (\ref{eq:devBogdpsi}) in $\hat{H}_3$ leads to the reduced Beliaev-Landau coupling amplitude in the Cartesian
$\delta\hat{\psi}$ point of view:
\be
\mathcal{A}_{\delta\psi}^{2\leftrightarrow 1}(\qq_1,\qq_2;\qq_3)= \frac{3}{4}s_1 s_2 s_3+\frac{1}{4}(s_1 d_2 d_3+s_2 d_1 d_3- s_3 d_1 d_2)
\label{eq:A2d1dpsi}
\ee
with $s_i=s_{\qq_i}$, $s_\qq=U_\qq+V_\qq$, $d_i=d_{\qq_i}$ and $d_\qq=U_\qq-V_\qq$. 

In the modulus-phase point of view, 
adapting reference \cite{CastinMora2003} to a symmetry breaking approach,
we expand $\hat{\psi}(\rr)$ in powers of the density fluctuation $\delta\hat{\rho}(\rr)= \hat{\rho}(\rr)-\psi_0^2$ and
of the phase $\hat{\theta}(\rr)$ so that $\delta\hat{\psi}(\rr)=\delta\hat{\psi}_1(\rr)+\delta\hat{\psi}_2(\rr)+\ldots$,
with
\bea
\delta\hat{\psi}_1(\rr) &=& \frac{\delta\hat{\rho}(\rr)}{2\psi_0} + \ii \psi_0 \hat{\theta}(\rr) \\
\delta\hat{\psi}_2(\rr) &=& -\frac{1}{2} \psi_0 \hat{\theta}^2(\rr) -\frac{\delta\hat{\rho}^2(\rr)}{8\psi_0^3} 
+ \ii \frac{\hat{\theta}(\rr)\delta\hat{\rho}(\rr)}{2\psi_0}
\eea
To quadratic order in $\delta\hat{\rho}$ and $\hat{\theta}$, one replaces $\delta\hat{\psi}$ in $\hat{H}_2$ with $\delta\hat{\psi}_1$; 
the normal mode expansion of the corresponding quadratic Hamiltonian in $(\delta\hat{\rho},\hat{\theta})$ gives (again neglecting
for simplicity the symmetry-breaking related anomalous modes):
\be
\delta\hat{\psi}_1(\rr)=
\frac{1}{L^{3/2}}\sum_{\qq\ne\mathbf{0}} U_\qq \eee^{\ii\qq\cdot\rr} \hat{b}^{\rho\theta}_\qq + 
V_\qq \eee^{-\ii\qq\cdot\rr} (\hat{b}^{\rho\theta}_\qq)^\dagger
\label{eq:devmodbogphmod}
\ee
with the same amplitudes $(U_\qq,V_\qq)$ as in equation (\ref{eq:devBogdpsi}) but with operator-valued coefficients 
$\hat{b}^{\rho\theta}_\qq,(\hat{b}^{\rho\theta}_\qq)^\dagger$ that coincide with 
$\hat{b}^{\delta\psi}_\qq,(\hat{b}^{\delta\psi}_\qq)^\dagger$, as $\delta\hat{\psi}_1$ and $\delta\hat{\psi}$ do,
only to first order in the fluctuations.
To cubic order in $\delta\hat{\rho}$ and $\hat{\theta}$, two different contributions are obtained: $(i)$ the obvious one coming 
from the replacement of $\delta\hat{\psi}$ with $\delta\hat{\psi}_1$ in $\hat{H}_3$, which leads to the Beliaev-Landau coupling
amplitude $\mathcal{A}_{\delta\psi}^{2\leftrightarrow 1}$, this time among the $\hat{b}^{\rho\theta}_\qq$, and 
$(ii)$ the more indirect one, coming from the replacement of
$\delta\hat{\psi}$ with $\delta\hat{\psi}_1+\delta\hat{\psi}_2$ in $\hat{H}_2$ and extraction of the corresponding cubic contributions
of the type $\delta\hat{\psi}_2^\dagger\delta\hat{\psi}_1$ and hermitian conjugate, which brings an extra term
$\mathcal{A}^{2\leftrightarrow 1}_{\mathrm{via}\, H_2}$ to the Beliaev-Landau coupling amplitude. At the end, the Beliaev-Landau coupling
amplitude in the $(\hat{\rho},\hat{\theta})$ point of view reads:
\be
\mathcal{A}_{\rho\theta}^{2\leftrightarrow 1}= \mathcal{A}_{\delta\psi}^{2\leftrightarrow 1} + \mathcal{A}^{2\leftrightarrow 1}_{\mathrm{via}\, H_2}
\ee
with
\begin{multline}
\mathcal{A}^{2\leftrightarrow 1}_{\mathrm{via}\, H_2}(\qq_1,\qq_2;\qq_3)= 
\frac{\check{\omega}_3d_3}{8}(d_1d_2-s_1s_2) +\frac{\check{\omega}_3s_3}{8}(s_1d_2+s_2d_1)  \\
-\frac{\check{\omega}_2d_2}{8}(d_1d_3+s_1s_3) -\frac{\check{\omega}_1d_1}{8}(d_2d_3+s_2s_3) 
+\frac{d_3}{8}(s_1 s_2 \check{\omega}_2 + s_2 s_1 \check{\omega}_1) 
-\frac{s_3}{8}(d_2s_1 \check{\omega}_1 + d_1 s_2 \check{\omega}_2)
\end{multline}
with $\check{\omega}_i=\hbar\omega_{\qq_i}/mc^2$, knowing that $mc^2=\mu$ at this order. From the Bogoliubov expressions
$\omega_i=[\check{E}_i(\check{E}_i+2)]^{1/2}$ and $s_i=1/d_i=[\check{E}_i/(\check{E}_i+2)]^{1/4}$, one has
$\check{\omega}_i d_i =(\check{E}_i+2)s_i$
and $\check{\epsilon}_i s_i=\check{E}_i d_i$, where $\check{E}_i=E_{\qq_i}/mc^2$ and $E_\qq=\hbar^2 q^2/2m$. Using as well
the conservation of momentum $\qq_3=\qq_1+\qq_2$ and its squares $q_3^2=(\qq_1+\qq_2)^2$, 
$q_2^2=(\qq_3-\qq_1)^2$ and $q_1^2=(\qq_3-\qq_2)^2$, we obtain
\be
\mathcal{A}_{\rho\theta}^{2\leftrightarrow 1}(\qq_1,\qq_2;\qq_3)=\\
\frac{1}{8}(s_3 d_1 d_2 \check{\qq}_1\cdot\check{\qq}_2+
s_2 d_1 d_3 \check{\qq}_1\cdot\check{\qq}_3 + 
s_1 d_2 d_3 \check{\qq}_2\cdot\check{\qq}_3)\\
-\frac{1}{16} s_1 s_2 s_3 (\check{q}_1^2+\check{q}_2^2+\check{q}_3^2)
\label{eq:ampcouplbellanrhoth}
\ee
with $\check{\qq}_i=\hbar\qq_i/mc$. 

We can now compare the Beliaev-Landau coupling amplitudes $\mathcal{A}_{\delta\psi}^{2\leftrightarrow 1}$ and
$\mathcal{A}_{\rho\theta}^{2\leftrightarrow 1}$ among themselves and to the quantum hydrodynamic coupling amplitude.
With the expressions of $s_i$ and $d_i$ given above, it is found from equation (\ref{eq:A2d1dpsi}) that
$\mathcal{A}_{\delta\psi}^{2\leftrightarrow 1}$ diverges when all the $q_i$ tend to zero (out of the energy shell).
On the contrary, even out of the energy shell, 
$\mathcal{A}_{\rho\theta}^{2\leftrightarrow 1}$ becomes equivalent to the quantum hydrodynamic result (\ref{eq:Ahydro21})
(here $\Lambda_{\rm F}=0$) that tends to zero, but it does not coincide with it (it differs by relative corrections $O(q_i^2)$)
due to the difference between the microscopic 
Hamiltonian (\ref{eq:hamilmicrobosons}) and the quantum hydrodynamic Hamiltonian.

As a final check, one can also directly insert the expansion (\ref{eq:devmodbogphmod}) (turned into an expansion
of $\delta\hat{\rho}$ and $\hat{\theta}$) in the terms of the Hamiltonian (40) of reference 
\cite{CastinMora2003} that are cubic in $\delta\hat{\rho}$ and $\hat{\theta}$.  One then directly recovers the Beliaev-Landau
coupling amplitude (\ref{eq:ampcouplbellanrhoth}) among the $\hat{b}_\qq^{\rho\theta}$.  Furthermore, if one  neglects everywhere
in this Hamiltonian (40) the kinetic energy stored in the density-variation $\delta\hat{\rho}$ of the field, 
at the origin of the so-called quantum pressure term, one recovers exactly the quantum hydrodynamic Beliaev-Landau coupling
(\ref{eq:Ahydro21}) (with here $\Lambda_{\rm F}=0$).
}

\section{{\vers Beliaev coupling beyond hydrodynamics}}
\label{appen:SonetWingate}

{\vers We give here details of the calculation allowing one to obtain, for the unitary Fermi gas, the first beyond-hydrodynamic correction
to the Beliaev phonon coupling amplitude from the Son and Wingate Lagrangian \cite{SonWingate2006}, see equation (\ref{eq:couplA}).}

For convenience we move to a Hamiltonian formalism, introducing the field $\Pi$ {\bleu that is} canonically conjugate {\bleu to the phase field $\phi$ and that represents} (up to a sign) density fluctuations, to 
obtain (here $\hbar=m=1$)
\bea
\!\!\!\!\!\!\!\!\!{\cal H}_2^{(0)} &=& \frac{2 \mu^{-1/2}}{15 c_0} \Pi^2 + \frac{5}{2} c_0 \mu^{3/2} \frac{1}{2} (\mathbf{grad}\, \phi )^2 \\
\!\!\!\!\!\!\!\!\!{\cal H}_3^{(0)} &=& \frac{4 \mu^{-2}}{3 (15 c_0)^2} \Pi^3 - \frac{1}{2} \Pi  (\mathbf{grad}\, \phi )^2 \\
\!\!\!\!\!\!\!\!\!{\cal H}_2^{(2)} &=& - \frac{16 c_1 \mu^{-3/2}}{(15 c_0)^2} (\mathbf{grad} \, \Pi)^2 - c_2 \mu^{1/2} (\Delta \phi )^2 \\
\!\!\!\!\!\!\!\!\!{\cal H}_3^{(2)} &=& - \frac{96 c_1 \mu^{-3}}{(15 c_0)^3} \Pi (\mathbf{grad} \, \Pi)^2 + \frac{2 c_2 \mu^{-1}}{15 c_0} 
\Pi (\Delta \phi )^2 
\eea
where the index $2$ or $3$ refers to the expansion order of the Hamiltonian in $\Pi$ and $\phi$ while
the exponent $(0)$ or $(2)$ refers to the expansion order in powers of spatial gradients,
the zeroth order being the standard hydrodynamics. Note that $\Pi$ and $\mathbf{grad} \, \phi$ are of the same order. At the hydrodynamic level, the Hamiltonian depends on a single constant $c_0$ that must be determined from a microscopic theory. It is linked as follows to the Bertsch parameter 
{\vers (\ref{eq:Bertsch})}:
\be
c_0=\frac{2^{5/2}}{15 \pi^2 \xi_B^{3/2}} 
\ee
The first correction to hydrodynamics involves two other {\vers dimensionless} constants $c_1$ and $c_2$.
By following the procedure already used in this paper, we first use the quadratized Hamiltonian ${\cal H}_2^{(0)}+{\cal H}_2^{(2)}$ to determine the excitation spectrum as in \cite{SonWingate2006}: 
\be
\label{eq:spectrum_SW}
\omega_\qq = \left( \frac{2\mu}{3} \right)^{1/2} q \left[1-\pi^2 (2\xi_B)^{1/2} \left(c_1+ \frac{3}{2}c_2\right) \left(\frac{q}{k_{\rm F}}\right)^2 
+o(q^2)\right]  
\equiv cq \left[ 1 + \frac{\gamma}{8} \left( \frac{\hbar q}{mc}\right)^2 + o(q^2) \right] 
\ee
and the modal expansion of the fields in the quantization volume $L^3$:
\bea
\Pi(\rr)&=&\frac{1}{L^{3/2}} \sum_{\qq \neq \zero} \Pi_\qq (b_\qq + b_{-\qq}^\ast) \eee^{\ii \qq \cdot \rr}\\
\phi(\rr)&=&\frac{1}{L^{3/2}} \sum_{\qq \neq \zero} \phi_\qq (b_\qq - b_{-\qq}^\ast) \eee^{\ii \qq \cdot \rr}
\eea
with the amplitudes
\bea
 \!\!\!\!\!\!\!\!\!\!\Pi_\qq  &=& \frac{1}{\sqrt{2}} \left( \frac{A_\qq}{B_\qq}\right)^{1/4} \\
 \!\!\!\!\!\!\!\!\!\! \phi_\qq &=& \frac{\ii}{\sqrt{2}} \left( \frac{A_\qq}{B_\qq}\right)^{-1/4} \\
 \!\!\!\!\!\!\!\!\!\! \frac{A_\qq}{B_\qq} \!\!\!&=&\!\!\! \frac{75}{8}(c_0 \mu q)^2 \!\bbcro{ 1 + \frac{8 \mu^{-1}}{15 c_0} \left(c_1 -\frac{3}{2}c_2\right)q^2
\!+ o(q^2) } 
 \label{eq:modal_amplitudes}
\eea
Indeed $\left(\begin{array}{c} \Pi_\qq \\ \phi_\qq \end{array}\right)$ is an eigenvector 
{\violet of eigenvalue $- \ii \omega_\qq$} of {\violet the}
matrix $\left(\begin{array}{cc} 0 & -A_\qq \\ B_\qq & 0 \end{array}\right)$ {\violet that appears in the
Hamiltonian equations of motion.} 
Note that different linear combinations of the constants $c_1$ and $c_2$ appear in the spectrum (\ref{eq:spectrum_SW})
and in the modal amplitudes (\ref{eq:modal_amplitudes}). By inserting the modal decomposition in the cubic Hamiltonian
${\cal H}_3^{(0)}+{\cal H}_3^{(2)}$, and isolating the Beliaev terms $2\leftrightarrow1$ as we did in (\ref{eq:H3hydro}), we obtain
the on-shell Beliaev coupling amplitude to first order beyond hydrodynamics {\vers as given in (\ref{eq:couplA})}.

\section{{\vers Three-phonon and four-phonon processes in a bosonic model}}
\label{appen:modelebosons}

{\vers Here we describe and we implement the microscopic bosonic model used in section \ref{sec:modelebosons} to test the $2\leftrightarrow 2$ effective
phonon coupling predicted by quantum hydrodynamics.}
{\vers We consider spinless bosons on a cubic lattice of lattice constant $l$, with a large enough interaction range $b$ so that their
excitation spectrum is concave at low wave number $q$.} The Hamiltonian of the lattice model reads
\be
\hat{H}_{\rm B}= l^3  \sum_{\rr} \hat{\psi}^\dagger(\rr) \left( - \frac{\hbar^2}{2m_{\rm B}}  \Delta_{\rr}\right)  \hat{\psi}(\rr)
+ \frac{l^6}{2} \sum_{\rr,\rr'} V(\rr-\rr')  \hat{\psi}^\dagger(\rr) \hat{\psi}^\dagger(\rr') \hat{\psi}(\rr'){\hat{\psi}}(\rr)
\label{eq:hamiltonienB}
\ee
with the interaction potential
\be
V(\rr) = V_0 \eee^{-r^2/2b^2}
\ee
of Fourier transform
\be
\tilde{V}(\qq) = \tilde{V}_0 \eee^{-q^2b^2/2}\quad\mbox{with}\quad\tilde{V}_0=(2\pi)^{3/2}b^3V_0
\ee

The Bose gas is in the weakly interacting regime $(\rho_{\rm B} a_{\rm B}^3)^{1/2} \ll 1$, where $\rho_{\rm B}=N_{\rm B}/L^3$ is the density of the
bosons and {the} $s$-wave scattering length $a_{\rm B}$ is given by $4\pi\hbar^2 a_{\rm B}/m_{\rm B}=\tilde{V}_0$
{in the Born approximation}.
Following the Bogoliubov theory \cite{Bogolioubov1958} in its $U(1)$ symmetry preserving version \cite{Gardiner1997,CastinDum1998,CastinLesHouches2001,CastinSinatra2009}, we split the bosonic field operator as
\be
\hat{\psi}(\rr)=\eee^{\ii\hat{\theta}_0}\bbcro{\hat{n}_0^{1/2} \phi_0(\rr)+\hat{\Lambda}(\rr)}
\ee
where $\hat{\theta}_0$ is the condensate phase operator, $\hat{n}_0$ is the number of bosons in the condensate mode $\phi_0(\rr)=1/L^{3/2}$
and the non-condensed field operator $\hat{\Lambda}(\rr)$, orthogonal to the condensate mode, conserves the number of particles.
Within the subspace with fixed total number of bosons $N_{\rm B}$, we eliminate $\hat{n}_0$ through the relation
\be
\hat{n}_0 = N_{\rm B}- l^3  \sum_{\rr} \hat{\Lambda}^\dagger(\rr) \hat{\Lambda}(\rr)
\label{eq:devn0}
\ee
To describe the  $2\leftrightarrow2$ processes, we expand the Hamiltonian in powers of $\hat{\Lambda}$ up to order $4$:
\be
\hat{H}_{\rm B}=\hat{H}_{\rm B0}+\hat{H}_{\rm B2}+\hat{H}_{\rm B3}+\hat{H}_{\rm B4}+\ldots
\ee
We obtain \footnote{We are here in the large $N_{\rm B}$ limit and we neglect $1$ as compared to $\hat{n}_0$ and to $N_{\rm B}$.
In the equations \eqref{eq:H2B} and \eqref{eq:H4B}, there was a cancellation of the Hartree contribution with a chemical potential-type
contribution originating from the expansion of the condensate interaction energy $\tilde{V}_0 {\hat{n}_0^2}/{2L^3}$ in powers of
the number of non-condensed particles.}
\bea
\hat{H}_{\rm B0} &=& \tilde{V}_0 \frac{N_{\rm B}^2}{2L^3} \label{eq:H0B} \\
\hat{H}_{\rm B2} &=& l^3 \sum_\rr \hat{\Lambda}^\dagger(\rr) \left( - \frac{\hbar^2}{2m_{\rm B}}  \Delta_{\rr}\right) \hat{\Lambda}(\rr) + \rho_{\rm B} l^6 \sum_{\rr,\rr'} V(\rr-\rr') \bb{\hat{\Lambda}^\dagger(\rr) \hat{\Lambda}(\rr')  + \frac{1}{2}\bbcro{\hat{\Lambda}^\dagger(\rr) \hat{\Lambda}^\dagger(\rr')+\hat{\Lambda}(\rr) \hat{\Lambda}(\rr')}}
\label{eq:H2B} \\
\hat{H}_{\rm B3} &=& \frac{\rho_{\rm B}^{1/2}}{2} l^6 \sum_{\rr,\rr'} V(\rr-\rr') \bb{ \bbcro{\hat{\Lambda}^\dagger(\rr)+\hat{\Lambda}^\dagger(\rr')} \hat{\Lambda}(\rr') \hat{\Lambda}(\rr) + \hat{\Lambda}^\dagger(\rr') \hat{\Lambda}^\dagger(\rr) \bbcro{\hat{\Lambda}(\rr)+\hat{\Lambda}(\rr')}} \label{eq:H3B}\\
\hat{H}_{\rm B4} &=& \frac{l^6}{2}  \sum_{\rr,\rr'} V(\rr-\rr')  \hat{\Lambda}^\dagger(\rr) \hat{\Lambda}^\dagger(\rr') \hat{\Lambda}(\rr'){\hat{\Lambda}}(\rr)
-\frac{\tilde{V}_0}{2L^3} \bb{l^3  \sum_{\rr} \hat{\Lambda}^\dagger(\rr) \hat{\Lambda}(\rr)}^2 \notag \\
&&-\frac{1}{L^3} \bb{l^3  \sum_{\rr} \hat{\Lambda}^\dagger(\rr) \hat{\Lambda}(\rr)} \bb{l^6\sum_{\rr,\rr'} V(\rr-\rr') \bbcro{\hat{\Lambda}^\dagger(\rr) \hat{\Lambda}(\rr') + \frac{1}{2}\bbcro{\hat{\Lambda}^\dagger(\rr) \hat{\Lambda}^\dagger(\rr')+\hat{\Lambda}(\rr) \hat{\Lambda}(\rr')}} } \label{eq:H4B}
\eea

\paragraph{$0$th order}
From $\hat{H}_{\rm B 0}$ we get the equation of state to leading order:
\be
\mu_{\rm B} = \rho_{\rm B} \tilde{V}_0 \equiv \frac{\hbar^2}{2m\xi^2}
\label{eq:etatB}
\ee
where we introduced the healing length $\xi$ and the chemical potential $\mu_{\rm B}$ of the Bose gas.
We then apply relation \eqref{eq:vitesseson} to obtain the sound velocity
\be
m_{\rm B} c_{\rm B}^2 = \mu_{\rm B}
\ee
\paragraph{$2$nd order}
$\hat{H}_{\rm B 2}$ can be diagonalised by a Bogoliubov transformation:
\be
\hat{\Lambda}(\rr)= \frac{1}{L^{3/2}} \sum_{\qq\neq\zero} \bb{U_\qq^{\rm B} \hat{b}_\qq \eee^{\ii\qq\cdot\rr}+ V_\qq^{\rm B} \hat{b}_\qq^\dagger \eee^{-\ii\qq\cdot\rr} }
\label{eq:fleurB}
\ee
where the $\hat{b}_\qq$ are annihilation operators of bosonic quasiparticles and the amplitudes $U^{\rm B}_\qq$ and $V^{\rm B}_\qq$
are given by
\bea
U^{\rm B}_\qq+V^{\rm B}_\qq &\equiv& S_\qq =\bb{\frac{\frac{\hbar^2q^2}{2m_{\rm B}}}{\frac{\hbar^2q^2}{2m_{\rm B}}+2\rho_{\rm B}\tilde{V}(\qq)}  }^{1/4} \\
U^{\rm B}_\qq-V^{\rm B}_\qq &\equiv& D_\qq = \frac{1}{S_\qq}
\eea
We have introduced the angular eigenfrequencies of the Bogoliubov quasiparticles
\be
\hbar\omega^{\rm B}_\qq=\bbcro{\frac{\hbar^2q^2}{2m_{\rm B}}\bb{\frac{\hbar^2q^2}{2m_{\rm B}}+2\rho_{\rm B}\tilde{V}(\qq)}}^{1/2}
\ee
The dispersion relation $q\mapsto\omega_\qq$ is then concave in the vicinity of $q=0$ under the condition, {assumed to be satisfied} in what follows,
\be
b>\xi
\ee
\paragraph{$3$rd order}
We insert the modal expansion \eqref{eq:fleurB} in the cubic Hamiltonian \eqref{eq:H3B} that we write in the form \eqref{eq:H3hydro}
with the constants $m$, $c$ and $\rho$ replaced with $m_{\rm B}$, $c_{\rm B}$ and $\rho_{\rm B}$ respectively,
and with coupling amplitudes now given by
\be
\mathcal{A}_{\rm B}^{2\leftrightarrow1}(\qq_1,\qq_2;\qq_3)  =  \frac{1}{4\tilde{V}_0} \left[ \tilde{V}(\qq_3)S_{3}(S_{1} S_{2}-D_{1}D_{2}) 
+\tilde{V}(\qq_1)S_{1}(S_{2} S_{3}+D_{2}D_{3}) +\tilde{V}(\qq_2)S_{2}(S_{1} S_{3}+D_{1}D_{3})\right] \label{eq:A2donne1B} 
 \ee
 \be
\mathcal{A}_{\rm B}^{3\leftrightarrow0}(\qq_1,\qq_2,\qq_3)  =  \frac{1}{12\tilde{V}_0} \left[\tilde{V}(\qq_1)S_{1}(S_{2} S_{3}-D_{2}D_{3})  
+\tilde{V}(\qq_2)S_{2}(S_{1} S_{3}-D_{1}D_{3})+\tilde{V}(\qq_3)S_{3}(S_{1} S_{2}-D_{1}D_{2}) \right]\label{eq:A3donne0B}
 \ee
with the notations $S_i\equiv S_{\qq_i}$ and $D_i\equiv D_{\qq_i}$.
\paragraph{$4$th order}
We insert the modal expansion \eqref{eq:fleurB} in the quartic Hamiltonian \eqref{eq:H4B} that we write in the form \eqref{eq:H2donne2hydro}
with $(m,c,\rho)\to (m_{\rm B},c_{\rm B},\rho_{\rm B})$ and a direct $2\leftrightarrow2$ coupling amplitude given by
\begin{multline}
\mathcal{A}_{\rm B}^{2\leftrightarrow2,{\rm dir}}(\qq_1,\qq_2;\qq_3,\qq_4) = \frac{1}{32\tilde{V}_0}
\bbcrol{ \bb{\tilde{V}(\qq_1+\qq_2)+\tilde{V}(\qq_3+\qq_4)}(S_{1} S_{2}-D_{1} D_{2})(S_{3} S_{4}-D_{3}D_{4}) } 
+\bb{\tilde{V}(\qq_3-\qq_1)+\tilde{V}(\qq_2-\qq_4)} \\
\times (S_{1} S_{3}+D_{1} D_{3})(S_{2} S_{4}+D_{2}D_{4})
\bbcror{+\bb{\tilde{V}(\qq_4-\qq_1)+\tilde{V}(\qq_2-\qq_3)} (S_{1} S_{4}+D_{1} D_{4})(S_{2} S_{3}+D_{2}D_{3})} 
\label{eq:A2donne2Bdir}
\end{multline}
We considered here the general case where the $\qq_i$ and their opposite are two-by-two distinct, in which case only the first term of $\hat{H}_{\rm B4}$ contributes. {The other terms} of $\hat{H}_{\rm B4}$, that originate from the expansion \eqref{eq:devn0} of $\hat{n}_0$
in powers of the number of non-condensed particles, contribute to the equation of state beyond Bogoliubov theory \cite{CastinMora2003}.

\paragraph{Effective coupling amplitude}
We obtain the on-shell effective coupling amplitude of the bosonic model $\mathcal{A}_{\rm B, {OnS}}^{2\leftrightarrow2,{\rm eff}}$
from the amplitudes (\ref{eq:A2donne1B},\ref{eq:A3donne0B},\ref{eq:A2donne2Bdir}) as prescribed by the equation \eqref{eq:A22eff}.
We then expand it in the limit of small wave vectors and express the result in terms of the angular frequencies $\omega_{\qq_i}^{\rm B}$, $i=1,2,3,4$
and $\omega_{\qq_1+\qq_2}^{\rm B}$, $\omega_{\qq_1-\qq_3}^{\rm B}$, $\omega_{\qq_1-\qq_4}^{\rm B}$. 
{\vers We then recover the hydrodynamic prediction corrected by the Landau and Khalatnikov prescription, as explained in
section \ref{sec:modelebosons}.}

{
\section{Contribution of the Beliaev and higher order processes to the $T=0$ decay of the single phonon state}
\label{appen:1d3}

\subsection{Presentation of the problem and link with the resolvent of the Hamiltonian}

We consider the damping rate $\Gamma_\qq$ of a phonon with wave vector $\qq$ prepared at zero temperature in a spatially homogeneous gas of fermions. We assume that the phononic excitation branch $q\mapsto \omega_\qq$ is convex in vicinity of $q=0$. The general problem is to determine the behavior of $\Gamma_\qq$ in the limit $q\to 0$, where one can use an effective low energy theory to describe the coupling among phonons.
The dominant process is of course the $1\to 2$ Beliaev one. By introducing a dimensionless parameter $\gamma >0$
describing both the correction in $q^3$ to the hydrodynamic linear excitation spectrum, and the first correction to the hydrodynamic coupling amplitude $1\to 2$, we have obtained the provisional result (\ref{eq:Bel_result_prov}) up to
the sub-leading order $q^7$. In this appendix, we give some details on the obtention of the final result at the Beliaev order (\ref{eq:Bel_result}), and we examine the possible contributions of order $q^7$ of all the higher order processes, such as 
the cascade process $1\to 2\to 3$  in figure \ref{fig:1d3}, leading to the correction (\ref{eq:Gamma1d3}) and to the final result (\ref{eq:amorunit}).

In the subspace with a fixed total momentum $\hbar\qq$, the one-phonon state $|\qq\rangle$ is the only discrete state, as it is the only one completely characterized by a single wave vector value. It is an eigenstate of $\hat{H}_2$, the part of the 
Hamiltonian that is quadratic in the phonon creation $\hat{b}^\dagger$ and annihilation $\hat{b}$ operators.
It is nevertheless coupled to the two-phonon, three-phonon, etc.\ continua by the rest of the Hamiltonian $\hat{V}=\hat{H}_3+\hat{H}_4+\ldots$, that contains cubic, quartic, etc.\ terms, when written in the normal order for the $\hat{b}$.  

As a consequence, the discrete state will in general get diluted in the continua, giving rise to a complex pole $z_\qq$ 
in the analytic continuation of the resolvent
$\hat{G}(z)=(z-\hat{H})^{-1}$ of the full Hamiltonian \cite{Cohen}. This pole can be written as 
\be
z_\qq =\hbar\omega_\qq-\ii \frac{\hbar\Gamma_\qq}{2}
\label{eq:decompzq}
\ee
where $\omega_\qq$ is the angular eigenfrequency of the phonon and $\Gamma_\qq$ its damping rate
\footnote{We cannot calculate here the corrections to $\omega_\qq$ due to the coupling with the continua, even
at the order $2$ in $\hat{V}$, because we rely on an effective Hamiltonian. If we try, we would encounter principal part integrals with ultraviolet divergences that, once inserted in the unperturbed value in $\hat{H}_2$ would give, following the ideas of renormalisation, the true value, which would however remain uncalculated and unknown in the absence of a microscopic model \cite{KCS2015}. 
Within this context, we {\rouge do not} find it useful to distinguish between the non perturbed phonon eigenenergy for $\hat{H}_2$,  
which we should rigorously note $\epsilon_\qq^{(2)}$, and the true energy  $\epsilon_\qq$. By the way, in order to obtain the 
scaling laws with $q$ of the perturbative terms of order $n$ in $\hat{V}$, we assume that the wave numbers of the virtual phonons are $O(q)$, hence that the ultraviolet cutoff is set at a wave number $Aq$, $A\gg 1$. This has the big advantage that  the real correction to the angular eigenfrequency $\epsilon_\qq^{(2)}$ due to $\hat{H}_3$ to the leading order is 
$O(q^5)$ and it affects neither the $\gamma$ parameter in equation (\ref{eq:fleur}) nor the damping rate $\Gamma_\qq$
to the order $q^7$.
}.
\setcounter{noterenormalisation}{\thefootnote}

\subsection{Perturbative calculation and power counting}
\label{subsec:app_comptage}

The $q\to 0$ limit corresponds to the weak coupling limit. This is apparent in the quantum hydrodynamic theory as
the modal expansion of the velocity and density fluctuations $\hat{\mathbf{v}}$ and $\delta\hat{\rho}$ involves coefficients 
that tend to zero as $q^{1/2}$, so that the matrix elements of  $\hat{H}_p$  between phonon Fock states behave as 
$q^{p/2}$:
\be
\hat{H}_p \underset{q\to 0}{\approx} q^{p/2}
\label{eq:odgHn}
\ee
One can then attempt a perturbative calculation of $\Gamma_\qq$ starting from the exact expression obtained by the projector method \cite{Cohen}
\be
\langle \qq| \hat{G}(z) |\qq\rangle = \frac{1}{z-\langle \qq| \hat{H}_{\rm eff}(z) |\qq\rangle}
\ee
then expanding in powers of $\hat{V}$ the matrix element of the effective Hamiltonian in the state $|\qq\rangle$:
\be
\langle \qq| H_{\rm eff}(z) |\qq\rangle = \langle \qq|\hat{H}_2|\qq\rangle +\langle \qq|\hat{V}\hat{Q}\frac{\hat{Q}}{z\hat{Q}-\hat{Q}\hat{H}_2\hat{Q}}
\hat{Q} \hat{V} |\qq\rangle + \langle \qq|\hat{V} \hat{Q} \frac{\hat{Q}}{z\hat{Q}-\hat{Q}\hat{H}_2\hat{Q}}\hat{Q}\hat{V}\hat{Q}
\frac{\hat{Q}}{z\hat{Q}-\hat{Q}\hat{H}_2\hat{Q}}\hat{Q}\hat{V}|\qq\rangle + \ldots
\label{eq:HeffdevpuisV}
\ee
where $\hat{Q}=1-|\qq\rangle \langle \qq|$ projects orthogonally to $|\qq\rangle$. The result (\ref{eq:Bel_result}) 
corresponds to the second term on the right-hand side of the equation, the Beliaev term, in which the contribution of $\hat{V}$ reduces to that of  $\hat{H}_3$.
In all the following terms of order  $3$ in $\hat{V}$ or higher, $z$ can be approximated in the denominator by
\be
z=\hbar\omega_\qq+\ii \eta, \quad \eta\to 0^+, 
\ee
the displacement of $z$ by an imaginary part that is $O(q^5)$ in these terms does not contribute to $\Gamma_\qq$
at the order $q^7$. The same conclusion does not hold for the Beliaev term, as we shall see in subsection 
 \ref{subsec:app_ordreBel}.

Let us consider in (\ref{eq:HeffdevpuisV}) the term $\mathcal{T}_n$ of order $n$ in $\hat{V}$ and let us try to give an upper bound to its order in $q$ as done in the simple reasoning below equation (\ref{eq:Bel_result}).  

Taking (\ref{eq:odgHn}) into account, we must retain in the $\hat{V}$  factors as many as possible contributions of 
$\hat{H}_3$ and, if not, those of  $\hat{H}_4$. As $\hat{H}_3$ changes the parity of the phonon number, while $\hat{H}_4$ preserves it, we will retain only $\hat{H}_3$ if $n$ is even, but it will be necessary to keep at least one factor $\hat{H}_4$
if $n$ is odd. The minimal number of independent wave vectors of the virtual phonons is  $n/2$ for $n$ even, $(n+1)/2$ for $n$ odd, and the integration over each independent wave vector provides a factor {\bleu $q^3$.}
Finally, one has to count a factor  $1/q$ for each of the $n-1$ energy denominators. All together, this leads to 
\bea
\mathcal{T}_n=O(q^{2n+1}) & \mbox{if} & n \ \mbox{even} \\
\mathcal{T}_n=O(q^{2n+3}) & \mbox{if} & n \ \mbox{odd}
\eea
One would then conclude with no surprise, that the processes of higher order than Beliaev, that is of order $n\geq 3$ in $\hat{V}$,
give a contribution to  $\Gamma_\qq$ that is $O(q^9)$ hence negligible.

The simple reasoning above {\violet forgets} the possibility {\violet that} the energy denominators {\violet are}
of order $q^3$. For this case to occur, {\violet the contribution to the energy denominator
of the linear part of the spectrum must have a chance to cancel out}
so the considered processes {\violet must have the possibility}
to be resonant. {\violet One} must then keep  in $\hat{H}_3$ only the terms in 
 $\hat{b}^\dagger \hat{b}^\dagger \hat{b}$, that {\violet constitute} the raising part  $\hat{H}_3^{(+)}$ of the cubic Hamiltonian, or the terms in $\hat{b}^\dagger \hat{b}\hat{b}$, that {\violet constitute} its lowering part $\hat{H}_3^{(-)}$. The other terms in $\hat{b}^\dagger
\hat{b}^\dagger\hat{b}^\dagger$ and $\hat{b}\hat{b}\hat{b}$ induce nonresonant processes.
In $\hat{H}_4$, we shall only keep terms  $\hat{b}^\dagger \hat{b}^\dagger \hat{b}\hat{b}$, that are the most favorable potentially resonant terms, as they add a single independent wave vector. 

In order {\violet to have} all the energy denominators {\violet of order} $q^3$, 
all the independent wave vectors must be emitted forward, within a narrow cone of angular aperture $O(q)$ with respect to the direction of $\qq$. The angular integration then {\violet brings} a factor $q^2$ for each independent wave vector, that is $q^n$ for $n$ even and  $q^{n+1}$ for $n$ odd, but we gain a factor $q^{-2(n-1)}$ thanks to $n-1$ small denominators.

This leads to the refined upper bounds
\bea
\label{eq:majaffinpair}
\mathcal{T}_n=O(q^{n+3}) & \mbox{\bleu if} & n \ \mbox{even} \\
\mathcal{T}_n=O(q^{n+6}) & \mbox{\bleu if} & n \ \mbox{odd}
\eea
The conclusion {\bleu of} this discussion is that only the order $4$ in $\hat{V}$ has a chance to {\violet produce} 
a correction in $q^7$
to the result $\Gamma_\qq^{\rm Bel}$ in equation (\ref{eq:Bel_result}). We analyze it in more detail in subsection
\ref{subsec:app_ordre4}.
We can also confirm that {\vers the} order $2$ in $\hat{V}$ gives the leading contribution to $\Gamma_\qq$ in $q^5$. This 
however does not exonerate us from a rigorous study of its sub-leading contributions up to the order $q^7$, which is done in subsection \ref{subsec:app_ordreBel}.

\subsection{Study to second order in $\hat{V}$}
\label{subsec:app_ordreBel}

In this subsection we restrict the effective Hamiltonian in equation  (\ref{eq:HeffdevpuisV}) to second order in $\hat{V}$ 
and calculate the resulting value of $\Gamma_\qq$ up to the order $q^7$.

\begin{figure}[t] 
\begin{center}
\includegraphics[width=0.15\textwidth,clip=]{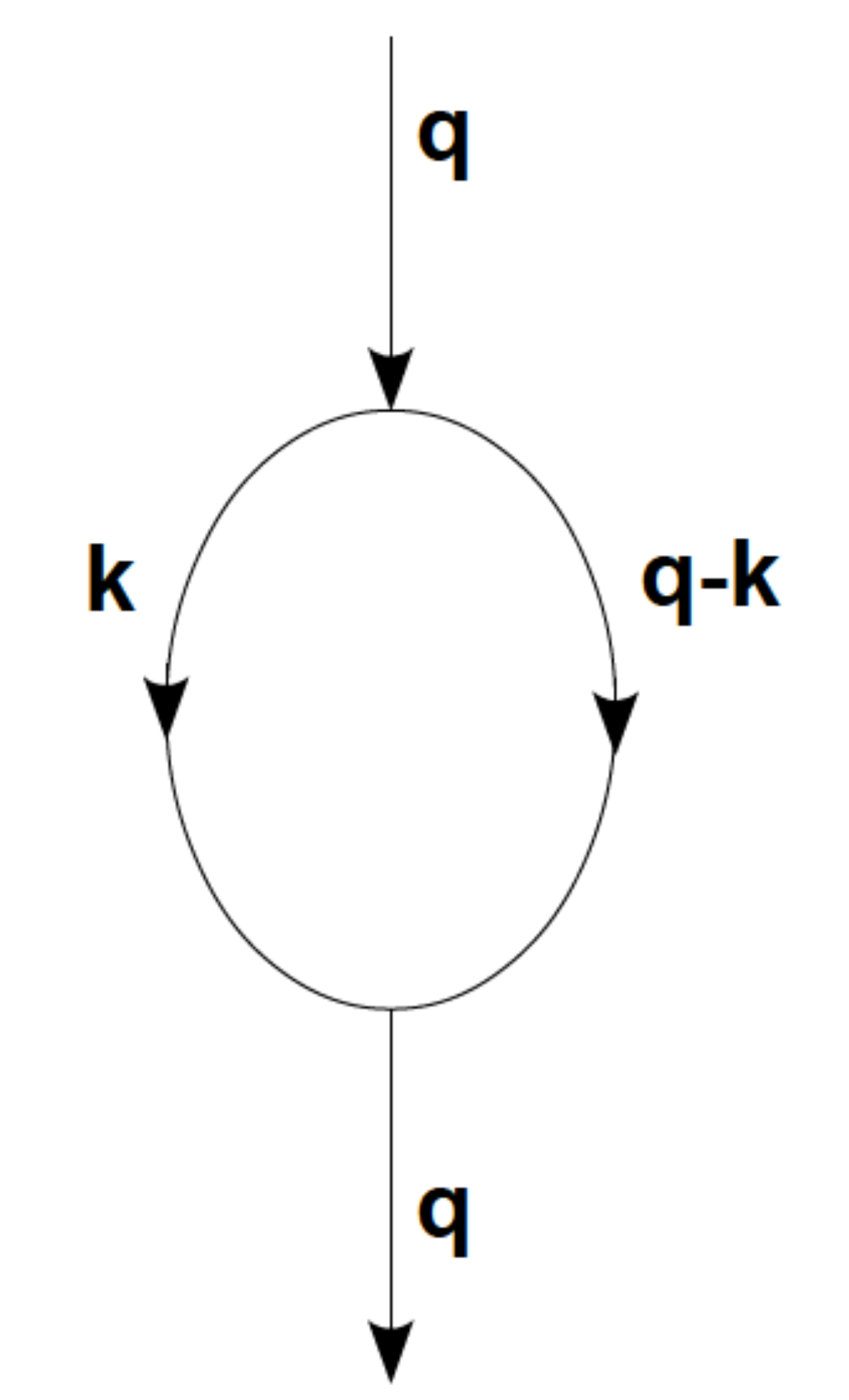}
\end{center}
\caption{\label{fig:diag2} Diagram of order 2 in $\hat{V}$ that contributes to the damping rate  $\Gamma_\qq$ of the phonon $\qq$ to the order $q^7$. It represents the Beliaev process whose leading order is $q^5$. On the figure, each vertex represents an action of  $\hat{H}_3^{(\pm)}$.
}
\end{figure} 

To this order in $\hat{V}$ several different diagrams actually contribute, as $\hat{V}=\hat{H}_3+\hat{H}_4+\hat{H}_5+\ldots$. The leading order in $q$ can be estimated by power counting. If the two factors $\hat{V}$ are both equal to $\hat{H}_3$, one has to integrate over at least one independent wave vector and the leading order is  $q^5$ as in equation  (\ref{eq:majaffinpair}). If they are both equal to $\hat{H}_4$, one has to integrate over at least two independent wave vectors
 and the leading order is  $q^9$ without the \g{small denominators} effect, but not any better with the \g{small denominators}.
 If one of the factors $\hat{V}$ involves $\hat{H}_p$, with $p\geq 5$, the contribution is even smaller and negligible. We can then restrict to $\hat{V}=\hat{H}_3$, and furthermore to $\hat{V}=\hat{H}_3^{(\pm)}$ as the non resonant terms in $\hat{b}^\dagger\hat{b}^\dagger\hat{b}^\dagger$ {\vers and} $\hat{b}\hat{b}\hat{b}$
of $\hat{H}_3$ involve two independent wave vectors and contribute in $O(q^8)$. 

There remains then a single diagram, the Beliaev one, represented in figure \ref{fig:diag2}. Let us introduce the corresponding 
Beliaev self-energy, for any complex number $z$ with $\im z >0$:
\be
\Delta \epsilon_\qq^{\rm Bel}(z)=\frac{1}{2} \sum_{\kk} \frac{|\langle\kk,\qq-\kk|\hat{H}_3^{(+)}|\qq\rangle|^2}
{z-(\epsilon_\kk+\epsilon_{\qq-\kk})}=\frac{2(mc^2)^2}{\rho} \int \frac{\dd^3k}{(2\pi)^3} 
\frac{[\mathcal{A}^{2\leftrightarrow 1}_{\rm hydro}(\kk,\qq-\kk;\qq)]^2}{z-(\epsilon_\kk+\epsilon_{\qq-\kk})}
\label{eq:DeQBelZ}
\ee
where we considered the thermodynamic limit and we have used the form  (\ref{eq:H3hydro}) of the hydrodynamic Hamiltonian. To this order of approximation, the pole  $z_\qq$ associated to the phonon $\qq$ is the solution of the implicit equation 
\be
z_\qq^{\rm Bel} = \epsilon_\qq + \Delta \epsilon_\qq^{\rm Bel \downarrow }(z_\qq^{\rm Bel})
\label{eq:impli}
\ee
where the downward arrow denotes the analytic continuation of the self-energy from the upper half-plane 
 $\im z>0$ to the lower one  $\im z < 0$ through the {\bleu branch} cut on the real positive axis. Let us separate $z^{\rm Bel}_\qq$
 into the real and imaginary parts as in (\ref{eq:decompzq}) and let us expand the right-hand side in powers of  $\Gamma_\qq$:
\be
z_\qq^{\rm Bel} = \epsilon_\qq + \Delta \epsilon_\qq^{\rm Bel\downarrow}(\epsilon_\qq) -\frac{\ii\hbar\Gamma_\qq}{2} 
\frac{\dd}{\dd z} \Delta \epsilon_\qq^{\rm Bel\downarrow}(\epsilon_\qq) + O({\bleu q^9})
\ee
It is enough here to truncate the Taylor expansion to the order one; on the other hand the zeroth order is not enough 
\footnote{The Taylor contribution of order $n$ is  $O(q^{2n+5})$ as the $n$th order derivative of the self-energy, that involves an energy denominator to the power $n+1$, is of order $q^8/q^{3(n+1)}$ accounting for the \g{small denominators} effect.}. The values of the analytic continuation and of its derivative in $\epsilon_\qq$ are obtained as the limits when 
$\eta\to 0^+$ of the noncontinuated functions in $z=\epsilon_\qq+\ii\eta$:
\be
z_\qq^{\rm Bel} = \epsilon_\qq + \Delta \epsilon_\qq^{\rm Bel} (\epsilon_\qq+\ii\eta) -\frac{\ii\hbar\Gamma_\qq}{2} 
\frac{\dd}{\dd z} \Delta \epsilon_\qq^{\rm Bel} (\epsilon_\qq+\ii\eta) + O({\bleu q^9})
\label{eq:taylore}
\ee
The second term is the ordinary perturbative result. It leads to the damping rate $\Gamma_{\qq,{\rm pert}}^{\rm Bel}$
given by the zero temperature limit of equation (\ref{eq:GamqBel}), 
\be
\frac{\hbar \Gamma_{\qq,{\rm pert}}^{\rm Bel}}{2} = -\im \Delta \epsilon_\qq^{\rm Bel} (\epsilon_\qq+\ii\eta)
\ee
and its value for the unitary gas has been calculated up to the order $q^7$ in equation (\ref{eq:Bel_result_prov}).
Once the imaginary part of equation (\ref{eq:taylore}) has been taken, we find a nonnegligible correction  to this order, resulting from a first self consistent iteration of $z_\qq$ in the implicit equation (\ref{eq:impli}):
\be
\Gamma_\qq^{\rm Bel} = \Gamma_{\qq,{\rm pert}}^{\rm Bel} \left[1+\re \frac{\dd}{\dd z} \Delta \epsilon_\qq^{\rm Bel} (\epsilon_\qq+\ii\eta)
+O({\bleu q}^4)\right]
\ee
with
\be
\frac{\dd}{\dd z} \Delta \epsilon_\qq^{\rm Bel} (\epsilon_\qq+\ii\eta) = -\frac{2(mc^2)^2}{\rho} 
\int \frac{\dd^3k}{(2\pi)^3} \frac{[\mathcal{A}^{2\leftrightarrow 1}_{\rm hydro}(\kk,\qq-\kk;\qq)]^2}{(\epsilon_\qq-\epsilon_\kk-\epsilon_{\qq-\kk}+\ii\eta)^2}
\ee
In order to obtain an equivalent of this derivative when $\check{q}{\bleu\equiv \hbar q/mc}\to 0$, we apply the same technique as in section 
\ref{subsec:ccadlk}.
We use spherical coordinates of polar axis  $\qq$ and perform the rescalings 
$k=\bar{k}q$ and $\theta=\gamma^{1/2}\check{q}\check{\theta}$ on the modulus and polar angle of the vector $\kk$, with $\bar{k}<1$ to profit from the \g{small denominators} effect. In the integrand and hence in the coupling amplitude 
(\ref{eq:Ahydro21}),  we take the limit $\check{q}\to 0$ with $\bar{k}$ and $\check{\theta}$ fixed, with the intermediate results
\bea
|\qq-\kk|=q(1-\bar{k}) \left[1+\frac{\gamma\bar{k}\check{\theta}^2\check{q}^2}{2(1-\bar{k})^2} + O(\check{q}^4)\right] && \\
\label{eq:intermdeux}
\epsilon_\qq-\epsilon_\kk-\epsilon_{\qq-\kk} \sim \frac{\gamma\check{q}^3 mc^2}{8}\left[3\bar{k}(1-\bar{k})-\frac{4\bar{k}\check{\theta}^2}{1-\bar{k}}
\right] &&  \\
{[\mathcal{A}^{2\leftrightarrow 1}_{\rm hydro}(\kk,\qq-\kk;\qq)]}^2 \sim \frac{2}{9} \check{q}^3 \bar{k}(1-\bar{k}) &&
\eea
taking into account the fact that $1+\Lambda_\mathrm{F}=8/9$ at unitarity.
The integration with respect to  $\check{\theta}$ over $[0,+\infty[$ is simple after the change of variable 
 $X=\check{\theta}^2$, and we are left with
\be
\frac{\dd}{\dd z} \Delta \epsilon_\qq^{\rm Bel} (\epsilon_\qq+\ii\eta) \sim \frac{8 \check{q}^2}{9\pi^2\gamma}  \left(\frac{mc}{\hbar\rho^{1/3}}\right)^{3}
\int_0^{1} \dd\bar{k} \frac{\bar{k}^2(1-\bar{k})^2}{3\bar{k} (1-\bar{k}) +\ii\eta}
\label{eq:derivexplic}
\ee
The dominated convergence theorem allows us to take the $\eta \to 0$ limit in the integrand and to obtain the result 
(\ref{eq:Bel_result}), knowing that $\rho(\hbar/mc)^3=\sqrt{3}/(\pi^2\xi_B^{3/2})$.
 As a side remark,  the fact that the derivative with respect to $z$ (\ref{eq:derivexplic}) is real in the limit 
$\eta\to 0^+$, means that a change in the real dispersion relation  $q\mapsto\omega_\qq$ of order $q^5$ does not affect 
 $\Gamma_\qq$ to the order $q^7$, which {\bleu the calculation technique of} subsection
\ref{subsec:tadpdlguat0} {\bleu also shows}.

\subsection{Study to the order 4 in $\hat{V}$}
\label{subsec:app_ordre4}

We have seen in subsection  \ref{subsec:app_comptage} that the order 4 in $\hat{V}$ may contribute to $\Gamma_\qq$
to the order $q^7$.
Let us write the corresponding correction $\delta\Gamma_\qq$ restricting to the leading order, that is keeping 
$\hat{H}_3^{(-)}$ in the first two factors $\hat{V}$ and $\hat{H}_3^{(+)}$ in the last two, this sequence being imposed by the presence of the projectors $\hat{Q}$:
\be
-\frac{\hbar\delta\Gamma_\qq}{2}=\im \langle\qq| \hat{W}^{(-)} \frac{\hat{Q}}{\epsilon_\qq+\ii\eta-\hat{H}_2} \hat{W}^{(+)} |\qq\rangle
\ee
where we introduced, for  $\varepsilon=\pm$, the effective coupling operator $1\leftrightarrow 3$ to second order 
\be
\hat{W}^{(\varepsilon)} \equiv \hat{H}_3^{(\varepsilon)} \frac{\hat{Q}}{\epsilon_\qq+\ii\eta-\hat{H}_2} \hat{H}_3^{(\varepsilon)}
\ee
In these expressions, one can use for $\hat{H}_3$ the quantum hydrodynamic approximation (\ref{eq:H3hydro}), provided one includes in $\epsilon_\qq=\hbar\omega_\qq$ the cubic correction (\ref{eq:fleur}) to the linear hydrodynamic spectrum.
As the matrix elements of $\hat{H}_3$ are real in the Fock basis, the matrices representing $\hat{W}^{(\pm)}$ are obtained 
one from the other by transposition. After the insertion of a closure relation in the three-phonon subspace, one obtains 
\be
-\frac{\hbar\delta\Gamma_\qq}{2}=\im \frac{1}{3!} \sum_{\qq_1,\qq_2,\qq_3} 
\delta_{\qq_1+\qq_2+\qq_3,\qq} \frac{\bleu (\langle\qq|\hat{W}^{(-)}|\qq_1,\qq_2,\qq_3\rangle)^2}
{\epsilon_\qq+\ii\eta-(\epsilon_{\qq_1}+\epsilon_{\qq_2}+\epsilon_{\qq_3})}
\label{eq:slfdus}
\ee
where, {\violet by virtue of} equation (\ref{eq:H3hydro}),
\begin{multline}
\langle\qq|\hat{W}^{(-)}|\qq_1,\qq_2,\qq_3\rangle=\frac{4(mc^2)^2}{\rho L^3} 
\left[\frac{\mathcal{A}_{\rm hydro}^{2\leftrightarrow 1}(\qq_1,\qq_2;\qq_1+\qq_2)\mathcal{A}_{\rm hydro}^{2\leftrightarrow 1}(\qq_1+\qq_2,\qq_3;\qq)}
{\epsilon_\qq+\ii\eta-(\epsilon_{\qq_1+\qq_2}+\epsilon_{\qq_3})} \right.\\
\left.+ \frac{\mathcal{A}_{\rm hydro}^{2\leftrightarrow 1}(\qq_2,\qq_3;\qq_2+\qq_3)\mathcal{A}_{\rm hydro}^{2\leftrightarrow 1}(\qq_2+\qq_3,\qq_1;\qq)}
{\epsilon_\qq+\ii\eta-(\epsilon_{\qq_2+\qq_3}+\epsilon_{\qq_1})}
+ \frac{\mathcal{A}_{\rm hydro}^{2\leftrightarrow 1}(\qq_1,\qq_3;\qq_1+\qq_3)\mathcal{A}_{\rm hydro}^{2\leftrightarrow 1}(\qq_1+\qq_3,\qq_2;\qq)}
{\epsilon_\qq+\ii\eta-(\epsilon_{\qq_1+\qq_3}+\epsilon_{\qq_2})}
\right]
\end{multline}
The contribution to the matrix element of the first term between square brackets, that {\violet we note}
$P(\qq_1,\qq_2|\qq_3)$, 
corresponds to the process in figure \ref{fig:1d3}. The other two {\violet terms}, 
$P(\qq_2,\qq_3|\qq_1)$ and $P(\qq_1,\qq_3|\qq_2)$, 
are deduced from the first one by circular permutation. As the rest of the summand in  (\ref{eq:slfdus}) is invariant 
by permutation of the three wave vectors, we can replace its numerator  $[P(\qq_1,\qq_2|\qq_3)+P(\qq_2,\qq_3|\qq_1)
+P(\qq_1,\qq_3|\qq_2)]^2$
by $3 [P(\qq_1,\qq_2|\qq_3)]^2+ 6 P(\qq_1,\qq_2|\qq_3) P(\qq_1,\qq_3|\qq_2)$, which amounts to considering the 
two diagrams in figure \ref{fig:diag4} and the integral expression 

\begin{multline}
-\frac{\hbar\delta\Gamma_\qq}{2} =\lim_{\eta\to 0^+} \im \frac{8(mc^2)^4}{\rho^2}
\int\frac{\dd^3q_2\dd^3q_3}{(2\pi)^6}  \left\{
\frac{[\mathcal{A}_{\rm hydro}^{2\leftrightarrow 1}(\qq_1,\qq_2;\qq_1+\qq_2)\mathcal{A}_{\rm hydro}^{2\leftrightarrow 1}(\qq_1+\qq_2,\qq_3;\qq)]^2}
{[\epsilon_\qq+\ii\eta-(\epsilon_{\qq_1}+\epsilon_{\qq_2}+\epsilon_{\qq_3})] 
[\epsilon_\qq+\ii\eta-(\epsilon_{\qq_1+\qq_2}+\epsilon_{\qq_3})]^2} \right. \\
\left.
+\frac{2\mathcal{A}_{\rm hydro}^{2\leftrightarrow 1}(\qq_1,\qq_2;\qq_1+\qq_2)\mathcal{A}_{\rm hydro}^{2\leftrightarrow 1}(\qq_1+\qq_2,\qq_3;\qq)
\mathcal{A}_{\rm hydro}^{2\leftrightarrow 1}(\qq_1,\qq_3;\qq_1+\qq_3)\mathcal{A}_{\rm hydro}^{2\leftrightarrow 1}(\qq_1+\qq_3,\qq_2;\qq)}
{[\epsilon_\qq+\ii\eta-(\epsilon_{\qq_1}+\epsilon_{\qq_2}+\epsilon_{\qq_3})][\epsilon_\qq+\ii\eta-(\epsilon_{\qq_1+\qq_2}+\epsilon_{\qq_3})]
[\epsilon_\qq+\ii\eta-(\epsilon_{\qq_1+\qq_3}+\epsilon_{\qq_2})]}
\right\}
\end{multline}
where we eliminated the sum over $\qq_1$ thanks to momentum conservation,
$\qq_1=\qq-(\qq_2+\qq_3)$, and took the thermodynamic limit.

\begin{figure}[t] 
\begin{center}
\includegraphics[width=7cm,clip=]{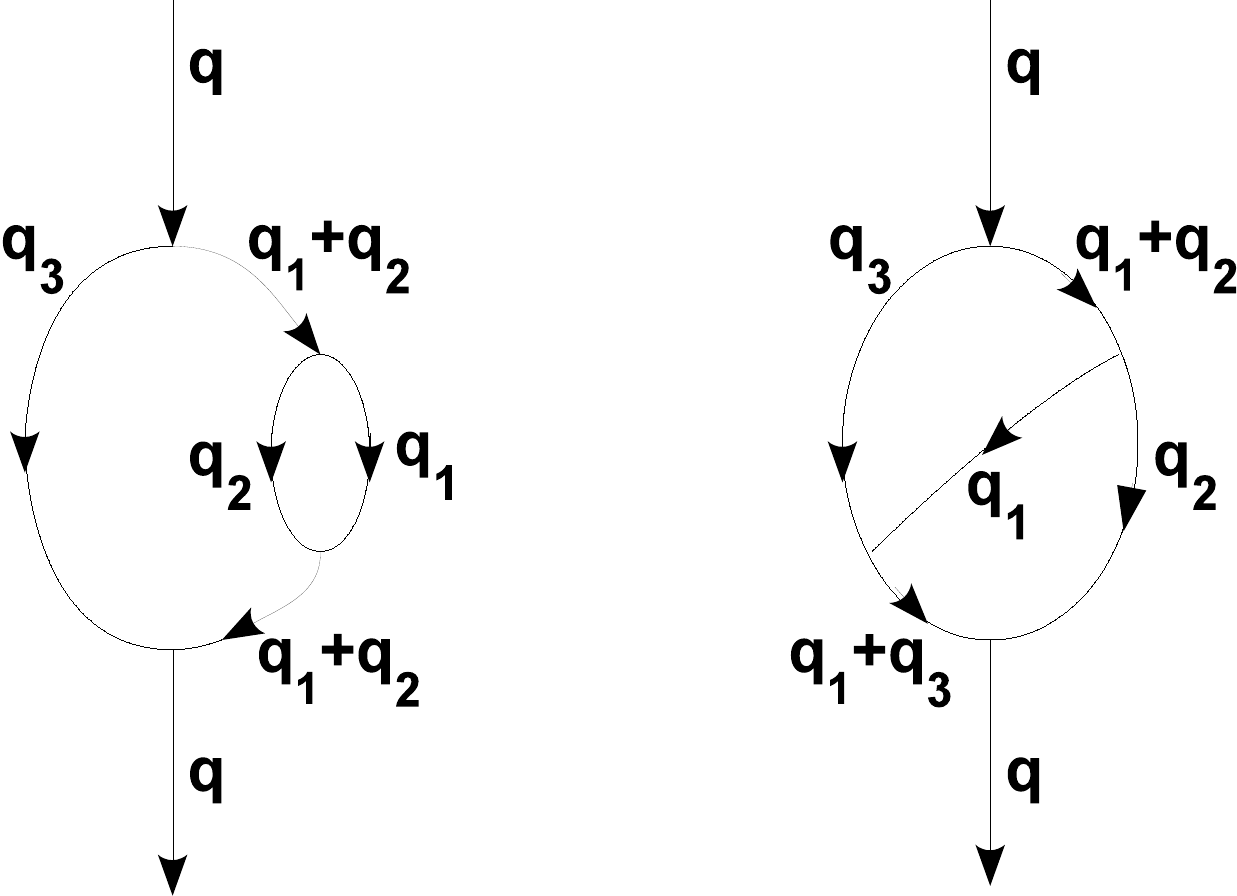}
\end{center}
\caption{\label{fig:diag4} 
The two diagrams of order 4 in $\hat{V}$ that may contribute to the damping rate 
$\Gamma_\qq$ of the phonon $\qq$ to the order $q^7$ thanks to the \g{small denominator} effect. 
The one on the left (I) is a Beliaev process with a single loop correction (itself of the Beliaev nature) to the virtual phonons angular eigenfrequency. The one on the right (II) is a Beliaev process enriched by an interaction between the virtual phonons.
Each vertex corresponds to an action of $\hat{H}_3^{(\pm)}$.}
\end{figure} 

In practice, the integration over  $\qq_2$ {\vers and} $\qq_3$ is performed in spherical coordinates of polar axis $\qq$.
To evaluate the contribution to $\delta\Gamma_\qq$ of the emission cones of $\qq_2$ and $\qq_3$ 
{\violet of angular aperture} $O(q)$ {\violet around $\qq$,} we rescale the polar angles $\theta_i$ as follows:
\be
\check{\theta}_i = \frac{\theta_i}{\gamma^{1/2} \check{q}}
\label{eq:thetaitildeq}
\ee
then we let $\check{q}=\hbar q/mc$ tend to zero with $\check{\theta}_i$ fixed. The calculation is similar to the one in section
 \ref{subsec:ccadlk}.
One {\violet introduces} the polar representation as in (\ref{eq:reppolth}): 
\be
{\violet \check{\theta}_2 = R \cos\alpha  \qquad \qquad \check{\theta}_3 = R \sin\alpha}
\ee
The wave numbers $q_i$ must be rescaled 
by $q$, meaning that $\bar{q}_i \equiv \frac{q_i}{q}$ are kept fixed when taking the {\violet $\check{q}\to 0$}
limit, with the constraint 
\be
\bar{q}_2+\bar{q}_3\leq 1
\ee
{\violet following} from (\ref{eq:qq1q2q3}) and {\bleu from} the positivity of $q_1$.

We give as an intermediate result the expression of an energy denominator in the $\hat{W}^{(-)}$ matrix elements,
\be
\frac{\epsilon_\qq-(\epsilon_{\qq-\qq_3}+\epsilon_{\qq_3})}{mc^2}\!\underset{\check{q}\to 0}{=}\!\frac{\gamma \check{q}^3}{2} 
\left[\frac{3}{4}\bar{q}_3 (1-\bar{q}_3) -\frac{\bar{q}_3 R^2\sin^2\alpha}{1-\bar{q}_3}\right] + O(\check{q}^5)
\ee
and the energy denominator in the 3 phonon subspace,
\be\frac{\epsilon_\qq-(\epsilon_{\qq_1}+\epsilon_{\qq_2}+\epsilon_{\qq_3})}{mc^2} \underset{\check{q}\to 0}{=} \frac{\gamma \check{q}^3}{2}
(v-u R^2) +O(\check{q}^5)
\ee
whose manifestly positive coefficients (once they are written in the proper form) look like those of equations 
 (\ref{eq:defualign}) and (\ref{eq:defvalign}):
\begin{multline}
u\equiv\frac{\bar{q}_2\cos^2\alpha+\bar{q}_3\sin^2\alpha-\bar{q}_2\bar{q}_3(1-\cos\phi \sin 2\alpha)}{1-\bar{q}_2-\bar{q}_3} \\
= \frac{(\bar{q}_2\cos\alpha-\bar{q}_3\sin\alpha)^2+\bar{q}_2\bar{q}_3 (1+\cos\phi)\sin 2\alpha} {1-\bar{q}_2-\bar{q}_3} 
+\bar{q}_2\cos^2\alpha+\bar{q}_3\sin^2\alpha 
\end{multline}
and
\be
v\equiv\frac{1}{4}[1-\bar{q}_2^3-\bar{q}_3^3-(1-\bar{q}_2-\bar{q}_3)^3]=\frac{3}{4}(1-\bar{q}_2)(1-\bar{q}_3)(\bar{q}_2+\bar{q}_3)
\ee
One finally obtains
\be
-\frac{\hbar \delta\Gamma_\qq}{2}\!\underset{\check{q}\to 0}{=}\! mc^2\frac{[3(1+\Lambda_{\rm F})]^4}{3\times 2\pi} 
\frac{\check{q}^7}{2^8\pi^4\gamma} \left(\frac{mc}{\hbar\rho^{1/3}}\right)^6 (\im I^{(\mathrm{I})}+\im I^{(\mathrm{II})}) + o(\check{q}^7)
\ee
with the parameter $\Lambda_{\rm F}$ defined in equation (\ref{eq:lambdaF}) 
{
and the quintuple integrals, coming from the left (type I) diagram and the right (type II) diagram of figure \ref{fig:diag4}, that 
should be evaluated in the limit $\eta\to 0^+$:
\begin{multline}
I^{(\mathrm{I})}=\lim_{\eta\to 0^+} \int_0^1 \dd\bar{q}_2 \int_0^{1-\bar{q}_2} \dd\bar{q}_3 \int_{-\pi}^{\pi} \dd\phi
\int_0^{\pi/2} \dd\alpha \int_0^{A^2/\check{q}^2}\frac{\dd X}{2} 
\frac{3X\sin\alpha\cos\alpha (1-\bar{q}_3)^2 (1-\bar{q}_2-\bar{q}_3) \bar{q}_2^3 \bar{q}_3^3}
{(v-uX+\ii\eta)[\frac{3}{4}\bar{q}_3(1-\bar{q}_3)-\frac{\bar{q}_3 X\sin^2\alpha}{1-\bar{q}_3}+\ii\eta]^2}
\label{eq:IexpI}
\end{multline}
\begin{multline}
I^{(\mathrm{II})}=\lim_{\eta\to 0^+} \int_0^1 \dd\bar{q}_2 \int_0^{1-\bar{q}_2} \dd\bar{q}_3 \int_{-\pi}^{\pi} \dd\phi \int_0^{\pi/2} \dd\alpha 
\\ \times \int_0^{+\infty}\frac{\dd X}{2}
\frac{6X\sin\alpha\cos\alpha (1-\bar{q}_2)(1-\bar{q}_3)(1-\bar{q}_2-\bar{q}_3) \bar{q}_2^3 \bar{q}_3^3}
{(v-uX+\ii\eta)[\frac{3}{4}\bar{q}_2(1-\bar{q}_2)-\frac{\bar{q}_2 X\cos^2\alpha}{1-\bar{q}_2}+\ii\eta]
[\frac{3}{4}\bar{q}_3(1-\bar{q}_3)-\frac{\bar{q}_3 X\sin^2\alpha}{1-\bar{q}_3}+\ii\eta]}
\label{eq:IexpII}
\end{multline}
In the integration over the radius $R$ we have performed the change of variables $X=R^2$. This allows to see that 
the integral over $R$, that is over $X$, in (\ref{eq:IexpII}) has a zero imaginary part in the limit 
$\eta\to 0^+$, and hence
\be
\im I^{(\mathrm{II})}=0
\label{eq:IexpIInul}
\ee
Indeed all the factors in the integrand denominator in (\ref{eq:IexpII}) have the form $\alpha-\beta X +\ii \eta$,
with $\alpha>0$ and $\beta>0$. As a consequence, the integral over $X\in [0,+\infty[$ can be evaluated using the following elementary theorem:

\begin{figure}[t] 
\begin{center}
\includegraphics[width=0.3\textwidth,clip=]{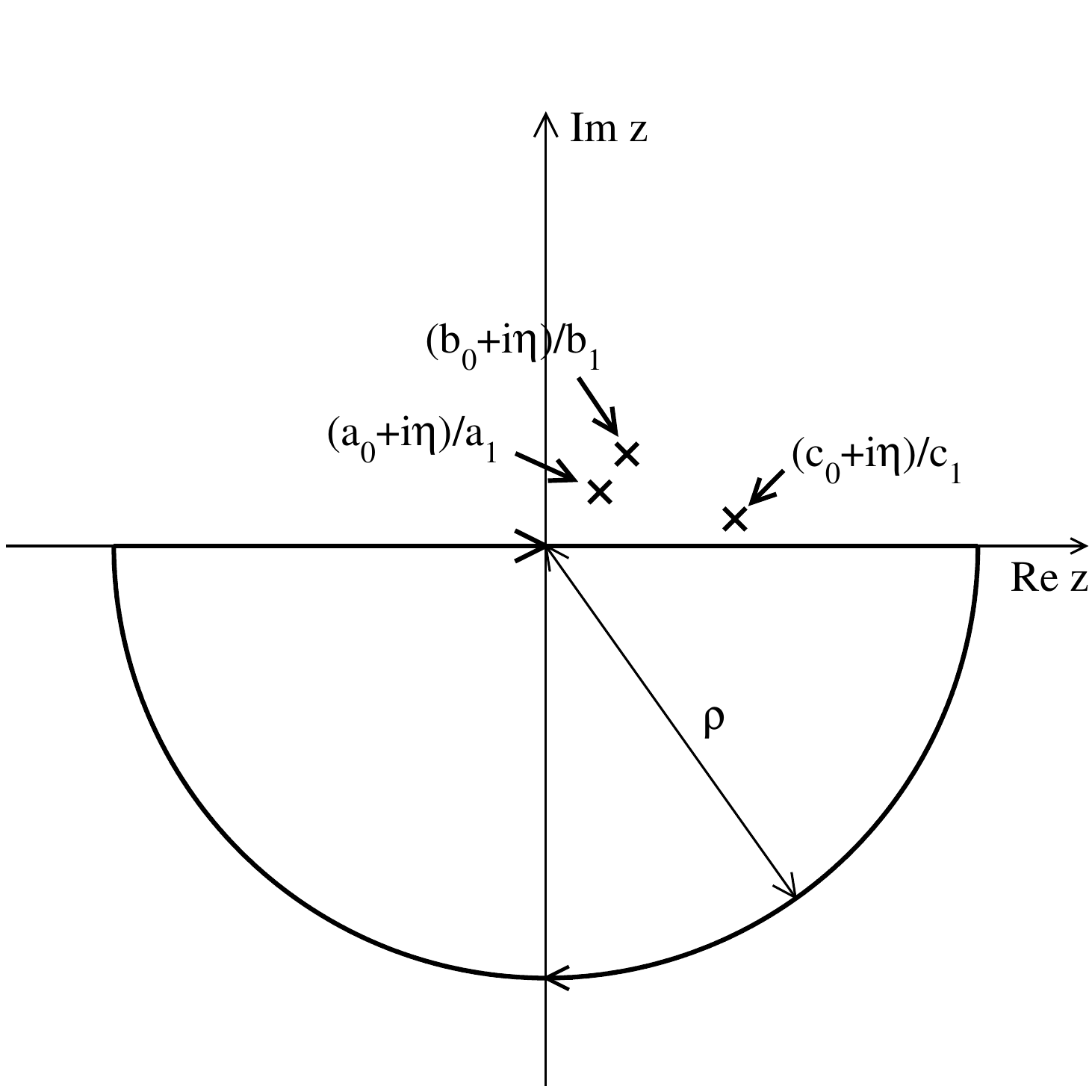}
\end{center}
\caption{Thick line: contour integration in the complex plane that should be used to apply the residue theorem, in order to obtain the identity (\ref{eq:resid}). The real numbers $a_0,a_1,b_0,b_1,c_0,c_1$ and $\eta$ are positive.
The radius $\rho$ of the half-circle tends to infinity. The crosses correspond to the poles of the integrand in the case they are all simple, but the result holds also in the general case.}
\label{fig:contour}
\end{figure}

\noindent{\bf Theorem:} {\sl We define $a(X)=a_0-a_1X$, $b(X)=b_0-b_1X$ and $c(X)=c_0-c_1X$ for all $X\in\mathbb{R}$.
If all the coefficients $a_0, a_1, b_0, b_1, c_0, c_1$ are $>0$ and
$\eta\in \mathbb{R}$, then}
\begin{equation}
\lim_{\eta\to 0^+} \int_{\mathbb{R}^+} \dd X \im \frac{X}{[a(X)+\ii \eta] [b(X)+\ii \eta] [c(X)+\ii \eta]} = 0 \label{eq:theo}
\end{equation}

\noindent{\bf Proof:}
We first note that the theorem would be trivially proved if the integral over $X$ was taken over $\mathbb{R}^-$. 
Indeed, for all $\eta >0$ and for all $X<0$,
\begin{equation}
\left|\frac{X}{[a(X)+\ii \eta] [b(X)+\ii \eta] [c(X)+\ii \eta]}\right| < \frac{X}{a(X) b(X) c(X)}
\end{equation}
where the upper bound has a finite integral over $\mathbb{R}^-$, so that the dominated 
convergence theorem {\bleu can be used} to exchange the order of the integral and limit $\eta\to 0^+$.
We can then extend the integration domain in (\ref{eq:theo}) to the whole $\mathbb{R}$ without changing the result.
On the other hand we can use the following identity valid for any $\eta>0$ not necessarily infinitesimal, concerning both the
real and the imaginary parts: 
\begin{equation}
\int_{\mathbb{R}} \dd X \frac{X}{[a(X)+\ii \eta] [b(X)+\ii \eta] [c(X)+\ii \eta]} = 0 \label{eq:resid}
\end{equation}
To show it, {\bleu we} use the residue theorem by closing downwards the integration contour following a half-circle 
of radius $\rho\to +\infty$ in the lower half-plane; the behavior of the integrand as $1/X^2$  for large $|X|$ allows for it,
which leads to a {\bleu vanishing} $O(1/\rho)$ contribution {\bleu of} the half-circle. 
As the integrand poles are all in the upper half-plane,
they are not enclosed by the contour, see the figure \ref{fig:contour}, hence (\ref{eq:resid}) and (\ref{eq:theo})
\footnote{We supposed here that the leading contribution to $\delta\Gamma_\qq$ comes from the
\g{bicone} configuration in which both $\qq_2$ and $\qq_3$ are in the forward emission cone of angular aperture $O(\check{q})$ with respect to the direction of $\qq$.
One can imagine a more subtle scenario, called \g{unicone}, in which only the vector $\qq_3$ would be in this cone,
while $\qq_2$ would be at an angle $\approx \check{q}^0$ 
with $\qq$. In this case, only the denominator of $P(\qq_1,\qq_2|\qq_3)$ is $\approx \check{q}^3$, while that of
$P(\qq_1,\qq_3|\qq_2)$ is $\approx \check{q}$. The crossed term $P(\qq_1,\qq_3|\qq_2) P(\qq_1,\qq_2|\qq_3)$ is then 
negligible with respect to $[P(\qq_1,\qq_2|\qq_3)]^2$. The global energy denominator
$\epsilon_\qq+\ii\eta-(\epsilon_{\qq_1}+\epsilon_{\qq_2}+\epsilon_{\qq_3})$ of (\ref{eq:slfdus}) is also $\approx \check{q}$, 
which makes us lose a factor $\check{q}^2$, but this is exactly compensated by the loss of a factor $\check{q}^2$ in the numerator in the polar integral $\int \dd \theta_2 \sin\theta_2$. The ensemble seems then to contribute to the same order 
$\check{q}^7$ as the bicone configuration. The integration over the polar angle $\theta_3$ rescaled as in 
(\ref{eq:thetaitildeq}), with the inclusion of the three-dimensional Jacobian $\bar{q}_3^2$ from the integration over 
$\bar{q}_3$, however leads to
\[
\int_0^{+\infty}\!\!\frac{\bar{q}_3^2 \check{\theta}_3\dd\check{\theta}_3}
{\left[\frac{3}{4} \bar{q}_3(1-\bar{q}_3) -\frac{\bar{q}_3\check{\theta}_3^2}{1-\bar{q}_3}+\ii\eta\right]^2}
=-\frac{\bar{q}_3(1-\bar{q}_3)/2}{\frac{3}{4} \bar{q}_3(1-\bar{q}_3)+\ii\eta} 
\]
which has a real limit $-\frac{2}{3}$ when $\eta\to 0^+$.
Hence the sought imaginary part in $\delta\Gamma_\qq$ may only come from the global energy denominator,
which would provide a Dirac $\delta[\epsilon_\qq-(\epsilon_{\qq_1}+\epsilon_{\qq_2}+\epsilon_{\qq_3})]$; 
however, to the leading order in $\check{q}$ that we consider here, $\qq_3=\bar{q}_3 \qq$, with $0<\bar{q}_3<1$, and the dispersion relation is linear, so that the argument of the Dirac {\bleu distribution cannot} be zero unless $\qq_2$ and $\qq$ are collinear {\bleu and of} the same direction, which is in contradiction with the hypothesis of having a 
 $\qq_2$ outside the forward emission cone. The same arguments hold for the damping rate 
$\Gamma_\qq^{2\leftrightarrow 2}$ of section \ref{subsec:ccadlk}, and justifies that we only have considered there  the 
\g{bicone} configuration.}. \qed

One might think that the same reasoning applies to the contribution (\ref{eq:IexpI}) and that $\im I^{(\mathrm{I})}=0$, 
in which case 
$\delta\Gamma_\qq=o(\check{q}^7)$ and there would be no correction to add to the result  (\ref{eq:Bel_result}).
This is not the case because, in order to obtain a finite value of $I^{(\mathrm{I})}$, we {\violet have} this time to keep in the integral over $X$ a finite value $A^2/\check{q}^2$ of the upper bound, which prevents the application of the theorem.
Here, $A>0$ is a cut-off constant whose precise value is not relevant, and the power law in $\check{q}^{-2}$ comes from the fact that $\theta_i\leq \pi$ and then $\check{\theta}_i\leq \pi/(\gamma^{1/2}\check{q})$ in (\ref{eq:thetaitildeq}), 
which implies $X=R^2=\check{\theta}_2^2+\check{\theta}_3^2=O(\check{q}^{-2})$. 
If one simply replaces $A^2/\check{q}^2$ by
$+\infty$ in equation (\ref{eq:IexpI}), one finds indeed that the integral over $X$ diverges as $\alpha^{-2}$ when $\alpha$
tends to zero, which leads to an integral over $\alpha$ that diverges logarithmically for $\alpha=0$.
\footnote{This phenomenon does not occur in the contribution (\ref{eq:IexpII}) because neither the factor containing 
the term $X\sin^2\alpha$, dangerous when $\alpha\to 0$, nor the one containing the term
$X\cos^2\alpha$, dangerous when $\alpha\to \pi/2$, are taken to the square.
As a consequence, the integral over $X$ from $0$ to $+\infty$ diverges only as $\ln \alpha$ or $\ln(\frac{\pi}{2}-\alpha)$ when $\alpha\to 0$ or $\alpha\to \pi/2$.
Mathematically, by replacing in equation (\ref{eq:IexpII}) the integral $\int_0^{+\infty} \dd X f(X)$ by $-\int_{-\infty}^{0} \dd X f(X)$ from the identity (\ref{eq:resid}), then by using the dominated convergence theorem,
we can justify the exchange of the limit  $\eta\to 0^+$ and the integration over $\bar{q}_2, \bar{q}_3, \phi$ and $\alpha$, and prove the result (\ref{eq:IexpIInul}).}
To progress, we cut the integration interval over $\alpha$ into two, a  sub-interval $[0,\nu]$ for which the cut-off $A^2/\check{q}^2$ in the integration over $X$ is necessary, and a sub-{\bleu interval} $[\nu,\pi/2]$ for which the integration over $X$ can be extended to $+\infty$:
\be
\int_0^{\pi/2} \dd \alpha \int_0^{A^2/\check{q}^2} \!\!\!\!\dd X f(\alpha, X) \to \int_0^{\nu} \dd \alpha \int_0^{A^2/\check{q}^2}\!\!\!\! \dd X f(\alpha, X) + \int_{\nu}^{\pi/2} \dd \alpha  \int_0^{+\infty} \dd X f(\alpha, X) 
\ee
The contribution of the second bit to $\im I^{(\mathrm{I})}$ is zero, because of the theorem.
The contribution of the first bit does not depend on the value of $\nu$. We can then calculate it for a value of 
$\nu$ sufficiently small to approximate each contribution depending on $\alpha$ in the integrand by its leading order in 
 $\alpha$, that is $\sin\alpha \simeq \alpha$, $\cos\alpha\simeq 1$,
$u\simeq u_0=\bar{q}_2 (1-\bar{q}_3)/(1-\bar{q}_2-\bar{q}_3)$. The dependence in $\phi$ then disappears and the integration over $\phi$ simply gives a factor $2\pi$. After a simple calculation of the integral over $\alpha$:
\be
\int_0^\nu \frac{\alpha \, \dd\alpha}{[\frac{3}{4}\bar{q}_3(1-\bar{q}_3)-\frac{\bar{q}_3 X\alpha^2}{1-\bar{q}_3}+\ii\eta]^2}= \left[\frac{1-\bar{q}_3}{2\bar{q}_3X [\frac{3}{4}\bar{q}_3(1-\bar{q}_3)-\frac{\bar{q}_3 X\alpha^2}{1-\bar{q}_3}+\ii\eta]}\right]_{\alpha=0}^{\alpha=\nu}
\label{eq:apresintegalpha}
\ee
we shall concentrate on the integration over $X$. The fully integrated term $\alpha=\nu$ of equation 
(\ref{eq:apresintegalpha}), after a multiplication by the factor $X$ in the numerator of (\ref{eq:IexpI}) and a division 
by the denominator $(v-u_0X+\ii\eta)$, gives an integrand {\bleu that is} $O(1/X^2)$; one can in this case replace the upper bound $A^2/\check{q}^2$ {\bleu in the integration over} $X$ by $+\infty$ and use the following variant of the theorem, in order to show that its contribution to $\im I^{(\mathrm{I})}$ is exactly zero:
\be
\lim_{\eta\to 0^+} \int_{\mathbb{R}^+} \dd X \im \frac{1}{[a(X)+\ii \eta] [b(X)+\ii \eta]} = 0 \label{eq:vartheo}
\ee
The fully integrated  term $\alpha=0$ of equation (\ref{eq:apresintegalpha}) leads on the contrary to an integral over $X$ of nonzero imaginary part in the limit $\eta\to 0^+$:
\begin{multline}
\im \int_0^{A^2/\check{q}^2} \frac{(1-\bar{q}_3)\dd X}{2\bar{q}_3(v-u_0 X+\ii\eta)[\frac{3}{4}\bar{q}_3(1-\bar{q}_3)+\ii\eta]} =-\frac{1-\bar{q}_3}{2\bar{q}_3u_0} \im \frac{\left[\ln(v-u_0A^2\check{q}^{-2}+\ii\eta) - \ln(v+\ii\eta)\right]}{\frac{3}{4}\bar{q}_3(1-\bar{q}_3)+\ii\eta} \\
\underset{\eta\to 0^+}{\to} -\frac{\pi}{u_0} \frac{{\vers\Theta}(u_0A^2\check{q}^{-2}-v)}{\frac{3}{2}\bar{q}_3^2}
\end{multline}
where {\bleu $\Theta$ is the Heaviside function and} $\ln$ is the usual branch of the complex logarithm, with a branch cut along the real negative axis.
We are then left with
\be
\im I^{(\mathrm{I})} = 2\pi^2 \int_0^1 \dd\bar{q}_2 \int_0^{1-\bar{q}_2} \!\!\!\!\dd\bar{q}_3 (1-\bar{q}_3) (1-\bar{q}_2-\bar{q}_3)^2 \bar{q}_2^2 \bar{q}_3 {\vers\Theta}(u_0A^2\check{q}^{-2}-v)  \underset{\check{q}\to 0}{\to}
\frac{\pi^2}{840}
\ee
hence the correction (\ref{eq:Gamma1d3}) to be added to the result (\ref{eq:Bel_result}).
}

To conclude, one might ask whether there is a physical interpretation to the contributing diagram of order $4$ in $\hat{V}$,
which is the left one (I) in figure \ref{fig:diag4}. In the calculation of the complex energy shift $\Delta \epsilon_\qq^{\rm Bel}$ of a phonon $\qq$ induced by a Beliaev process, this diagram takes into account the effect of a modification of the angular eigenfrequency of the virtual phonons, modification itself induced by a Beliaev process at this order.
More quantitatively, let us introduce the Beliaev self-energy (\ref{eq:DeQBelZ}),
where we note now more rigorously $\epsilon_\kk^{(2)}$ the unperturbed eigenenergy of a phonon with wave vector $\kk$
in $\hat{H}_2$, to distinguish it from the exact energy $\epsilon_\qq$ 
(see the note \thenoterenormalisation). 
We verified that the change in the complex energy of the phonon $\qq$ originating from the left diagram (I) is exactly
\be
\delta z_\qq^{\rm (I)}= \frac{1}{2} \sum_{\qq_3} \frac{(\langle \qq_3,\qq-\qq_3|\hat{H}_3|\qq\rangle)^2}{[\epsilon_\qq+\ii\eta
-(\epsilon_{\qq_3}^{(2)}+\epsilon_{\qq-\qq_3}^{(2)})]^2} \left[\Delta \epsilon_{\qq_3}^{\rm Bel}(\epsilon_\qq+\ii\eta-\epsilon_{\qq-\qq_3}^{(2)}) 
+\Delta \epsilon_{\qq-\qq_3}^{\rm Bel}(\epsilon_\qq+\ii\eta-\epsilon_{\qq_3}^{(2)})\right]
\label{eq:reformul}
\ee
The reader will notice that the argument $z$ of $\Delta \epsilon^{\rm Bel}_{\qq-\qq_3}$ is here the exact energy 
$\epsilon_\qq$ minus the non perturbed energy of the virtual phonon that is spectator in the process of diagram (I), that is the one that does not participate to the loop.
The sum of the correction (I) and of the usual perturbative Beliaev shift of the $\qq$ phonon energy can then be written,
to the order $4$ in $\hat{H}_3$, as a Beliaev shift for a renormalized complex dispersion relation:
\be
\Delta \epsilon_\qq^{\rm Bel}(\epsilon_\qq+\ii\eta)+\delta z_\qq^{\rm (I)}\simeq \frac{1}{2} \sum_{\qq_3,\qq_4} 
\frac{(\langle\qq_3,\qq_4|\hat{H}_3|\qq\rangle)^2 \delta_{\qq_3+\qq_4,\qq}}
{\epsilon_\qq+\ii\eta-[\epsilon_{\qq_3}^{(2)}+\Delta\epsilon_{\qq_3}^{\rm Bel}(\epsilon_\qq+\ii\eta-\epsilon_{\qq_4}^{(2)})
+\epsilon_{\qq_4}^{(2)}+\Delta\epsilon_{\qq_4}^{\rm Bel}(\epsilon_\qq+\ii\eta-\epsilon_{\qq_3}^{(2)})]}
\label{eq:jolimaislong}
\ee
By the way, the formulation (\ref{eq:reformul}) provides a second calculation method, alternative to the one of 
equation (\ref{eq:IexpI}). With the rescalings of subsection  \ref{subsec:ccadlk} suitable to the 
\g{small denominators}, and by using (\ref{eq:intermdeux})
with $(\qq,\kk)$ successively equal to $(\qq_3,\kk)$ and $(\qq,\qq_3)$, where $\kk$ is the integration variable
appearing in the expression taken from (\ref{eq:DeQBelZ}) of $\Delta \epsilon_{\qq_3}^{\rm Bel}(\epsilon_\qq+\ii\eta-\epsilon_{\qq-\qq_3}^{(2)})$, we find
\begin{multline}
\frac{\Delta \epsilon_{\qq_3}^{\rm Bel}(\epsilon_\qq+\ii\eta-\epsilon_{\qq-\qq_3}^{(2)})}{mc^2}=
-\frac{9(1+\Lambda_{\rm F})^2}{64\pi^2} \left(\frac{mc}{\hbar\rho^{1/3}}\right)^3
\bar{q}_3^5 \check{q}^5 \\ \times \int_0^1 \dd\bar{k}\, \bar{k}^2 (1-\bar{k})^2 
\left[\ii\pi +\ln \left(\frac{4\bar{q}_3\bar{k}A^2}{(1-\bar{k})\check{q}^2}\right)-\ln\left(\ii\eta+3\bar{q}_3(1-\bar{q}_3)+3\bar{q}_3^3\bar{k}(1-\bar{k})
-\frac{4\bar{q}_3\check{\theta_3}^2}{1-\bar{q}_3}\right)\right]
\label{eq:debelexpli}
\end{multline}
Here $\bar{q}_3=q_3/q$, $\bar{k}=k/q_3$, $\check{\theta}_3=\theta_3/(\gamma^{1/2}\check{q})$, $\theta_3$ is the non oriented angle between the vectors 
$\qq$ and $\qq_3$. The integral over $\check{\theta}=\theta/(\gamma^{1/2}\check{q}_3)$, where $\theta$ is the angle 
between $\kk$ and $\qq$, has been explicitly performed over the interval  $[0,A/\check{q}_3]$ with the same cut-off parameter $A$ as in equation (\ref{eq:IexpI}).
After the insertion of (\ref{eq:debelexpli}) in equation (\ref{eq:reformul}), we obtain 
\begin{multline}
\frac{\im \delta z^{\rm (I)}_\qq}{mc^2}=-\frac{9^2(1+\Lambda_{\rm F})^4}{64\pi^4\gamma} \left(\frac{mc}{\hbar\rho^{1/3}}\right)^6
\check{q}^7 \int_0^1 \dd\bar{q}_3\, \bar{q}_3^8 (1-\bar{q}_3) \int_0^1 \dd\bar{k}\, \bar{k}^2(1-\bar{k})^2  \\
\times \im \int_0^{+\infty} \dd X_3 \frac{\ii\pi-\ln\left(\ii\eta+3\bar{q}_3(1-\bar{q}_3)+3\bar{q}_3^3\bar{k}(1-\bar{k})
-\frac{4\bar{q}_3X_3}{1-\bar{q}_3}\right)}{\left(\ii\eta+3\bar{q}_3(1-\bar{q}_3)-\frac{4\bar{q}_3 X_3}{1-\bar{q}_3}\right)^2}
\end{multline}
We omitted here the $\check{\theta}_3$-independent logarithmic real term, in between square brackets in 
equation (\ref{eq:debelexpli}), because its contribution to the final result is clearly real in the limit 
$\eta\to 0^+$. The dependence on the cut-off parameter $A$ then {\bleu disappears}.
The integral over $X_3=\check{\theta}_3^2$ of the bit with the logarithm is calculated by parts, by taking the derivative of the logarithm;
one finds that its contribution is zero in the limit  $\eta\to 0^+$ by using the identity (\ref{eq:vartheo}).
Finally, only the term $\ii\pi$ matters, and it leads exactly to the result (\ref{eq:Gamma1d3}).

}

\providecommand*\hyphen{-}

\end{document}


\selectlanguage{francais}
\title{Processus d’interaction \`a trois et quatre phonons dans un gaz de fermions condens\'e par paires}
\author{H. Kurkjian, Y. Castin, A. Sinatra}
\affiliation{Laboratoire Kastler Brossel, ENS-PSL, CNRS, UPMC-Sorbonne Universit\'es et Coll\`ege de France, Paris, France}

\begin{abstract}
Nous effectuons une \'etude d\'etaill\'ee des processus d'interaction et de la dur\'ee de vie des phonons dans un gaz de fermions
condens\'e par paires dans la zone de raccordement CBE-BCS, dans le r\'egime faiblement collisionnel. Nous d\'eterminons les amplitudes de couplage phonon-phonon par une th\'eorie 
microscopique s'appuyant sur un ansatz variationnel BCS g\'en\'eralis\'e au cas de paires en mouvement, qui se pr\^ete \`a un d\'eveloppement
syst\'ematique autour de l'approximation de champ moyen BCS de l'\'etat fondamental.
Nous trouvons que les amplitudes de couplage d\'eduites de l'hydrodynamique quantique par Landau et Khalatnikov ne sont valables que
sur la couche de masse, c'est-\`a-dire pour des processus r\'esonnants, qui conservent l'\'energie.
Notre th\'eorie microscopique conduit au m\^eme spectre d'excitation que la RPA ({\sl Random Phase Approximation}), avec aux faibles nombres
d'onde une relation de dispersion de d\'epart lin\'eaire (phononique) qui passe de convexe à concave lorsqu'on passe 
de la limite CBE \`a la limite BCS. 
Dans le cas convexe, le m\'ecanisme d'amortissement dominant \`a basse temp\'erature est celui de Beliaev-Landau 2 phonons $\leftrightarrow$ 1 phonon.
Dans le cas concave, c'est le processus de Landau-Khalatnikov 2 phonons $\leftrightarrow$ 2 phonons. Dans les deux cas, en sortant la d\'ependance
en la force des interactions par un adimensionnement astucieux des vecteurs d'onde, nous obtenons une expression universelle du
taux d'amortissement. Dans le cas des processus $2\leftrightarrow2$, notre expression universelle corrige et étend les r\'esultats 
analytiques de Landau et Khalatnikov [ZhETF {\bf 19}, 637 (1949)]. {Dans le cas de l'amortissement de Beliaev $1\to 2$ pour le gaz
unitaire à temp\'erature nulle, nous calculons 
la première correction en $q^7$ au taux d'amortissement hydrodynamique en $q^5$, ce qui n'avait pas \'et\'e fait de manière syst\'ematique.}
\end{abstract}
\maketitle

\section{Introduction}
Dans de nombreux syst\`emes \`a $N$ corps, les excitations collectives de basse \'energie sont des phonons.
\`A basse temp\'erature, les interactions entre phonons d\'eterminent leur dur\'ee de vie, leur temps de corr\'elation et leur libre parcours moyen.
Elles jouent donc un r\^ole central dans les ph\'enom\`enes de transport, tels que la conduction thermique dans les solides di\'electriques et
les propri\'et\'es hydrodynamiques d\'ependant de la temp\'erature comme la viscosit\'e et l'att\'enuation du son dans l'h\'elium liquide 
\cite{Khalatnikov1949,Khalatnikov1966,PhysRevLett.25.220,Sherlock1975,Smith2005,1367-2630-9-3-052}, mais aussi dans les propri\'et\'es de coh\'erence macroscopique des gaz d\'eg\'en\'er\'es, puisqu'elles
d\'eterminent le temps de coh\'erence intrins\`eque du condensat de bosons ou de paires de fermions \cite{CastinSinatra2009,KCS2015}.
Le processus dominant dans la d\'ecroissance des phonons varie suivant le syst\`eme physique consid\'er\'e, 
et d\'epend en particulier de la concavit\'e 
de la branche d'excitation phononique \cite{Wyatt1992,Wyatt2009,KCS2016}. Lorsque la relation de dispersion est convexe aux faibles
nombres d'onde, le processus dominant est celui de Beliaev-Landau $2 \leftrightarrow 1$ \cite{Beliaev1958,Salasnich2015}, qui met en jeu
trois quasi-particules. Lorsqu'elle est concave, c'est celui de Landau-Khalatnikov $2 \leftrightarrow 2$ \`a quatre quasi-particules,
puisque celui de Beliaev-Landau n'est plus r\'esonnant.

Parmi les syst\`emes \`a $N$ corps, les gaz d'atomes froids offrent la possibilit\'e unique de contr\^oler et d'ajuster la valeur de certains 
param\`etres microscopiques, en particulier la force des interactions. Dans les gaz de fermions froids \`a deux \'etats de spin $\uparrow$
et $\downarrow$, les interactions se produisent seulement dans l'onde $s$ entre fermions de spin oppos\'e. Nous les consid\'erons ici dans le r\'egime universel o\`u leur port\'ee $b$ est n\'egligeable. Elles sont alors caract\'eris\'ees
par un seul param\`etre, la longueur de diffusion $a$ dans l'onde $s$, $|a|\gg b$, ajustable {\sl via} un champ magn\'etique ext\'erieur au voisinage
d'une r\'esonance de Feshbach. Ceci a permis l'\'etude dans ces syst\`emes du raccordement entre la limite de Bardeen-Cooper-Schrieffer (BCS)
$1/a\to -\infty$, o\`u les paires condens\'ees de fermions $(\kk\uparrow,-\kk\downarrow)$ sont localis\'ees dans l'espace des impulsions au voisinage
de la surface de Fermi, et la limite de la condensation de Bose-Einstein (CBE) $1/a\to +\infty$, 
o\`u les dim\`eres de fermions fortement li\'es se comportent
comme des particules bosoniques \cite{Thomas2002,Salomon2003,Grimm2004,Grimm2004b,Ketterle2004,Ketterle2005,Salomon2010,Zwierlein2012,Grimm2013}.
Dans le cas non polaris\'e, c'est-\`a-dire avec des populations \'egales dans les deux \'etats de spin, le spectre d'excitation est compos\'e de deux
branches~: une branche fermionique \`a seuil non nul donc avec bande interdite, 
qui d\'ecrit l'excitation des degr\'es de libert\'e internes des paires $\uparrow\downarrow$,
et une branche bosonique de d\'epart phononique aux faibles nombres d'onde, qui d\'ecrit l'excitation du mouvement du centre de masse
des paires \cite{Anderson1958,Strinati1998,CKS2006,Tempere2011,Randeria2014,KCS2016}. Quand elles ont une densit\'e suffisamment faible, 
ces excitations \'el\'ementaires constituent des quasi-particules en interaction faible et \`a longue dur\'ee de vie.

La relation de dispersion des phonons dans les gaz de fermions froids de spin $1/2$ passe, près de la limite unitaire $1/a=0$, 
de convexe du c\^ot\'e CBE \cite{CKS2006} à concave du c\^ot\'e BCS \cite{Strinati1998}; \`a la limite unitaire, en l'absence de verdict exp\'erimental,
on s'attend \`a ce qu'elle soit convexe aux faibles nombres d'onde \cite{Salasnich2008,Manes2009,Rupak2009,Salasnich2011,Tempere2011,KCS2016}, m\^eme si les valeurs pr\'edites du param\`etre de courbure varient significativement. Ceci veut dire en tout cas que ce syst\`eme physique permet, \`a lui seul, 
d'\'etudier exp\'erimentalement l'interaction entre phonons aussi bien dans le cas concave que dans le cas convexe. En particulier, en passant
du premier au second cas par changement de la longueur de diffusion, on devrait observer une augmentation brutale de la dur\'ee de vie
des phonons. Les excitations collectives peuvent \^etre \'etudi\'ees exp\'erimentalement aussi bien dans le domaine spectral 
par diffusion de Bragg \cite{Davidson2002,Davidson2002b}
que dans le domaine temporel \cite{Dalibard2002,Kinast2005}. Si l'on ajoute \`a cela la possibilit\'e de cr\'eer des gaz spatialement homog\`enes dans
des potentiels de pi\'egeage \`a fond plat \cite{Hadzibabic2013}, la mesure de la d\'ecroissance des phonons due aux
processus $2 \leftrightarrow1$ ou $2 \leftrightarrow 2$ devrait \^etre r\'ealisable \`a court terme.

En attendant, il manque toujours une \'etude th\'eorique exhaustive des interactions entre phonons et une expression g\'en\'erale des 
taux d'amortissement \`a basse temp\'erature pour une interaction de force quelconque.
L'exemple le plus imm\'ediat d'une relation de dispersion convexe est la branche d'excitation de Bogolioubov du gaz de bosons
en interaction faible. Sur cet exemple, Beliaev a obtenu les amplitudes de couplage dans une th\'eorie microscopique et en a d\'eduit
les taux d'amortissement dus aux processus $1\to2$ \cite{Beliaev1958}. La contribution des processus de Landau  $2\to1$,
pr\'esents seulement \`a temp\'erature non nulle, a \'et\'e calcul\'ee plus tard \cite{Martin1965,Kondor1974,Stringari1997}.
Dans le cas qui nous int\'eresse des gaz superfluides de fermions de spin $1/2$, la r\'ef\'erence \cite{Salasnich2015} a calcul\'e
le taux d'amortissement de Beliaev $1\to2$  \`a temp\'erature nulle \`a l'aide d'une th\'eorie effective
de basse \'energie ph\'enom\'enologique allant un cran au-del\`a de l'hydrodynamique quantique, afin de prendre en compte la courbure de la relation
de dispersion des phonons. Cependant, les auteurs ont omis d'inclure les corrections du m\^eme ordre \`a l'amplitude de couplage 
entre phonons, ce qui rend leur traitement inconsistant.
Dans le cas bosonique de la phase II de l'$^4{\rm He}$, dont on croyait \`a l'origine la relation de dispersion concave, 
Landau et Khalatnikov ont obtenu l'amplitude de couplage des processus $2\leftrightarrow2$ en tenant compte des processus virtuels 
non r\'esonnants \`a trois phonons et en introduisant \`a la main un terme correctif cubique dans le spectre d'excitation hydrodynamique.
Cependant, ils ont calcul\'e les taux d'amortissement seulement dans la limite des petits et des grands nombres d'onde, ce qui nous a incit\'es
\`a reprendre et compl\'eter leur travail, d'autant plus que l'\'etude des processus $2\leftrightarrow2$ \`a basse \'energie dans l'$^4{\rm He}$
a \'et\'e suspendue vers 1970, quand il a \'et\'e d\'ecouvert que la relation de dispersion y est en fait convexe aux faibles nombres d'onde \cite{PhysRevLett.25.220,Sherlock1975}, du moins \`a pression pas trop \'elev\'ee \cite{Sherlock1975,Smith2005}.
Les processus $2\leftrightarrow2$ dans l'$^4{\rm He}$
font toujours l'objet de recherches dans l'intervalle de nombre d'onde o\`u la relation de dispersion
redevient concave, ce que la pr\'esence d'une petite r\'egion convexe rend assez difficile \cite{Wyatt2009}. \`A notre connaissance,
les \'etudes existantes des processus $2\leftrightarrow2$ dans les gaz de fermions froids se restreignent \`a la limite unitaire
et au r\'egime hydrodynamique fortement collisionnel, o\`u l'indicateur de la dissipation est la viscosit\'e de cisaillement du gaz \cite{Schafer2007,Zwerger2011,Manuel2013}.
Dans cet article, nous pr\'esentons au contraire une \'etude compl\`ete des processus d'interaction entre phonons dans un gaz de fermions
froid \`a suffisamment basse temp\'erature pour \^etre dans le r\'egime faiblement collisionnel, et ceci pour une interaction de force quelconque dans la zone de raccordement CBE-BCS,
ce qui inclut les cas concave et convexe.

Nous d\'ecrivons pour cela le gaz de fermions non polaris\'e condens\'e par paires \`a l'aide d'un mod\`ele microscopique semi-classique, s'appuyant
sur un \'etat variationnel comportant des paires en mouvement. Ce mod\`ele permet de d\'evelopper de fa\c{c}on syst\'ematique
le hamiltonien en puissances de deux champs canoniquement conjugu\'es $\beta$ et $\beta^\ast$, qui sont de faible amplitude \`a basse densit\'e
d'excitations du gaz. Le mod\`ele microscopique et le principe du d\'eveloppement sont expos\'es dans la section \ref{sec:methode}.

Dans la section \ref{sec:quadratique} nous d\'eveloppons le hamiltonien jusqu'\`a l'ordre deux en les champs $\beta$ et $\beta^\ast$,
et nous diagonalisons cette approximation quadratique pour obtenir le spectre d'excitation 
\cite{Anderson1958,Strinati1998,CKS2006,Tempere2011,Randeria2014,KCS2016}.
\`A part ces r\'esultats spectraux d\'ej\`a connus, nous obtenons les modes propres de quasi-particule et les amplitudes correspondantes
de cr\'eation et d'annihilation, que nous quantifions ensuite. Le tout proc\`ede d'une certaine \'el\'egance gr\^ace \`a la nature
hamiltonienne des \'equations du mod\`ele semi-classique.

Dans la section \ref{sec:interactions}, nous calculons les amplitudes de couplage des processus d'interaction \`a trois et \`a quatre phonons.
Sur la couche de masse, nous comparons avec succ\`es les pr\'edictions de l'hydrodynamique quantique \cite{Khalatnikov1949} 
sur les amplitudes des processus $2 \leftrightarrow 1$ et $2 \leftrightarrow 2$ aux r\'esultats 
d'une th\'eorie microscopique, th\'eorie qui est pour $2 \leftrightarrow 1$
celle du mod\`ele fermionique de la section \ref{sec:quadratique}, et pour $2 \leftrightarrow 2$ celle d'un mod\`ele de bosons en interaction
de port\'ee ajust\'ee pour que la relation de dispersion soit concave aux faibles nombres d'onde. 
Le test microscopique des pr\'edictions de l'hydrodynamique que cela constitue
est particuli\`erement int\'eressant pour $2 \leftrightarrow 2$, puisque $(i)$ ce cas met en jeu des processus non
r\'esonnants $2 \leftrightarrow 1$ et $3 \leftrightarrow 0$, dont les amplitudes ne sont pas correctement donn\'ees par l'hydrodynamique,
$(ii)$ l'hydrodynamique de base conduit \`a une divergence de l'amplitude $2 \leftrightarrow 2$ que Landau et Khalatnikov ont
d\^u r\'egulariser en mettant \`a la main un terme de courbure dans la relation de dispersion des phonons.

Dans la section \ref{sec:amortissement}, nous appliquons directement les r\'esultats pr\'ec\'edents au calcul du taux d'amortissement
des phonons dans un gaz de fermions dans tout le raccordement CBE-BCS dans le r\'egime faiblement collisionnel. Pour ce faire, nous incorporons dans une \'equation pilote
les amplitudes de couplage des processus $2 \leftrightarrow 1$ et $2 \leftrightarrow 2$ de la section \ref{sec:interactions}. 
Annon\c{c}ons d\'ej\`a les principaux r\'esultats de cette section \ref{sec:amortissement}~:  
$(i)$ pour une relation de dispersion aussi bien convexe que concave,
gr\^ace \`a l'introduction de quantit\'es soigneusement adimensionn\'ees, nous obtenons une courbe universelle donnant le taux d'amortissement
en fonction du nombre d'onde, pour une interaction de force quelconque entre les limites CBE et BCS, $(ii)$ nous donnons analytiquement
tous les comportements aux limites des taux d'amortissement, pour $\hbar cq \ll k_{\rm B}T$ et $\hbar cq \gg k_{\rm B}T$ o\`u $c$ est la 
vitesse du son et $T$ la temp\'erature, ce qui, dans le cas concave,
nous conduit \`a corriger les r\'esultats de Landau et Khalatnikov \cite{Khalatnikov1949}, 
enfin $(iii)$ dans le cas particulier du gaz unitaire \`a temp\'erature nulle, nous calculons
la premi\`ere correction \`a la pr\'ediction hydrodynamique pour le taux d'amortissement des phonons, en utilisant la th\'eorie des champs effective
de Son et Wingate \cite{SonWingate2006,Manuel2010}, ce qui nous permet de compl\'eter le r\'esultat de la r\'ef\'erence \cite{Salasnich2015}.


\section{Principe de l'approche microscopique}
\label{sec:methode}

\subsection{Hamiltonien d'interaction dans l'onde $s$}


Nous considérons un gaz de fermions dans deux états internes notés $\uparrow$ et $\downarrow$ évoluant sur un réseau cubique de pas $l$ avec des conditions aux limites périodiques dans un volume $[0,L]^3$ de taille finie. Les fermions de spin opposé ont une interaction sur site caractérisée par une constante de couplage $g_0$
\be
V(\rr,\rr')=g_0\frac{\delta_{\rr,\rr'}}{l^3}
\label{eq:potentiel}
\ee
La discr\'etisation de l'espace introduit automatiquement une coupure ultraviolette, puisque les vecteurs d'onde \`a une particule sont restreints \`a la premi\`ere zone de Brillouin $\mathrm{PZB}=[-\pi/l,\pi/l[^3$ du r\'eseau, tout en pr\'eservant la simplicit\'e d'une interaction de contact.
Le hamiltonien du système  dans l'ensemble grand canonique de potentiel chimique $\mu$ s'écrit~:
\begin{multline}
\hat{H}= l^3  \sum_{\rr, \sigma=\uparrow/\downarrow} \hat{\psi}_\sigma^\dagger(\rr) \left( - \frac{\hbar^2}{2m}  \Delta_{\rr} -\mu\right)  \hat{\psi}_\sigma(\rr) \\
+ g_0 l^3 \sum_{\rr} \hat{\psi}_\uparrow^\dagger(\rr) \hat{\psi}_\downarrow^\dagger(\rr)\hat{\psi}_\downarrow(\rr){\hat{\psi}_\uparrow}(\rr)
\label{eq:hamiltonien}
\end{multline}
où l'opérateur laplacien discret $\Delta_\rr$ admet les fonctions $\eee^{\ii\kk \cdot \rr}$ 
comme fonctions propres de valeur propre $-k^2$ 
et l'opérateur champ des fermions possède les relations d'anticommutation discrètes $\{ \hat{\psi}_\sigma(\rr) , \hat{\psi}_{\sigma'}^\dagger({\rr'}) \} = \delta_{\sigma \sigma'} \delta_{\rr \rr'}/l^3$, 
avec $\sigma, \sigma'=\uparrow {\rm ou} \downarrow$. 
Le r\'egime d'interactions de port\'ee nulle s'obtient en passant \`a la limite d'un espace continu $l\to 0$
\`a valeur fix\'ee de la longueur de diffusion $a$. 
Il est alors d'usage d'éliminer la constante de couplage $g_0$, qui tend vers $0$ lorsque $l\to0$, 
au profit de la longueur de diffusion $a$ mesurable expérimentalement.
On utilise pour cela la relation issue de la théorie de la diffusion dans l'espace libre ($L=\infty$)
appliquée au potentiel \eqref{eq:potentiel} \cite{CastinLesHouches2004,Varenna}~:
\begin{equation}
\label{eq:g0}
\frac{1}{g_0} = \frac{m}{4 \pi \hbar^2 a} - \int_{\rm PZB} \frac{\dd^3k}{(2\pi)^3} \frac{m}{\hbar^2 k^2} 
\end{equation}

\subsection{Approximation BCS de l'état fondamental}

L'idée de la théorie BCS \cite{BCS1957} est de chercher  une expression approchée de l'état fondamental du hamiltonien \eqref{eq:hamiltonien} parmi la famille d'états
\begin{equation}
\ket{\psi^{\rm BCS}}=  \prod_{\kk\in\mathcal{D}}  \bb{ U_\kk - V_\kk \hat{a}_{\kk \uparrow}^\dagger \hat{a}_{-\kk \downarrow}^\dagger} \ket{0}
\label{eq:ansatzBCS2}
\end{equation}
où $\mathcal{D}=\frac{2\pi}{L}\mathbb{Z}^{3}\cap[-\pi/{l},\pi/{l}[^3$ est l'ensemble des vecteurs de la première zone de Brillouin compatibles avec les conditions aux limites périodiques, l'opérateur $\hat{a}_{\kk \sigma}$, coefficient de Fourier de l'opérateur champ $\hat{\psi}_\sigma(\rr)$, crée un fermion de vecteur d'onde $\kk$ et de spin $\sigma$, $V_\kk$ est l'amplitude de probabilité de présence d'un fermion $\kk\sigma$ dans $\ket{\psi^{\rm BCS}}$ 
et $U_\kk=\sqrt{1-|V_\kk|^2}$. Les minimiseurs de la fonctionnelle énergie classique $\bra{\psi^{\rm BCS}} \hat{H} \ket{\psi^{\rm BCS}}$ sont répartis sur un cercle
\be
U_\kk=U_\kk^0 \qquad V_\kk=V_\kk^0 \eee^{\ii \phi}
\label{eq:paramBCS}
\ee
où $\phi\in[0,2\pi[$ et les valeurs de référence $U_\kk^0$ et $V_\kk^0$ sont choisies réelles.

Pour étudier le comportement du système à basse énergie, nous choisissons un état donné du cercle, celui de phase nulle
\be
\ket{\psi^{\rm BCS}_0}=  \prod_\kk  \bb{ U_\kk^0 - V_\kk^0 \hat{a}_{\kk \uparrow}^\dagger \hat{a}_{-\kk \downarrow}^\dagger} \ket{0}
\label{eq:psiBCS0}
\ee
comme origine du développement. 
Dans cet état à symétrie brisé, on définit le \textit{param\`etre d'ordre} de la théorie BCS
\be
\Delta\equiv g_0 \bra{\psi^{\rm BCS}_0}{\hat{\psi}_\downarrow \hat{\psi}_\uparrow}\ket{\psi^{\rm BCS}_0} = - \frac{g_0}{L^3} \sum_{\kk\in\mathcal{D}} U_\kk^0 V_\kk^0,
\label{eq:defDelta}
\ee
quantité qui remplace $g_0$ ou $a$ comme paramètre naturel de la force des interactions dans la théorie BCS. Il permet d'écrire les coefficients $U_\kk^0,V_\kk^0$ sous la forme
\be
V_\kk^0=\sqrt{\frac{1}{2}\bb{1-\frac{\xi_\kk}{\epsilon_\kk}}} \quad \mbox{et} \quad U_\kk^0=\sqrt{\frac{1}{2}\bb{1+\frac{\xi_\kk}{\epsilon_\kk}}}
\label{eq:Vk0}
\ee
avec les énergies
\bea
\xi_\kk &=& \frac{\hbar^2k^2}{2m}-\mu+ \frac{g_0{\rho}}{2} \label{eq:defxik} \\
\epsilon_\kk &=& \sqrt{\Delta^2+\xi_\kk^2} \label{eq:defepsilonk}
\eea
\`A l'aide de la relation fort utile $U_\kk^0 V_\kk^0= \Delta/2\epsilon_\kk$, 
l'\'equation (\ref{eq:defDelta}) peut \^etre mise sous la forme plus habituelle
\be
\frac{1}{g_0} = - \frac{1}{L^3}\sum_{\kk\in\mathcal{D}} \frac{1}{2\epsilon_\kk}
\label{eq:gap}
\ee
La densité totale moyenne ${\rho}$, nombre total moyen $\meanv{\hat{N}}$ de particules par unité de volume, vaut dans l'état \eqref{eq:psiBCS0}
\be
{\rho}\equiv\frac{\meanv{\hat{N}}}{L^3}\equiv\frac{k_{\rm F}^3}{3\pi^2}=\frac{2}{L^3}\sum_{\kk\in\mathcal{D}} (V_\kk^0)^2
\label{eq:rhobar}
\ee
où $k_{\rm F}$ est le nombre d'onde de Fermi du gaz parfait de densité $\rho$.
Associé à la relation \eqref{eq:Vk0} ceci conduit conduit à l'équation d'état BCS du gaz reliant ${\rho}$ à $\mu$ et $\Delta$.

On définit enfin des opérateurs de création et d'annihilation d'excitations fermioniques, obtenus à partir des opérateurs de création et d'annihilation de particules par la transformation de Bogolioubov~:
\bea
\hat{{\gamma}}_{\kk\uparrow}     &=& U_\kk^0 \hat{a}_{\kk\uparrow} + V_\kk^0 \hat{a}_{-\kk\downarrow}^\dagger \label{eq:bup} \\
\hat{{\gamma}}_{-\kk\downarrow} &=& -V_\kk^0 \hat{a}_{\kk\uparrow}^\dagger + U_\kk^0 \hat{a}_{-\kk\downarrow}  \label{eq:bdown}
\eea
Ce sont les opérateurs dont l'état \eqref{eq:psiBCS0} est le vide. Ils annihilent des quasi-particules $\kk\uparrow$ et $-\kk\downarrow$ d'énergie $\epsilon_\kk$.
Pour \^etre complets, nous donnons le d\'eveloppement modal correspondant 
des champs fermioniques~:
\be
\begin{pmatrix}\psi_\uparrow(\rr) \\ \psi_\downarrow^\dagger(\rr) \end{pmatrix}=
\frac{1}{L^{3/2}} \sum_{\kk\in\mathcal{D}} \hat{\gamma}_{\kk\uparrow} \begin{pmatrix} U_\kk^0 \\ V_\kk^0\end{pmatrix}
\eee^{\ii\kk\cdot\rr} + \hat{\gamma}_{\kk\downarrow}^\dagger \begin{pmatrix} -V_{\kk}^0 \\ U_\kk^0 \end{pmatrix}
\eee^{-\ii\kk\cdot\rr}
\label{eq:devmodal}
\ee


\subsection{Ansatz de paires en mouvement}


Nous paramétrons les fluctuations autour du fondamental BCS à symétrie brisée (le vide de quasi-particules)
par l'état cohérent de quasi-particules ou vide comprim\'e~:
\be
\ket{\psi}(t)=\mathcal{N}(t) \exp\bb{\sum_{\kk,\qq} z_{\kk+\qq,\kk}(t) \hat{\gamma}_{\kk+\qq\uparrow}^\dagger \hat{\gamma}_{-\kk\downarrow}^\dagger } \ket{\psi_{\rm BCS}^0}
\label{eq:Ansatz}
\ee
o\`u les amplitudes complexes $z_{\kk+\qq,\kk}$ sont pris comme les param\`etres variationnels ind\'ependants et $\mathcal{N}(t)$ est un facteur de normalisation. Cet ansatz est inspir\'e de la section 9.9b de la r\'ef\'erence \cite{Ripka1985}.
Contrairement au fondamental BCS \eqref{eq:psiBCS0} dans lequel toutes les paires de particules sont au repos, les paires de quasi-particules ont maintenant un vecteur d'onde du centre de masse $\qq$ non nul. 
{\rosso Ceci va nous permettre d'acc\'eder aux modes collectifs phononiques (sans bande interdite) du syst\`eme et de d\'eterminer leurs interactions. Nous n'incluons pas dans l'ansatz les excitations fermioniques
\'el\'ementaires, que produirait l'action directe des op\'erateurs $\hat{\vers\gamma}_{\kk\sigma}^\dagger$
sur l'\'etat BCS fondamental $\ket{\psi_{\rm BCS}^0}$~; elles pr\'esentent en effet une bande d'\'energie
interdite, de largeur bien plus grande que les \'energies d'excitation qui nous int\'eressent ici.
Remarquons cependant que les op\'erateurs $\hat{\vers\gamma}_{\kk\sigma}$ et $\hat{\vers\gamma}_{\kk\sigma}^\dagger$ d'annihilation et de cr\'eation de quasi-particules fermioniques sont des combinaisons lin\'eaires 
des op\'erateurs $\hat{a}_{\kk\sigma}$ et $\hat{a}_{\kk\sigma}^\dagger$ d'annihilation et de cr\'eation 
de particules,
voir les \'equations ~(\ref{eq:bup},\ref{eq:bdown}), si bien que notre \'etat coh\'erent de paires
de quasi-particules (\ref{eq:Ansatz}) peut \^etre vu aussi comme un \'etat coh\'erent de paires
de particules, c'est-\`a-dire {\sl in fine} comme un ansatz BCS d\'ependant du temps.
Parmi tous ces ansatz \'equivalents, et gaussiens en les champs fermioniques, la forme (\ref{eq:Ansatz})
nous para\^\i t conduire aux calculs les plus simples dans le r\'egime de faible excitation, 
puisqu'elle fait appara\^\i tre directement les amplitudes $z_{\kk+\qq,\kk}$ comme petits param\`etres
du d\'eveloppement.}

Les références \cite{Ripka1985} et \cite{TheseHK} expliquent comment appliquer le principe variationnel dans l'état \eqref{eq:Ansatz}. On introduit les variables
\be
\beta_{\kk,\kk'}=-\bb{z(1+z^\dagger z)^{-1/2}}_{\kk,\kk'}
\label{eq:defbeta}
\ee
où $z$ est la matrice $(z_{\kk,\kk'})_{\kk,\kk'\in\mathcal{D}}$ afin de mettre les équations du mouvement sous la forme hamiltonienne
\bea
\ii\hbar\frac{\dd\beta_{\kk',\kk}}{\dd t} &=& \frac{\partial E}{ \partial \beta_{\kk',\kk}^* } \label{eq:hamiltonienne1}\\
-\ii\hbar\frac{\dd\beta_{\kk',\kk}^*}{\dd t} &=& \frac{\partial E}{\partial \beta_{\kk',\kk}}  \label{eq:hamiltonienne2}
\eea
Le hamiltonien classique correspondant
\be
E\equiv\bra{\psi}\hat{H}\ket{\psi}
\label{eq:defE}
\ee
s'évalue grâce au théorème de Wick, qui relie $E$ \`a des moyennes d'op\'erateurs 
bilin\'eaires en les $\hat{\gamma}_{\kk\sigma}$. Ces moyennes s'obtiennent par décomposition de Schmidt de la matrice $z$ et s'expriment naturellement en fonction du champ $\beta$ avec une notation
matricielle~:
\bea
\meanv{\hat{\gamma}_{-\kk\downarrow} \hat{\gamma}_{\kk'\uparrow}} &=& - \bb{\beta(1-\beta^\dagger\beta)^{1/2}}_{\kk',\kk} \label{eq:kappa} \\
\meanv{\hat{\gamma}_{\kk\uparrow}^\dagger \hat{\gamma}_{-\kk'\downarrow}^\dagger} &=& - \bb{\beta(1-\beta^\dagger\beta)^{1/2}}_{\kk,\kk'}^* \label{eq:kappa_star} \\
\meanv{\hat{\gamma}_{\kk'\uparrow}^\dagger \hat{\gamma}_{\kk\uparrow}}&=& \bb{\beta\beta^\dagger}_{\kk,\kk'} \label{eq:rho_up} \\
\meanv{\hat{\gamma}_{-\kk'\downarrow}^\dagger \hat{\gamma}_{-\kk\downarrow}} &=& \bb{\beta^\dagger\beta}_{\kk',\kk}\label{eq:rho_down}
\eea


\section{Hamiltonien classique à l'ordre quadratique : variables normales de la branche bosonique et opérateurs de phonons}
\label{sec:quadratique}


\subsection{\'Equations du mouvement linéarisées}


Pour linéariser les équations du mouvement \eqref{eq:hamiltonienne1} et \eqref{eq:hamiltonienne2}, nous supposons que l'état \eqref{eq:Ansatz} diffère peu du fondamental BCS et donc que
\be
\forall \kk,\kk',\,|z_{\kk,\kk'}|\ll 1 \quad \mbox{et} \quad |\beta_{\kk,\kk'}|\ll 1
\label{eq:petitesamplitudes}
\ee
Puis nous développons la fonctionnelle énergie en puissances du champ $\beta$
\be
E=E_0+E_2+O(\beta^3)
\label{eq:devE}
\ee
où $E_0=\bra{\psi^{\rm BCS}_0}\hat{H}\ket{\psi^{\rm BCS}_0}$ est une constante et $E_2$ est bilinéaire en $\beta$. Le développement est dépourvu de termes linéaires puisque le fondamental BCS (qui correspond à $\beta=0$) est un minimiseur de $E$. Comme dans la RPA d'Anderson \cite{Anderson1958}, les équations du mouvement linéarisées qui découlent de ce développement par remplacement de $E$ par $E_2$ dans \eqref{eq:hamiltonienne1} et \eqref{eq:hamiltonienne2} sont découplées par valeur du vecteur d'onde total $\qq$. Pour cette raison, nous récrivons les coordonnées du champ $\beta_{\kk_1,\kk_2}$ sous une forme indice-exposant à la Anderson, en mettant en indice le vecteur d'onde relatif $(\kk_1+\kk_2)/2$ et en exposant le vecteur d'onde du centre de masse $\kk_1-\kk_2$ de la paire physique $(\kk_1\uparrow, -\kk_2\downarrow)$, voir l'\'equation (\ref{eq:kappa}),
\be
\beta_{\kk}^\qq\equiv\beta_{\kk+\qq/2,\kk-\qq/2}
\label{eq:notationsRPA}
\ee
et nous définissons les vecteurs $\beta^\qq=(\beta_\kk^\qq)_{\kk\in\mathcal{D}}$ 
et $\bar{\beta}^\qq=((\beta_{\kk}^{-\qq})^*)_{\kk\in\mathcal{D}}$.
Les équations du mouvement se mettent alors sous la forme matricielle
\be
\ii\hbar \frac{\dd}{\dd t} \begin{pmatrix}
\beta^\qq \\
\bar{\beta}^\qq
\end{pmatrix}=
\mathcal{L}^\qq
 \begin{pmatrix}
\beta^\qq \\
\bar{\beta}^\qq
\end{pmatrix},
\label{eq:systvaria}
\ee
avec un opérateur d'évolution $\mathcal{L}^\qq$ symplectique
\be
\sigma_z \mathcal{L}^\qq \sigma_z = (\mathcal{L}^\qq)\dagger
\label{eq:symplectique}
\ee
et présentant une symétrie particule-trou
\be
\sigma_x \mathcal{L}^\qq \sigma_x = -(\mathcal{L}^\qq)^*
\label{eq:renversementtemps}
\ee
avec $\sigma_x=\begin{pmatrix}0&1\\1&0\end{pmatrix}$ et $\sigma_z=\begin{pmatrix}1&0\\0&-1\end{pmatrix}$ en notation par blocs. La relation (\ref{eq:renversementtemps}) est une sym\'etrie particule-trou au sens
o\`u l'action de $\sigma_x$ \'echange le champ $\beta$ et son complexe conjugu\'e $\bar{\beta}$,
ce qui \'evoque bien le fait que, dans le formalisme quantique, l'op\'erateur de champ des trous est
l'hermitien conjugu\'e de l'op\'erateur de champ des particules.

Pour écrire explicitement l'opérateur $ \mathcal{L}^\qq$, il est plus aisé de faire un changement de base \g{somme et différence} en définissant
\bea
y_{\kk}^\qq=\beta_{\kk}^\qq - \bar{\beta}_{\kk}^{\qq} \\
s_{\kk}^\qq=\beta_{\kk}^\qq + \bar{\beta}_{\kk}^{\qq}
\eea
pour obtenir
\be
\ii\hbar\frac{\dd y_{\kk}^\qq }{\dd t}  =  \epsilon_{\kk\qq} s_{\kk}^\qq    +  \frac{g_0}{L^3} \sum_{\kk'\in\mathcal{D}} \bb{W_{\kk\qq}^-  W_{\kk'\qq}^- + w_{\kk\qq}^+  w_{\kk'\qq}^+} {s}_{\kk'}^\qq \label{eq:systvaria1} 
\ee
\be
\ii\hbar\frac{\dd{s_{\kk}^\qq }}{\dd t}  =  \epsilon_{\kk\qq} y_{\kk}^\qq    +   \frac{g_0}{L^3} \sum_{\kk'\in\mathcal{D}} \bb{W_{\kk\qq}^+ W_{\kk'\qq}^+ - w_{\kk\qq}^- w_{\kk'\qq}^-} y_{\kk'}^\qq 
\label{eq:systvaria2}
\ee
Les coefficients $W_{\kk,\qq}^\pm$ et $w_{\kk,\qq}^\pm$ sont des regroupements astucieux des coefficients $U_\kk^0$ et $V_\kk^0$
\bea
W_{\kk\qq}^\pm=U_{\kk+\qq/2}^0U_{\kk-\qq/2}^0\pm V_{\kk+\qq/2}^0V_{\kk-\qq/2}^0 \label{eq:defW} \\
w_{\kk\qq}^\pm=U_{\kk+\qq/2}^0V_{\kk-\qq/2}^0\pm U_{\kk-\qq/2}^0V_{\kk+\qq/2}^0 \label{eq:defw} 
\eea
et les énergies $\epsilon_{\kk\qq}$ sont celles du continuum à deux excitations fermioniques
\be
\epsilon_{\kk\qq}=\epsilon_{\kk+\qq/2}+\epsilon_{\kk-\qq/2}
\ee
Les équations du mouvement (\ref{eq:systvaria1}) et (\ref{eq:systvaria2}) contiennent deux contributions : premièrement, une partie {individuelle}, qui ne couple que les amplitudes de même vecteur d'onde relatif $\kk$ et qui contient l'évolution triviale des opérateurs $\hat{\gamma}_{\kk\sigma}$ sous le hamiltonien BCS $\hat{H}_{\rm BCS}=E_0+\sum_{\kk\sigma} \epsilon_\kk \hat{\gamma}_{\kk\sigma}^\dagger \hat{\gamma}_{\kk\sigma} $, et deuxièmement, une partie collective qui couple les amplitudes normales $ y_{\kk}^\qq$ et $ s_{\kk}^\qq$ à des amplitudes collectives de même impulsion totale $\qq$.
Notons aussi que ces équations semi-classiques coïncident avec les équations de la RPA d'Anderson \cite{Anderson1958} prises en valeur moyenne dans l'état \eqref{eq:Ansatz}. Pour s'en convaincre, il faut exprimer les équations (78a--d) de la référence \cite{Anderson1958} en termes des opérateurs de quasi-particules grâce aux relations \eqref{eq:bup} et \eqref{eq:bdown} et remarquer que les moyennes des opérateurs $\hat{\gamma}_{\kk\sigma}^\dagger \hat{\gamma}_{\kk'\sigma}$ sont négligeables à l'ordre linéaire de la théorie variationnelle d'après \eqref{eq:rho_up} et \eqref{eq:rho_down}.


\subsection{Branche d'excitation collective}


Nous cherchons maintenant les modes propres du système d'équations (\ref{eq:systvaria1},\ref{eq:systvaria2}) d'énergie $\hbar\omega_\qq$ positive mais inférieure au continuum $k\mapsto\epsilon_{\kk+\qq/2}+\epsilon_{\kk-\qq/2}$ à deux excitation fermioniques, que l'on obtient en diagonalisant la partie individuelle de (\ref{eq:systvaria1},\ref{eq:systvaria2})
\be
0<\hbar\omega_\qq<\inf_{\kk}  (\epsilon_{\kk+\qq/2}+\epsilon_{\kk-\qq/2})
\ee
L'équation aux valeurs propres associée au système (\ref{eq:systvaria1},\ref{eq:systvaria2}) est \'ecrite dans le cas général dans la référence \cite{TheseHK}. 
Nous donnons dans l'annexe \ref{app:RPA} les grandes étapes de son obtention, 
dans la limite simplificatrice d'un espace continu $l\to0$. Nous obtenons ainsi l'équation implicite
\be
I_{++}(\omega_\qq,q) I_{--}(\omega_\qq,q) = {\hbar^2\omega^2_\qq} \left[I_{+-}(\omega_\qq,q)\right]^2
\label{eq:dedispersion}
\ee
o\`u l'on a introduit les int\'egrales
\bea
\label{eq:intpp}
I_{++}(\omega,q) \!\!&=&\!\!\int_{\mathbb{R}^3} \!\!\dd^3k\left[\frac{\epsilon_{\kk\qq}(W_{\kk\qq}^+)^2}{(\hbar\omega)^2
-(\epsilon_{\kk\qq})^2}+\frac{1}{2\epsilon_\kk}\right]  \\
\label{eq:intmm}
I_{--}(\omega,q) \!\!&=&\!\! \int_{\mathbb{R}^3} \!\!\dd^3k\left[\frac{\epsilon_{\kk\qq}(W_{\kk\qq}^-)^2}
{(\hbar\omega)^2-(\epsilon_{\kk\qq})^2}+\frac{1}{2\epsilon_\kk}\right]  \\
\label{eq:intpm}
\!\!\!\! I_{+-}(\omega,q) \!\!&=&\!\! \int_{\mathbb{R}^3} \!\!\dd^3k\frac{W_{\kk\qq}^+ W_{\kk\qq}^-}{(\hbar\omega)^2-(\epsilon_{\kk\qq})^2}
\eea
La même équation \eqref{eq:dedispersion} peut \^etre obtenue par la RPA, comme il a \'et\'e fait par Anderson \cite{Anderson1958} dans la limite de couplage faible et par la r\'ef\'erence \cite{TheseHK} dans le cas g\'en\'eral. Il est \`a noter que le syst\`eme lin\'eaire obtenu par la RPA contient des termes suppl\'ementaires proportionnels aux op\'erateurs $\hat{\vers\gamma}_{\kk+\qq\uparrow}^\dagger \hat{\vers\gamma}_{\kk\uparrow}$ et $\hat{\vers\gamma}_{-\kk\downarrow}^\dagger \hat{\vers\gamma}_{-\kk-\qq\downarrow}$. Ces termes ne peuvent contribuer aux \'equations (\ref{eq:systvaria1},\ref{eq:systvaria2}) puisque leur valeur moyenne dans l'ansatz \eqref{eq:Ansatz} est quadratique en le champ $\beta$. De plus, ils jouent le r\^ole de termes sources et n'affectent pas le spectre d'excitation. L'\'equation (\ref{eq:dedispersion}) s'obtient par d'autres approches encore : une approximation gaussienne de l'action dans un formalisme d'intégrale de chemin
\cite{Strinati1998,Randeria2014} et
une m\'ethode de fonctions de Green combinée à une approximation diagrammatique
\cite{CKS2006}.
Les conditions sur $q$ d'existence de la solution $\omega_\qq$ sont discutées dans la référence \cite{CKS2006}, et la concavité du spectre est étudiée dans la référence \cite{KCS2016}. Au-delà des travaux cités, nous construisons ici les opérateurs associés aux modes collectifs.


\subsection{Construction des variables normales de la branche bosonique}
\label{sec:variablesnormales}


\subsubsection{Cas général}


Grâce aux symétries \eqref{eq:symplectique} et \eqref{eq:renversementtemps} de l'opérateur d'évolution, l'obtention des amplitudes normales s'effectue de manière entièrement analogue au cas bosonique par la théorie de Bogolioubov \cite{CastinDum1998}.
La première étape est d'écrire le mode propre $\vec{e}_{+}$ d'énergie $\hbar\omega_\qq>0$
\be
\vec{{e}}_{+}(\qq) =
\begin{pmatrix} 
M_{\kk}^{\qq} \vspace{2mm}\\
N_{\kk}^{\qq}
\end{pmatrix}_{\kk\in\mathcal{D}}
\ee
Pour obtenir l'expression analytique des coefficients $M$ et $N$ dans la limite continue $l\to0$, nous résolvons le système (\ref{eq:systvaria1},\ref{eq:systvaria2}) avec $\dd/\dd t \to - \ii \omega_\qq$ et nous utilisons les équations \eqref{eq:yqsq} et \eqref{eq:proprecollective} pour éliminer les amplitudes collectives, ce qui conduit à
\bea
\!\!\!\!\!\!M_{\kk}^{\qq}-N_{\kk}^{\qq} &=& \frac{2\Delta\bb{ \epsilon_{\kk\qq} W_{\kk\qq}^+  - W_{\kk\qq}^- \frac{I_{++}(\omega_\qq,q)}{I_{+-}(\omega_\qq,q)}   }
}{ \sqrt{\norm_\qq}(\epsilon_{\kk\qq}^2 - \hbar^2\omega_\qq^2 )} \label{eq:Ykq} \\
\!\!\!\!\!\!M_{\kk}^{\qq}+N_{\kk}^{\qq} &=&  \frac{2\Delta \bb{ \hbar^2\omega_\qq^2 W_{\kk\qq}^+  -  \epsilon_{\kk\qq} W_{\kk\qq}^- \frac{I_{++}(\omega_\qq,q)}{I_{+-}(\omega_\qq,q)}   }}{ \hbar\omega_\qq \sqrt{\norm_\qq}(\epsilon_{\kk\qq}^2 - \hbar^2\omega_\qq^2)} 
\label{eq:Skq}
\eea
où $\norm_\qq$ est une constante de normalisation sans dimension, déterminée plus bas par l'équation \eqref{eq:normalisation}. Relevons une propriété importante de $M$ et $N$~: leur invariance aussi bien par parité interne ($\kk\to-\kk$) qu'externe ($\qq\to-\qq$) vis-à-vis des paires, qui résulte de l'invariance par parité du problème et de l'appariement dans l'onde $s$,
\bea
M_{\kk}^{\qq} &=& M_{-\kk}^{\qq}   = M_{\kk}^{-\qq} \label{eq:invartranspo1}\\
N_{\kk}^{\qq} &=& N_{-\kk}^{\qq}   = N_{\kk}^{-\qq} \label{eq:invartranspo2}
\eea
En vertu de la symétrie particule-trou \eqref{eq:renversementtemps},
 le vecteur propre $\vec{e}_{+}$ possède un vecteur propre associé $\vec{e}_{-}$ d'énergie $-\hbar\omega_\qq$  qui s'obtient par simple multiplication par $\sigma_x$
\be
\vec{{e}}_{-}(\qq) =
\sigma_x
\begin{pmatrix} 
M_{\kk}^{\qq} \vspace{2mm}\\
N_{\kk}^{\qq}
\end{pmatrix}_{\kk\in\mathcal{D}}
=
\begin{pmatrix} 
N_{\kk}^{\qq} \vspace{2mm}\\
M_{\kk}^{\qq}
\end{pmatrix}_{\kk\in\mathcal{D}}
\ee
Nous cherchons ensuite les vecteurs duaux qui servent à calculer la projection du champ $\beta$ sur les modes propres $\vec{e}_{+}$ et $\vec{e}_{-}$. En vertu de la symétrie symplectique \eqref{eq:symplectique}, ils s'obtiennent par multiplication par $\sigma_z$
\bea
\vec{{d}}_{+}(\qq)= \sigma_z
\begin{pmatrix} 
M_{\kk}^{\qq} \vspace{2mm}\\
N_{\kk}^{\qq}
\end{pmatrix}_{\kk\in\mathcal{D}} &=& 
 \begin{pmatrix} 
M_{\kk}^{\qq} \vspace{2mm}\\
-N_{\kk}^{\qq}
\end{pmatrix}_{\kk\in\mathcal{D}} \label{eq:dual+}
\\
\vec{{d}}_{-}(\qq)= - \sigma_z
\begin{pmatrix} 
N_{\kk}^{\qq} \vspace{2mm}\\
M_{\kk}^{\qq}
\end{pmatrix}_{\kk\in\mathcal{D}} &=&
 \begin{pmatrix} 
-N_{\kk}^{\qq} \vspace{2mm}\\
M_{\kk}^{\qq} 
\end{pmatrix}_{\kk\in\mathcal{D}} \label{eq:dual-}
\eea
Finalement, les vecteurs et leurs duaux sont normalisés en fixant la valeur de $\mathcal{N}_\qq$ de façon que
\be
\bb{\vec{{d}}_{\pm }}^* \cdot \vec{e}_{\pm } =  \sum_{\kk\in\mathcal{D}} 
\left[(M_\kk^\qq)^2-(N_\kk^\qq)^2\right] =1
\label{eq:normalisation}
\ee

Pour définir les amplitudes $b_\qq$ des modes collectifs, nous projetons le champ $\beta$~:
\bea
\!\!\!\!\!\!b_\qq=\vec{{d}}_{+}(\qq) \cdot \begin{pmatrix}  \beta^{\qq} \\ \bar{\beta}^{\qq}\end{pmatrix} \!\!\!&=&\!\!\! 
\sum_{\kk\in\mathcal{D}} \left[M_\kk^\qq \beta_\kk^\qq - N_{\kk}^\qq ({\beta}_\kk^{-\qq})^*\right] \\
\!\!\!\!\!\!b_{-\qq}^*=\vec{{d}}_{-}(\qq) \cdot \begin{pmatrix}  \beta^{\qq} \\ \bar{\beta}^{\qq}\end{pmatrix} \!\!\!&=&\!\!\! 
\sum_{\kk\in\mathcal{D}} \left[-N_\kk^\qq \beta_\kk^\qq +M_{\kk}^\qq ({\beta}_\kk^{-\qq})^*\right]
\label{eq:BSC}
\eea
ce qui constitue le premier résultat majeur de cet article.
L'égalité $b_{\qq}^*=(b_{\qq})^*$ suggérée par la notation est une conséquence de l'invariance par transposition (\ref{eq:invartranspo1}).
Pour exprimer en sens inverse le champ classique en fonction des amplitudes de phonons, nous le développons sur les vecteurs propres~:
\be
\begin{pmatrix} 
\beta^{\qq} \\
\bar{\beta}^{\qq}
\end{pmatrix} = 
b_\qq \vec{e}_{+}(\qq) + b^*_{-\qq} \vec{e}_{-}(\qq) +\ldots
\label{eq:champsurmodes}
\ee
L'ellipse matérialisée par les points de suspension concerne ici la composante du champ sur les autres modes excités de vecteur d'onde total $\qq$. On montre \cite{TheseHK} qu'à la limite d'un espace continu ces modes excités constituent le continuum $\kk\mapsto\epsilon_{\kk+\qq/2}+\epsilon_{\kk-\qq/2}$ à deux quasi-particules fermioniques ayant un vecteur d'onde du centre de masse $\qq$.
En projetant l'équation vectorielle \eqref{eq:champsurmodes}, nous obtenons~:
\bea
\label{eq:devmodbetajoli1}
\beta_{\kk+\qq/2,\kk-\qq/2} &=& M_{\kk}^{\qq}   b_\qq + N_{\kk}^{\qq} b^*_{-\qq}   + \ldots  \\
{\beta}_{\kk-\qq/2,\kk+\qq/2}^* &=& N_{\kk}^{\qq}   b_\qq + M_{\kk}^{\qq} b^*_{-\qq}   + \ldots
\label{eq:devmodbetajoli2}
\eea
Ces dernières relations permettent d'interpréter $M$ et $N$
comme les coefficients d'une nouvelle transformation de Bogolioubov, qui, après quantification, réarrange cette fois les opérateurs bilinéaires de quasi-particules fermioniques des \'equations (\ref{eq:kappa},\ref{eq:kappa_star}) en opérateurs de quasi-particules bosoniques, et qui se superpose à celle induite par les coefficients $U_\kk^0$ et $V_\kk^0$, qui réarrange les opérateurs de particules en opérateurs de quasi-particules fermioniques. Cette transformation agit au niveau des paires, d'où le fait que les coefficients $M$ et $N$ aient deux indices.


\subsubsection{Limite des grandes longueurs d'onde}
 

\`A la limite des grandeurs longueurs d'onde $q\to 0$, nous rappelons le développement de l'énergie $\hbar\omega_\qq$ qu'obtient la référence \cite{KCS2016}~:
\be
\hbar \omega_\qq \underset{q\to 0}{=} \hbar c q \left[1+\frac{\gamma}{8} \bb{\frac{\hbar q}{mc}}^2+ O\bb{\frac{\hbar q}{mc}}^4
\right]. \label{eq:fleur}
\ee
Dans cette expression, $c$ est la vitesse du son, reliée, comme pour tout superfluide, à l'équation d'état par la formule hydrodynamique
\be
mc^2=\rho \frac{\dd\mu}{\dd\rho},\label{eq:vitesseson}
\ee
où la dérivée est prise à longueur de diffusion $a$ fixée, et le coefficient $\gamma$ de l'ordre suivant du développement est donné comme une fraction rationnelle en les variables
\be
x = \frac{\Delta}{\mu} \ \ \ \ \mbox{et}\ \ \ \ 
y = \frac{\dd\Delta}{\dd\mu}=\frac{\int_{\mathbb{R}^3}\dd^3k \frac{\xi_\kk}{\epsilon_\kk^3}}{\int_{\mathbb{R}^3}\dd^3k \frac{\Delta}{\epsilon_\kk^3}}
\label{eq:xety}
\ee
reliées entre elles dans la dernière égalité par l'équation d'état BCS \eqref{eq:rhobar}.

Les combinaisons \eqref{eq:Ykq} et \eqref{eq:Skq} de $M_\kk^\qq$ et $N_\kk^\qq$ introduites dans le présent article possèdent alors les développements suivants en $q=0$~:
\bea
\sqrt{\norm_\qq} (M_{\kk}^{\qq}-N_{\kk}^{\qq}) &=&  \frac{\Delta}{\epsilon_\kk} + O(q^2) \label{eq:devYkq} \\
\sqrt{\norm_\qq} (M_{\kk}^{\qq}+N_{\kk}^{\qq}) &=&  -\frac{\hbar c q \epsilon_\kk}{2\Delta} \frac{\dd W_{\kk\zero}^-}{\dd \mu}  +O(q^3) \label{eq:devSkq} \\
\norm_\qq/L^3 &=&  \frac{\hbar c q}{2}  \frac{\dd{\rho}}{\dd\mu}  +O(q^3) \label{eq:devnorm}
\eea
Notons que ces expressions 
se raccordent à l'ordre le plus bas avec les coefficients des modes d'énergie nul $\vec{e}_n$ et anormal $\vec{e}_a$ de l'opérateur d'évolution $\mathcal{L}^\zero$ du sous-espace d'impulsion nul qu'obtient la référence \cite{KCS2013}.

{\rosso Comme exemple d'application du d\'eveloppement modal (\ref{eq:devmodbetajoli1},\ref{eq:devmodbetajoli2})
et de la limite \`a faible $q$ (\ref{eq:devYkq},\ref{eq:devSkq}) des amplitudes correspondantes,
d\'eterminons le d\'eveloppement modal de la densit\'e du gaz
\be
\rho(\rr,t)=\sum_\sigma \langle \hat{\psi}_\sigma^\dagger(\rr) \hat{\psi}_\sigma(\rr)\rangle
\label{eq:defrhomicro}
\ee
et de la phase $\theta(\rr,t)$ du param\`etre d'ordre d\'efinie par
\be
g_0\langle\hat{\psi}_\downarrow(\rr)\hat{\psi}_\uparrow(\rr)\rangle =
|g_0\langle\hat{\psi}_\downarrow(\rr)\hat{\psi}_\uparrow(\rr)\rangle|
\eee^{\ii \theta(\rr,t)}, 
\label{eq:defthetamicro}
\ee
o\`u la valeur moyenne est prise dans l'ansatz variationnel (\ref{eq:Ansatz}) d\'ependant du temps.
En d\'eveloppant l'ansatz au premier ordre en les amplitudes $z_{\kk+\qq,\kk}(t)\simeq -\beta_{\kk+\qq,\kk}(t)$,
en utilisant le th\'eor\`eme de Wick pour calculer les valeurs moyennes dans l'\'etat BCS fondamental
et en rempla\c{c}ant les champs fermioniques $\hat{\psi}_\sigma(\rr)$ par leur d\'eveloppement modal
(\ref{eq:devmodal}), nous mettons les premiers \'ecarts aux valeurs BCS fondamentales sous la forme
\bea
\delta\rho(\rr,t) &\!\!\!=\!\!\!& \frac{1}{L^3}\sum_{\kk,\qq} w_{\kk\qq}^+ [\beta_{\kk+\frac{\qq}{2},\kk-\frac{\qq}{2}}(t)\eee^{\ii\qq\cdot\rr}+\mbox{c.c.}]+O(\beta^2) \\
\delta\theta(\rr,t)&\!\!\!=\!\!\!&\frac{-g_0}{2\ii\Delta L^3} \sum_{\kk,\qq} W_{\kk\qq}^+ [\beta_{\kk+\frac{\qq}{2},\kk-\frac{\qq}{2}}(t)\eee^{\ii\qq\cdot\rr}
-\mbox{c.c.}]+O(\beta^2) \nonumber \\
&&
\eea
o\`u les coefficients $w_{\kk\qq}^+$ et $W_{\kk\qq}^+$ sont ceux des \'equations (\ref{eq:defW},\ref{eq:defw}).
Nous rempla\c{c}ons ensuite le champ $\beta$ par son d\'eveloppement (\ref{eq:devmodbetajoli1},\ref{eq:devmodbetajoli2}) sur les modes collectifs et utilisons la propri\'et\'e de sym\'etrie (\ref{eq:invartranspo2})
pour obtenir
\bea
\label{eq:devmodrho}
\delta\rho(\rr,t)   &=& \frac{1}{L^{3/2}}     \sum_\qq \mathcal{B}_\qq (b_\qq+b_{-\qq}^*) \eee^{\ii\qq\cdot\rr}+O(\beta^2)\\
\delta\theta(\rr,t) &=& \frac{-\ii}{L^{3/2}} \sum_\qq \mathcal{C}_\qq (b_\qq-b_{-\qq}^*) \eee^{\ii\qq\cdot\rr}+O(\beta^2)
\label{eq:devmodtheta}
\eea
avec les coefficients
\bea
\mathcal{B}_\qq &=& \frac{1}{L^{3/2}} \sum_\kk w_{\kk\qq}^+ (M_\kk^\qq+N_\kk^\qq) \\
\mathcal{C}_\qq &=& \frac{g_0}{2\Delta L^{3/2}} \sum_\kk W_{\kk\qq}^+ (N_\kk^\qq-M_\kk^\qq)
\eea
Dans la limite des faibles $q$, compte tenu de la valeur (\ref{eq:rhobar}) de la densit\'e, 
de l'\'equation (\ref{eq:gap}) sur le param\`etre d'ordre et de l'expression thermodynamique (\ref{eq:vitesseson}) de la vitesse du son, nous obtenons les lois d'\'echelle simples suivantes~:
\bea
\mathcal{B}_\qq &=& \left(\frac{\hbar q\rho}{2mc}\right)^{1/2}[1 + O(q^2)] \\
\mathcal{C}_\qq &=&  \left(\frac{2mc}{\hbar q\rho}\right)^{1/2}[1 + O(q^2)]
\eea
Ces r\'esultats font l'objet d'une comparaison \`a l'hydrodynamique quantique dans la section \ref{subsubsec:approchehydro}.
}


\subsection{Quantification des variables normales}
\label{sec:phonons_quantification}

Pour quantifier les amplitudes des modes bosoniques obtenus par notre approche semi-classique, nous nous souvenons qu'au premier ordre de l'approximation des petites amplitudes \eqref{eq:petitesamplitudes}, le champ $\beta_\kk^\qq$ se confond avec $-\meanv{\hat{\gamma}_{-\kk+\qq/2\downarrow} \hat{\gamma}_{\kk+\qq/2 \uparrow}}$, c'est-\`a-dire avec la valeur moyenne d'un op\'erateur de paire de quasi-particules, voir l'\'equation (\ref{eq:kappa}).  Nous sommes donc tentés de faire la substitution
\be
\beta_{\kk}^\qq=\beta_{\kk+\qq/2,\kk-\qq/2}\rightarrow - {\hat{\gamma}_{-\kk+\qq/2\downarrow} \hat{\gamma}_{\kk+\qq/2 \uparrow}}. \label{eq:substitution}
\ee
En soumettant l'amplitude normale $b_\qq$ à cette substitution, nous obtenons l'opérateur quantique~:
\be
{
\hat{b}_\qq= - \sum_{\kk\in\mathcal{D}} \bb{M_{\kk}^{\qq} \hat{\gamma}_{-\kk+\qq/2\downarrow} \hat{\gamma}_{\kk+\qq/2 \uparrow} -  N_{\kk}^{\qq} {\hat{\gamma}_{\kk-\qq/2 \uparrow}^\dagger \hat{\gamma}_{-\kk-\qq/2\downarrow}^\dagger } } \label{eq:BqQuantique}}
\ee
Ce geste intuitif peut en fait être justifié par plusieurs arguments.
Le premier est que l'opérateur quantique \eqref{eq:BqQuantique} coïncide avec celui que nous aurions obtenu dans le cadre de la RPA d'Anderson en diagonalisant le système homogène pour les opérateurs de création et d'annihilation de paires de quasi-particules $\hat{\gamma}_{\kk\uparrow} \hat{\gamma}_{\kk'\downarrow}$ et $\hat{\gamma}_{\kk'\downarrow}^\dagger \hat{\gamma}_{\kk\uparrow}^\dagger$, les opérateurs $\hat{\gamma}_{\kk\sigma}^\dagger\hat{\gamma}_{\kk'\sigma}$, dont la dynamique par la RPA est triviale et dont la valeur moyenne est d'ordre supérieur pour la théorie variationnelle, étant traités comme des termes source. Le second est l'application
de la procédure de quantification par images bosoniques du chapitre 11 de la référence \cite{Ripka1985} où appara\^\i t un \'equivalent quantique du champ $\beta$, le champ d'opérateurs bosoniques $\mathbb{B}$ (à ne pas confondre avec les opérateurs de création d'excitations bosoniques $\hat{b}_\qq$). L'expression de l'\textit{image bosonique} des opérateurs fermioniques à deux corps tel $\hat{\gamma}_{-\kk+\qq/2\downarrow} \hat{\gamma}_{\kk+\qq/2 \uparrow}$, en termes de ce champ bosonique $\mathbb{B}$, qui permettrait d'envisager une substitution telle que \eqref{eq:substitution}, n'est pas simple dans le cas général car elle implique une série infinie en puissances de $\mathbb{B}$.  Mais, dans limite où le gaz est faiblement excité, l'opérateur population des images bosoniques $\mathbb{B}\mathbb{B}^\dagger$ peut être négligé, ce qui autorise la substitution \eqref{eq:substitution}, où nous confondons les opérateurs fermioniques à deux corps et leur image bosonique.
Enfin, le dernier argument  est le caractère quasi-bosonique \eqref{eq:BqQuantique} de l'opérateur $\hat{b}_\qq$ lorsque le gaz est faiblement excité. En effet, calculons le commutateur de $\hat{b}_\qq$ et $\hat{b}^\dagger_\qq$~:
\begin{multline}
\bbcro { \hat{b}_\qq,\hat{b}_\qq^\dagger   }-1= \\ \sum_{\kk\in\mathcal{D}} \bbcrol{ \bb{{N}_\kk^\qq}^2 
\bb{  \hat{\gamma}_{\kk-\qq/2\uparrow}^\dagger \hat{\gamma}_{\kk-\qq/2\uparrow} + \hat{\gamma}_{-\kk-\qq/2\downarrow}^\dagger \hat{\gamma}_{-\kk-\qq/2\downarrow}  } } \\ \bbcror{
- \bb{{M}_\kk^\qq}^2 \bb{  \hat{\gamma}_{\kk+\qq/2\uparrow}^\dagger \hat{\gamma}_{\kk+\qq/2\uparrow} + \hat{\gamma}_{-\kk+\qq/2\downarrow}^\dagger \hat{\gamma}_{-\kk+\qq/2\downarrow}  } }
\label{eq:commbq}
\end{multline}
Les termes par lesquels ce commutateur diffère de l'unité sont des opérateurs de populations de quasi-particules fermioniques. Ils sont donc exactement nuls dans le fondamental BCS et du second ordre en le champ $z=O(\beta)$ dans un état cohérent \eqref{eq:Ansatz} de quasi-particules.


\section{Au-delà de l'ordre quadratique : Interactions entre phonons et comparaison à l'hydrodynamique}
\label{sec:interactions}

{\rosso
L'interaction entre les phonons, responsable de leur amortissement, est au c\oe ur de ce travail.
Elle est obtenue dans cette section, sous la forme d'amplitudes de couplage \`a trois ou quatre phonons
apparaissant dans des hamiltoniens d'interaction phononique cubiques ou quartiques en le champ des phonons.
Bien que les expressions obtenues soient un peu longues, les calculs sont conceptuellement tr\`es
simples~: on part soit d'un mod\`ele microscopique, la fonctionnelle \'energie (\ref{eq:defE}) 
des fermions ou le hamiltonien bosonique (\ref{eq:hamiltonienB}) plus simple \`a manipuler,
soit du hamiltonien (\ref{eq:hamiltonienhydro}) de l'hydrodynamique quantique, puis on d\'eveloppe
cette fonctionnelle ou ces hamiltoniens \`a l'ordre trois ou quatre en les fluctuations des champs, 
et enfin on injecte dans les termes cubiques ou quartiques obtenus le d\'eveloppement lin\'eaire des fluctuations
sur les modes normaux phononiques, dont les coefficients sont justement les op\'erateurs de cr\'eation
et d'annihilation de phonons. Dans le cas des processus \`a trois phonons, 
trait\'es dans la section \ref{sec:3phonons}, cette proc\'edure suffit. Pour les processus
\`a quatre phonons de la section \ref{sec:4phonons}, il y a une \'etape suppl\'ementaire~:
si l'on veut obtenir la v\'eritable amplitude de couplage \`a quatre phonons, dite {\sl effective},
il faut ajouter \`a l'amplitude du couplage direct \`a quatre phonons celle des processus
\`a trois phonons trait\'es au second ordre de la th\'eorie des perturbations, comme l'avait
d\'ej\`a compris la r\'ef\'erence \cite{Khalatnikov1949}.
Un r\'esultat marquant de cette section sera que, dans la limite des faibles nombres d'onde,
les amplitudes de couplage \`a trois ou quatre phonons issues
des mod\`eles microscopiques co\"\i ncident avec celles de l'hydrodynamique quantique sur la couche
de masse (processus conservant l'\'energie des phonons non perturb\'ee), 
mais s'en \'ecartent violemment en dehors.
}


\subsection{Processus à trois phonons}
\label{sec:3phonons}
Nous étudions dans cette sous-section les processus à trois quasi-particules bosoniques, ce qui inclut les processus $2\leftrightarrow1$ dits de Beliaev-Landau et les processus non résonnants $3\leftrightarrow0$. Notre objectif est de faire apparaître ces processus dans le développement d'un hamiltonien en puissances d'un champ d'excitations, et d'extraire les éléments de matrice associés. Pour cela nous utiliserons deux théories que nous comparerons. Premièrement, nous utiliserons la théorie variationnelle microscopique que nous venons d'exposer en prenant la fonctionnelle énergie $E$ \eqref{eq:defE} comme hamiltonien classique, en la développant maintenant jusqu'à l'ordre trois en $\beta$,
\be
E=E_0+E_2+E_3+O(\beta^4)
\label{eq:devE3}
\ee
et en injectant le développement \eqref{eq:champsurmodes} du champ sur les modes propres collectifs dans le terme trilinéaire $E_3$. Ce calcul microscopique sera mis en \oe uvre seulement dans le cas $2\leftrightarrow1$. Deuxièmement, nous utiliserons l'hydrodynamique quantique de Landau et Khalatnikov \cite{Khalatnikov1949}, théorie mésoscopique qui traite les paires de fermions à grande échelle comme des champs bosoniques mais qui a l'avantage de reposer sur l'équation d'état exacte. Elle sera appliqu\'ee aux processus $2\leftrightarrow1$ et $3\leftrightarrow0$. La comparaison des deux théories nous permettra de discuter la validité de l'hydrodynamique quantique.

\subsubsection{Approche microscopique}


L'esprit du calcul microscopique de l'amplitude de couplage est simple~: il faut exprimer le hamiltonien classique \eqref{eq:defE} en termes des amplitudes $b_\qq$ associées au modes propres collectifs de la théorie linéarisée (\textit{cf}. la sous-section \ref{sec:variablesnormales}), isoler les termes contenant les amplitudes de création ou d'annihilation $b_\qq^*$ et $b_\qq$ correspondant aux processus étudiés et extraire leur coefficient. Pour les processus à trois phonons, ce sont les termes cubiques du hamiltonien \eqref{eq:devE}, rassemblés dans $E_3$, qu'il faut exprimer en fonction des amplitudes $b_\qq$.

\paragraph{Cas général}


La partie cubique de $E$ se met sous la forme
\begin{multline}
E_3=\frac{g_0}{L^3}\!\!\sum_{\kk,\kk',\qq\in\mathcal{D}}\!\!\mathcal{T}_{\kk,\kk'}^{\qq}
 \bbcrol{ \bbl{\beta_{\kk'+\qq/2,\kk'-\qq/2}
{\rosso \meanv{\hat{\gamma}_{\kk+\qq/2\uparrow}^\dagger\hat{\gamma}_{\kk-\qq/2\uparrow}}}}} \\
\bbcror{\bbr{+ {\beta}_{\kk'-\qq/2,\kk'+\qq/2}
{\rosso \meanv{\hat{\rosso\gamma}_{-\kk-\qq/2\downarrow}^\dagger \hat{\rosso\gamma}_{-\kk+\qq/2\downarrow}} }} + {\rm c.c.}} \label{eq:E3}
\end{multline}
où nous avons introduit le tenseur
\be
\mathcal{T}_{\kk,\kk'}^\qq = \frac{\rosso W_{\kk\qq}^-w_{\kk'\qq}^+ - w_{\kk\qq}^+W_{\kk'\qq}^- - w_{\kk\qq}^-W_{\kk'\qq}^+ - W_{\kk\qq}^+w_{\kk'\qq}^-}{2}
\label{eq:TkkPq}
\ee 
Nous utilisons les relations (\ref{eq:rho_up},\ref{eq:rho_down}) et injectons le développement \eqref{eq:champsurmodes} du champ sur les modes propres collectifs dans l'expression \eqref{eq:E3} pour obtenir le résultat suivant \footnote{Tout au long de cet article, les amplitudes de couplage $\mathcal{A}^{m\leftrightarrow n}$ seront des fonctions symétriques de leurs $m$ premiers arguments, les vecteurs d'onde $\qq_1,\qq_2,\ldots,\qq_m$ des phonons entrants, ainsi que de leurs $n$ derniers arguments, les vecteurs d'onde $\qq_{m+1},\qq_{m+2},\ldots,\qq_{m+n}$ des phonons sortants. Cette propriété de symétrie détermine les amplitudes de manière unique.}~:
\begin{multline}
E_{\rm 3}= \frac{m c^2} {(\rho L^3)^{1/2}} \sum_{\qq_1,\qq_2,\qq_3\in\mathcal{D}} \delta_{\qq_1+\qq_2,\qq_3} \\ \times\bb{\mathcal{A}_{\rm micro}^{2\leftrightarrow1} (\qq_1,\qq_2;\qq_3) b^*_{\qq_1} b_{\qq_2}^* b_{\qq_3} + {\rm c.c.}} + \ldots
\label{eq:E3Beliaev}
\end{multline}
Les vecteurs $\qq_1$, $\qq_2$ et $\qq_3$ sont les trois vecteurs d'onde des phonons impliqués dans le processus de Beliaev-Landau, $\qq_3$ étant le vecteur d'onde du phonon qui se désintègre  ou se forme à partir des phonons $\qq_1$ et $\qq_2$. 
L'ellipse $\ldots$ dans \eqref{eq:E3Beliaev} concerne les processus à trois excitations impliquant des excitations autres que bosoniques
 ainsi que les processus non résonnants $3\leftrightarrow0$ qui seraient décrits par les termes $b_{\qq_1} b_{\qq_2} b_{\qq_3}$ et $b_{\qq_1}^* b_{\qq_2}^* b_{\qq_3}^*$ mais que nous n'étudions pas par la théorie microscopique. 
Nous avons mis en facteur $m c^2/(\rho L^3)^{1/2}$, pour faire apparaître l'amplitude de couplage adimensionnée $\mathcal{A}_{\rm micro}^{2\leftrightarrow1} (\qq_1,\qq_2;\qq_3)$ du processus $b^*_{\qq_1} b^*_{\qq_2} b_{\qq_3}$, finie et non nulle à la limite thermodynamique~:
\begin{multline}
\frac{mc^2}{(\rho L^3)^{1/2}}  \mathcal{A}_{\rm micro}^{2\leftrightarrow1} (\qq_1,\qq_2;\qq_3) = \frac{g_0}{L^3} \sum_{\kk,\kk'\in\mathcal{D}} \\ \left[ \mathcal{T}_{\kk,\kk'}^{\qq_1} {M}_{\kk'}^{\qq_1} \bb{ {M}_{\kk-\qq_2/2}^{\qq_3}  {M}_{\kk-\qq_3/2}^{\qq_2} + {N}_{\kk+\qq_3/2}^{\qq_2}  {N}_{\kk+\qq_2/2}^{\qq_3} } \right.\\ \left.
+\mathcal{T}_{\kk,\kk'}^{\qq_1} {N}_{\kk'}^{\qq_1} \bb{ {M}_{\kk+\qq_3/2}^{\qq_2}  {M}_{\kk+\qq_2/2}^{\qq_3} + {N}_{\kk-\qq_2/2}^{\qq_3}  {N}_{\kk-\qq_3/2}^{\qq_2} } \right.  \\ 
\left. +\mathcal{T}_{\kk,\kk'}^{\qq_3} \bb{{N}_{\kk'}^{\qq_3} {N}_{\kk+\qq_1/2}^{\qq_2} {M}_{\kk-\qq_2/2}^{\qq_1} 
 + {M}_{\kk'}^{\qq_3} {M}_{\kk+\qq_1/2}^{\qq_2} {N}_{\kk-\qq_2/2}^{\qq_1}} \right. \\ \left. + \qq_1 \leftrightarrow \qq_2 \vphantom{\mathcal{T}_{\kk,\kk'}^{\qq_1} } \right]
\label{eq:amplitudemicro}
\end{multline}
où la notation $+\qq_1 \leftrightarrow \qq_2$ signifie qu'il faut ajouter aux termes d\'ej\`a présents dans l'équation \eqref{eq:amplitudemicro} ceux obtenus en échangeant $\qq_1$ et $\qq_2$ et en laissant $\qq_3$ inchangé.
{\rosso 
Malgr\'e la simplicit\'e de notre m\'ethode de calcul microscopique de l'amplitude de couplage
$2\leftrightarrow1$, l'expression (\ref{eq:amplitudemicro}) obtenue est assez lourde, \`a cause
de la resommation sur les vecteurs d'onde internes $\kk$ et $\kk'$ des paires en interaction. Une
expression beaucoup plus maniable peut \^etre \'ecrite dans la limite des grandes longueurs d'onde, comme
nous le voyons maintenant.}


\begin{figure}[htb] 
\begin{center}
\includegraphics[width=0.52\textwidth,clip=]{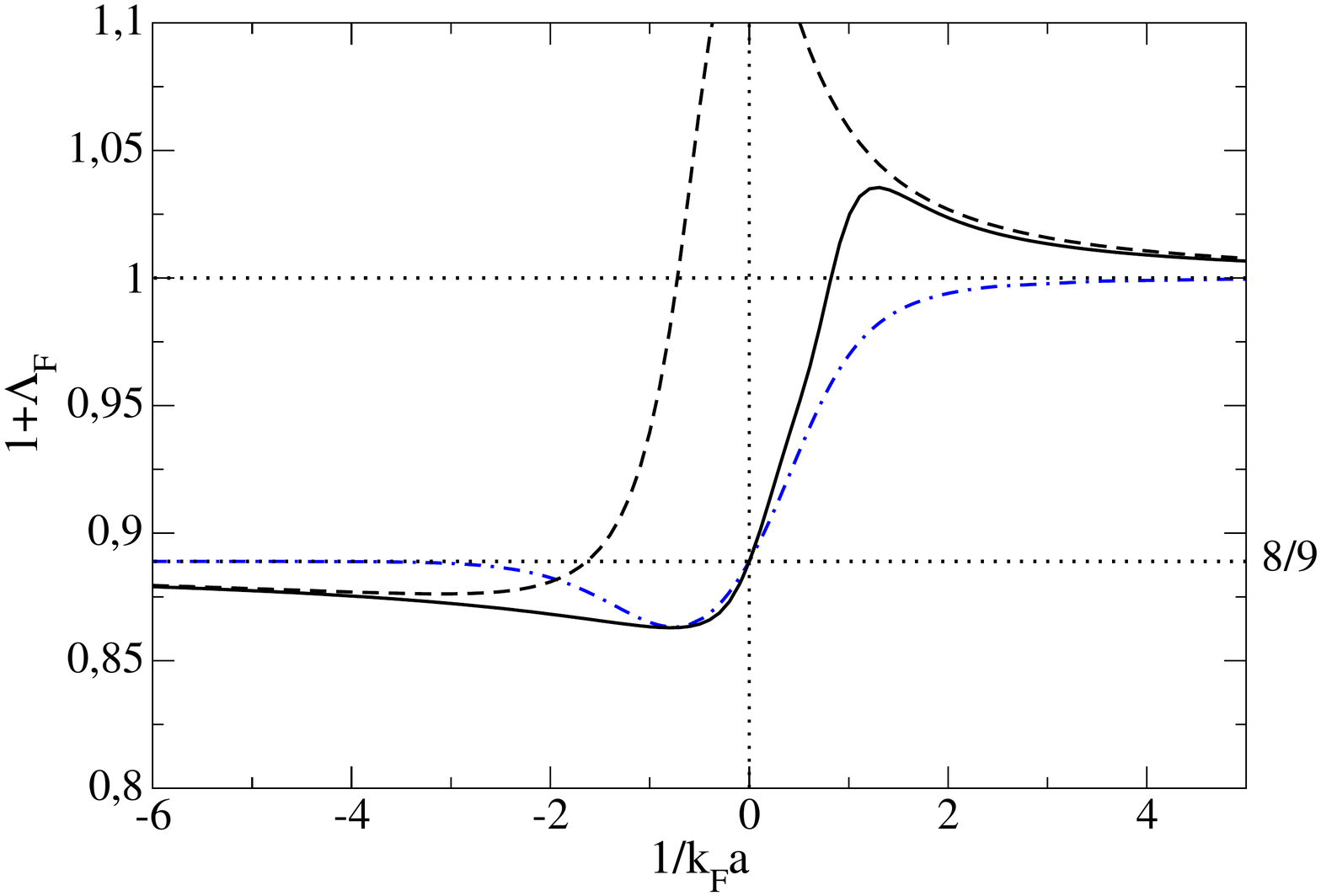} 
\end{center}
\caption{ \label{fig:lambdaF} Le facteur $1+\Lambda_{\rm F}$, contribution de l'équation d'état du gaz de fermions à l'amplitude de couplage 
à trois phonons $2\leftrightarrow1$ (\cf l'équation \eqref{eq:lambdaF}), est représenté en fonction de la force des interactions $1/k_{\rm F} a$. La courbe en trait noir plein provient de l'équation d'état mesurée expérimentalement dans la référence \cite{Salomon2010}. Dans les limites CBE ($k_{\rm F} a \to 0^+$) et BCS ($k_{\rm F} a \to 0^-$) nous donnons en tireté noir ses comportements asymptotiques obtenus à partir des corrections dites de Lee-Huang-Yang \cite{Yang1957,Khalatnikov1957,Galitskii1958} (inclusion de la première correction à l'énergie de champ moyen, en $(k_{\rm F} a)^2$ du côté BCS et en $\rho a\sqrt{\rho a^3}$ du côté CBE). La courbe prédite par l'équation d'état BCS \eqref{eq:rhobar} est donnée à fin de comparaison en tireté-pointillé bleu. Relevons que le facteur $1+\Lambda_{\rm F}$ tend vers $1$ à la limite CBE, comme pour un gaz de bosons en interaction faible. Il vaut $8/9$ à la limite unitaire, et tend vers cette valeur dans le régime BCS, puisque dans les deux cas on a $\mu\propto{\rho}^{2/3}$.}
\end{figure} 

\paragraph{Limite des grandes longueurs d'onde}


\`A basse température, ce sont les phonons de faibles vecteurs d'onde qui dominent la cinétique du gaz, ce qui motive l'étude de l'amplitude $ \mathcal{A}^{2\leftrightarrow1} $ dans la limite $q_1,q_2,q_3\to0$. En outre, c'est seulement dans cette limite que nous pouvons espérer comparer au résultat hydrodynamique. Nous effectuons donc le développement limité de l'expression \eqref{eq:amplitudemicro}, en nous plaçant à la limite d'un espace continu $l\to 0$, ce qui permet d'effectuer les intégrales sur les degrés de liberté internes $\kk$ et $\kk'$ des paires et d'obtenir une expression qui dépend uniquement des vecteurs $\qq_1$, $\qq_2$ et $\qq_3$ du processus. Le calcul microscopique, qui ne sera pas détaillé ici, figure dans la référence \cite{TheseHK}. C'est le caractère fermionique sous-jacent du problème qui lui donne tout son sel. 

Plutôt que d'utiliser les vecteurs d'onde, nous choisissons de paramétrer le résultat par les énergies $\hbar\omega_\qq$, en nous autorisant le raccourci ${\omega}_i \equiv \omega_{\qq_i}, \ i=1,2,3$.
Le passage d'un jeu de variables à l'autre se fait grâce aux relations de dispersion \eqref{eq:fleur} et de conservation de l'impulsion. Nous obtenons dans la limite $\omega_i\to 0$~:
\begin{widetext}
\begin{align}
&\mathcal{A}_{\rm micro}^{2\leftrightarrow1} (\qq_1,\qq_2;\qq_3) = \bb{\frac{\hbar}{mc^2}}^{-1/2} \frac{1}{2^{3/2} \sqrt{{\omega}_1 {\omega}_2 {\omega}_3} }  \Bigg[   2J(x,y)    \bb{{\omega}_1 +{\omega}_2 - {\omega}_3}   \notag
  \\    
   &+\bb{\frac{\hbar}{mc^2}}^{2} \bbaco{A(x,y) {({\omega}_1 +{\omega}_2-{\omega}_3)({\omega}_1^2 + {\omega}_2^2+{\omega}_3^2) }   + B(x,y) \bb{{\omega}_3^3-{\omega}_1^3-{\omega}_2^3  }  
  + C(x,y) {\omega}_1 {\omega}_2 {\omega}_3} + O(\omega^5)   \Bigg]  \label{eq:finaleA}
\end{align}
\end{widetext}
où les fractions rationnelles $A$, $B$, $C$ et $J$ en les variables $x$ et $y$ de l'équation \eqref{eq:xety} sont données dans l'annexe \ref{ann:fractionsrationnelles}. Cette amplitude de couplage contient un dénominateur en $({\omega}_1 {\omega}_2 {\omega}_3)^{1/2}$ qui peut la faire diverger à la limite des grandes longueurs d'onde si le processus considéré n'est pas sur la couche de masse, c'est-à-dire ne vérifie pas la relation de conservation de l'énergie
\be
\omega_{3}=\omega_{1}+\omega_{2}.
\label{eq:cons_energie}
\ee
Nous avons vérifié que la limite CBE de notre résultat \eqref{eq:finaleA} coïncide avec l'expression prédite par la théorie de Bogolioubov \cite{Giorgini1998,CastinSinatra2009} pour un gaz de dimères bosoniques faiblement interagissant. \footnote{Prendre la limite CBE revient à faire \cite{KCS2016} $x\to0^+$ avec $y\sim-4/x$~; nous donnons dans l'annexe \ref{ann:fractionsrationnelles} les expressions \eqref{eq:limBEC} des fractions rationnelles dans cette limite. Grâce à elles, on vérifie sans mal que l'expression de $\mathcal{A}_{\rm micro}^{2\leftrightarrow1}$ dans la limite CBE coïncide avec l'équation (16) de la référence \cite{CastinSinatra2009} développée jusqu'à l'ordre $3$ en $q$ et écrite en termes des énergies des excitations élémentaires de Bogolioubov pour un gaz de $\meanv{\hat{N}_{\rm B}}=\meanv{\hat{N}}/2$ dimères bosoniques de masse $m_{\rm B}=2m$, et de vitesse du son $c$.}


\paragraph{Processus résonnants}


Nous concluons cette étude microscopique en calculant l'amplitude de couplage sur la couche de masse, c'est-à-dire pour les processus qui vérifient la conservation de l'énergie \eqref{eq:cons_energie}, ce qui n'est possible dans la présente limite $\omega_3\to0$ que si le paramètre $\gamma$ est positif, c'est-à-dire si la relation de dispersion $q\mapsto\omega_\qq$ est convexe aux faibles $q$. Les fractions rationnelles $B$ et $C$ se regroupent alors pour former la quantité thermodynamique
\be
2B(x,y)+\frac{2}{3}C(x,y) =  1+\frac{{\rho}}{3} {\frac{\dd^2\mu}{\dd{\rho}^2}} \bb{\frac{\dd\mu}{\dd{\rho}}}^{-1} \equiv 1+\Lambda_{\rm F}  \label{eq:lambdaF}
\ee
dont l'expression en termes de $x$ et $y$ s'obtient en dérivant deux fois l'équation d'état BCS \eqref{eq:rhobar} par rapport à $\mu$ et est donnée en annexe \ref{ann:fractionsrationnelles}. Ceci conduit à la très jolie formule de l'amplitude de couplage sur la couche de masse~:
\be
\mathcal{A}_{\rm Cdm}^{2\leftrightarrow1}(\qq_1,\qq_2;\qq_3) = 3 \bb{1+\Lambda_{\rm F}} \sqrt{\frac{\hbar^3 {q}_1{q}_2 {q}_3}{32m^3c^3}}   + O\bb{q_3^{7/2}} \label{eq:ACM}
\ee
L'apparition du facteur thermodynamique $1+\Lambda_{\rm F}$, reli\'e au param\`etre de
Gr\"uneisen $u=\frac{\partial \ln c}{\partial\ln\rho}$ par $1+\Lambda_{\rm F}=\frac{2}{3} (1+u)$,
est la seule différence entre cette formule (\ref{eq:ACM}) et son équivalent pour un gaz de bosons en interaction faible ($|\Lambda_{\rm F}|\ll 1$) d\'ecoulant de la théorie de Bogolioubov (à ce sujet, voir les équations (D8) et (D9) de la référence \cite{CastinSinatra2009}). Les variations de ce facteur dans la zone de raccordement CBE-BCS sont représentées sur la figure \ref{fig:lambdaF}.
Notons qu'il vaut 1 dans la limite CBE ($x\to0^+$, $y\sim-4/x$) comme dans un gaz de bosons en interaction faible où $\mu\propto{\rho}$. Il vaut $8/9$ tant à la limite unitaire ($x=y$) que dans la limite BCS ($x\to0$, $y\to0$), où $\mu\propto{\rho}^{2/3}$. 


\subsubsection{Approche hydrodynamique}
\label{subsubsec:approchehydro}


Nous cherchons maintenant à comparer notre résultat microscopique \eqref{eq:finaleA} à la prédiction de l'hydrodynamique quantique irrotationnelle de Landau et Khalatnikov. Cette théorie décrit le gaz aux grandes échelles spatiales par deux champs quantiques hermitiens $\hat{\rho}(\rr,t)$ et $\hat{\vv}(\rr,t)$, en omettant directement la densité de quasi-particules fermioniques, exponentiellement petite en $1/T$.
En le supposant irrotationnel, nous écrivons le champ de vitesse $\hat{\vv}(\rr,t)$
comme le gradient (discret) d'un opérateur champ de phase $\hat{\phi}(\rr,t)$,
\be
\hat{\vv}(\rr,t)=\frac{\hbar}{m}\nabla \hat{\phi}(\rr,t),
\ee
canoniquement conjugué 
à l'opérateur champ de densité $\hat{\rho}(\rr,t)$~:
\be
[\hat{\rho}(\rr,t),\hat{\phi}(\rr',t)]=\ii\frac{\delta_{\rr,\rr'}}{{l}^3}.
\label{eq:rhophi}
\ee
La dynamique de ces deux champs est alors donnée par le hamiltonien hydrodynamique
\begin{equation}
\hat{H}_{\rm hydro} = l^3\sum_{\rr} \left[\frac{\hbar^2}{2m} (\nabla\hat{\phi})\cdot \hat{\rho}\ (\nabla \hat{\phi})
+ e_{0,0}(\hat{\rho})\right],
\label{eq:hamiltonienhydro}
\end{equation}
où $e_{0,0}$ est la densité volumique d'énergie nue, qui sera renormalisée par l'énergie de point zéro des modes selon la procédure décrite dans la référence \cite{KCS2015} pour donner la densité volumique d'énergie vraie $e_0$ dans l'état fondamental. Cette dernière est reliée au potentiel chimique à température nulle par $\mu=\dd e_0/\dd \rho$.

La suite est classique et similaire à l'approche microscopique. Nous linéarisons les équations du mouvement pour de faibles fluctuations spatiales du champ de densité et de phase
\bea
\hat{\rho}(\rr,t) & =&  \hat{\rho}_0 +\delta{\hat{\rho}}(\rr,t)  \label{eq:fluc1} \\ 
\hat{\phi}(\rr,t) & =&  \hat{\phi}_0(t) +\delta{\hat{\phi}}(\rr,t) \label{eq:fluc2}
\eea
où nous nous autoriserons à remplacer $\hat{\rho}_0$ par la densité moyenne $\rho$.
Nous développons ensuite les champs sur les modes propres de la dynamique linéarisée
\begin{eqnarray}
\label{eq:devmod1}
\!\!\!\!\!\!\delta{\hat{\rho}}(\rr,t)&=&\frac{{\rho}^{1/2}}{L^{3/2}}  \!\!
\sum_{\qq \in \mathcal{D}^*} \!\!
\left(\frac{\hbar q}{2 m {c}}\right)^{1/2} \!\! (\hat{b}_\qq+\hat{b}_{-\qq}^\dagger) \,\mathrm{e}^{\ii \qq\cdot \rr} \\
\label{eq:devmod2}
\!\!\!\!\!\!\delta{\hat{\phi}}(\rr,t)&=&\frac{-\ii}{{\rho}^{1/2} L^{3/2}} \!\!
\sum_{\qq \in \mathcal{D}^*} \!\!
\left(\frac{m {c}}{2\hbar q}\right)^{1/2} \!\! (\hat{b}_\qq-\hat{b}_{-\qq}^\dagger) \,\mathrm{e}^{\ii \qq\cdot \rr}
\end{eqnarray}
ce qui fait apparaître des opérateurs bosoniques $\hat{b}_\qq$ qui sont le pendant hydrodynamique de ceux que nous avons définis à l'équation \eqref{eq:BqQuantique}. Le spectre de ces excitations bosoniques est purement linéaire $\hbar \omega^{\rm hydro}_\qq = \hbar c q$ et la vitesse du son $c$ à la densité $\rho$ s'obtient par la même formule hydrodynamique que plus haut à l'équation \eqref{eq:vitesseson}.
Nous isolons ensuite la partie cubique du hamiltonien \eqref{eq:hamiltonienhydro}
\begin{equation}
\hat{H}_{\rm hydro}^{(3)} = l^3\sum_{\rr} \left[ \frac{\hbar^2}{2m} \nabla\delta\hat{\phi}\cdot \delta\hat{\rho} \nabla\delta\hat{\phi}
+ \frac{1}{6} \frac{\dd^2\mu}{\dd\rho^2} (\delta\hat{\rho})^3\right],
\label{eq:hamiltonienhydro3}
\end{equation}
et y injectons les développements (\ref{eq:devmod1}) et (\ref{eq:devmod2}). Ceci conduit au résultat
\begin{widetext}
\begin{multline}
\hat{H}_{\rm hydro}^{(3)} = \frac{m{c}^2}{\bb{{\rho}L^3}^{1/2}} 
\sum_{\qq_1,\qq_2,\qq_3\in\mathcal{D}^*}  \bbcrol{ \delta_{\qq_1 + \qq_2,\qq_3} \mathcal{A}^{2\leftrightarrow 1}_{\rm hydro}({\qq_1,\qq_2;\qq_3}) 
\left(\hat{b}_{\qq_1}^\dagger \hat{b}_{\qq_2}^\dagger \hat{b}_{\qq_3} +\mbox{h.c.}\right) }  \\
+\bbcror{  \delta_{\qq_1 +\qq_2 + \qq_3,\zero}\mathcal{A}^{3\leftrightarrow 0}_{\rm hydro} ({\qq_1,\qq_2,\qq_3})   \left(\hat{b}^\dagger_{\qq_1}\hat{b}^\dagger_{\qq_2}\hat{b}_{\qq_3}^\dagger +\mbox{h.c.}\right) }
\label{eq:H3hydro}
\end{multline}
\end{widetext}
où les amplitude de couplage des processus $2\leftrightarrow1$ et $3\leftrightarrow0$ valent respectivement
\be
\mathcal{A}^{2\leftrightarrow 1}_{\rm hydro}({\qq_1,\qq_2;\qq_3})  \!=\!  \sqrt{\frac{\hbar^3 {q}_1{q}_2 {q}_3}{32m^3c^3}} \bb{3\Lambda_{\rm F}\!+\!u_{12}\!+\!u_{13}\!+\!u_{23} } \label{eq:Ahydro21} 
\ee
\be
\mathcal{A}^{3\leftrightarrow0}_{\rm hydro}({\qq_1,\qq_2,\qq_3})  \!=\!  \frac{1}{3} \sqrt{\frac{\hbar^3 {q}_1{q}_2 {q}_3}{32m^3c^3}} \bb{3\Lambda_{\rm F}  \!+\!u_{12}\!+\!u_{13}\!+\!u_{23} }
\label{eq:Ahydro30}
\ee
en termes du paramètre $\Lambda_{\rm F}$ défini à l'équation \eqref{eq:lambdaF} et du cosinus de l'angle entre les vecteurs $\qq_i$ et $\qq_j$,
\be
u_{ij}=\frac{\qq_i \cdot \qq_j}{q_i q_j}
\ee
L'amplitude $2\leftrightarrow1$ est manifestement différente de l'amplitude \eqref{eq:finaleA} obtenue par l'approche microscopique (en particulier elle ne diverge pas à la limite des grandes longueurs d'onde). 
{\rosso 
Ceci peut surprendre \`a premi\`ere vue. On pourrait s'attendre en effet \`a ce que les composantes
du champ sur les modes collectifs dans la th\'eorie microscopique soient l'\'equivalent exact des champs
de l'hydrodynamique quantique~; ainsi, le d\'eveloppement modal des champs de phase et de densit\'e
(\ref{eq:devmodrho},\ref{eq:devmodtheta}) de la th\'eorie microscopique semble reproduire,
apr\`es quantification des variables normales $b_\qq$, celui (\ref{eq:devmod1},\ref{eq:devmod2})
de l'hydrodynamique quantique (\`a un facteur global $2$ pr\`es dans l'\'equation (\ref{eq:devmodtheta})
d\^u au fait que la phase $\theta(\rr,t)$ du champ d'appariement est conjugu\'e \`a la densit\'e
de paires plut\^ot qu'\`a la densit\'e de particules), si bien que les op\'erateurs $\hat{b}_\qq$
dans l'\'equation (\ref{eq:BqQuantique}) devraient \^etre les \'equivalents exacts de ceux des \'equations
(\ref{eq:devmod1},\ref{eq:devmod2}).

Le d\'esaccord constat\'e sur les amplitudes de couplage \`a trois phonons peut recevoir deux
explications~: $(i)$ la plus naturelle, l'\'ecart entre les hamiltoniens du mod\`ele microscopique
et de l'hydrodynamique, $(ii)$ la moins \'evidente, le fait que les d\'eveloppements modaux (\ref{eq:devmodrho},\ref{eq:devmodtheta}) du mod\`ele microscopique sont valables seulement au premier ordre en $\beta$, 
alors que ceux (\ref{eq:devmod1},\ref{eq:devmod2}) de l'hydrodynamique quantique sont exacts, ce qui
fait que les op\'erateurs $\hat{b}_\qq$ des deux mod\`eles ne co\"\i ncident qu'au premier ordre en $\beta$.
La premi\`ere explication repose sur le fait que l'hydrodynamique quantique n'est qu'une th\'eorie
effective de basse \'energie qui ne vaut qu'aux grandes longueurs d'onde. 
La seconde explication repose sur le fait que l'hydrodynamique quantique introduit directement les
champs de densit\'e $\hat{\rho}(\rr,t)$ et de phase $\hat{\phi}(\rr,t)$ comme variables canoniquement
conjugu\'ees, en une repr\'esentation phase-module, alors que la th\'eorie microscopique introduit
les champs de paires $\beta_{\kk,\kk'}(t)$ et $\beta^*_{\kk,\kk'}(t)$ comme variables conjugu\'ees,
si bien que les champs de densit\'e $\rho(\rr,t)$ et de phase $\theta(\rr,t)$ ne sont que des champs
d\'eriv\'es fonctions non lin\'eaires des champs primitifs $\beta$ et $\beta^*$, voir les \'equations
(\ref{eq:defrhomicro},\ref{eq:defthetamicro}).
Pour d\'eterminer qualitativement la contribution relative de ces deux explications au d\'esaccord 
observ\'e sur les amplitudes de couplage, le plus simple est refaire toute l'\'etude pour un
gaz de bosons en interaction faible, ce qui conduit \`a des calculs moins lourds et rejoint le
cas fermionique dans la limite CBE $k_\mathrm{F} a\to 0^+$, voir l'annexe~\ref{appen:diffmicrohydro}.
Ceci nous permet de conclure que l'explication $(ii)$ est la bonne~: le faible \'ecart entre les
op\'erateurs $\hat{b}_\qq$ dans les th\'eories microscopique et hydrodynamique explique la divergence
de l'amplitude de couplage dans le mod\`ele microscopique hors de la couche de masse dans la limite
des faibles nombres d'onde $q_i$. Au contraire, l'explication $(i)$ ne rend compte
que d'une correction $O(q_i^{7/2})$ n\'egligeable \`a l'amplitude de couplage $\approx q_i^{3/2}$
de l'hydrodynamique.}

On vérifie cependant aisément que les résultats de la th\'eorie microscopique et de l'hydrodynamique quantique pour l'amplitude de couplage $2\leftrightarrow 1$
sont équivalents lorsque l'énergie est conservée, \textit{cf}. l'équation \eqref{eq:ACM} et l'\'equation \eqref{eq:Ahydro21} \'ecrite sur la couche de masse comme suit, 
\be
\mathcal{A}^{2\leftrightarrow 1}_{\rm hydroCdM}({\qq_1,\qq_2;\qq_3})  \!=\!
3(1+\Lambda_{\rm F}) \sqrt{\frac{\hbar^3 {q}_1{q}_2 {q}_3}{32m^3c^3}}
\label{eq:Ahydro21surlacouche}
\ee
En effet, dans la théorie hydrodynamique, les trois vecteurs d'onde $\qq_1$, $\qq_2$ et $\qq_3$ sont colinéaires et de même sens sur la couche de masse, en vertu du cas d'\'egalit\'e $|\qq_1+\qq_2|=q_1+q_2$ dans l'inégalité triangulaire.
Cette \'equivalence des r\'esultats est fort satisfaisante pour la suite puisque, comme nous le verrons, les taux d'amortissement $\Gamma_\qq$ de Beliaev-Landau des phonons dans la section \ref{sec:ccbld} font 
intervenir les amplitudes de couplage seulement sur la couche de masse. De ce point de vue, 
la th\'eorie microscopique et l'hydrodynamique quantique 
conduisent \`a la m\^eme physique de basse \'energie, comme on pouvait s'y attendre.

\subsection{Processus à quatre phonons}
\label{sec:4phonons}


Nous nous intéressons maintenant au processus à quatre phonons $2\leftrightarrow2$ lorsque la branche d'excitation $q\mapsto\omega_\qq$ est concave aux faibles nombres d'onde.
Dans ce cas, c'est en effet le processus à plus petit nombre de phonons qui puisse être résonnant, les processus $1\leftrightarrow 2$ et $1\leftrightarrow 3$ étant interdits \cite{KCS2016} par la conservation de l'énergie. Ce processus est aussi plus riche du point de vue théorique puisqu'il s'obtient entre autres par des processus intermédiaires $1\leftrightarrow 2$ ou $3\leftrightarrow0$ virtuels et non résonnants. Pour cette raison, l'équivalence de l'hydrodynamique, qui décrit mal les processus $1\leftrightarrow 2$ hors de la couche de masse, et de l'approche microscopique est tout sauf évidente.
Dans cette section, nous donnons la prédiction hydrodynamique pour l'amplitude de couplage $2\leftrightarrow2$ effective, incluant les processus virtuels puis nous validons ce résultat par un modèle microscopique. Plutôt que le modèle microscopique fermionique particulièrement lourd, nous optons pour un modèle de bosons avec des interactions de portée non nulle, ajustées pour qu'il présente une branche d'excitation concave aux faibles $q$.


\subsubsection{Amplitude de transition}


Nous cherchons l'amplitude de transition entre un état initial d'énergie $E_i$, état de Fock quelconque de quasi-particules bosoniques
\be
\ket{i}=\ket{(n_\qq)_{\qq\in \mathcal{D}}}
\ee
et un état final d'énergie $E_f$ où deux phonons de vecteur d'onde $\qq_1$ et $\qq_2$ ont été annihilés et remplacés par des phonons de vecteur d'onde $\qq_3$ et $\qq_4$~:
\be
\ket{f}= \frac{\hat{b}_{\qq_3}^\dagger \hat{b}_{\qq_4}^\dagger \hat{b}_{\qq_1} \hat{b}_{\qq_2}}{\sqrt{n_{\qq_1} n_{\qq_2} (1+n_{\qq_3})(1+n_{\qq_4})}} \ket{i}
\ee
Quel que soit le modèle considéré, le hamiltonien se met sous  la forme
\be
\hat{H}=E_0+\hat{H}_2+\hat{H}_3+\hat{H}_4+\ldots
\label{eq:hamiltoniendev}
\ee
où $E_0$ est une constante, $\hat{H}_2$ le hamiltonien quadratique libre des quasi-particules et $\hat{H}_3$ et $\hat{H}_4$ les développements aux ordres suivants.
Comme $\hat{H}_3$ ne couple pas $\ket{i}$ à $\ket{f}$ directement, nous calculons le couplage au second ordre de la théorie des perturbations, ce qui revient à traiter $\hat{H}_4$ au premier ordre et  $\hat{H}_3$ au second ordre pour construire un hamiltonien effectif \cite{Cohen}~:
\be
\bra{f} \hat{H}^{2\leftrightarrow2,\rm eff} \ket{i} = \bra{f} \hat{H}_4 \ket{i}+\sum_\lambda \frac{\bra{f} \hat{H}_3 \ket{\lambda} \bra{\lambda} \hat{H}_3 \ket{i}}{E_i-E_\lambda}\equiv\mathcal{A}_{i\to f}
\label{eq:Aif}
\ee
Il y a 6 états intermédiaires $\ket{\lambda}$, numérotés de I à VI, accessibles à température nulle c'est-\`a-dire
lorsque tous les modes $\qq\neq\qq_1,\qq_2$ sont initialement vides. Ce sont des états où un phonon virtuel est créé et réabsorbé par un processus à trois phonons non résonnant. Nous les représentons par les diagrammes de gauche sur la figure \ref{fig:diagrammes}, avec une ligne tiretée pour figurer le phonon intermédiaire virtuel. \`A ces six états intermédiaires $\lambda$=I--VI correspondent les diagrammes $\lambda$=I'--VI', représentés à droite sur la figure \ref{fig:diagrammes}, où le phonon intermédiaire a le même vecteur d'onde mais est détruit puis recréé plutôt que créé puis détruit.
 Ces états intermédiaires ne sont possibles qu'à température non nulle puisque le phonon intermédiaire doit être initialement présent dans $\ket{i}$.

\tikzset{->-/.style={decoration={
  markings,
  mark=at position .5 with {\arrow{>}}},postaction={decorate}}}
\tikzset{-<-/.style={decoration={
  markings,
  mark=at position .5 with {\arrow{<}}},postaction={decorate}}}

\setlength{\tabcolsep}{8pt}
\begin{figure}[htb]
\begin{tabular}{V{0.44\linewidth}V{0.44\linewidth}}
{
\begin{tikzpicture}[scale=1.45]
\coordinate (A) at (-0.5,0);
\coordinate (B) at (0.5,0);
\def\xA{-0.5}
\def\yA{0}
\def\xB{0.5}
\def\yB{0}
\draw (A) node{\textbullet};
\draw (B) node{\textbullet};
\draw[->-,dashed] (A) -- node[above]{$\qq_{\rm S}$} (B);
\draw[->-] (\xA-0.5,\yA+0.5) -- node[above=0.1cm]{$\qq_{1}$} (A);
\draw[->-] (\xA-0.5,\yA-0.5) -- node[below=0.1cm]{$\qq_{2}$} (A);
\draw[->-] (B) -- node[above=0.1cm]{$\qq_{3}$} (\xB+0.5,\yB+0.5);
\draw[->-] (B) -- node[below=0.1cm]{$\qq_{4}$} (\xB+0.5,\yB-0.5);
\draw (\xA*0.5+\xB*0.5,\yA*0.5+\yB*0.5-0.6) node{I};
\end{tikzpicture}
} &
{\centering \begin{tikzpicture}[scale=1.45]
\coordinate (A) at (-0.6,0);
\coordinate (B) at (0.6,0);
\def\xA{-0.6}
\def\yA{0}
\def\xB{0.6}
\def\yB{0}
\draw (A) node{\textbullet};
\draw (B) node{\textbullet};
\draw[->-] (\xB-0.5,\yB+0.5) -- node[above=0.1cm]{$\qq_{1}$} (B);
\draw[->-] (\xB-0.5,\yB-0.5) -- node[below=0.1cm]{$\qq_{2}$} (B);
\draw[->-,double] (B)  -- node[above]{$\qq_{\rm S}$} (\xB+0.5,\yB);
\draw[->-,double] (\xA-0.5,\yA) -- node[above]{$\qq_{\rm S}$} (A);
\draw[->-] (A)  -- node[above=0.1cm]{$\qq_{3}$} (\xA+0.5,\yA+0.5) ;
\draw[->-] (A)  -- node[below=0.1cm]{$\qq_{4}$} (\xA+0.5,\yA-0.5);
\draw (\xA*0.5+\xB*0.5,\yA*0.5+\yB*0.5-0.7) node{I'};
\end{tikzpicture}} \tabularnewline
{
\begin{tikzpicture}[scale=1.2]
\coordinate (A) at (-0.4,0.3);
\coordinate (B) at (0.4,-0.3);
\def\xA{-0.4}
\def\yA{0.3}
\def\xB{0.4}
\def\yB{-0.3}
\draw (A) node{\textbullet};
\draw (B) node{\textbullet};
\draw[->-,dashed] (A) -- node[right=0.1cm]{$\qq_{\rm t1}$} (B);
\draw[->-] (\xA-1,\yA) -- node[above]{$\qq_{1}$} (A);
\draw[->-]  (A) -- node[above left]{$\qq_{3}$} (\xA+0.5,\yA+0.7);
\draw[->-] (\xB-0.5,\yB-0.7) -- node[below right]{$\qq_{2}$} (B);
\draw[->-] (B) -- node[below]{$\qq_{4}$} (\xB+1,\yB);
\draw (\xA*0.5+\xB*0.5-0.7,\yA*0.5+\yB*0.5-0.7) node{II};
\end{tikzpicture}
} &
{\begin{tikzpicture}[scale=1.45]
\coordinate (A) at (0.6,0);
\coordinate (B) at (-0.6,0);
\def\xA{0.6}
\def\yA{0}
\def\xB{-0.6}
\def\yB{0}
\draw (A) node{\textbullet};
\draw (B) node{\textbullet};
\draw[->-,double] (\xB-0.5,\yB+0.5) -- node[above=0.1cm]{$\qq_{\rm t1}$} (B);
\draw[->-] (\xB-0.5,\yB-0.5) -- node[below=0.1cm]{$\qq_{2}$} (B);
\draw[->-] (B)  -- node[above]{$\qq_{\rm 4}$} (\xB+0.5,\yB);
\draw[->-,] (\xA-0.5,\yA) -- node[above]{$\qq_{1}$} (A);
\draw[->-] (A)  -- node[above=0.1cm]{$\qq_{3}$} (\xA+0.5,\yA+0.5) ;
\draw[->-,double] (A)  -- node[below=0.1cm]{$\qq_{\rm t1}$} (\xA+0.5,\yA-0.5);
\draw (\xA*0.5+\xB*0.5,\yA*0.5+\yB*0.5-0.5) node{II'};
\end{tikzpicture}} \tabularnewline
{
\begin{tikzpicture}[scale=1.2]
\coordinate (A) at (-0.4,0.3);
\coordinate (B) at (0.4,-0.3);
\def\xA{-0.4}
\def\yA{0.3}
\def\xB{0.4}
\def\yB{-0.3}
\draw (A) node{\textbullet};
\draw (B) node{\textbullet};
\draw[->-,dashed] (A) -- node[right=0.1cm]{$-\qq_{\rm t1}$} (B);
\draw[->-] (\xA-1,\yA) -- node[above]{$\qq_{2}$} (A);
\draw[->-]  (A) -- node[above left]{$\qq_{4}$} (\xA+0.5,\yA+0.7);
\draw[->-] (\xB-0.5,\yB-0.7) -- node[below right]{$\qq_{1}$} (B);
\draw[->-] (B) -- node[below]{$\qq_{3}$} (\xB+1,\yB);
\draw (\xA*0.5+\xB*0.5-0.7,\yA*0.5+\yB*0.5-0.7) node{III};
\end{tikzpicture}
} &
{\begin{tikzpicture}[scale=1.45]
\coordinate (A) at (0.6,0);
\coordinate (B) at (-0.6,0);
\def\xA{0.6}
\def\yA{0}
\def\xB{-0.6}
\def\yB{0}
\draw (A) node{\textbullet};
\draw (B) node{\textbullet};
\draw[->-,double] (\xB-0.5,\yB+0.5) -- node[above=0.1cm]{$-\qq_{\rm t1}$} (B);
\draw[->-] (\xB-0.5,\yB-0.5) -- node[below=0.1cm]{$\qq_{1}$} (B);
\draw[->-] (B)  -- node[above]{$\qq_{\rm 3}$} (\xB+0.5,\yB);
\draw[->-,] (\xA-0.5,\yA) -- node[above]{$\qq_{2}$} (A);
\draw[->-] (A)  -- node[above=0.1cm]{$\qq_{4}$} (\xA+0.5,\yA+0.5) ;
\draw[->-,double] (A)  -- node[below=0.1cm]{$-\qq_{\rm t1}$} (\xA+0.5,\yA-0.5);
\draw (\xA*0.5+\xB*0.5,\yA*0.5+\yB*0.5-0.5) node{III'};
\end{tikzpicture}} \tabularnewline
{\begin{tikzpicture}[scale=1.2]
\coordinate (A) at (-0.4,0.3);
\coordinate (B) at (0.4,-0.3);
\def\xA{-0.4}
\def\yA{0.3}
\def\xB{0.4}
\def\yB{-0.3}
\tikzset{->-/.style={decoration={
  markings,
  mark=at position .5 with {\arrow{>}}},postaction={decorate}}}
\tikzset{-<-/.style={decoration={
  markings,
  mark=at position .5 with {\arrow{<}}},postaction={decorate}}}
\draw (A) node{\textbullet};
\draw (B) node{\textbullet};
\draw[->-,dashed] (A) -- node[right=0.1cm]{$\qq_{\rm t2}$} (B);
\draw[->-] (\xA-1,\yA) -- node[above]{$\qq_{1}$} (A);
\draw[->-]  (A) -- node[above left]{$\qq_{4}$} (\xA+0.5,\yA+0.7);
\draw[->-] (\xB-0.5,\yB-0.7) -- node[below right]{$\qq_{2}$} (B);
\draw[->-] (B) -- node[below]{$\qq_{3}$} (\xB+1,\yB);
\draw (\xA*0.5+\xB*0.5-0.7,\yA*0.5+\yB*0.5-0.7) node{IV};
\end{tikzpicture}} &
{\begin{tikzpicture}[scale=1.45]
\coordinate (A) at (0.6,0);
\coordinate (B) at (-0.6,0);
\def\xA{0.6}
\def\yA{0}
\def\xB{-0.6}
\def\yB{0}
\draw (A) node{\textbullet};
\draw (B) node{\textbullet};
\draw[->-,double] (\xB-0.5,\yB+0.5) -- node[above=0.1cm]{$\qq_{\rm t2}$} (B);
\draw[->-] (\xB-0.5,\yB-0.5) -- node[below=0.1cm]{$\qq_{2}$} (B);
\draw[->-] (B)  -- node[above]{$\qq_{\rm 3}$} (\xB+0.5,\yB);
\draw[->-] (\xA-0.5,\yA) -- node[above]{$\qq_{1}$} (A);
\draw[->-] (A)  -- node[above=0.1cm]{$\qq_{4}$} (\xA+0.5,\yA+0.5) ;
\draw[->-,double] (A)  -- node[below=0.1cm]{$\qq_{\rm t2}$} (\xA+0.5,\yA-0.5);
\draw (\xA*0.5+\xB*0.5,\yA*0.5+\yB*0.5-0.5) node{IV'};
\end{tikzpicture}} \tabularnewline
{\begin{tikzpicture}[scale=1.2]
\coordinate (A) at (-0.4,0.3);
\coordinate (B) at (0.4,-0.3);
\def\xA{-0.4}
\def\yA{0.3}
\def\xB{0.4}
\def\yB{-0.3}
\tikzset{->-/.style={decoration={
  markings,
  mark=at position .5 with {\arrow{>}}},postaction={decorate}}}
\tikzset{-<-/.style={decoration={
  markings,
  mark=at position .5 with {\arrow{<}}},postaction={decorate}}}
\draw (A) node{\textbullet};
\draw (B) node{\textbullet};
\draw[->-,dashed] (A) -- node[right=0.1cm]{$-\qq_{\rm t2}$} (B);
\draw[->-] (\xA-1,\yA) -- node[above]{$\qq_{2}$} (A);
\draw[->-]  (A) -- node[above left]{$\qq_{3}$} (\xA+0.5,\yA+0.7);
\draw[->-] (\xB-0.5,\yB-0.7) -- node[below right]{$\qq_{1}$} (B);
\draw[->-] (B) -- node[below]{$\qq_{4}$} (\xB+1,\yB);
\draw (\xA*0.5+\xB*0.5-0.7,\yA*0.5+\yB*0.5-0.7) node{V};
\end{tikzpicture}} &
{\begin{tikzpicture}[scale=1.45]
\coordinate (A) at (0.6,0);
\coordinate (B) at (-0.6,0);
\def\xA{0.6}
\def\yA{0}
\def\xB{-0.6}
\def\yB{0}
\draw (A) node{\textbullet};
\draw (B) node{\textbullet};
\draw[->-,double] (\xB-0.5,\yB+0.5) -- node[above=0.1cm]{$-\qq_{\rm t2}$} (B);
\draw[->-] (\xB-0.5,\yB-0.5) -- node[below=0.1cm]{$\qq_{1}$} (B);
\draw[->-] (B)  -- node[above]{$\qq_{\rm 4}$} (\xB+0.5,\yB);
\draw[->-] (\xA-0.5,\yA) -- node[above]{$\qq_{2}$} (A);
\draw[->-] (A)  -- node[above=0.1cm]{$\qq_{3}$} (\xA+0.5,\yA+0.5) ;
\draw[->-,double] (A)  -- node[below=0.1cm]{$-\qq_{\rm t2}$} (\xA+0.5,\yA-0.5);
\draw (\xA*0.5+\xB*0.5,\yA*0.5+\yB*0.5-0.5) node{V'};
\end{tikzpicture}} \tabularnewline
{\begin{tikzpicture}[scale=1.45]
\coordinate (A) at (-0.5,-0.3);
\coordinate (B) at (0.5,0.3);
\def\xA{-0.5}
\def\yA{-0.3}
\def\xB{0.5}
\def\yB{0.3}
\draw (A) node{\textbullet};
\draw (B) node{\textbullet};
\draw[->-,dashed] (A) -- node[left=0.1cm]{$-\qq_{\rm S}$} (B);
\draw[->-] (A) -- node[below]{$\qq_{3}$} (\xA+1,\yA);
\draw[->-] (A) -- node[below]{$\qq_{4}$} (\xA+0.7,\yA-0.7);
\draw[->-] (\xB-1,\yB) -- node[above]{$\qq_{1}$} (B);
\draw[->-] (\xB-0.7,\yB+0.7) -- node[above]{$\qq_{2}$} (B);
\draw (\xA*0.5+\xB*0.5-0.7,\yA*0.5+\yB*0.5-0.7) node{VI};
\end{tikzpicture}} &
{\begin{tikzpicture}[scale=1.45]
\coordinate (A) at (-0.3,0);
\coordinate (B) at (0.3,0);
\def\xA{-0.3}
\def\yA{0}
\def\xB{0.3}
\def\yB{0}
\draw (A) node{\textbullet};
\draw (B) node{\textbullet};
\draw[->-] (\xA-0.5,\yA) -- node[above left]{$\qq_{1}$} (A);
\draw[->-] (\xA-0.5,\yA+0.5) -- node[above=0.2cm]{$\qq_{2}$} (A);
\draw[->-,double] (\xA-0.5,\yA-0.5) -- node[left=0.1cm]{$-\qq_{\rm S}$} (A);
\draw[->-] (B) -- node[above right]{$\qq_{4}$} (\xB+0.5,\yB);
\draw[->-] (B) -- node[above=0.2cm]{$\qq_{3}$} (\xB+0.5,\yB+0.5);
\draw[->-,double] (B) -- node[right=0.1cm]{$-\qq_{\rm S}$} (\xB+0.5,\yB-0.5);
\draw (\xA*0.5+\xB*0.5,\yA*0.5+\yB*0.5-0.5) node{VI'};
\end{tikzpicture}} \tabularnewline
\end{tabular}
\caption{Diagrammes du second ordre pour le processus à $4$ phonons $(\qq_1,\qq_2)\to(\qq_3,\qq_4)$ avec $\qq_1+\qq_2=\qq_3+\qq_4$ vu comme  deux processus à trois phonons successifs. Les vecteurs incidents $\qq_1$ et $\qq_2$ et émergents $\qq_3$ et $\qq_4$ sont représentés par un trait plein fléché. Sur la colonne de gauche les diagrammes incluent un phonon intermédiaire virtuel, représenté par un tiret\'e fléché. Sur la colonne de droite ils incluent un phonon intermédiaire réel, représenté par un trait double fléché. Les deux diagrammes d'une même ligne possèdent le même phonon intermédiaire : en I et I' $\qq_{\rm S}=\qq_1+\qq_2=\qq_3+\qq_4$, en II et II' $\qq_{\rm t1}=\qq_1-\qq_3=\qq_4-\qq_2$, en III et III' $-\qq_{\rm t1}$, en IV et IV' $\qq_{\rm t2}=\qq_1-\qq_4=\qq_3-\qq_2$, en V et V' $-\qq_{\rm t2}$, en VI et VI' $-\qq_{\rm S}$.\label{fig:diagrammes}}
\end{figure}


\subsubsection{Amplitude de couplage $2\leftrightarrow2$ effective}


Nous définissons l'amplitude effective $\mathcal{A}^{2\leftrightarrow2,{\rm eff}}$ du couplage du processus $2\leftrightarrow2$ à partir de l'écriture suivante du hamiltonien effectif 
\begin{multline}
\hat{H}^{2\leftrightarrow2,{\rm eff}} = \frac{mc^2}{\rho L^3} \sum_{\qq_1,\qq_2,\qq_3,\qq_4\in\mathcal{D}}  \delta_{\qq_1+\qq_2,\qq_3+\qq_4}\\  \times \mathcal{A}^{2\leftrightarrow2,{\rm eff}}(\qq_1,\qq_2;\qq_3,\qq_4)   \hat{b}_{\qq_3}^\dagger \hat{b}_{\qq_4}^\dagger \hat{b}_{\qq_1} \hat{b}_{\qq_2}
\label{eq:H2donne2eff}
\end{multline}
dont, par construction, l'élément de matrice entre $\ket{i}$ et $\ket{f}$ est l'amplitude de transition $\mathcal{A}_{i\to f}$. Ceci conduit à poser
\be
\mathcal{A}_{i\to f} = \sqrt{n_{\qq_1} n_{\qq_2} (1+n_{\qq_3})(1+n_{\qq_4})} \frac{4mc^2}{\rho L^3} \mathcal{A}^{2\leftrightarrow2,{\rm eff}}
\ee
où le facteur $4$ est un simple facteur de comptage. 

Nous envisageons à partir de maintenant un processus sur la couche de masse, c'est-à-dire vérifiant la relation
\be
\omega_1+\omega_2=\omega_3+\omega_4,
\ee
où nous nous permettons à nouveau le raccourci $\omega_i=\omega_{\qq_i}$, $i=1,2,3,4$.
Dans ce cas, une simplification intervient entre chaque diagramme $\lambda$=I,II,III,IV,V,VI et son correspondant $\lambda'$=I',II',III',IV',V',VI' sur la colonne de droite de la figure \ref{fig:diagrammes}, qui permet de se ramener au cas à température nulle. En effet, d'un diagramme à l'autre les dénominateurs d'énergie sont opposés tandis que les numérateurs ne diffèrent que par un facteur de population du phonon intermédiaire $\qq$, facteur qui vaut $(1+n_\qq)$ pour les  diagrammes où le phonon intermédiaire est d'abord virtuellement créé et $n_\qq$ pour ceux où il est d'abord détruit. Par exemple pour I et I' nous avons $E_i-E_{\rm I}=\omega_1+\omega_2-\omega_{\qq_1+\qq_2}$ tandis que $E_i-E_{\rm I'}=\omega_{\qq_3+\qq_4}-\omega_3-\omega_4=-(E_i-E_{\rm I})$ ; dans l'élément de matrice $\bra{f} \hat{H}_3 \ket{{\rm I}} \bra{{\rm I}} \hat{H}_3 \ket{i}$ nous avons le facteur $(1+n_{\qq_1+\qq_2})$  tandis que dans $\bra{f} \hat{H}_3 \ket{{\rm I'}} \bra{{\rm I'}} \hat{H}_3 \ket{i}$ nous avons le facteur $n_{\qq_1+\qq_2}$. Ainsi, en regroupant les diagrammes deux à deux, nous obtenons une amplitude de couplage effective identique à celle que l'on aurait à température nulle, soit avec $n_\qq=0$ pour $\qq\neq\qq_1,\qq_2$.
En fonction de l'amplitude de couplage $2\leftrightarrow2$ directe sur la couche de masse $\mathcal{A}^{2\leftrightarrow2,{\rm dir}}_{\rm Cdm}$ (reliée à $\hat{H}_4$ comme $\mathcal{A}^{2\leftrightarrow2,{\rm eff}}$ est reliée à $\hat{H}^{2\leftrightarrow2,{\rm eff}}$) et des amplitudes $\mathcal{A}^{2\leftrightarrow1}$ et $\mathcal{A}^{3\leftrightarrow0}$ déjà introduites dans la section \ref{sec:3phonons}, cette amplitude effective sur la couche de masse s'écrit alors
\begin{multline}
\!\!\!\!\mathcal{A}^{2\leftrightarrow2,{\rm eff}}_{\rm Cdm}(\qq_{1},\qq_{2};\qq_{3},\qq_{4}) = \mathcal{A}^{2\leftrightarrow2,{\rm dir}}_{\rm Cdm} (\qq_{1},\qq_{2};\qq_{3},\qq_{4}) \\
+ \frac{\mathcal{A}^{2\leftrightarrow1}(\qq_1,\qq_2;\qq_{1}+\qq_{2}) \mathcal{A}^{2\leftrightarrow1}(\qq_3,\qq_4;\qq_{1}+\qq_{2}) }{\check{\omega}_1+\check{\omega}_2-\check{\omega}  _{1+2}}  \\
+ \frac{9\mathcal{A}^{3\leftrightarrow0}(\qq_3,\qq_4,-\qq_{1}-\qq_{2})  \mathcal{A}^{3\leftrightarrow0}(\qq_1,\qq_2,-\qq_{1}-\qq_{2}) }{-(\check{\omega}_1+\check{\omega}_2+\check{\omega}_{1+2})} \\  
+\frac{\mathcal{A}^{2\leftrightarrow1}(\qq_3,\qq_{1}-\qq_3;\qq_{1}) \mathcal{A}^{2\leftrightarrow1}(\qq_{1}-\qq_3,\qq_2;\qq_{4}) }{\check{\omega}_1-\check{\omega}_3-\check{\omega}_{1-3}} \\
+ \frac{\mathcal{A}^{2\leftrightarrow1}(\qq_4,\qq_{3}-\qq_1;\qq_{2}) \mathcal{A}^{2\leftrightarrow1}(\qq_{3}-\qq_1,\qq_1;\qq_{3}) }{\check{\omega}_3-\check{\omega}_1-\check{\omega}_{3-1}}\\
+ \frac{\mathcal{A}^{2\leftrightarrow1}(\qq_4,\qq_{1}-\qq_4;\qq_{1}) \mathcal{A}^{2\leftrightarrow1}(\qq_{1}-\qq_4,\qq_2;\qq_{3}) }{\check{\omega}_1-\check{\omega}_4-\check{\omega}_{1-4}} \\
+ \frac{\mathcal{A}^{2\leftrightarrow1}(\qq_3,\qq_{4}-\qq_1;\qq_{2}) \mathcal{A}^{2\leftrightarrow1}(\qq_{4}-\qq_1,\qq_1;\qq_{4}) }{\check{\omega}_4-\check{\omega}_1-\check{\omega}_{4-1}}
\label{eq:A22eff}
\end{multline}
où nous avons adimensionné les énergies par $mc^2$,
\be
\check{\omega}\equiv\frac{\hbar\omega}{mc^2} \quad \mbox{avec la convention} \quad \omega_{i\pm j}\equiv\omega_{\qq_i\pm\qq_j}.
\ee
Notons que le couplage du diagramme VI (qu'en changeant légèrement l'ordre de la figure \ref{fig:diagrammes} nous écrivons ici juste après celui du diagramme I) présente un facteur de comptage égal à 9.


\subsubsection{Amplitude effective dans l'hydrodynamique}
\label{sec:A22effhydro}

\`A basse température nous cherchons la limite des grandes longueurs d'onde de l'expression \eqref{eq:A22eff} sur la couche de masse. Si l'on veut pouvoir utiliser l'hydrodynamique quantique pour ce faire, il est impératif de la corriger d'abord par analogie avec une approche microscopique. En effet, le spectre obtenu par l'hydrodynamique est purement linéaire ce qui conduit les dénominateurs d'énergie dans \eqref{eq:A22eff} à s'annuler lorsque les vecteurs d'onde sont colinéaires, et fait diverger l'amplitude de couplage effective. Cette divergence est un artefact de la théorie hydrodynamique : en réalité le spectre possède un paramètre de courbure $\gamma$ non nul, ici négatif, qui empêche les dénominateurs de s'annuler. Lorsque les vecteurs d'onde sont presque colinéaires, nous régularisons donc la divergence de l'amplitude de couplage en remplaçant la relation de dispersion hydrodynamique $\hbar\omega_\qq^{\rm hydro}=\hbar c q$ par le développement \eqref{eq:fleur} comme l'ont fait à l'origine Landau et Khalatnikov \cite{Khalatnikov1949}. \`A l'argument d'autorité, nous ajouterons une justification microscopique : pour des vecteurs d'onde colinéaires, notre modèle microscopique bosonique (\textit{cf}. la sous-section \ref{sec:modelebosons}) donne un \'equivalent aux grandes longueurs d'onde de la formule \eqref{eq:A22eff} 
qui coïncide avec celui obtenu par l'hydrodynamique modifiée de Landau et Khalatnikov. Ajouter à la main la courbure du spectre est donc une façon de réconcilier l'approche hydrodynamique avec le résultat microscopique.

\'Etablissons d'abord l'amplitude directe du processus $2\leftrightarrow2$, celle qui s'obtient à partir des termes quartiques du hamiltonien
\begin{equation}
\hat{H}_{\rm hydro}^{(4)} = \frac{1}{24} \frac{\dd^3\mu}{\dd\rho^3} l^3\sum_{\rr}  \delta\hat{\rho}^4
\label{eq:H4hydro}
\end{equation}
Comme précédemment, nous injectons les développements (\ref{eq:devmod1}) et (\ref{eq:devmod2}) dans $\hat{H}_{\rm hydro}^{(4)}$.
Nous ne gardons que les termes $2\leftrightarrow2$ $\hat{b}_{\qq_3}^\dagger  \hat{b}_{\qq_4}^\dagger \hat{b}_{\qq_1} \hat{b}_{\qq_2}$~:
\begin{multline}
\hat{H}_{\rm hydro}^{2\leftrightarrow2,{\rm dir}} = \frac{mc^2}{\rho L^3} \sum_{\qq_1,\qq_2,\qq_3,\qq_4\in\mathcal{D}}  \delta_{\qq_1+\qq_2,\qq_3+\qq_4}\\  \times \mathcal{A}^{2\leftrightarrow2,{\rm dir}}_{\rm hydro}(\qq_1,\qq_2;\qq_3,\qq_4)   \hat{b}_{\qq_3}^\dagger \hat{b}_{\qq_4}^\dagger \hat{b}_{\qq_1} \hat{b}_{\qq_2}
\label{eq:H2donne2hydro}
\end{multline}
L'amplitude de couplage $2\leftrightarrow2$ directe adimensionnée est donnée par
\be
\mathcal{A}^{2\leftrightarrow2,{\rm dir}}_{\rm hydro}(\qq_1,\qq_2;\qq_3,\qq_4)  = \frac{{\rm \Sigma_F}}{16} \sqrt{\frac{\hbar^4 q_1 q_2 q_3 q_4}{m^4c^4}}
\label{eq:A22dirhydro}
\ee
où nous avons défini
\be
{\rm \Sigma_F} \equiv \frac{\rho^3}{mc^2}\frac{\dd^3\mu}{\dd\rho^3}
\ee

Nous combinons alors l'amplitude \eqref{eq:A22dirhydro} et nos expressions précédentes des amplitudes $2\leftrightarrow1$ \eqref{eq:Ahydro21} et $3\leftrightarrow0$ \eqref{eq:Ahydro30} pour obtenir
\bea
\mathcal{A}^{2\leftrightarrow2,{\rm eff}}_{\rm hydro\ corr, Cdm}&&\!\!\!\!\!\!\!\!\!(\qq_1,\qq_2;\qq_3,\qq_4) =\frac{1}{16}\sqrt{\frac{\hbar^4\omega_1\omega_2\omega_3\omega_4}{m^4c^8}} \bbl{\vphantom{\frac{1}{2}}{\rm \Sigma_F} } \notag\\
&+&\frac{(\omega_1+\omega_2)^2 A_{1234}+\omega_{1+2}^2 B_{1234}}{(\omega_1+\omega_2)^2-\omega_{1+2}^2} \notag \\
&+&\frac{(\omega_1-\omega_3)^2 A_{1324}+\omega_{1-3}^2 B_{1324}}{(\omega_1-\omega_3)^2-\omega_{1-3}^2} \notag \\
&+&\bbr{\frac{(\omega_1-\omega_4)^2 A_{1423}+\omega_{1-4}^2 B_{1423}}{(\omega_1-\omega_4)^2-\omega_{1-4}^2}} \label{eq:A2donne2effhydro}
\eea
où l'indice \g{corr} signifie que l'on va au-delà de l'approximation hydrodynamique dans la relation de dispersion au dénominateur de l'expression \eqref{eq:A2donne2effhydro}, c'est-à-dire qu'on utilise l'approximation cubique \eqref{eq:fleur}. Nous avons introduit les coefficients
\bea
A_{ijkl}&=&(3\Lambda_{\rm F}+u_{ij})(1+u_{kl})+(3\Lambda_{\rm F}+u_{kl})(1+u_{ij})\notag\\
&&+(1+u_{ij})(1+u_{kl}) \\
B_{ijkl}&=&(3\Lambda_{\rm F}+u_{ij}) (3\Lambda_{\rm F}+u_{kl})
\eea


\subsubsection{Amplitude effective dans un gaz de bosons en interaction faible de portée non nulle}

\label{sec:modelebosons}
Pour aborder les processus $2\leftrightarrow2$ par une approche microscopique, et comprendre comment l'hydrodynamique doit être corrigée pour éliminer sa divergence, nous choisissons, plutôt que la théorie variationnelle fermionique de la partie \ref{sec:methode}, rendue particulièrement lourde par la prise en compte des degrés de liberté internes des paires, un modèle de bosons possédant une branche d'excitation concave aux faibles $q$. L'idée est d'utiliser un potentiel d'interaction entre bosons
de portée $b$ suffisamment grande pour donner un d\'epart concave \`a 
la branche d'excitation de Bogolioubov. 
Dans l'annexe \ref{appen:modelebosons}, nous d\'ecrivons le mod\`ele et calculons
l'amplitude de couplage effective $2\leftrightarrow 2$  dans la limite d'interactions faibles
avec la th\'eorie de Bogolioubov. Sur la couche de masse, dans la limite des faibles nombres
d'onde, nous retrouvons alors la pr\'ediction \eqref{eq:A2donne2effhydro} de l'hydrodynamique
quantique corrig\'ee, pour peu qu'on la sp\'ecialise \`a l'\'equation d'\'etat 
\eqref{eq:etatB} des bosons en prenant $\Lambda_{\rm F}={\rm \Sigma_F}=0$ et en effectuant la
substitution $(m,c,\rho)\to(m_{\rm B},c_{\rm B},\rho_{\rm B})$, o\`u $m_{\rm B}$ est la masse
d'un boson et $c_{\rm B}$ est la vitesse du son dans le gaz de bosons de densit\'e
$\rho_{\rm B}$.
La courbure de la branche d'excitation, qui supprime la divergence de l'amplitude de couplage
effective $2\leftrightarrow2$, appara\^\i t naturellement dans le mod\`ele microscopique
et n'a pas besoin d'\^etre ajout\'ee \`a la main. Ceci fournit une justification 
microscopique \`a la prescription de Landau et Khalatnikov \cite{Khalatnikov1949} d\'ecrite au début de la section \ref{sec:A22effhydro}.
Ceci n'avait rien d'évident \textit{a priori} car le terme de couplage $2\leftrightarrow2$ direct
\eqref{eq:A2donne2Bdir} diverge à la limite des grandes longueurs d'onde  d'une façon qui compense subtilement la divergence du terme de couplage virtuel issu du second ordre de la théorie des perturbations. La physionomie du calcul microscopique est donc fort différente de celle du calcul hydrodynamique, dans lequel le terme de couplage direct est nul pour l'équation d'état \eqref{eq:etatB} considérée.


\section{Application : amortissement des phonons dans tout le raccordement CBE-BCS}
\label{sec:amortissement}


\subsection{Expression générale des taux par équation pilote}

Pour calculer le taux d'amortissement des phonons du mode $\qq$, nous voyons ce mode comme un oscillateur harmonique couplé au réservoir que  constitue l'ensemble des autres modes de quasi-particules \cite{CastinSinatra2007}, supposé être à l'équilibre thermique à la température $T$. Nous récrivons donc le hamiltonien effectif de basse énergie sous la forme
\be
\hat{H} = \hbar\omega_\qq \hat{b}_\qq^\dagger \hat{b}_\qq + \bb{\sum_{\qq'\neq\qq} \hbar\omega_{\qq'} \hat{b}_{\qq'}^\dagger \hat{b}_{\qq'}}+(\hat{R}^\dagger \hat{b}_\qq +  \hat{b}_\qq^\dagger \hat{R})+\ldots
\ee
Le premier et le second terme, tous deux issus de $\hat{H}_2$, décrivent l'évolution libre respectivement du mode $\qq$ et du réservoir. Le troisième terme, simple récriture des termes de $\hat{H}_3$ ou de $\hat{H}^{2\leftrightarrow2,\rm eff}$ contenant $\hat{b}_\qq$ ou $\hat{b}_\qq^\dagger$, décrit le couplage entre le réservoir et le mode $\qq$. L'ellipse $\ldots$ concerne les processus non linéaires d'ordre plus élevé et tous ceux qui n'impliquent pas le mode $\qq$, ainsi que les termes de déplacement d'énergie du mode $\qq$, de la forme $\hat{R}' \hat{b}_\qq^\dagger\hat{b}_\qq$ avec $\hat{R}'$ un opérateur du réservoir. Dans l'approximation de Born-Markov,
valable dans la limite de couplage faible et faiblement collisionnelle d'une temp\'erature $T$ et d'un nombre d'onde $q$ tendant vers z\'ero \`a $\hbar c q/k_B T$ non nul fix\'e,
\footnote{{\rosso Lorsque $T$ et $q$ tendent vers z\'ero, les nombres d'onde $q_i$ des phonons interm\'ediaires
contribuant \`a $\Gamma_\qq$  tendent vers z\'ero, ainsi que les amplitudes de couplage phonon-phonon
qui interviennent dans l'op\'erateur $\hat{R}$. Pourtant, le r\'esultat (\ref{eq:Gammaq_epil}) reste
soumis \`a la condition de validit\'e $\omega_\qq \gg \Gamma_{\qq_{\rm th}}$, 
avec $\hbar c q_{\rm th}=k_B T$~: la pulsation $\omega_\qq$ du mode consid\'er\'e doit \^etre bien plus
grande que le taux de relaxation typique des phonons thermiques, ce qui d\'efinit le r\'egime faiblement
collisionnel (par opposition au r\'egime hydrodynamique).}}
l'équation d'évolution du nombre moyen d'excitations $\meanv{\hat{n}_\qq}$ dans le mode $\qq$ s'écrit \cite{Cohen}
\be
\frac{\dd}{\dd t}\meanv{\hat{n}_\qq}=-\Gamma_\qq (\meanv{\hat{n}_\qq}-\bar{n}_\qq)
\ee
où $\bar{n}_{\qq}$ désigne la population du mode $\qq$ à l'équilibre thermique
\be
\bar{n}_{\qq} = \frac{1}{\exp\bb{\frac{\hbar\omega_{\qq}}{k_{\rm B} T}}-1} \label{eq:moyennecanB}
\ee
Le taux d'amortissement $\Gamma_\qq$ est donné par
\be
\Gamma_\qq = \int_{-\infty}^{+\infty} \frac{\dd t}{\hbar^2} \eee^{-\ii\omega_\qq t}\mbox{Tr}_{\rm R}\bb{[\hat{R},\hat{R}^\dagger(t)]\sigma_{\rm R}^{\rm \acute{e}q}(t)}
\label{eq:Gammaq_epil}
\ee
où $\mbox{Tr}_{\rm R}$ désigne la trace sur tous les états possibles du réservoir, $\sigma_{\rm R}^{\rm \acute{e}q}$ est l'opérateur densité du réservoir à l'équilibre thermique, $[\hat{A},\hat{B}]$ est le commutateur des opérateurs $\hat{A}$ et $\hat{B}$ et l'évolution de l'opérateur $\hat{R}^\dagger(t)$ du réservoir est celle en représentation d'interaction donnée par le hamiltonien $\sum_{\qq'\neq\qq} \hbar\omega_{\qq'} \hat{b}_{\qq'}^\dagger \hat{b}_{\qq'}$.

\subsection{Cas convexe : amortissement de Beliaev-Landau}
\label{sec:ccbld}

\begin{figure}[htb] 
\begin{center}
\includegraphics[width=0.49\textwidth,clip=]{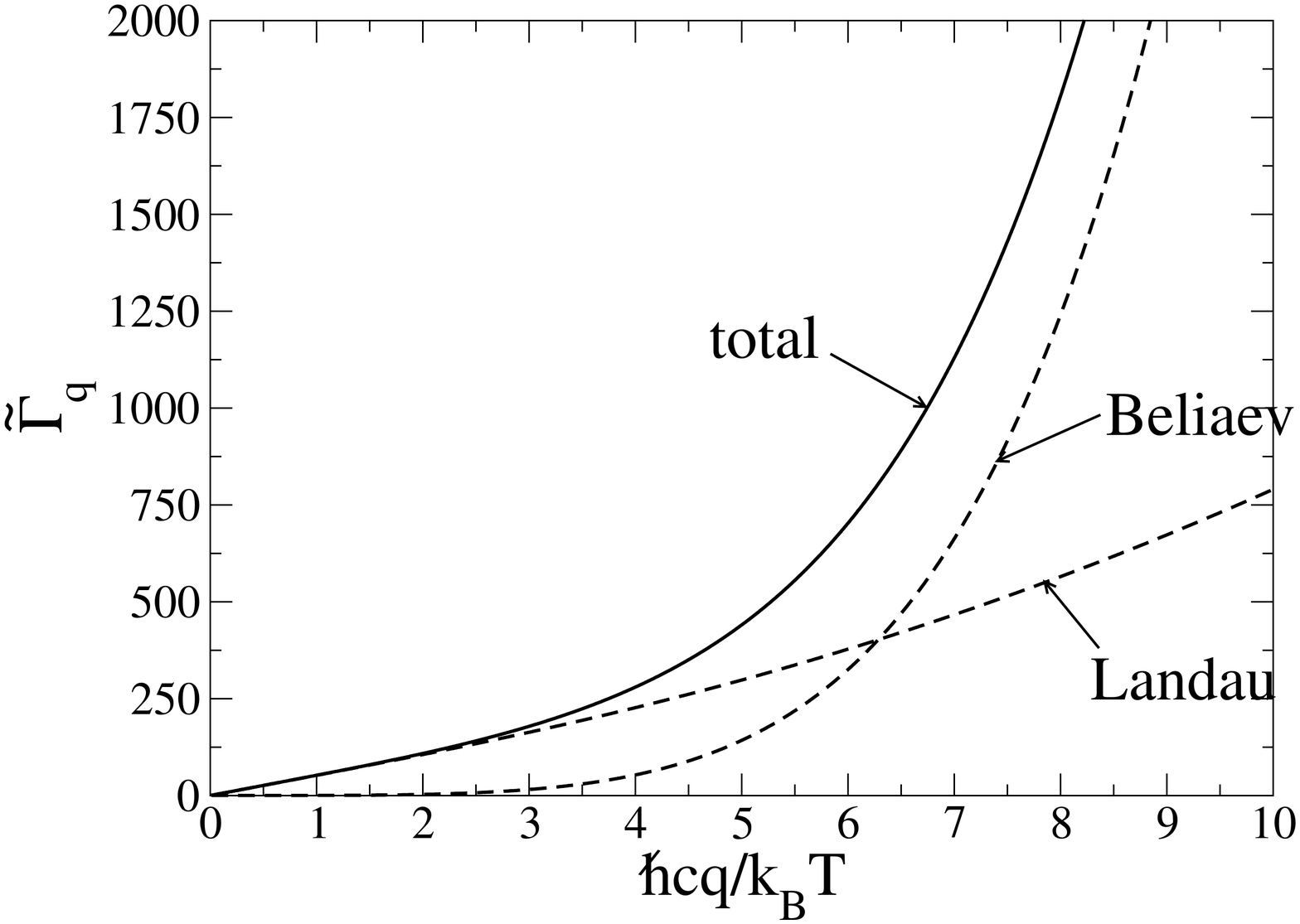}
\end{center}
\caption{ \label{fig:GammaBelLan} Comportement limite à basse température du taux total d'amortissement des phonons en fonction du nombre d'onde adimensionné $\tilde{q}=\hbar c q /k_{\rm B} T$ dans le cas d'une relation de dispersion convexe, en trait plein, somme des taux de Beliaev \eqref{eq:GammaBel} et de Landau \eqref{eq:GammaLan} représentés en tireté. Les taux sont adimensionnés comme dans les équations \eqref{eq:GammaBeladim} et \eqref{eq:GammaLanadim}.}
\end{figure}

Dans le cas d'une relation de dispersion convexe, les termes bilinéaires présents dans $\hat{R}$ conduisent à des processus résonnants et sont donc ceux qui déterminent l'expression de $\Gamma_\qq$ à l'ordre le plus bas en la température. 
{\rouge Il faut donc utiliser dans l'\'equation (\ref{eq:Gammaq_epil}) l'expression suivante de $\hat{R}^\dagger$:
\begin{multline}
\hat{R}^\dagger=\frac{mc^2}{(\rho L^3)^{1/2}} \left[\sum_{\qq_2,\qq_3}\mathcal{A}^{2\leftrightarrow 1}(\qq_2,\qq_3;\qq)
\hat{b}_{\qq_2}^\dagger \hat{b}_{\qq_3}^\dagger \delta_{\qq,\qq_2+\qq_3} \right.\\
\left. +\sum_{\qq_1,\qq_2} 2\mathcal{A}^{2\leftrightarrow 1}(\qq_2,\qq;\qq_1) \hat{b}_{\qq_1}^\dagger \hat{b}_{\qq_2} \delta_{\qq_2+\qq,\qq_1}\right]
\label{eq:Rbellan}
\end{multline}
}
Nous séparons la contribution du processus de Beliaev $(\qq',(\qq-\qq'))\leftrightarrow\qq$ (premier terme dans (\ref{eq:Rbellan})) et de Landau $(\qq,\qq')\leftrightarrow\qq+\qq'$ (second terme dans (\ref{eq:Rbellan}))~:
\bea
\Gamma_q^{\rm Bel}&=&\frac{(mc^2)^2}{2\pi^2\hbar^2 \rho} \int_{\mathbb{R}^3} \dd^3 q' |\mathcal{A}^{2\leftrightarrow1}_{\rm Cdm}(\qq',\qq-\qq';\qq)|^2  \notag \\ 
&\times&\!\!\delta(\omega_{\qq'}+\omega_{\qq-\qq'}-\omega_{\qq})(1+\bar{n}_{\qq-\qq'}+\bar{n}_{\qq'}) \label{eq:GamqBel}\\
\Gamma_q^{\rm Lan}&=&\frac{(mc^2)^2}{\pi^2\hbar^2 \rho} \int_{\mathbb{R}^3} \dd^3 q'  |\mathcal{A}^{2\leftrightarrow1}_{\rm Cdm}(\qq',\qq;\qq'+\qq)|^2 \notag \\ 
&\times& \delta(\omega_{\qq+\qq'}-\omega_{\qq'}-\omega_{\qq})(\bar{n}_{\qq'}-\bar{n}_{\qq'+\qq})
\label{eq:GamqLan}
\eea
\`A basse température, nous pouvons utiliser l'expression \`a l'ordre dominant \eqref{eq:ACM} de l'amplitude de couplage $2\leftrightarrow1$ sur la couche de masse, ou de mani\`ere \'equivalente l'expression hydrodynamique (\ref{eq:Ahydro21surlacouche}) sur la couche de masse. Après intégration sur le vecteur d'onde $\qq'$ en coordonnées sphériques d'axe polaire $\qq$, nous obtenons les équivalents\footnote{Si l'on utilise directement la relation de dispersion lin\'eaire $\omega_\qq=cq$ de l'hydrodynamique quantique dans l'argument de la distribution de Dirac des \'equations (\ref{eq:GamqBel},\ref{eq:GamqLan}), on tombe sur l'intégrale polaire $\int_{-1}^1 \dd u \delta(u-1)$ ($u$ est le cosinus de l'angle entre $\qq$ et $\qq'$, l'\'energie n'est conserv\'ee que lorsque $u=1$) dont on pourrait penser à tort qu'elle vaut $1/2$. En tenant compte de la stricte convexité de la relation de dispersion au moyen du param\`etre de courbure $\gamma>0$, on tombe sur la valeur correcte $\int_{-1}^1 \dd u \delta(u-u_0)=1$, où la racine $u_0\in]-1,1[$ assurant la conservation de l'\'energie tend vers $1$ \`a basse temp\'erature. 
Ceci montre bien pourquoi les r\'esultats (\ref{eq:GammaBeladim},\ref{eq:GammaLanadim}) 
ne d\'ependent pas de la valeur 
de $\gamma$, un Dirac ayant toujours une masse totale unit\'e. Ils d\'ependent cependant, si l'on veut, du signe de $\gamma$~: si $\gamma$ \'etait n\'egatif, il n'y aurait dans l'intervalle
d'int\'egration aucune racine $u_0$ assurant la conservation de l'\'energie et l'int\'egrale
polaire serait nulle, ce qui signifierait l'absence d'amortissement Beliaev-Landau pour une relation
de dispersion concave. }~:
\setcounter{notepasundemi}{\thefootnote}
\bea
\!\!\!\!\!\!\!\!\!\!\!\!\Gamma_q^{\rm Bel} \!\!\!&\underset{T\to0}{\sim}& \!\!\! \frac{9(1+\Lambda_{\rm F})^2}{32\pi} \frac{mc^2}{\hbar\rho} \bb{\frac{mc}{\hbar}}^3 \bb{\frac{k_{\rm B}T}{mc^2}}^5 \tilde{\Gamma}^{\rm Bel} (\tilde{q}) \label{eq:GammaBeladim}\\
\!\!\!\!\!\!\!\!\!\!\!\!\Gamma_q^{\rm Lan} \!\!\!&\underset{T\to0}{\sim}&\!\!\! \frac{9(1+\Lambda_{\rm F})^2}{32\pi} \frac{mc^2}{\hbar\rho} \bb{\frac{mc}{\hbar}}^3 \bb{\frac{k_{\rm B}T}{mc^2}}^5 \tilde{\Gamma}^{\rm Lan} (\tilde{q}) \label{eq:GammaLanadim}
\eea
où nous avons adimensionné les nombres d'onde par le nombre d'onde thermique typique comme suit~:
\be
\tilde{q}=\frac{\hbar c q}{k_{\rm B}T}
\label{eq:tildeq}
\ee
et où la limite $T\to0$ est prise à $\tilde{q}$ fixé.
Les fonctions $\tilde{\Gamma}^{\rm Bel}$ et $\tilde{\Gamma}^{\rm Lan}$ sont des fonctions universelles de $\tilde{q}$ s'exprimant en termes des fonctions de Bose $g_\alpha(z)=\sum_{n=1}^{+\infty} z^n/n^\alpha$, aussi appelées fonctions polylogarithmes ${\rm Li}_\alpha(z)$,
 et de la fonction $\zeta$ de Riemann, $\zeta(\alpha)=g_\alpha(1)$,\footnote{Il suffit de développer $1/(\eee^{\tilde{\qq}'}-1)$ et $1/(\eee^{\tilde{\qq}'+\tilde{\qq}}-1)$ en série entière de la variable $\eee^{-\tilde{\qq}'}$ et d'échanger la sommation et l'intégration. On rappelle que $\zeta(4)=\pi^4/90$.}
\bea
\tilde{\Gamma}^{\rm Bel} (\tilde{q})&=& \frac{\tilde{q}^5}{30}- \frac{4\pi^4}{15} \tilde{q}  +48\bbcro{\zeta(5)-g_5(\eee^{-\tilde{q}})} \notag \\ 
&-&24\tilde{q}\,g_4(\eee^{-\tilde{q}}) + 4\tilde{q}^2[\zeta(3)-g_3(\eee^{-\tilde{q}})] \label{eq:GammaBel}\\
\tilde{\Gamma}^{\rm Lan}(\tilde{q}) &=& \tilde{\Gamma}^{\rm Bel}(\tilde{q}) - \frac{\tilde{q}^5}{30} + \frac{8\pi^4}{15} \tilde{q} \label{eq:GammaLan}
\eea
On en déduit les comportements aux limites suivants~:
\bea
\!\!\!\!\!\!\!\!\!\!\!\!\tilde{\Gamma}^{\rm Bel} \!\!\!&\underset{\tilde{q}\to0}{=}& \!\!\!\frac{\tilde{q}^4}{6} + \frac{\tilde{q}^6}{360} + O(\tilde{q}^8) \\ 
\!\!\!\!\!\!\!\!\!\!\!\!\tilde{\Gamma}^{\rm Bel} \!\!\!&\underset{\tilde{q}\to+\infty}{=}& \!\!\!\frac{\tilde{q}^5}{30} \!+\! 4\zeta(3)\tilde{q}^2\!-\!\frac{4\pi^4}{15}\tilde{q}+48\zeta(5) \!+\! O(\tilde{q}^2\eee^{-\tilde{q}}) \label{eq:GamqBel_lim}
\eea
ceux de $\tilde{\Gamma}^{\rm Lan}$ s'en déduisant trivialement par l'équation \eqref{eq:GammaLan}.
Enfin, les variations des deux taux et de leur somme entre ces deux extrêmes sont représentées sur la figure \ref{fig:GammaBelLan}.

\subsection{Cas concave : amortissement de Landau-Khalatnikov}
\label{subsec:ccadlk}

Dans le cas d'une relation de dispersion concave, le mode $\qq$ est amorti à suffisamment basse température par des processus $2\leftrightarrow2$~; nous retenons donc dans l'op\'erateur $\hat{R}^\dagger$ de l'\'equation (\ref{eq:Gammaq_epil}) les termes en $\hat{b}_{\qq_3}^\dagger \hat{b}_{\qq_4}^\dagger \hat{b}_{\qq_2}$, issus du hamiltonien quartique \eqref{eq:H2donne2eff}, c'est-\`a-dire
\be
\hat{R}^\dagger = \frac{mc^2}{\rho L^3}\!\!\!\sum_{\qq_2,\qq_3,\qq_4}\!\!\! 2\mathcal{A}_{\rm eff}^{2\leftrightarrow 2}(\qq,\qq_2;\qq_3,\qq_4)
\hat{b}_{\qq_3}^\dagger \hat{b}_{\qq_4}^\dagger \hat{b}_{\qq_2} \delta_{\qq+\qq_2,\qq_3+\qq_4}
\ee
Ceci conduit à
\begin{multline}
\Gamma_\qq^{2\leftrightarrow2}=\frac{(mc^2)^2}{4\pi^5\hbar^2\rho^2}  \int_{\mathbb{R}^6} \dd^3q_2 \dd^3q_3 |\mathcal{A}_{\rm Cdm}^{2\leftrightarrow2,{\rm eff}}(\qq,\qq_2;\qq_3,\qq_4)|^2 \\
\times\delta(\omega_{\qq_3}+\omega_{\qq_4}-\omega_{\qq_2}-\omega_{\qq})\bbcrol{\bar{n}_{\qq_2}(1+\bar{n}_{\qq_3})(1+\bar{n}_{\qq_4})} \\\bbcror{-(1+\bar{n}_{\qq_2})\bar{n}_{\qq_3} \bar{n}_{\qq_4}}
\label{eq:Gamma2d2_epil}
\end{multline}
où le vecteur $\qq_4$ est relié aux autres vecteurs d'onde par la conservation de l'impulsion 
\be
\qq_4=\qq+\qq_2-\qq_3
\label{eq:impulsion}
\ee 
Il nous faut maintenant obtenir, \`a $\tilde{q}=\hbar c q/k_B T$ fix\'e,
un \'equivalent du taux $\Gamma_\qq^{2\leftrightarrow2}$ \`a basse temp\'erature.
Choisissons l'axe $Oz$ du repère sphérique selon le vecteur $\qq$~; les vecteurs $\qq_i$ ont alors pour coordonnées $(q_i,\theta_i,\phi_i)$. Plaçons-nous 
à température $T$, contrôlée par le petit paramètre
\be
\epsilon\equiv\frac{k_{\rm B} T}{mc^2} \ll 1,
\ee
suffisamment basse pour que les nombres d'onde typiques soient petits devant $mc/\hbar$, de sorte que la branche bosonique n'est peuplée que dans la partie où elle est quasiment linéaire. Dans ce cas, l'amplitude de couplage $\mathcal{A}_{\rm Cdm}^{2\leftrightarrow2}$, qui serait divergente aux angles nuls pour une relation de dispersion linéaire comme nous l'avons déjà signalé dans la section \ref{sec:A22effhydro}, 
est extrêmement piquée autour de $\theta_2=\theta_3=0$, avec, comme nous le verrons, un pic de largeur d'ordre $\epsilon$ en $\theta_2$ et $\theta_3$,\footnote{La largeur du pic est d'ordre $\epsilon$ car le changement d'\'energie des phonons virtuels interm\'ediaires lorsque les $\theta_i$ varient 
de $0$ \`a $\epsilon$ est du m\^eme ordre que la correction cubique \`a la relation de dispersion hydrodynamique, sachant que les nombres d'onde
typiques sont $\approx k_B T/\hbar c$. 
En prenant comme exemple le phonon interm\'ediaire $\qq_{\rm S}=\qq+\qq_2$, on trouve que $\hbar c[|\qq+\qq_2|-(q+q_2)]
\sim -\frac{\hbar c q q_2}{q+q_2} \theta_2^2 \approx k_B T \theta_2^2$~; cet \'ecart est du m\^eme ordre que le terme cubique 
$\approx k_B T |\gamma| \epsilon^2$ dans l'\'equation (\ref{eq:fleur}) lorsque $\theta_2^2 \approx |\gamma| \epsilon^2$, d'o\`u le changement de variable
(\ref{eq:tildetheta}).}
s'élevant d'un facteur $1/\epsilon^2$ au-dessus de la valeur typique de l'amplitude à angles non nuls. Compte tenu de la conservation de l'impulsion \eqref{eq:impulsion} et de l'énergie, on trouve aussi que $\theta_4=O(\epsilon)$ sur la largeur du pic. Nous redéfinissons alors les nombres d'onde comme dans l'équation \eqref{eq:tildeq} et les angles polaires par
\be
\tilde{\theta}_i = \frac{\theta_i}{\epsilon |\gamma|^{1/2}}
\label{eq:tildetheta}
\ee
avec $\gamma<0$ le paramètre de courbure \eqref{eq:fleur}, puis nous effectuons un développement limité de l'amplitude de couplage \eqref{eq:A2donne2effhydro} lorsque $\epsilon\to0$ à valeurs fixées des quantités redimensionnées:
\begin{multline}
\mathcal{A}_{\rm Cdm}^{2\leftrightarrow2}(\qq,\qq_2;\qq_3,\qq_4) \underset{\epsilon\to0}{=} \bb{\frac{3(1+\Lambda_{\rm F})}{4}}^2 \frac{(\tilde{q} \tilde{q}_2 \tilde{q}_3 \tilde{q}_4)^{1/2}}{|\gamma|} \\ \times
\mathcal{A}_{\rm r\text{é}d}^{2\leftrightarrow2}(\tilde{q},\tilde{q}_2,\tilde{q}_3,\tilde{\theta}_2,\tilde{\theta}_3) + O(\epsilon^2)
\end{multline}
La contribution en $\Sigma_{\rm F}$ de l'amplitude directe $\mathcal{A}^{2\leftrightarrow2,{\rm dir}}$ disparaît à cet ordre du développement en $\epsilon$ ce qui, combiné à notre redéfinition astucieuse \eqref{eq:tildetheta} des angles polaires, nous a permis de sortir les facteurs $1+\Lambda_{\rm F}$ et $\gamma$ dépendant de la force des interactions. On obtient ainsi une amplitude réduite $\mathcal{A}_{\rm r\text{é}d}^{2\leftrightarrow2}$ universelle~:
\begin{multline}
\mathcal{A}_{\rm r\text{é}d}^{2\leftrightarrow2}(\tilde{q},\tilde{q}_2,\tilde{q}_3,\tilde{\theta}_2,\tilde{\theta}_3) = 
\frac{1}{\tilde{q} \tilde{q}_2\bb{\frac{\tilde{\theta}_2^2}{(\tilde{q} +\tilde{q}_2)^2}+\frac{3}{4}}} \\ - \frac{1}{\tilde{q} \tilde{q}_3\bb{\frac{\tilde{\theta}_3^2}{(\tilde{q} -\tilde{q}_3)^2}+\frac{3}{4}}}  - \frac{1}{\tilde{q} (\tilde{q}+\tilde{q}_2-\tilde{q}_3) \bb{\frac{\tilde{\theta}_4^2}{(\tilde{q}_3 -\tilde{q}_2)^2}+\frac{3}{4}}}
\label{eq:Ared}
\end{multline}
Les premier, deuxième et troisième termes de \eqref{eq:Ared} proviennent respectivement des deuxième, troisième et quatrième termes de \eqref{eq:A2donne2effhydro}. Dans ces deux derniers, on a pris soin de distinguer les cas $q>q_3$ et $q<q_3$, $q>q_4$ et $q<q_4$ avant de passer \`a la limite $\epsilon\to 0$.
En combinant la relation implicite issue de la conservation de l'énergie,
\be
\tilde{q}_4 = \tilde{q}+ \tilde{q}_2 - \tilde{q}_3 - \frac{\epsilon^2 |\gamma|}{8}\bb{\tilde{q}^3+\tilde{q}_2^3-\tilde{q}_3^3-\tilde{q}_4^3}+O(\epsilon^4)
\label{eq:tildeq4}
\ee
itérée une fois, à un calcul de géométrie sphérique consistant à projeter la relation \eqref{eq:impulsion} sur $\qq$, on obtient l'expression de $\tilde{\theta}_4$~:
\begin{multline}
\tilde{\theta}_4^2= \frac{\tilde{q}_2 \tilde{\theta}_2^2- \tilde{q}_3 \tilde{\theta}_3^2-\frac{1}{4}\bbcro{\tilde{q}^3+\tilde{q}_2^3-\tilde{q}_3^3-(\tilde{q}+\tilde{q}_2-\tilde{q}_3)^3}}{\tilde{q}+ \tilde{q}_2 - \tilde{q}_3} \\ + O(\epsilon^2)
\end{multline}
Notons que $\tilde{q}_3<\tilde{q}+\tilde{q}_2$ d'après \eqref{eq:tildeq4} et que l'intégrande de \eqref{eq:Gamma2d2_epil} dépend azimutalement seulement de $\phi\equiv\phi_2-\phi_3$ par invariance par rotation autour de $\qq$.

\begin{figure}[htb] 
\begin{center}
\includegraphics[width=0.49\textwidth,clip=]{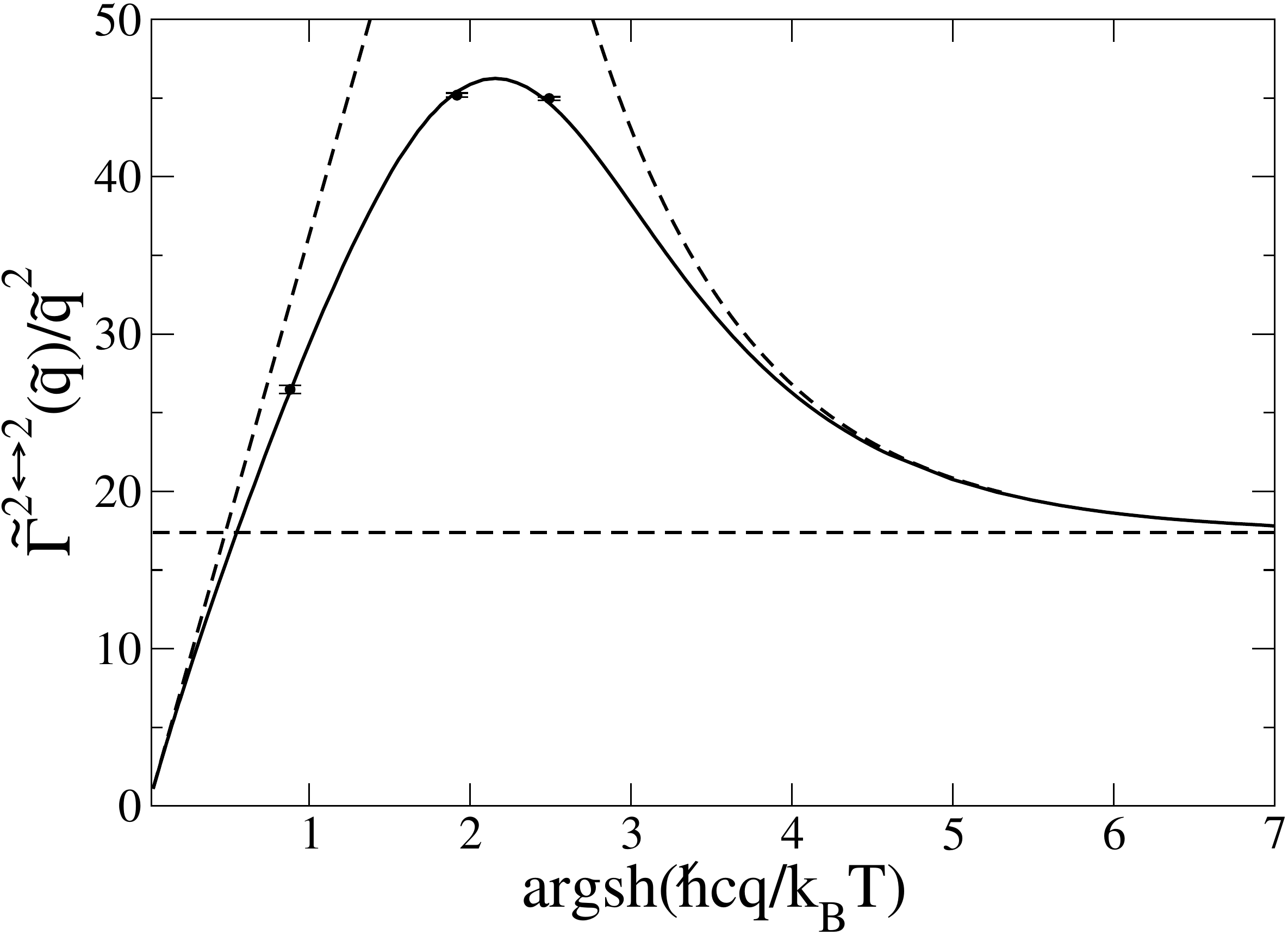}
\end{center}
\caption{ \label{fig:F} Taux d'amortissement $2\leftrightarrow2$ en fonction du nombre d'onde adimensionné $\tilde{q}=\hbar cq/k_B T$. Afin de faire ressortir les comportements aux limites, nous divisons le taux par $\tilde{q}^2$ et traçons en fonction de $\mbox{argsh}(\tilde{q}){=\ln(\tilde{q}+\sqrt{1+\tilde{q}^2})}$. La tangente oblique en $\tilde{q}=0$ figure le comportement limite à petits $\tilde{q}$ (\textit{cf}. l'équation \eqref{eq:petitsq}). La droite horizontale est la limite de $\tilde{\Gamma}^{2\leftrightarrow2}{(\tilde{q})}/\tilde{q}^2$ lorsque $\tilde{q}\to\infty$ (\textit{cf}. l'équation \eqref{eq:grandsq}) et la courbe asymptote 
aux grands $\tilde{q}$ est un ajustement de $\tilde{\Gamma}^{2\leftrightarrow2}{(\tilde{q})}/\tilde{q}^2$ affine en $1/\tilde{q}$. 
{Les disques noirs proviennent d'un calcul num\'erique de l'int\'egrale (\ref{eq:Gamma2d2_epil}) pour la relation de dispersion bogolioubovienne 
(\ref{eq:formedeBogolioubov}) tronquée à un intervalle $[0,q_{\rm max}]$ sur lequel sa d\'eriv\'ee est strictement positive~; les valeurs repr\'esent\'ees 
correspondent à une extrapolation lin\'eaire à $\epsilon=0$, et la mi-largeur des barres d'erreur à l'écart à une extrapolation quadratique.}}
\end{figure}

Il nous reste à intégrer le Dirac de conservation de l'énergie. Pour cela, nous passons en représentation polaire pour les angles redimensionnés
\be
\tilde{\theta}_2 = R \cos\alpha  \qquad \qquad \tilde{\theta}_3 = R \sin\alpha
\label{eq:reppolth}
\ee
et mettons la différence d'énergie entre état initial et final sous la forme
\begin{multline}
\omega_{\qq_3}+\omega_{\qq_4}-\omega_{\qq_2}-\omega_{\qq} = \frac{mc^2}{\hbar}  \frac{\epsilon^3|\gamma|}{2} \bb{u R^2 + v} +O(\epsilon^5) \label{eq:dansDirac}
\end{multline}
avec 
\begin{align}
&u=\frac{\tilde{q}(\tilde{q}_3 \sin^2\alpha - \tilde{q}_2 \cos^2\alpha)+\tilde{q}_2 \tilde{q}_3 (1-\sin2\alpha\cos\phi)}{\tilde{q}+ \tilde{q}_2 - \tilde{q}_3} \label{eq:defualign}\\
&v= \frac{1}{4} \bbcro{\tilde{q}^3+\tilde{q}_2^3-\tilde{q}_3^3-(\tilde{q}+\tilde{q}_2-\tilde{q}_3)^3} \label{eq:defvalign}
\end{align}
\'Ecrit sous la forme \eqref{eq:dansDirac}, le Dirac de conservation de l'énergie est immédiat à intégrer sur $R$. Nous  exprimons alors le taux d'amortissement $2\leftrightarrow2$ en termes d'une fonction universelle $\tilde{\Gamma}^{2\leftrightarrow2}$ dépendant uniquement du vecteur d'onde adimensionné $\tilde{q}$~:
\be
\frac{\hbar \Gamma_{\qq}^{2\leftrightarrow2}}{mc^2}  \!\!\underset{\epsilon\to0}{\sim} \!\! \frac{81(1+\Lambda_{\rm F})^4}{256\pi^4|\gamma|} \bb{\frac{k_{\rm B} T}{mc^2}}^7 \bb{\frac{mc}{\hbar\rho^{1/3}}}^6  \tilde{\Gamma}^{2\leftrightarrow2}(\tilde{q})
\label{eq:Gamma2d2}
\ee
ce qui constitue un autre résultat majeur de cet article. La fonction $\tilde{\Gamma}^{2\leftrightarrow2}(\tilde{q})$ est donnée par une intégrale quadruple \footnote{Précisons les intervalles d'intégration $[\alpha_{\rm min},\alpha_{\rm max}]$ et $[\phi_{\rm min},\phi_{\rm max}]$ imposés par la fonction de Heaviside. On définit $\chi=\frac{{q}_2 {q}_3+{q} ({q}_3\sin^2\alpha-{q}_2\cos^2\alpha) }{{q}_2 {q}_3\sin2\alpha}$, $A=\frac{{q}}{2{q}_3} +\frac{{q}}{2{q}_2}$, $C=1-\frac{{q}}{2{q}_3}+\frac{{q}}{2{q}_2}$, $A'=\frac{A}{(1+A^2)^{1/2}}$ et $C'=\frac{C}{(1+A^2)^{1/2}}$. 
{$(i)$  Si $v>0$ et $q<q_3$}, 
$2\alpha_{\rm min}=\mbox{acos}\, A' - \mbox{acos}\, C' $ et $2\alpha_{\rm max}=\mbox{acos}\, A'  + \mbox{acos}\, C' $.
{$(ii)$  Si $v>0$ et $q>q_3$},
$\alpha_{\rm min}=0$ et $2\alpha_{\rm max}=\mbox{acos}\, A'  + \mbox{acos}\, C'$.
{$(iii)$  Si $v<0$ et $q<q_3$},
$\alpha_{\rm min}=0$ et $\alpha_{\rm max}=\pi/2$.
{$(iv)$  Si $v<0$ et $q>q_3$},
$2\alpha_{\rm min}=\mbox{acos}\, C' - \mbox{acos}\, A' $ et $\alpha_{\rm max}=\pi/2$.
Dans les cas $(i)$ et $(ii)$ ($v>0$), $\phi_{\rm min}=0$, $\phi_{\rm max}=\pi$ si $\chi<-1$ et $\phi_{\rm max}=\mbox{acos}\, \chi$ sinon.
Dans les cas $(iii)$ et $(iv)$ ($v<0$), $\phi_{\rm max}=\pi$, $\phi_{\rm min}=0$ si $\chi>1$ et $\phi_{\rm min}=\mbox{acos}\, \chi$ sinon.
L'intégrale sur $\phi$ est en fait calculable analytiquement. On se ramène à $\int_0^{\Phi} \dd\phi\bb{\sum_i \frac{b_i}{a_i+\cos\phi}}^2$ qui s'exprime en termes de la primitive $F(\Phi)=\int_0^{\Phi} \dd\phi\frac{1}{a+\cos\phi}$ et de sa dérivée par rapport à $a$. Si $a\in]-1,1[$, $F(\Phi)=\frac{2}{(1-a^2)^{1/2}} \mbox{argth}\,\bbcro{\bb{\frac{1-a}{1+a}}^{1/2} \tan\frac{\Phi}{2}  }$. Si $|a|>1$, $F(\Phi)=\frac{2}{(a^2-1)^{1/2}} \mbox{atan}\,\bbcro{\frac{a-1}{(a^2-1)^{1/2}} \tan\frac{\Phi}{2}  }$. Si $a=1$, $F(\Phi)=\tan \frac{\Phi}{2}$.}
\begin{multline}
\tilde{\Gamma}^{2\leftrightarrow2}(\tilde{q}) =  \int_0^{+\infty} \dd\tilde{q}_2  \int_0^{\tilde{q} + \tilde{q}_2} \dd\tilde{q}_3 \frac{\tilde{q} \tilde{q}_2^3\tilde{q}_3^3(\tilde{q}+\tilde{q}_2-\tilde{q}_3)}{|v|} \\
\times \frac{[1+f(\tilde{q}_2)] f(\tilde{q}_3) f(\tilde{q}+\tilde{q}_2-\tilde{q}_3)}{f(\tilde{q})}    \\
\times \int_{0}^{\pi/2} \dd\alpha  \int_{0}^\pi \dd\phi \sin\alpha\cos\alpha \, \Theta\bb{-\frac{v}{u}} \\
 \times\left\vert \frac{v}{u}\mathcal{A}_{\rm r\text{é}d}^{2\leftrightarrow2}\bb{\tilde{q},\tilde{q}_2,\tilde{q}_3,\left\vert\frac{v}{u}\right\vert^{1/2}\!\!\cos\alpha,\left\vert\frac{v}{u}\right\vert^{1/2}\!\!\sin\alpha} \right\vert^2
\end{multline}
expression dans laquelle nous avons introduit la fonction de Heaviside $\Theta(x\geq0)=1,\ \Theta(x<0)=0$ et la fonction de Bose réduite $f(x)=1/(\eee^x-1)$. Nous avons récrit les nombres d'occupation dans \eqref{eq:Gamma2d2_epil} grâce à la propriété $(1+\bar{n}_{\qq_i})/\bar{n}_{\qq_i}=\eee^{\hbar\omega_{\qq_i}/k_{\rm B}T}$ et à la conservation de l'énergie. La fonction $\tilde{\Gamma}^{2\leftrightarrow2}$ est représentée sur la figure \ref{fig:F}. 
Ses comportements aux faibles et aux grands $\tilde{q}$ peuvent être obtenus analytiquement~:
\bea
\tilde{\Gamma}^{2\leftrightarrow2}(\tilde{q}) &\underset{\tilde{q}\to0}{=}& \frac{16\pi^5}{135} \tilde{q}^3 +O(\tilde{q}^4) \label{eq:petitsq}\\
\tilde{\Gamma}^{2\leftrightarrow2}(\tilde{q}) &\underset{\tilde{q}\to\infty}{=}& \frac{16\pi\zeta(5)}{3} \tilde{q}^2 + O(\tilde{q}) \label{eq:grandsq}
\eea
{Ces comportements aux limites diffèrent de ceux obtenus par Landau et Khalatnikov dans la référence \cite{Khalatnikov1949} (voir leurs équations (7.6) et (7.12) dans la version \cite{LandauCollectedPapers}), y compris sur l'ordre en $\tilde{q}$ des termes dominants. La raison de ce désaccord est que ces auteurs négligent dans l'amplitude de couplage, aussi bien à faible $\tilde{q}$ qu'à grand $\tilde{q}$, la contribution des diagrammes II à V de la figure \ref{fig:diagrammes} (le diagramme VI est non r\'esonnant), ce qui n'est pas justifié, comme l'avait d\'ejà remarqu\'e la r\'ef\'erence \cite{Wyatt1992} dans un calcul num\'erique.
En r\'ealit\'e,} ceux-ci interfèrent destructivement à l'ordre dominant avec le diagramme I, et rendent les résultats exacts (\ref{eq:petitsq},\ref{eq:grandsq}) sous-dominants par rapport aux prédictions de Landau et Khalatnikov de deux ordres en $\tilde{q}$.

\subsection{Amortissement des phonons au-delà de l'hydrodynamique}

Cette sous-section est consacr\'ee au gaz unitaire de fermions \`a temp\'erature nulle. Nous y calculons le taux d'amortissement des phonons
au-del\`a du r\'esultat hydrodynamique (\ref{eq:GamqBel_lim}) que l'on peut r\'ecrire sous la forme
\begin{equation}
\left(\Gamma_q^{\rm Bel}\right)_{\rm hydro}^{\rm gaz\ unitaire} = \left( \frac{2}{9\pi}\right) 
\left( \frac{mc^2}{\hbar}\right) \left( \frac{mc}{\hbar \rho^{1/3}}\right)^3 \frac{\check{q}^5}{30} 
\label{eq:GamqBelhydro}
\end{equation}
avec les notations
\be
\check{q} \equiv \frac{\hbar q}{mc} \quad \quad {\rm et} \quad \quad \check{\omega}_q \equiv \frac{\hbar \omega_\qq}{mc^2}
\ee
\`A cette fin, nous calculons d'abord l'amplitude de couplage ${\cal A}^{2 \leftrightarrow 1}_{\rm Cdm}(\qq_1,\qq_2;\qq_3)$ au premier ordre 
en le param\`etre de courbure $\gamma$ du spectre d'excitation,
ce qui nous donne dans l'expression (\ref{eq:ACM}) la premi\`ere correction au r\'esultat hydrodynamique, puis nous d\'eduisons le taux d'amortissement de Beliaev de l'\'equation 
(\ref{eq:GamqBel}), sans oublier d'inclure la correction d'ordre un en $\gamma$ au spectre d'excitation lui-m\^eme.
Nous trouvons aussi qu'il faut aller au-del\`a d'une simple distribution de Dirac dans la r\`egle d'or de Fermi
(\ref{eq:GamqBel}) 
et qu'il faut prendre en compte des processus d'ordre plus \'elev\'e que celui de Beliaev.


\subsubsection{Amplitude du processus de Beliaev $2\leftrightarrow 1$ pour le gaz unitaire}

Nous calculons ici l'amplitude de couplage ${\cal A}^{2 \leftrightarrow 1}_{\rm Cdm}(\qq_1,\qq_2;\qq_3)$ à l'ordre un en $\gamma$,
\`a l'aide de la th\'eorie des champs effective de Son et Wingate d\'eduite de l'invariance conforme du gaz unitaire \cite{SonWingate2006}.
Ces auteurs ont construit un lagrangien incluant la premi\`ere correction \`a l'hydrodynamique, fonction d'un \g{champ de phase} $\phi$. Le lecteur trouvera tous les d\'etails sur le formalisme
dans leur publication \cite{SonWingate2006}, ainsi que dans la r\'ef\'erence \cite{Manuel2010}
qui l'a utilis\'e pour calculer des corrections \`a la vitesse du son \`a temp\'erature non nulle.

En bref, dans un premier temps, nous quadratisons le lagrangien pour obtenir le spectre d'excitation
phononique $\omega_\qq$ et les amplitudes modales correspondantes. Le spectre
est de la forme cubique (\ref{eq:fleur}) avec un param\`etre de courbure $\gamma=-\frac{64}{45 c_0}\left(c_1+\frac{3}{2}c_2\right)$~; ici,
la constante sans dimension $c_0$ appara\^\i t dans le lagrangien 
\`a l'ordre dominant (hydrodynamique), et les constantes sans dimension $c_1$ et $c_2$
\`a l'ordre sous-dominant (premi\`ere correction \`a l'hydrodynamique).
La constante $c_0$ s'exprime en termes du param\`etre de Bertsch $\xi_B$, 
mesur\'e avec pr\'ecision \cite{Zwierlein2012}, et qui relie le potentiel chimique $\mu$ au nombre d'onde de Fermi $k_{\rm F}$ ou \`a la densit\'e totale $\rho$ du gaz unitaire~:
\be
\mu=\xi_B \frac{\hbar^2k_{\rm F}^2}{2m} \quad \mbox{et} \quad \rho=\frac{k_{\rm F}^3}{3\pi^2}
\label{eq:Bertsch}
\ee
Les constantes $c_1$ et $c_2$ n'ont pas encore \'et\'e mesur\'ees, ni m\^eme calcul\'ees avec 
pr\'ecision \`a partir d'une th\'eorie microscopique. On ne dispose que d'estimations,
voir la r\'ef\'erence \cite{Manuel2013} et les articles qu'elle cite. Dans un deuxi\`eme
temps, nous d\'eduisons des termes cubiques du lagrangien le couplage entre les phonons puis
l'amplitude des processus de Beliaev sur la couche de masse au premier ordre au-del\`a de l'hydrodynamique~:
\begin{multline}
{\cal A}^{2 \leftrightarrow 1}_{\rm {Cdm}}(\kk,\kk';\qq) =-\frac{\sqrt{2}}{3}(\check{\omega}_q\check{\omega}_k\check{\omega}_{k'})^{1/2} \\
\times \bbcro{1-\frac{7\gamma}{32}(\check{\omega}_q^2+\check{\omega}_k^2+\check{\omega}_{k'}^2)
+o(\check{\omega}_q^2)}
\label{eq:couplA}
\end{multline}
De fa\c{c}on remarquable, ce r\'esultat fait intervenir exactement la m\^eme combinaison
lin\'eaire $\gamma=-\frac{64}{45 c_0}\left(c_1+\frac{3}{2}c_2\right)$ de $c_1$ et $c_2$
que dans le spectre d'excitation. Les d\'etails de notre calcul ont \'et\'e rel\'egu\'es dans
l'annexe~\ref{appen:SonetWingate}.


\subsubsection{Taux d'amortissement des phonons dans le gaz unitaire \`a $T=0$}
\label{subsec:tadpdlguat0}

\`A l'ordre trois en $q$, nous pouvons mettre le spectre d'excitation sous une forme de Bogolioubov~:
\begin{equation}
\check{\omega}_q=\check{q}\left(1+\frac{\gamma}{4}\check{q}^2\right)^{1/2}
\label{eq:formedeBogolioubov}
\end{equation}
et donc r\'eutiliser le r\'esultat (A14) de l'annexe A de la r\'ef\'erence \cite{CastinSinatra2009} pour effectuer l'int\'egration
angulaire dans l'\'equation (\ref{eq:GamqBel}),
\begin{multline}
\left(\Gamma_q^{\rm Bel}\right)^{\rm gaz\, unitaire}= 
\left( \frac{mc^2}{\pi \hbar }\right) \left( \frac{mc}{\hbar \rho^{1/3}}\right)^3 \\
 \times\int_0^{\check{q}} \dd \check{k} 
|{\cal A}^{2 \leftrightarrow 1}_{\rm Cdm}(k,k';q)|^2 \frac{\check{k}}{\check{q}} 
\frac{\check{\omega}_q-\check{\omega}_k}{\left[ 1 + \gamma (\check{\omega}_q-\check{\omega}_k)^2 \right]^{1/2}}
\end{multline}
o\`u nous avons tenu compte, dans notre notation, du fait que l'amplitude de couplage (\ref{eq:couplA}), 
dans laquelle $\check{\omega}_{k'}=\check{\omega}_q-\check{\omega}_k$, d\'epend seulement du module des vecteurs d'onde.
\`A l'aide du changement de variable $\kappa\equiv \check{k}/\check{q}$, nous mettons $[\left(\Gamma_q^{\rm Bel}\right)^{\rm Unitary \, gas} \gamma^2 /\check{q}]$ sous la forme d'une int\'egrale sur $\kappa \in [0,1]$ d'une fonction de $\kappa$ et du petit param\`etre $\gamma \check{q}^2$.
En d\'eveloppant cette fonction en puissances de $\gamma \check{q}^2$ jusqu'\`a l'ordre trois sous-dominant, puis en effectuant
l'int\'egration sur $\kappa$, nous obtenons {le r\'esultat provisoire}
\begin{multline}
\left(\Gamma_q^{\rm Bel}\right)^{\rm gaz\, unitaire}_{\rm prov} \underset{q\to0}{=} \left( \frac{2}{9\pi}\right) 
\left( \frac{mc^2}{\hbar}\right) \left( \frac{mc}{\hbar \rho^{1/3}}\right)^3 
\\\times \frac{\check{q}^5}{30}\left[{1 - \frac{25}{112}\gamma \check{q}^2+ o(\check{q}^2)} \right] 
\label{eq:Bel_result_prov} 
\end{multline}
Le terme sous-dominant de notre r\'esultat {provisoire} (\ref{eq:Bel_result_prov}) diff\`ere, m\^eme en signe, 
de celui de la r\'ef\'erence \cite{Salasnich2015}~: {il correspond à une r\'eduction du taux d'amortissement 
par rapport au r\'esultat hydrodynamique plut\^ot qu'à une augmentation.}
Ce d\'esaccord {est surprenant car notre calcul et celui de \cite{Salasnich2015} 
ont \'et\'e effectu\'es dans le m\^eme esprit et avec la m\^eme m\'ethodologie.
Il pourrait} 
r\'esulter du fait que la d\'ependance de ${\cal A}^{2 \leftrightarrow 1}$ en $\gamma$ a \'et\'e {en d\'efinitive}
n\'eglig\'ee dans
la r\'ef\'erence \cite{Salasnich2015}, alors qu'elle donne une contribution du m\^eme ordre que la d\'ependance du spectre en $\gamma$.

{Avant d'accepter le r\'esultat (\ref{eq:Bel_result_prov}), il est prudent de se demander quelle correction y apporterait le fait 
que l'\'etat à un phonon consid\'er\'e est instable, de largeur $\hbar \Gamma_\qq/2$, si bien que l'\'energie des phonons libres, lors du processus
de d\'ecroissance \`a deux phonons, n'est certainement pas conserv\'ee de mani\`ere exacte, contrairement \`a ce que suppose
la distribution de Dirac dans l'\'equation (\ref{eq:GamqBel}). Afin d'estimer l'ordre de grandeur de cet effet, nous rempla\c{c}ons
le Dirac par une lorentzienne de mi-largeur $\Gamma_\qq/2$,
\be
\pi\delta(\omega_{\qq'}+\omega_{\qq-\qq'}-\omega_{\qq}) \to \frac{\Gamma_\qq/2}{(\omega_{\qq'}+\omega_{\qq-\qq'}-\omega_{\qq})^2+\Gamma_\qq^2/4}
\label{eq:dirdevlor}
\ee
où $\Gamma_\qq$  peut \^etre assimil\'e à son terme dominant en $q^5$. Nous trouvons alors\footnote{Instruits par la note \thenotepasundemi, nous consid\'erons l'int\'egrale polaire $J=\int_{-1}^1 \dd u \frac{\eta'/\pi}{\eta'^2+(u-u_0)^2}$ o\`u $\eta'\approx \Gamma_\qq/\check{q}\approx \check{q}^4$
et $1-u_0=1-\cos\theta_0 \approx \check{q}^2$
comme dans (\ref{eq:thetaitildeq}). Alors $J-1\approx \check{q}^2$, d'où une correction $O(\check{q}^7)$ non n\'egligeable \`a
$\Gamma_\qq^{\rm Bel}$ due \`a la largeur en \'energie de l'\'etat initial. La correction en $\check{q}^7$ dans l'approche
heuristique (\ref{eq:dirdevlor}) peut \^etre calcul\'ee
jusqu'au bout. De manière amusante, elle co\"\i ncide avec le r\'esultat exact au signe près.
On le voit bien d'ailleurs dans le mod\`ele simple d'une fonction \'energie propre $\Sigma(z)=
\int_{-\Omega}^{\Omega}\dd\omega \frac{\hbar A}{z-\hbar\omega}$ o\`u $\Omega>0$ est fix\'e et $A>0$ est le petit
param\`etre du d\'eveloppement. La r\'esonance $z_0=\hbar\omega_0-\ii\hbar \Gamma_0/2$ dont il faut calculer
le taux de d\'ecroissance est solution de $z_0=\Sigma^{\downarrow}(z_0)$, o\`u $\Sigma^{\downarrow}(z)$ est
le prolongement analytique de $\Sigma(z)$ du demi-plan complexe sup\'erieur au demi-plan inf\'erieur. Le
d\'eveloppement perturbatif exact comme dans l'\'equation (\ref{eq:respersigmaprime}) donne $\hbar\Gamma_0/2 
=\pi A + \frac{2\pi A^2}{\hbar\Omega} + O(A^3)$, puisque $\Sigma^{\downarrow}(0)=-\ii \pi A$
et $\frac{\dd}{\dd z}\Sigma^{\downarrow}(0)=\frac{2A}{\hbar\Omega}$. Au contraire, l'approche heuristique
$\hbar\Gamma/2=\int_{-\Omega}^{\Omega} \dd\omega \frac{A\Gamma/2}{\omega^2+(\Gamma/2)^2}$ conduit
au d\'eveloppement incorrect $\hbar\Gamma/2 =
\pi A - \frac{2\pi A^2}{\hbar\Omega} + O(A^3)$.
} que ceci conduit, dans la limite $q\to 0$, \`a une modification de $\Gamma_\qq$ \`a l'ordre $q^7$.
Pour obtenir la correction exacte sur $\Gamma_\qq$, nous allons au-delà de ce calcul heuristique dans l'annexe \ref{appen:1d3}. \`A l'aide de la r\'esolvante du hamiltonien, nous \'ecrivons
\`a l'ordre de Beliaev une \'equation autocoh\'erente sur l'\'energie complexe $z_\qq=\hbar\omega_\qq-\ii\hbar\Gamma_\qq/2$
du phonon $\qq$. Si l'on remplace dans la partie implicite de l'\'equation $z_\qq$ par son approximation d'ordre z\'ero $\hbar\omega_\qq+\ii\eta$,
on retrouve exactement (\ref{eq:Bel_result_prov}). Si l'on effectue une it\'eration d'autocoh\'erence, c'est-\`a-dire que l'on prend pour $-2\im z_\qq/\hbar$
son approximation hydrodynamique habituelle (\ref{eq:GamqBelhydro}), on obtient le r\'esultat d\'efinitif \`a l'ordre de Beliaev~:
\begin{multline}
\left(\Gamma^{\rm Bel}_\qq\right)^{\rm gaz\, unitaire} = \left(\frac{2}{9\pi}\right) \left(\frac{mc^2}{\hbar}\right)
\left(\frac{mc}{\hbar\rho^{1/3}}\right)^3 \\
\times \frac{\check{q}^5}{30}
\left[1-\frac{25}{112}\gamma\check{q}^2 + \frac{4\sqrt{3}\xi_B^{3/2}}{243\gamma} \check{q}^2+o(\check{q}^2)\right]
\label{eq:Bel_result}
\end{multline}
o\`u $\xi_B$ est le param\`etre de Bertsch (\ref{eq:Bertsch}) du gaz unitaire.
}

\begin{figure}[htb] 
\begin{center}
\includegraphics[width=0.3\textwidth,clip=]{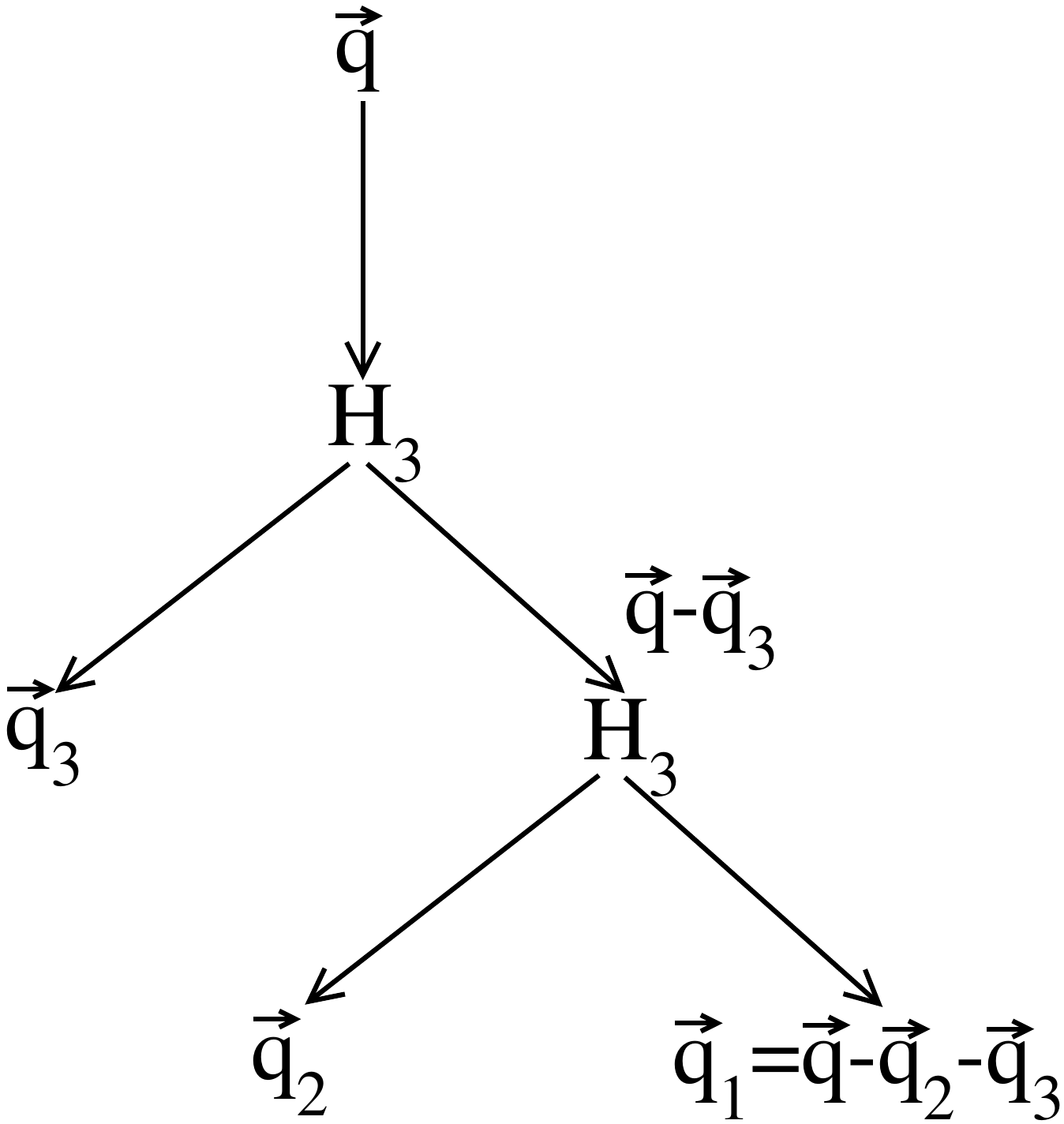}
\end{center}

\caption{\label{fig:1d3} Processus de couplage effectif $1\to 3$ du second ordre en $\hat{H}_3$ avec \'emission d'un phonon virtuel.
Dans le cas d'une relation de dispersion de d\'epart convexe, {il contribue  \`a l'ordre $(\hbar q/mc)^7$}
au taux de d\'ecroissance \`a $T=0$ d'un phonon de faible vecteur d'onde $\qq$
lorsque les phonons $\qq_1$, $\qq_2$ et $\qq_3$ sont \'emis avec un petit 
angle $O(\hbar q/mc)$ par rapport \`a $\qq$, comme le montre un simple comptage de puissances (voir le texte),
{et il doit \^etre ajout\'e au r\'esultat (\ref{eq:Bel_result}), comme on le v\'erifie dans l'annexe \ref{appen:1d3}
par un calcul explicite.}}
\end{figure} 

Pour terminer, et rendre le r\'esultat (\ref{eq:Bel_result}) rigoureux, il faut v\'erifier qu'aucun processus d'ordre plus \'elev\'e 
que le processus {de Beliaev} $1\to 2$ ne fournit de contribution en $\check{q}^7$.  Un candidat naturel est le processus de d\'ecroissance
en cascade repr\'esent\'e sur la figure \ref{fig:1d3}, qui encha\^ine deux processus $1\to 2$ induits par le hamiltonien cubique $\hat{H}_3$
et conduit \` a une amplitude de couplage effective $1\to 3$ du second ordre avec un phonon virtuel.
\footnote{Le lecteur attentif aura remarqu\'e que le hamiltonien quartique $\hat{H}_4$ donne directement un couplage
$1\to 3$ au premier ordre. L'amplitude correspondante est en $q^2$.  Ce couplage direct contribue \`a $\Gamma_\qq$ \`a l'ordre
$q^9$, n\'egligeable, m\^eme si l'on prend en compte comme dans ce qui suit l'effet d'un petit d\'enominateur en $q^3$. Aussi l'avons-nous 
pass\'e sous silence dans le corps du texte.}
Cherchons \`a estimer le changement correspondant
de l'\'energie complexe du phonon $\qq$ en traitant ce couplage effectif $1\to 3$ au second ordre de la th\'eorie des perturbations. 

\`A premi\`ere vue,
le r\'esultat est $O(\check{q}^9)$ donc n\'egligeable. Il faut en effet int\'egrer sur deux vecteurs d'onde de phonons \'emis ind\'ependants,
par exemple $\qq_2$ et $\qq_3$, le troisi\`eme $\qq_1$ \'etant impos\'e par la conservation de l'impulsion. Comme les nombres d'onde $q_i$
sont d'ordre $q$, ceci sort d\'ej\`a un facteur $q^6$. Vient ensuite le produit de quatre \'el\'ements de matrice de $\hat{H}_3$, puisque le couplage
effectif $1\to 3$ contient deux \'el\'ements de matrice en facteur et qu'il est trait\'e au second ordre~; voil\`a qui ajoute un facteur global
$(q^{3/2})^4=q^6$. Enfin, viennent en facteur trois d\'enominateurs d'\'energie, l'un issu de la formule des perturbations au second ordre et
les deux autres provenant de celui du couplage effectif $1\to 3$ trait\'e au second ordre~; comme les \'energies des phonons sont d'ordre $\hbar c q$,
il vient un facteur $q^3$ au d\'enominateur. Le tout est $O(\check{q}^9)$ comme nous l'annoncions. 

\begin{figure}[htb] 
\begin{center}
\includegraphics[width=0.49\textwidth,clip=]{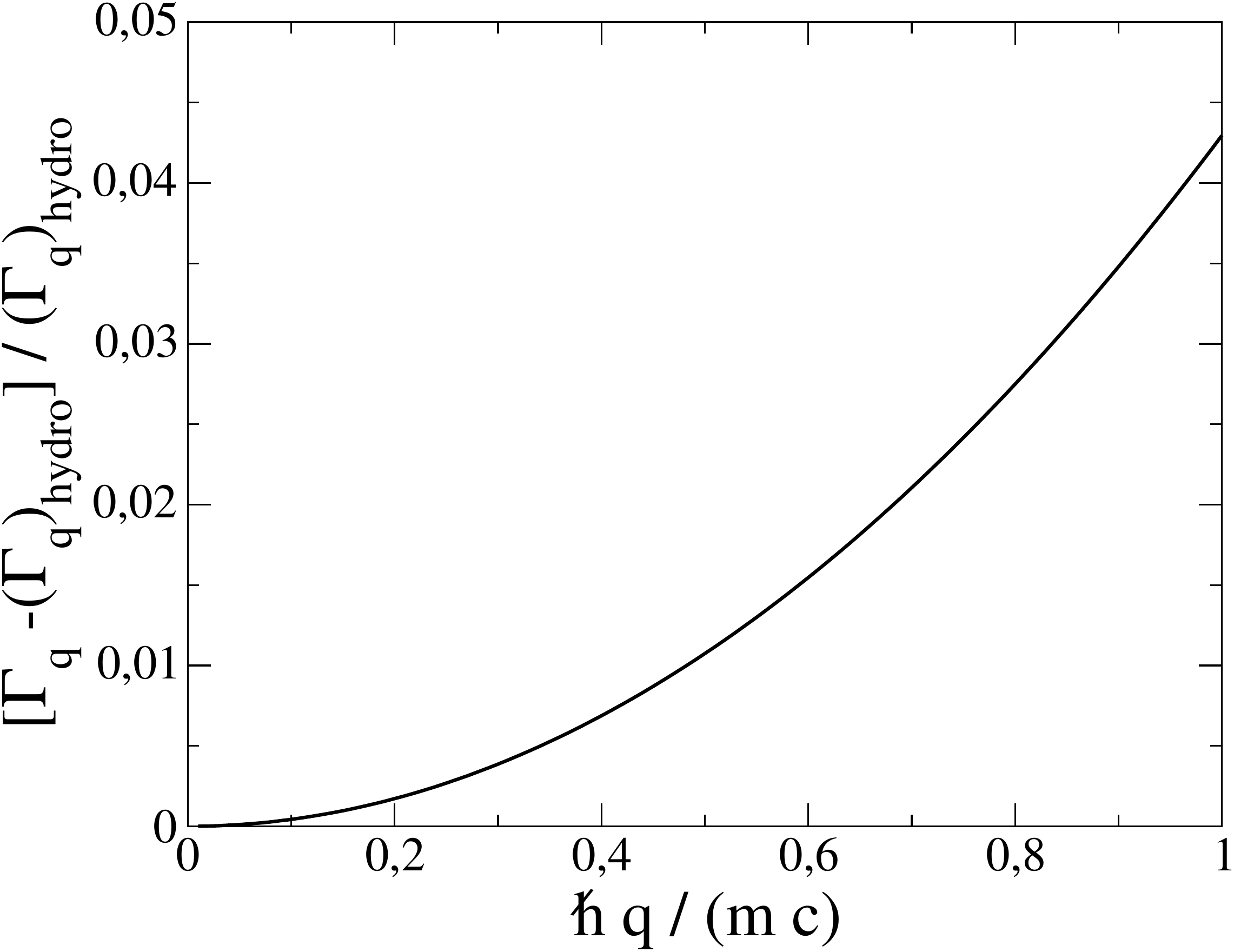}
\end{center}
\caption{\label{fig:BeyondHD}  {En valeur relative, première correction \eqref{eq:amorunit} \`a la prédiction hydrodynamique sur le taux d'amortissement des phonons du gaz unitaire à température nulle.}
On a pris pour le paramètre de courbure la prédiction de la RPA $\gamma\simeq {0,\! 083769}$
{et pour le param\`etre de Bertsch la valeur exp\'erimentale $\xi_B\simeq 0,\! 376$ de la r\'ef\'erence \cite{Zwierlein2012}.}}
\end{figure} 

Cependant, le raisonnement pr\'ec\'edent ignore l'exaltation due aux petits d\'enominateurs d'ordre $q^3$, qui se produit lorsque les vecteurs
d'onde $\qq_i$ sont \'emis vers l'avant, avec de petits angles $\theta_i$ par rapport \`a $\qq$, et qui joue un r\^ole crucial dans
la section \ref{subsec:ccadlk}.  Dans le cas limite o\`u $\qq$ et les $\qq_i$ sont tous colin\'eaires et de m\^eme sens, 
la conservation de l'impulsion impose 
\be
q=q_1+q_2+q_3
\label{eq:qq1q2q3}
\ee
si bien que la diff\'erence d'\'energie 
$\hbar\omega_\qq-(\hbar\omega_{\qq_1}+\hbar\omega_{\qq_2}+\hbar\omega_{\qq_3})$ n'est plus d'ordre $q$ mais d'ordre $q^3$, compte tenu
du terme cubique dans la relation de dispersion (\ref{eq:fleur}). Cette conclusion s'\'etend \`a tous les d\'enominateurs
d'\'energie et perdure tant que les angles d'\'emission sont $O(\check{q})$. 
En effet, $q_1=|\qq-\qq_2-\qq_3|$ varie en valeur relative au second ordre en les angles d'\'emission
$\theta_2$ et $\theta_3$ avec des coefficients d'ordre $\check{q}^0$, 
de m\^eme que la relation de dispersion vraie d\'evie en valeur relative de celle de l'hydrodynamique au second ordre en $\check{q}$.

Affinons donc l'estimation na\"\i ve $O(\check{q}^9)$ pr\'ec\'edente, en tenant compte de l'int\'egration sur $\qq_2$ et $\qq_3$
dans des c\^ones d'ouverture angulaire $O(\check{q})$ autour de $\qq$. Chaque c\^one occupe un angle
solide $O(\check{q}^2)$ donc on perd un facteur $q^4$ dans l'int\'egration sur les angles polaires. En revanche, on gagne un facteur $q^{-2}$
pour chaque d\'enominateur d'\'energie, donc un facteur global $q^{-6}$. On pr\'edit d\'esormais un changement d'\'energie complexe
d'ordre $\check{q}^7$, du m\^eme ordre que la correction \`a l'hydrodynamique dans l'\'equation (\ref{eq:Bel_result}). 

Dans l'annexe \ref{appen:1d3}, nous calculons donc explicitement la contribution au taux de d\'ecroissance $\Gamma_\qq$ des couplages effectifs
$1\to 3$ trait\'es au second ordre. Nous trouvons 
que son expression, une int\'egrale quintuple assez r\'ebarbative, 
conduit à
\begin{multline}
\left(\Gamma^{1\to 3}_\qq\right)^{\rm gaz\, unitaire} = \left(\frac{2}{9\pi}\right) \left(\frac{mc^2}{\hbar}\right)
\left(\frac{mc}{\hbar\rho^{1/3}}\right)^3 \\ \times\frac{\check{q}^5}{30}
\left[-\frac{2\sqrt{3}\xi_B^{3/2}}{567\gamma}\check{q}^2+o(\check{q}^2)\right]
\label{eq:Gamma1d3}
\end{multline}
Dans la m\^eme annexe, 
nous v\'erifions qu'aucun autre processus, d'ordre arbitrairement \'elev\'e en $\hat{H}_3$, $\hat{H}_4$, etc, n'est susceptible de contribuer
\`a l'ordre $\check{q}^7$ \`a $\Gamma_\qq$, m\^eme en tirant parti de petits d\'enominateurs en $q^3$. Aussi pouvons-nous
additionner les contributions (\ref{eq:Bel_result}) et (\ref{eq:Gamma1d3}) pour obtenir
\begin{multline}
\left(\Gamma_\qq\right)^{\rm gaz\, unitaire} \underset{q\to 0}{=} \left(\frac{2}{9\pi}\right) \left(\frac{mc^2}{\hbar}\right)
\left(\frac{mc}{\hbar\rho^{1/3}}\right)^3 \\ \times\frac{\check{q}^5}{30}
\left[1-\frac{25}{112}\gamma\check{q}^2 + \frac{22\sqrt{3}\xi_B^{3/2}}{1701\gamma} \check{q}^2+o(\check{q}^2)\right]
\label{eq:amorunit}
\end{multline}
un d\'eveloppement exact repr\'esent\'e sur la figure \ref{fig:BeyondHD} et \`a mettre au nombre des succ\`es de cet article.


\section{Conclusion}


Nous avons pr\'esent\'e dans cet article une \'etude compl\`ete des processus d'interaction entre phonons dans des gaz de fermions froids
\`a basse temp\'erature, pour une interaction entre fermions de port\'ee nulle mais de force quelconque, donc aussi bien pour une relation de dispersion des phonons
concave que convexe aux faibles nombres d'onde $q$.
Ceci nous a permis de clarifier les conditions de validit\'e des th\'eories effectives de basse \'energie, telles que l'hydrodynamique,
en les comparant \`a une approche microscopique tenant compte des degr\'es de libert\'e internes des paires. Ces th\'eories effectives ne pr\'edisent
correctement les amplitudes de couplage entre phonons que sur la couche de masse.

Un des principaux apports de cette \'etude est l'obtention microscopique de l'amplitude de couplage $2 \leftrightarrow 2$.
En effet, son expression au second ordre de la th\'eorie des perturbations inclut des processus non r\'esonnants $2 \leftrightarrow 1$
et $3 \leftrightarrow 0$. Comme ces processus peuvent en fait \^etre r\'esonnants dans un traitement 
d'hydrodynamique quantique, o\`u le spectre d'excitation
est lin\'eaire, Landau et Khalatnikov ont d\^u introduire \`a la main un terme de courbure dans la relation de dispersion, afin d'\'eviter 
une divergence de l'amplitude de couplage. \`A premi\`ere vue, cette proc\'edure est risqu\'ee puisque, comme nous l'avons dit, 
l'hydrodynamique quantique ne pr\'edit pas correctement l'amplitude des processus non r\'esonnants (donc hors de la couche de masse). 
Afin d'y apporter une justification microscopique, sans avoir
\`a effectuer les calculs tr\`es lourds que requerrait notre approche microscopique fermionique, nous nous sommes rabattus sur un mod\`ele de bosons 
en interaction faible, avec une port\'ee de l'interaction conduisant \`a une relation de dispersion concave. 
Bien que les expressions microscopiques des amplitudes de couplage diff\`erent de celles de l'hydrodynamique, nous trouvons, apr\`es sommation
au second ordre de tous les processus non r\'esonnants \`a trois phonons et sommation au premier ordre du processus $2 \leftrightarrow 2$ direct,
et apr\`es restriction \`a la couche de masse du couplage $2 \leftrightarrow 2$ effectif r\'esultant, que la prescription de Landau-Khalatnikov 
et la th\'eorie microscopique sont en accord.

Enfin, nous avons donn\'e des expressions universelles des taux d'amortissement pour les processus $2\leftrightarrow1$ et $2\leftrightarrow2$ \`a basse temp\'erature,
universelles au sens o\`u les fonctions r\'eduites $\tilde{q}\mapsto\tilde{\Gamma}(\tilde{q})$ apparaissant
dans les taux ne d\'ependent ni de l'esp\`ece atomique, ni de la force des interactions, ni de la temp\'erature. Le plus int\'eressant est l'obtention analytique, en fonction de $q$, du taux d'amortissement des phonons dans
le cas concave, d\'etermin\'e \`a l'ordre dominant en temp\'erature par les processus $2\leftrightarrow2$, voir l'\'equation \eqref{eq:Gamma2d2} et 
la Fig.~\ref{fig:F}, et qui a fait l'objet d'une lettre \cite{prepubamor}.
Dans les cas limites $\hbar c q \ll k_{\rm B} T$  et $\hbar c q \gg k_{\rm B} T$, notre r\'esultat est en d\'esaccord avec celui de la r\'ef\'erence
\cite{Khalatnikov1949} et lui est sous-dominant par deux ordres en $\hbar c q /k_{\rm B} T$. Ceci r\'esulte de l'hypoth\`ese erron\'ee faite dans 
\cite{Khalatnikov1949} que certains diagrammes d'interaction sont n\'egligeables, alors qu'en fait ils interf\`erent destructivement
avec le diagramme suppos\'e \^etre dominant. Nous avons aussi calcul\'e, pour le gaz unitaire \`a temp\'erature nulle, la premi\`ere 
correction {$\propto q^7$} \`a la pr\'ediction hydrodynamique {$\propto q^5$}
sur le taux d'amortissement d'un phonon unique, ce qui permet d'affiner la pr\'ediction
de \cite{Salasnich2015} par {$(i)$} inclusion {v\'eritable} d'une amplitude de couplage au-del\`a de l'hydrodynamique \cite{SonWingate2006},
{$(ii)$} prise en compte {d'une conservation de l'\'energie à $\hbar \Gamma_\qq$ pr\`es dans la règle d'or
de Fermi, c'est-à-dire d'un terme imaginaire pur d'ordre $q^5$ dans le d\'enominateur d'\'energie de la th\'eorie des perturbations},
et {$(iii)$ prise en compte des processus d'ordre plus \'elev\'e que Beliaev, en particulier du processus de couplage
effectif $1\to3$ trait\'e au second ordre. }

{Toutes nos pr\'edictions peuvent \^etre v\'erifi\'ees en laboratoire, dans l'\'etat de l'art des exp\'eriences sur les gaz d'atomes froids fermioniques. En particulier, une discussion pr\'ecise de l'observabilit\'e 
de l'amortissement de Landau-Khalatnikov $2\leftrightarrow 2$ avec des atomes froids pi\'eg\'es dans un potentiel
à fond plat est pr\'esent\'ee dans la r\'ef\'erence \cite{prepubamor}~; une m\'ethode d'excitation y est propos\'ee, qui doit permettre de peupler un mode d'onde stationnaire de phonons bien pr\'ecis
dans la bo\^\i te dans un r\'egime proche
de la limite thermodynamique, avec un nombre d'onde $q$ bien plus grand que la taille du syst\`eme mais
encore dans la partie lin\'eaire de la branche phononique.}

\appendix

\section{\'Equation implicite pour la branche d'excitation collective}
\label{app:RPA}

Nous expliquons ici comment passer du syst\`eme lin\'eaire (\ref{eq:systvaria1},\ref{eq:systvaria2}) \`a l'\'equation implicite (\ref{eq:dedispersion}) sur la pulsation
$\omega_\qq$ du mode collectif de vecteur d'onde $\qq$, dans la limite d'un pas
du r\'eseau tendant vers z\'ero, $l\to 0$. 
On introduit pour cela les amplitudes collectives
\bea
Y_\qq = \frac{g_0}{L^3} \sum_{\kk\in\mathcal{D}} W^+_{\kk\qq} y_\kk^\qq &\quad& S_\qq = \frac{g_0}{L^3} \sum_{\kk\in\mathcal{D}} W^-_{\kk\qq} s_\kk^\qq \label{eq:defcoll1}\\
y_\qq = \frac{g_0}{L^3} \sum_{\kk\in\mathcal{D}} w^-_{\kk\qq} y_\kk^\qq   &\quad& s_\qq = \frac{g_0}{L^3} \sum_{\kk\in\mathcal{D}} w^+_{\kk\qq} s_\kk^\qq \label{eq:defcoll2}
\eea
On résout le système linéaire $2\times2$ (\ref{eq:systvaria1},\ref{eq:systvaria2}) pour exprimer les inconnues $y_\kk^\qq$ et $s_\kk^\qq$ en fonction des amplitudes collectives. Puis on injecte les expressions obtenues dans les définitions (\ref{eq:defcoll1},\ref{eq:defcoll2}) pour aboutir au système linéaire homogène
\begin{multline}
\!\!\!\!\!\!\!\!\! \left( \begin{array}{llll}
\Sigma_{W^+ W^+}^\epsilon - 1 & \Sigma_{W^+ W^-}^\omega & -\Sigma_{W^+ w^-}^\epsilon & \Sigma_{W^+ w^+}^\omega \\
\Sigma_{W^- W^+}^\omega  & \Sigma_{W^- W^-}^\epsilon - 1 & -\Sigma_{W^- w^-}^\omega & \Sigma_{W^- w^+}^\epsilon \\
\Sigma_{w^- W^+}^\epsilon  & \Sigma_{w^- W^-}^\omega & -\Sigma_{w^- w^-}^\epsilon -1 & \Sigma_{w^- w^+}^\omega \\
\Sigma_{w^+ W^+}^\omega  & \Sigma_{w^+ W^-}^\epsilon & -\Sigma_{w^+ w^-}^\omega & \Sigma_{w^+ w^+}^\epsilon -1 \\
\end{array} \right) \\
\times\begin{pmatrix}
Y_\qq \\
S_\qq \\
y_\qq \\
s_\qq
\end{pmatrix}
=0
\label{eq:auxvalpropres}
\end{multline}
Dans les notations
\bea
\Sigma_{ab}^\epsilon &=& \frac{g_0}{L^3} \sum_{\kk\in\mathcal{D}} \frac{\epsilon_{\kk\qq} a_{\kk\qq} b_{\kk\qq}}{(\hbar\omega_\qq)^2-(\epsilon_{\kk\qq})^2} \\
\Sigma_{ab}^\omega &=& \frac{g_0}{L^3} \sum_{\kk\in\mathcal{D}} \frac{\hbar\omega_{\qq} a_{\kk\qq} b_{\kk\qq}}{(\hbar\omega_\qq)^2-(\epsilon_{\kk\qq})^2} 
\eea
les caractères $a$ et $b$ représentent l'un quelconque des $W^+,W^-,w^+,w^-$. 
Le système \eqref{eq:auxvalpropres} se simplifie à la limite continue $l\to0$. Comme $g_0\to0$ et $w_{\kk\qq}^\pm\underset{k\to+\infty}{=}O(1/k^2)$, on constate que dans les troisième et quatrième lignes tous les $\Sigma$ tendent vers $0$, ce qui conduit à 
\be
y_\qq=s_\qq=0
\label{eq:yqsq}
\ee
Nous divisons ensuite les deux premières lignes de \eqref{eq:auxvalpropres} par $g_0$. En vertu de l'équation \eqref{eq:gap}, on constate que tous les éléments de matrice divisés ont une limite finie et non nulle. Le système se réduit alors à son bloc supérieur gauche, que nous écrivons à la limite thermodynamique sous la forme~:
\be
\begin{pmatrix}
I_{++}(\omega_\qq,q) & \hbar\omega_\qq I_{+-}(\omega_\qq,q) \\
\hbar\omega_\qq I_{+-}(\omega_\qq,q) & I_{--}(\omega_\qq,q)
\end{pmatrix}
\begin{pmatrix}
Y_\qq \\
S_\qq 
\end{pmatrix}
=0
\label{eq:proprecollective}
\ee
avec des intégrales $I$, fonctions de la pulsation propre $\omega_\qq$ et du nombre d'onde $q$ d\'efinies par (\ref{eq:intpp},\ref{eq:intmm},\ref{eq:intpm}). 
L'\'equation implicite recherch\'ee sur la pulsation propre (\ref{eq:dedispersion}) exprime simplement l'annulation du d\'eterminant de la matrice $2\times 2$ dans l'\'equation (\ref{eq:proprecollective}).


\section{Fractions rationnelles en $x$ et $y$ apparaissant dans l'amplitude des processus à trois phonons}


\label{ann:fractionsrationnelles}
Les fractions rationnelles en $x$ et $y$ intervenant dans l'expression \eqref{eq:finaleA} de l'amplitude de couplage microscopique entre phonons valent
\bea
J(x,y) \!\!\!&=&\!\!\! \frac{3 x y}{2 x y+2}   \label{eq:polydexy} \\
A(x,y) \!\!\!&=&\!\!\! \frac{\sum_{i=0}^3 A_i(y) x^i}{360 \left(x^3+x\right) \left(y^2+1\right)^3}   \\
B(x,y) \!\!\!&=&\!\!\! \frac{\sum_{i=0}^3 B_i(y) x^i}{18 \left(x^3+x\right) \left(y^2+1\right)^2}   \\
C(x,y) \!\!\!&=&\!\!\! \frac{y (x y+1)}{6 \left(x^3+x\right) \left(y^2+1\right)^2}  \\
&&\times \left[x^2 \left(5 y^2+3\right)-x y \left(y^2-3\right)+2\left(y^2+2\right)\right] \notag
\label{eq:Axy}
\eea
avec
\bea
A_0(y) &=& -20 y \left(5 y^4+5 y^2+2\right)  \label{eq:polydey} \\
A_1(y) &=& 2 \left(10 y^6+109 y^4+63 y^2+24\right)  \\
A_2(y) &=& -y \left(53 y^4+186 y^2+13\right)  \\
A_3(y) &=& 25 y^6+116 y^4+167 y^2+36  \\
B_0(y) &=& -2 y \left(3 y^2+2\right)   \\
B_1(y) &=& 3 y^4+17 y^2+8  \\
B_2(y) &=& - 2 y \left(y^2+4\right)  \\
B_3(y) &=& 4 y^4+11 y^2+9  
\eea
Leurs valeurs dans la limite CBE $x\to0,y\sim-4/x$, utiles pour raccorder le résultat \eqref{eq:finaleA} à son équivalent dans la théorie de Bogolioubov sont
\bea
A(x,y) &\underset{\substack{x\to0\\y\sim-4/x}}{\to} \frac{1}{8}  \label{eq:limBEC} \qquad
B(x,y) &\underset{\substack{x\to0\\y\sim-4/x}}{\to} \frac{1}{4}  \\
J(x,y) &\underset{\substack{x\to0 \\ y\sim-4/x}}{\to} 2  \qquad
C(x,y) &\underset{\substack{x\to0\\y\sim-4/x}}{\to} \frac{3}{4} \notag 
\eea
Pour des processus résonnants, ces fractions rationnelles se combinent pour donner le coefficient thermodynamique
\be
1+\frac{\rho}{3}{\frac{\dd^2\mu}{\dd{\rho}^2}}\bb{\frac{\dd\mu}{\dd{\rho}}}^{-1}  
   = \frac{\sum_{i=0}^3 D_i(y) x^i}{9(x^3+x)(y^2+1)^2} \label{eq:unpluslambdaF}
\ee
avec
\bea
D_0(y) &=& -4 y^3 \\
D_1(y) &=& 4 \left(y^4+6 y^2+2\right) \\
D_2(y) &=& -y \left(y^4-6 y^2+5\right) \\
D_3(y) &=& 9 y^4+14 y^2+9 
\eea

{\rosso
\section{L'origine du d\'esaccord sur les amplitudes de couplage Beliaev-Landau aux faibles nombres d'onde
en dehors de la couche de masse, entre la th\'eorie microscopique et l'hydrodynamique quantique}
\label{appen:diffmicrohydro}

Dans cette annexe, comme nous l'annoncions dans l'avant-dernier paragraphe de la section \ref{sec:3phonons},
nous cherchons \`a percer \`a jour l'origine physique du d\'esaccord sur les amplitudes de couplage Beliaev-Landau aux faibles nombres d'onde
en dehors de la couche de masse, entre la th\'eorie microscopique et l'hydrodynamique quantique, 
les amplitudes divergeant dans le premier cas et tendant vers z\'ero dans le deuxi\`eme cas.
Pour nous \'epargner les calculs assez p\'enibles auxquels conduirait le mod\`ele fermionique microscopique,
nous utilisons ici un mod\`ele bosonique microscopique, celui du gaz spatialement uniforme de bosons 
en interaction faible, avec un hamiltonien $\hat{H}$ grand canonique au potentiel chimique $\mu$ de la forme
\be
\hat{H}=l^3\sum_\rr \hat{\psi}^\dagger \left(-\frac{\hbar^2}{2m}\Delta_\rr -\mu\right)\hat{\psi}
+\frac{g_0}{2} \sum_\rr l^3 \hat{\psi}^\dagger\hat{\psi}^\dagger\hat{\psi}\hat{\psi}
\label{eq:hamilmicrobosons}
\ee
Il est alors ais\'e de faire le lien entre cette th\'eorie microscopique et l'hydrodynamique quantique
au travers de la repr\'esentation module-phase de l'op\'erateur champ bosonique \cite{CastinMora2003},
\be
\hat{\psi}(\rr)=\eee^{\ii\hat{\theta}(\rr)} \hat{\rho}^{1/2}(\rr)
\ee
le champ de phase $\hat{\theta}(\rr)$ et le champ de densit\'e $\hat{\rho}(\rr)$  \'etant canoniquement
conjugu\'es.

Dans le formalisme de Bogolioubov brisant la sym\'etrie $U(1)$, on d\'ecompose le champ bosonique
comme suit, $\hat{\psi}(\rr)=\psi_0+\delta\hat{\psi}(\rr)$, o\`u $\psi_0>0$ minimise l'\'energie de champ classique, et l'on d\'eveloppe le hamiltonien en puissances de $\delta\hat{\psi}$, $\hat{H}=H_0+\hat{H}_2+\hat{H}_3+\ldots$. Le champ $\delta\hat{\psi}(\rr)$ est ensuite d\'evelopp\'e sur les modes normaux de $\hat{H}_2$ 
(d\'eveloppement de Bogolioubov)~: 
\be
\delta\hat{\psi}(\rr)=\frac{1}{L^{3/2}}\sum_{\qq\ne\mathbf{0}} U_\qq \eee^{\ii\qq\cdot\rr} \hat{b}^{\delta\psi}_\qq + 
V_\qq \eee^{-\ii\qq\cdot\rr} (\hat{b}^{\delta\psi}_\qq)^\dagger
\label{eq:devBogdpsi}
\ee
Pour simplifier, nous omettons ici la contribution des modes anormaux induits par la brisure de sym\'etrie.
Notre notation fait explicitement r\'ef\'erence au fait que les op\'erateurs de cr\'eation et d'annihilation
de quasi-particules r\'esultent d'un d\'eveloppement de $\hat{H}$ en puissances de $\delta\hat{\psi}$.
L'injection du d\'eveloppement (\ref{eq:devBogdpsi}) dans $\hat{H}_3$ conduit \`a l'amplitude
de couplage Beliaev-Landau r\'eduite dans la repr\'esentation cart\'esienne $\delta\hat{\psi}$~:
\be
\mathcal{A}_{\delta\psi}^{2\leftrightarrow 1}(\qq_1,\qq_2;\qq_3)= \frac{3}{4}s_1 s_2 s_3+\frac{1}{4}(s_1 d_2 d_3+s_2 d_1 d_3- s_3 d_1 d_2)
\label{eq:A2d1dpsi}
\ee
avec $s_i=s_{\qq_i}$, $s_\qq=U_\qq+V_\qq$, $d_i=d_{\qq_i}$ et $d_\qq=U_\qq-V_\qq$. 

Dans la repr\'esentation phase-module, en adaptant la r\'ef\'erence \cite{CastinMora2003} au cas \`a sym\'etrie
bris\'ee, nous d\'eveloppons $\hat{\psi}(\rr)$ en puissances des fluctuations de densit\'e $\delta\hat{\rho}(\rr)= \hat{\rho}(\rr)-\psi_0^2$ et de phase $\hat{\theta}(\rr)$ si bien que $\delta\hat{\psi}(\rr)=\delta\hat{\psi}_1(\rr)+\delta\hat{\psi}_2(\rr)+\ldots$, avec
\bea
\delta\hat{\psi}_1(\rr) &=& \frac{\delta\hat{\rho}(\rr)}{2\psi_0} + \ii \psi_0 \hat{\theta}(\rr) \\
\delta\hat{\psi}_2(\rr) &=& -\frac{1}{2} \psi_0 \hat{\theta}^2(\rr) -\frac{\delta\hat{\rho}^2(\rr)}{8\psi_0^3} 
+ \ii \frac{\hat{\theta}(\rr)\delta\hat{\rho}(\rr)}{2\psi_0}
\eea
\`A l'ordre quadratique en $\delta\hat{\rho}$ et $\hat{\theta}$, nous pouvons remplacer $\delta\hat{\psi}$
par $\delta\hat{\psi}_1$ dans $\hat{H}_2$~; le d\'eveloppement modal habituel du hamiltonian correspondant
quadratique en $(\delta\hat{\rho},\hat{\theta})$ donne (en omettant toujours les modes anormaux li\'es
\`a la brisure de sym\'etrie)~:
\be
\delta\hat{\psi}_1(\rr)=
\frac{1}{L^{3/2}}\sum_{\qq\ne\mathbf{0}} U_\qq \eee^{\ii\qq\cdot\rr} \hat{b}^{\rho\theta}_\qq + 
V_\qq \eee^{-\ii\qq\cdot\rr} (\hat{b}^{\rho\theta}_\qq)^\dagger
\label{eq:devmodbogphmod}
\ee
avec les m\^emes amplitudes $(U_\qq,V_\qq)$ que dans l'\'equation (\ref{eq:devBogdpsi}) 
mais avec des coefficients \`a valeurs op\'erateurs
$\hat{b}^{\rho\theta}_\qq,(\hat{b}^{\rho\theta}_\qq)^\dagger$ qui co\"\i ncident avec 
$\hat{b}^{\delta\psi}_\qq,(\hat{b}^{\delta\psi}_\qq)^\dagger$, tout comme $\delta\hat{\psi}_1$ co\"\i ncide
avec $\delta\hat{\psi}$, seulement au premier ordre en les fluctuations.
\`A l'ordre cubique en $\delta\hat{\rho}$ et $\hat{\theta}$, deux types de contributions apparaissent~:
$(i)$ une contribution \'evidente, provenant du remplacement de $\delta\hat{\psi}$ par $\delta\hat{\psi}_1$
dans $\hat{H}_3$, ce qui conduit \`a l'amplitude de couplage Beliaev-Landau $\mathcal{A}_{\delta\psi}^{2\leftrightarrow 1}$, cette fois entre les $\hat{b}^{\rho\theta}_\qq$, et $(ii)$ une contribution plus indirecte,
provenant du remplacement de $\delta\hat{\psi}$ par $\delta\hat{\psi}_1+\delta\hat{\psi}_2$ dans $\hat{H}_2$
et de l'extraction des contributions cubiques correspondantes, de la forme $\delta\hat{\psi}_2^\dagger\delta\hat{\psi}_1$ et hermitien conjugu\'e, ce qui donne naissance \`a un terme suppl\'ementaire $\mathcal{A}^{2\leftrightarrow 1}_{\mathrm{via}\, H_2}$ dans l'amplitude de couplage Beliaev-Landau. Finalement, l'amplitude
de couplage Beliaev-Landau en repr\'esentation $(\hat{\rho},\hat{\theta})$ s'\'ecrit~:
\be
\mathcal{A}_{\rho\theta}^{2\leftrightarrow 1}= \mathcal{A}_{\delta\psi}^{2\leftrightarrow 1} + \mathcal{A}^{2\leftrightarrow 1}_{\mathrm{via}\, H_2}
\ee
avec
\begin{multline}
\mathcal{A}^{2\leftrightarrow 1}_{\mathrm{via}\, H_2}(\qq_1,\qq_2;\qq_3)=  \\
\frac{\check{\omega}_3d_3}{8}(d_1d_2-s_1s_2) +\frac{\check{\omega}_3s_3}{8}(s_1d_2+s_2d_1)  \\
-\frac{\check{\omega}_2d_2}{8}(d_1d_3+s_1s_3) -\frac{\check{\omega}_1d_1}{8}(d_2d_3+s_2s_3)  \\
+\frac{d_3}{8}(s_1 s_2 \check{\omega}_2 + s_2 s_1 \check{\omega}_1) 
-\frac{s_3}{8}(d_2s_1 \check{\omega}_1 + d_1 s_2 \check{\omega}_2)
\end{multline}
et $\check{\omega}_i=\hbar\omega_{\qq_i}/mc^2$, sachant que $mc^2=\mu$ \`a cet ordre. Les expressions
de Bogolioubov $\omega_i=[\check{E}_i(\check{E}_i+2)]^{1/2}$ et $s_i=1/d_i=[\check{E}_i/(\check{E}_i+2)]^{1/4}$
conduisent \`a $\check{\omega}_i d_i =(\check{E}_i+2)s_i$ et $\check{\epsilon}_i s_i=\check{E}_i d_i$, 
o\`u $\check{E}_i=E_{\qq_i}/mc^2$ et $E_\qq=\hbar^2 q^2/2m$. \`A partir de la conservation de l'impulsion
$\qq_3=\qq_1+\qq_2$ et ses carr\'es $q_3^2=(\qq_1+\qq_2)^2$, 
$q_2^2=(\qq_3-\qq_1)^2$ et $q_1^2=(\qq_3-\qq_2)^2$, nous obtenons
\begin{multline}
\mathcal{A}_{\rho\theta}^{2\leftrightarrow 1}(\qq_1,\qq_2;\qq_3)=\\
\frac{1}{8}(s_3 d_1 d_2 \check{\qq}_1\cdot\check{\qq}_2+
s_2 d_1 d_3 \check{\qq}_1\cdot\check{\qq}_3 + 
s_1 d_2 d_3 \check{\qq}_2\cdot\check{\qq}_3)\\
-\frac{1}{16} s_1 s_2 s_3 (\check{q}_1^2+\check{q}_2^2+\check{q}_3^2)
\label{eq:ampcouplbellanrhoth}
\end{multline}
avec $\check{\qq}_i=\hbar\qq_i/mc$. 

Nous sommes d\'esormais en mesure de comparer les amplitudes de couplage Beliaev-Landau $\mathcal{A}_{\delta\psi}^{2\leftrightarrow 1}$ et $\mathcal{A}_{\rho\theta}^{2\leftrightarrow 1}$  entre elles et \`a 
l'hydrodynamique quantique. En utilisant les expressions de $s_i$ et $d_i$ ci-dessus, ainsi que l'\'equation (\ref{eq:A2d1dpsi}), nous trouvons que l'amplitude $\mathcal{A}_{\delta\psi}^{2\leftrightarrow 1}$ diverge lorsque
tous les $q_i$ tendant vers z\'ero en dehors de la couche de masse. 
Au contraire, m\^eme en dehors de la couche de masse, l'amplitude $\mathcal{A}_{\rho\theta}^{2\leftrightarrow 1}$ tend vers z\'ero dans cette limite et devient \'equivalente \`a l'amplitude (\ref{eq:Ahydro21}) de l'hydrodynamique quantique (ici $\Lambda_{\rm F}=0$)~; ces deux amplitudes ne co\"\i ncident pas, cependant, mais diff\`erent
par des corrections relatives $O(q_i^2)$, parce que les hamiltoniens microscopique (\ref{eq:hamilmicrobosons})
et de l'hydrodynamique quantique diff\`erent.

\`A titre d'ultime v\'erification, nous injectons directement le d\'eveloppement 
(\ref{eq:devmodbogphmod}) (r\'ecrit sous la forme d'un d\'eveloppement de $\delta\hat{\rho}$ et $\hat{\theta}$)
dans les termes du hamiltonien (40) de la r\'ef\'erence \cite{CastinMora2003} cubiques en $\delta\hat{\rho}$
et $\hat{\theta}$. Nous retrouvons alors directement l'amplitude de couplage Beliaev-Landau (\ref{eq:ampcouplbellanrhoth}) entre les $\hat{b}_\qq^{\rho\theta}$. De plus, en n\'egligeant partout dans ce hamiltonien (40)
l'\'energie cin\'etique contenue dans la variation spatiale de la fluctuation de densit\'e $\delta\hat{\rho}$
du champ (\`a l'origine du terme dit de pression quantique), nous retrouvons exactement
le couplage Beliaev-Landau (\ref{eq:Ahydro21}) de l'hydrodynamique (avec ici $\Lambda_{\rm F}=0$).
}

\section{Couplage de Beliaev au-del\`a de l'hydrodynamique}
\label{appen:SonetWingate}

Nous expliquons ici comment obtenir, pour le gaz unitaire, l'amplitude de couplage (\ref{eq:couplA})
des phonons par processus de Beliaev sur la couche de masse
au premier ordre au-del\`a de l'hydrodynamique quantique,
en partant du lagrangien de Son et Wingate \cite{SonWingate2006}.

Il nous est commode de passer au formalisme hamiltonien en introduisant le champ $\Pi$,
canoniquement conjugu\'e au champ de phase $\phi$ et qui repr\'esente les
fluctuations de densit\'e au signe pr\`es. Dans un syst\`eme d'unit\'es tel que $\hbar=m=1$, nous obtenons les contributions au hamiltonien~:
\bea
\!\!\!\!\!\!\!\!\!{\cal H}_2^{(0)} &=& \frac{2 \mu^{-1/2}}{15 c_0} \Pi^2 + \frac{5}{2} c_0 \mu^{3/2} \frac{1}{2} (\mathbf{grad}\, \phi )^2 \\
\!\!\!\!\!\!\!\!\!{\cal H}_3^{(0)} &=& \frac{4 \mu^{-2}}{3 (15 c_0)^2} \Pi^3 - \frac{1}{2} \Pi  (\mathbf{grad}\, \phi )^2 \\
\!\!\!\!\!\!\!\!\!{\cal H}_2^{(2)} &=& - \frac{16 c_1 \mu^{-3/2}}{(15 c_0)^2} (\mathbf{grad} \, \Pi)^2 - c_2 \mu^{1/2} (\Delta \phi )^2 \\
\!\!\!\!\!\!\!\!\!{\cal H}_3^{(2)} &=& - \frac{96 c_1 \mu^{-3}}{(15 c_0)^3} \Pi (\mathbf{grad} \, \Pi)^2 + \frac{2 c_2 \mu^{-1}}{15 c_0} 
\Pi (\Delta \phi )^2 
\eea
o\`u l'indice $2$ ou $3$ est l'ordre du d\'eveloppement du hamiltonien en puissances de $\Pi$ et $\phi$ et l'exposant $(0)$ ou $(2)$
est l'ordre du d\'eveloppement en puissances des gradients spatiaux, l'ordre $(0)$ \'etant l'hydrodynamique usuelle.
On aura compris que les champs $\Pi$ et $\mathbf{grad} \, \phi$ sont en fait du m\^eme ordre.
\`A l'ordre de l'hydrodynamique, le hamiltonien ne d\'epend que d'une seule constante $c_0$ qu'il faille d\'eduire d'une th\'eorie microscopique.
Elle s'exprime comme suit en fonction du param\`etre de Bertsch $\xi_B$ (\ref{eq:Bertsch})~:
\be
c_0=\frac{2^{5/2}}{15 \pi^2 \xi_B^{3/2}} 
\ee
La premi\`ere correction \`a l'hydrodynamique fait intervenir deux autres constantes sans dimension $c_1$ et $c_2$ qui, au contraire de $c_0$, n'ont pas encore
\'et\'e mesur\'ees ni m\^eme d\'eduites d'une th\'eorie microscopique. En suivant la proc\'edure d\'ej\`a utilis\'ee dans cet article,
nous obtenons \`a partir de l'approximation quadratique ${\cal H}_2^{(0)}+{\cal H}_2^{(2)}$ du hamiltonien le spectre d'excitation comme
dans la r\'ef\'erence \cite{SonWingate2006}~:
\begin{multline}
\label{eq:spectrum_SW}
\omega_\qq = \left( \frac{2\mu}{3} \right)^{1/2} q \bbcrol{ 1-\pi^2 (2\xi_B)^{1/2} \left(c_1+ \frac{3}{2}c_2\right) \left(\frac{q}{k_{\rm F}}\right)^2} \\
\bbcror{\vphantom{\frac{1}{1}}+ o(q^2)}\equiv cq \left[ 1 + \frac{\gamma}{8} \left( \frac{\hbar q}{mc}\right)^2 + o(q^2) \right] 
\end{multline}
et le d\'eveloppement modal des champs dans le volume de quantification $L^3$~:
\bea
\Pi(\rr)&=&\frac{1}{L^{3/2}} \sum_{\qq \neq \zero} \Pi_\qq (b_\qq + b_{-\qq}^\ast) \eee^{\ii \qq \cdot \rr}\\
\phi(\rr)&=&\frac{1}{L^{3/2}} \sum_{\qq \neq \zero} \phi_\qq (b_\qq - b_{-\qq}^\ast) \eee^{\ii \qq \cdot \rr}
\eea
avec les amplitudes
\bea
 \!\!\!\!\!\!\!\!\!\!\Pi_\qq  &=& \frac{1}{\sqrt{2}} \left( \frac{A_\qq}{B_\qq}\right)^{1/4} \\
 \!\!\!\!\!\!\!\!\!\! \phi_\qq &=& \frac{\ii}{\sqrt{2}} \left( \frac{A_\qq}{B_\qq}\right)^{-1/4} \\
 \!\!\!\!\!\!\!\!\!\! \frac{A_\qq}{B_\qq} \!\!\!&=&\!\!\! \frac{75}{8}(c_0 \mu q)^2 \!\bbcro{ 1 + \frac{8 \mu^{-1}}{15 c_0} \left(\!c_1\!-\!\frac{3}{2}c_2\!\right)q^2
\!+ o(q^2) } 
 \label{eq:modal_amplitudes}
\eea
En effet, $\left(\begin{array}{c} \Pi_\qq \\ \phi_\qq \end{array}\right)$ est un vecteur propre de la matrice
$\left(\begin{array}{cc} 0 & -A_\qq \\ B_\qq & 0 \end{array}\right)$ avec la valeur propre $- \ii \omega_\qq$,
cette matrice r\'esultant des \'equations du mouvement hamiltoniennes.
Soulignons que ce ne sont pas les m\^emes combinaisons lin\'eaires des constantes $c_1$ et $c_2$ qui apparaissent dans le spectre (\ref{eq:spectrum_SW})
et dans les amplitudes des modes (\ref{eq:modal_amplitudes}). Ensuite, en injectant la d\'ecomposition modale dans les termes cubiques
${\cal H}_3^{(0)}+{\cal H}_3^{(2)}$ du hamiltonien et en isolant les termes de Beliaev $2\leftrightarrow1$ comme nous l'avons fait dans 
l'\'equation (\ref{eq:H3hydro}), nous obtenons les amplitudes de couplage de Beliaev sur la couche de masse au premier ordre au-del\`a de l'hydrodynamique, sous la forme (\ref{eq:couplA}).

\section{Processus \`a trois et quatre phonons dans un mod\`ele de bosons}
\label{appen:modelebosons}

Nous d\'ecrivons et impl\'ementons ici le mod\`ele microscopique de bosons utilis\'e dans
la section \ref{sec:modelebosons} pour tester la valeur du couplage effectif $2\leftrightarrow 2$ 
entre phonons pr\'edite par l'hydrodynamique quantique. Nous consid\'erons donc des bosons 
sans spin sur un
r\'eseau cubique de pas $l$, avec une interaction de port\'ee $b$ suffisamment grande 
pour que leur spectre d'excitation soit concave aux faibles nombres d'onde $q$. 
Le hamiltonien du gaz dans le modèle sur réseau s'écrit
\begin{multline}
\hat{H}_{\rm B}= l^3  \sum_{\rr} \hat{\psi}^\dagger(\rr) \left( - \frac{\hbar^2}{2m_{\rm B}}  \Delta_{\rr}\right)  \hat{\psi}(\rr) \\
+ \frac{l^6}{2} \sum_{\rr,\rr'} V(\rr-\rr')  \hat{\psi}^\dagger(\rr) \hat{\psi}^\dagger(\rr') \hat{\psi}(\rr'){\hat{\psi}}(\rr)
\label{eq:hamiltonienB}
\end{multline}
avec le potentiel d'interaction
\be
V(\rr) = V_0 \eee^{-r^2/2b^2}
\ee
de transformée de Fourier
\be
\tilde{V}(\qq) = \tilde{V}_0 \eee^{-q^2b^2/2}\quad\mbox{avec}\quad\tilde{V}_0=(2\pi)^{3/2}b^3V_0
\ee

Le gaz de bosons est dans le régime d'interaction faible $(\rho_{\rm B} a_{\rm B}^3)^{1/2} \ll 1$, où $\rho_{\rm B}=N_{\rm B}/L^3$ est la densité de bosons et $a_{\rm B}$, la longueur de diffusion dans l'onde $s$ entre bosons, est donnée dans l'approximation de Born par $4\pi\hbar^2 a_{\rm B}/m_{\rm B}=\tilde{V}_0$.
Suivant la méthode habituelle de la théorie de Bogolioubov \cite{Bogolioubov1958} à symétrie $U(1)$ conservée \cite{Gardiner1997,CastinDum1998,CastinLesHouches2001,CastinSinatra2009}, nous écrivons l'opérateur champ bosonique sous la forme
\be
\hat{\psi}(\rr)=\eee^{\ii\hat{\theta}_0}\bbcro{\hat{n}_0^{1/2} \phi_0(\rr)+\hat{\Lambda}(\rr)}
\ee
où $\hat{\theta}_0$ est l'opérateur phase du condensat, $\hat{n}_0$ le nombre de bosons dans le mode $\phi_0(\rr)=1/L^{3/2}$ du condensat et l'opérateur champ non condensé $\hat{\Lambda}(\rr)$, orthogonal au mode du condensat $\phi_0(\rr)$, conserve le nombre de particules. Nous nous plaçons dans le sous-espace à nombre total de bosons $N_{\rm B}$ fixé et nous éliminons $\hat{n}_0$ grâce à la relation 
\be
\hat{n}_0 = N_{\rm B}- l^3  \sum_{\rr} \hat{\Lambda}^\dagger(\rr) \hat{\Lambda}(\rr)
\label{eq:devn0}
\ee
Pour décrire les processus $2\leftrightarrow2$, nous devons développer le hamiltonien en puissances de $\hat{\Lambda}$ jusqu'à l'ordre $4$
\be
\hat{H}_{\rm B}=\hat{H}_{\rm B0}+\hat{H}_{\rm B2}+\hat{H}_{\rm B3}+\hat{H}_{\rm B4}+\ldots
\ee
Nous avons \footnote{Nous sommes ici à la limite des grands $N_{\rm B}$ et nous négligeons $1$ devant $\hat{n}_0$ donc devant $N_{\rm B}$. Dans les équations \eqref{eq:H2B} et \eqref{eq:H4B}, nous avons avons simplifié la contribution de Hartree avec la contribution de type potentiel chimique provenant du développement de l'énergie d'interaction du condensat $\tilde{V}_0 {\hat{n}_0^2}/{2L^3}$ en puissances du nombre de particules non condensées.}
\bea
\hat{H}_{\rm B0} &=& \tilde{V}_0 \frac{N_{\rm B}^2}{2L^3} \label{eq:H0B} \\
\hat{H}_{\rm B2} &=& l^3 \sum_\rr \hat{\Lambda}^\dagger(\rr) \left( - \frac{\hbar^2}{2m_{\rm B}}  \Delta_{\rr}\right) \hat{\Lambda}(\rr) + \rho_{\rm B} l^6 \sum_{\rr,\rr'} V(\rr-\rr') \notag \\
&\times& \bb{\hat{\Lambda}^\dagger(\rr) \hat{\Lambda}(\rr')  + \frac{1}{2}\bbcro{\hat{\Lambda}^\dagger(\rr) \hat{\Lambda}^\dagger(\rr')+\hat{\Lambda}(\rr) \hat{\Lambda}(\rr')}}
\label{eq:H2B} \\
\hat{H}_{\rm B3} &=& \frac{\rho_{\rm B}^{1/2}}{2} l^6 \sum_{\rr,\rr'} V(\rr-\rr') 
\bbl{ \bbcro{\hat{\Lambda}^\dagger(\rr)+\hat{\Lambda}^\dagger(\rr')} \hat{\Lambda}(\rr') \hat{\Lambda}(\rr)}\notag \\
&+& \bbr{\hat{\Lambda}^\dagger(\rr') \hat{\Lambda}^\dagger(\rr) \bbcro{\hat{\Lambda}(\rr)+\hat{\Lambda}(\rr')}} \label{eq:H3B} \\
\hat{H}_{\rm B4} &=& \frac{l^6}{2}  \sum_{\rr,\rr'} V(\rr-\rr')  \hat{\Lambda}^\dagger(\rr) \hat{\Lambda}^\dagger(\rr') \hat{\Lambda}(\rr'){\hat{\Lambda}}(\rr) \notag \\
&-&\frac{\tilde{V}_0}{2L^3} \bb{l^3  \sum_{\rr} \hat{\Lambda}^\dagger(\rr) \hat{\Lambda}(\rr)}^2 
-\frac{1}{L^3} \bb{l^3  \sum_{\rr} \hat{\Lambda}^\dagger(\rr) \hat{\Lambda}(\rr)} \notag\\
&\times& \bbl{l^6\sum_{\rr,\rr'} V(\rr-\rr') \bbcrol{\hat{\Lambda}^\dagger(\rr) \hat{\Lambda}(\rr') + \frac{1}{2}\bbcrol{\hat{\Lambda}^\dagger(\rr) \hat{\Lambda}^\dagger(\rr')}}} \notag\\
&+&\bbr{\bbcror{\bbcror{\hat{\Lambda}(\rr) \hat{\Lambda}(\rr')}\vphantom{\frac{1}{1}}}\vphantom{\sum_{\rr,\rr'}} } 
\label{eq:H4B} 
\eea

\paragraph{Ordre $0$}
$\hat{H}_{\rm B 0}$ donne l'équation d'état à l'ordre le plus bas de la théorie
\be
\mu_{\rm B} = \rho_{\rm B} \tilde{V}_0 \equiv \frac{\hbar^2}{2m\xi^2}
\label{eq:etatB}
\ee
où nous avons introduit la longueur de relaxation $\xi$ et le potentiel chimique $\mu_{\rm B}$ du gaz de bosons.
En appliquant la relation \eqref{eq:vitesseson} à cette équation d'état, nous obtenons l'expression correspondante de la vitesse du son
\be
m_{\rm B} c_{\rm B}^2 = \mu_{\rm B}
\ee
\paragraph{Ordre $2$}
La diagonalisation de $\hat{H}_{\rm B 2}$ s'obtient par la transformation de Bogolioubov
\be
\hat{\Lambda}(\rr)= \frac{1}{L^{3/2}} \sum_{\qq\neq\zero} \bb{U_\qq^{\rm B} \hat{b}_\qq \eee^{\ii\qq\cdot\rr}+ V_\qq^{\rm B} \hat{b}_\qq^\dagger \eee^{-\ii\qq\cdot\rr} }
\label{eq:fleurB}
\ee
où les opérateurs $\hat{b}_\qq$ sont les opérateurs d'annihilation de quasi-particules bosoniques, et les amplitudes $U^{\rm B}_\qq$ et $V^{\rm B}_\qq$ sont données par
\bea
U^{\rm B}_\qq+V^{\rm B}_\qq &\equiv& S_\qq =\bb{\frac{\frac{\hbar^2q^2}{2m_{\rm B}}}{\frac{\hbar^2q^2}{2m_{\rm B}}+2\rho_{\rm B}\tilde{V}(\qq)}  }^{1/4} \\
U^{\rm B}_\qq-V^{\rm B}_\qq &\equiv& D_\qq = \frac{1}{S_\qq}
\eea
Nous avons introduit les pulsations propres des quasi-particules de Bogolioubov
\be
\hbar\omega^{\rm B}_\qq=\bbcro{\frac{\hbar^2q^2}{2m_{\rm B}}\bb{\frac{\hbar^2q^2}{2m_{\rm B}}+2\rho_{\rm B}\tilde{V}(\qq)}}^{1/2}
\ee
Nous voulons une relation de dispersion $q\mapsto\omega_\qq$ concave au voisinage de $q=0$, ce qui nous conduit à imposer la condition 
\be
b>\xi
\ee
\paragraph{Ordre $3$}
Nous injectons maintenant le développement modal \eqref{eq:fleurB} dans le hamiltonien cubique \eqref{eq:H3B} que nous mettons sous la forme \eqref{eq:H3hydro}
avec les constantes $m$, $c$ et $\rho$ remplacées respectivement par $m_{\rm B}$, $c_{\rm B}$ et $\rho_{\rm B}$ et des amplitudes de couplage cette fois-ci données par
\begin{multline}
\mathcal{A}_{\rm B}^{2\leftrightarrow1}(\qq_1,\qq_2;\qq_3)  =  \frac{1}{4\tilde{V}_0} \bbcrol{ \tilde{V}(\qq_3)S_{3}(S_{1} S_{2}-D_{1}D_{2})}  \\
\bbcror{+\tilde{V}(\qq_1)S_{1}(S_{2} S_{3}+D_{2}D_{3}) 
 +\tilde{V}(\qq_2)S_{2}(S_{1} S_{3}+D_{1}D_{3})} \label{eq:A2donne1B} 
 \end{multline}
 \begin{multline}
\mathcal{A}_{\rm B}^{3\leftrightarrow0}(\qq_1,\qq_2,\qq_3)  =  \frac{1}{12\tilde{V}_0} \bbcrol{\tilde{V}(\qq_1)S_{1}(S_{2} S_{3}-D_{2}D_{3})}  \\ 
\bbcror{+\tilde{V}(\qq_2)S_{2}(S_{1} S_{3}-D_{1}D_{3})+\tilde{V}(\qq_3)S_{3}(S_{1} S_{2}-D_{1}D_{2})} \label{eq:A3donne0B}
 \end{multline}
 avec la convention $S_i\equiv S_{\qq_i}$ et $D_i\equiv D_{\qq_i}$.
\paragraph{Ordre $4$}
Nous injectons le développement modal \eqref{eq:fleurB} dans le hamiltonien quartique \eqref{eq:H4B} que nous mettons sous la forme \eqref{eq:H2donne2hydro} avec la substitution $(m,c,\rho)\to (m_{\rm B},c_{\rm B},\rho_{\rm B})$ et
une amplitude de couplage $2\leftrightarrow2$ directe donnée par
\begin{multline}
\mathcal{A}_{\rm B}^{2\leftrightarrow2,{\rm dir}}(\qq_1,\qq_2;\qq_3,\qq_4) = \frac{1}{32\tilde{V}_0} \times \\  
\! \bbcrol{ \bb{\tilde{V}(\qq_1\!+\!\qq_2)\!+\!\tilde{V}(\qq_3\!+\!\qq_4)} \!(S_{1} S_{2}\!-\!D_{1} D_{2})(S_{3} S_{4}\!-\!D_{3}D_{4}) } \\ 
\!\!+\!\!\bb{\tilde{V}(\qq_3\!-\!\qq_1)\!+\!\tilde{V}(\qq_2\!-\!\qq_4)} \!(S_{1} S_{3}\!+\!D_{1} D_{3})(S_{2} S_{4}\!+\!D_{2}D_{4})\\
\bbcror{\!\!\!\!\!+\!\!\bb{\tilde{V}(\qq_4\!-\!\qq_1)\!+\!\tilde{V}(\qq_2\!-\!\qq_3)} \!(S_{1} S_{4}\!+\!D_{1} D_{4})(S_{2} S_{3}\!+\!D_{2}D_{3})} \!\!\!\!\!\!\!\!\!\!
\label{eq:A2donne2Bdir}
\end{multline}
Nous nous sommes placés ici dans le cas général où les $\qq_i$ et leurs opposés sont deux à deux distincts, auquel cas seul le premier terme de $\hat{H}_{\rm B4}$ contribue. En revanche les autres termes de $\hat{H}_{\rm B4}$, qui proviennent du développement \eqref{eq:devn0} de $\hat{n}_0$ en puissances du nombre de particules non condensées, contribuent à l'équation d'état au-delà de la théorie de Bogolioubov \cite{CastinMora2003}.
\paragraph{Amplitude de couplage effective}
\`A partir des amplitudes (\ref{eq:A2donne1B},\ref{eq:A3donne0B},\ref{eq:A2donne2Bdir}) nous formons l'amplitude de couplage effective du modèle bosonique sur la couche de masse, $\mathcal{A}_{\rm B, Cdm}^{2\leftrightarrow2,{\rm eff}}$ comme prescrit par l'équation \eqref{eq:A22eff}. Nous développons ensuite cette amplitude aux faibles vecteurs d'onde et l'exprimons en fonction des pulsations $\omega_{\qq_i}^{\rm B}$, $i=1,2,3,4$ ainsi que $\omega_{\qq_1+\qq_2}^{\rm B}$, $\omega_{\qq_1-\qq_3}^{\rm B}$ et $\omega_{\qq_1-\qq_4}^{\rm B}$. Nous retrouvons alors
la prédiction hydrodynamique \eqref{eq:A2donne2effhydro} corrig\'ee selon la prescription de Landau et Khalatnikov \cite{Khalatnikov1949}, comme il est dit au début de la section \ref{sec:A22effhydro}.

\section{Contribution de Beliaev et des processus d'ordre sup\'erieur dans le taux de d\'ecroissance d'un phonon \`a $T=0$}
\label{appen:1d3}

\subsection{Pr\'esentation du problème et lien avec la r\'esolvante}

On consid\`ere ici le taux de d\'ecroissance $\Gamma_\qq$ d'un phonon de vecteur d'onde $\qq$ pr\'epar\'e \`a temp\'erature nulle
dans un gaz spatialement homog\`ene de fermions pr\'esentant une branche d'excitation phononique $q\mapsto \omega_\qq$ convexe au voisinage de $q=0$.
La probl\'ematique g\'en\'erale est de d\'eterminer le comportement de $\Gamma_\qq$ dans la limite $q\to 0$, o\`u l'on
peut utiliser une th\'eorie effective de basse \'energie pour d\'ecrire le couplage entre les phonons. 
Le processus dominant est bien entendu celui $1\to 2$ de Beliaev~; en introduisant un param\`etre sans dimension
$\gamma >0 $ d\'ecrivant aussi bien la correction en $q^3$ au spectre lin\'eaire de l'hydrodynamique que la premi\`ere correction
\`a l'amplitude de couplage hydrodynamique $1\to 2$, nous avons obtenu le r\'esultat provisoire (\ref{eq:Bel_result_prov}) jusqu'\`a l'ordre
sous-dominant $q^7$. Dans cette annexe, il s'agit de donner quelques d\'etails sur l'obtention du r\'esultat
{complet (\ref{eq:Bel_result}) à l'ordre de Beliaev et de voir dans quelle mesure les processus d'ordre sup\'erieur \`a celui de Beliaev,
en réalité le processus en cascade $1\to 2\to 3$  de la figure \ref{fig:1d3}, contribuent \`a l'ordre $q^7$ pour conduire à la
correction (\ref{eq:Gamma1d3}) et au résultat final (\ref{eq:amorunit}).}

Dans le sous-espace de quantit\'e de mouvement totale fix\'ee $\hbar\qq$, l'\'etat \`a un phonon $|\qq\rangle$
est le seul \'etat discret, puisqu'il est le seul
à \^etre totalement caract\'eris\'e par la donn\'ee d'un unique vecteur d'onde.
C'est un \'etat propre de la partie $\hat{H}_2$ du hamiltonien quadratique en les op\'erateurs
de cr\'eation $\hat{b}^\dagger$ et d'annihilation $\hat{b}$ de phonons.
Cependant, il est coupl\'e aux continua \`a deux phonons et plus par le reste du hamiltonien $\hat{V}=\hat{H}_3+\hat{H}_4+\ldots$,
qui contient des termes cubiques, quartiques, etc, \'ecrits dans l'ordre normal vis-\`a-vis des $\hat{b}$.
L'\'etat discret va donc g\'en\'eriquement 
se diluer dans les continua et donner naissance \`a un p\^ole complexe $z_\qq$ dans le prolongement analytique de la r\'esolvante
$\hat{G}(z)=(z-\hat{H})^{-1}$ du hamiltonien complet \cite{Cohen}. Ce p\^ole s'\'ecrit
\be
z_\qq =\hbar\omega_\qq-\ii \frac{\hbar\Gamma_\qq}{2}
\label{eq:decompzq}
\ee
o\`u $\omega_\qq$ est la pulsation propre du phonon et $\Gamma_\qq$ son taux de d\'ecroissance. \footnote{On ne peut pas \`a proprement parler calculer ici des \g{corrections} à $\omega_\qq$ dues au couplage aux continua, 
m\^eme à l'ordre $2$ en $\hat{V}$, puisqu'on
se base sur un hamiltonien effectif. Si on cherchait \`a le faire, on tomberait sur des int\'egrales de parties
principales divergentes ultraviolettes, dont l'incorporation dans la valeur non perturb\'ee pr\'esente dans $\hat{H}_2$
donnerait, suivant les id\'ees de la renormalisation, la valeur vraie, qui resterait
non calcul\'ee et inconnue en l'absence d'un modèle microscopique \cite{KCS2015}. 
Dans ce contexte, nous n'avons pas jug\'e utile de distinguer l'\'energie propre non perturb\'ee des phonons 
dans $\hat{H}_2$, en tout rigueur $\epsilon_\qq^{(2)}$, de l'\'energie vraie $\epsilon_\qq$.
Par ailleurs, pour obtenir les lois d'\'echelle en $q$ 
des termes perturbatifs d'ordre $n$ en $\hat{V}$,
nous supposerons que les nombres d'onde des phonons virtuels sont $O(q)$ donc que la coupure ultraviolette est mise à un nombre d'onde
$Aq$, $A\gg 1$. Ceci a le grand avantage que la correction r\'eelle à la pulsation propre $\epsilon_\qq^{(2)}$, due \`a $\hat{H}_3$ à l'ordre dominant, est
un $O(q^5)$ et n'affecte ni le param\`etre $\gamma$ dans l'\'equation (\ref{eq:fleur}), ni le taux d'amortissement $\Gamma_\qq$ à l'ordre $q^7$.
}
\setcounter{noterenormalisation}{\thefootnote}

\subsection{Calcul perturbatif et comptage de puissances}
\label{subsec:app_comptage}

La limite $q\to 0$ correspond au r\'egime de couplage faible. On le voit bien dans la th\'eorie de de l'hydrodynamique quantique~:
le d\'eveloppement des fluctuations de densit\'e $\delta\hat{\rho}$ et de vitesse $\hat{\mathbf{v}}$ sur les modes de
phonons fait appara\^\i tre des coefficients s'annulant en $q^{1/2}$ si bien que les \'el\'ements de matrice
de $\hat{H}_p$ entre \'etats de Fock de phonons s'annulent en $q^{p/2}$~:
\be
\hat{H}_p \underset{q\to 0}{\approx} q^{p/2}
\label{eq:odgHn}
\ee
On peut donc tenter un calcul perturbatif de $\Gamma_\qq$, en partant de l'\'ecriture exacte obtenue par la m\'ethode des projecteurs
\cite{Cohen}
\be
\langle \qq| \hat{G}(z) |\qq\rangle = \frac{1}{z-\langle \qq| \hat{H}_{\rm eff}(z) |\qq\rangle}
\ee
puis en d\'eveloppant en puissances de $\hat{V}$ l'\'el\'ement de matrice du hamiltonien effectif dans l'\'etat $|\qq\rangle$~:
\begin{multline}
\langle \qq| H_{\rm eff}(z) |\qq\rangle = \langle \qq|\hat{H}_2|\qq\rangle +\langle \qq|\hat{V}\hat{Q}\frac{\hat{Q}}{z\hat{Q}-\hat{Q}\hat{H}_2\hat{Q}}
\hat{Q} \hat{V} |\qq\rangle \\
+ \langle \qq|\hat{V} \hat{Q} \frac{\hat{Q}}{z\hat{Q}-\hat{Q}\hat{H}_2\hat{Q}}\hat{Q}\hat{V}\hat{Q}
\frac{\hat{Q}}{z\hat{Q}-\hat{Q}\hat{H}_2\hat{Q}}\hat{Q}\hat{V}|\qq\rangle + \ldots
\label{eq:HeffdevpuisV}
\end{multline}
o\`u $\hat{Q}=1-|\qq\rangle \langle \qq|$ projette orthogonalement \`a $|\qq\rangle$. Le r\'esultat (\ref{eq:Bel_result}) correspond au second
terme du second membre, le terme de Beliaev, dans lequel la contribution de $\hat{V}$ se r\'eduit \`a celle de $\hat{H}_3$.
Dans tous les termes successifs d'ordre $>2$ en $\hat{V}$, $z$ peut \^etre approxim\'e au d\'enominateur par 
\be
z=\hbar\omega_\qq+\ii \eta, \eta\to 0^+, 
\ee
le d\'eplacement de $z$ par une partie imaginaire en $O(q^5)$ dans ces termes ne contribuant pas \`a $\Gamma_\qq$ à l'ordre $q^7$. Il n'en va pas de
m\^eme dans le terme de Beliaev, comme nous le verrons dans la sous-section \ref{subsec:app_ordreBel}.

Consid\'erons dans (\ref{eq:HeffdevpuisV}) le terme $\mathcal{T}_n$ d'ordre $n$ en $\hat{V}$ et cherchons \`a majorer son ordre en $q$, 
comme dans le raisonnement simple du paragraphe qui suit l'\'equation (\ref{eq:Bel_result}). 
Compte tenu de (\ref{eq:odgHn}), il faut garder autant que possible dans les facteurs $\hat{V}$   
la contribution de $\hat{H}_3$, \`a d\'efaut celle de $\hat{H}_4$. Comme $\hat{H}_3$ change la parit\'e du nombre de phonons, alors que
$\hat{H}_4$ la conserve, on ne gardera que des facteurs $\hat{H}_3$ si $n$ est pair, mais il faudra caser au moins un facteur $\hat{H}_4$
si $n$ est impair. 
Par ailleurs, le nombre minimal de vecteurs d'onde ind\'ependants caract\'erisant les phonons virtuels 
est $n/2$ pour $n$ pair, $(n+1)/2$ pour $n$ impair, l'int\'egration sur chaque vecteur d'onde ind\'ependant sortant un facteur $q^3$.
Enfin, on compte un facteur $1/q$ pour chacun des $n-1$ d\'enominateurs d'\'energie. Ceci conduit \`a
\bea
\mathcal{T}_n=O(q^{2n+1}) & \mbox{ si } & n \ \mbox{pair} \\
\mathcal{T}_n=O(q^{2n+3}) & \mbox{ si } & n \ \mbox{impair}
\eea
On tend alors \`a conclure sans surprise que les processus au-del\`a de Beliaev, c'est-\`a-dire d'ordre $n\geq 3$ en $\hat{V}$, 
ont une contribution $O(q^9)$ 
n\'egligeable à $\Gamma_\qq$.

Le raisonnement simple pr\'ec\'edent ne prend pas en compte la possibilit\'e que les d\'enominateurs d'\'energie soient d'ordre $q^3$.
Pour que ceci se produise, il faut que la contribution de la partie lin\'eaire du spectre aux d\'enominateurs  puisse s'annuler, ce qui implique que les processus consid\'er\'es puissent \^etre r\'esonnants. Il faut donc garder dans $\hat{H}_3$
seulement les termes en $\hat{b}^\dagger \hat{b}^\dagger \hat{b}$, qui en forment la partie ascendante $\hat{H}_3^{(+)}$,
ou les termes en $\hat{b}^\dagger \hat{b}\hat{b}$, qui en forment la partie descendante $\hat{H}_3^{(-)}$. Les autres termes en $\hat{b}^\dagger
\hat{b}^\dagger\hat{b}^\dagger$ et $\hat{b}\hat{b}\hat{b}$ induisent des processus non r\'esonnants.
Dans $\hat{H}_4$, on se limitera aux $\hat{b}^\dagger \hat{b}^\dagger \hat{b}\hat{b}$, qui sont 
les termes potentiellement r\'esonnants les plus favorables puisqu'ils n'ajoutent qu'un seul vecteur d'onde ind\'ependant.
Pour que tous les d\'enominateurs d'\'energie soient en $q^3$, il faut aussi que tous les vecteurs d'onde ind\'ependants soient \'emis
presque vers l'avant, dans un petit c\^one d'ouverture angulaire $O(q)$ autour de la direction de $\qq$. L'int\'egration sur les angles
solides fait sortir un facteur p\'enalisant $q^2$ par vecteur d'onde ind\'ependant, soit $q^n$ pour $n$ pair
et $q^{n+1}$ pour $n$ impair, mais l'on gagne un facteur $q^{-2(n-1)}$ gr\^ace aux $n-1$ petits d\'enominateurs. Ceci conduit
aux majorations affin\'ees
\bea
\label{eq:majaffinpair}
\mathcal{T}_n=O(q^{n+3}) & \mbox{ si } & n \ \mbox{pair} \\
\mathcal{T}_n=O(q^{n+6}) & \mbox{ si } & n \ \mbox{impair}
\eea
On en d\'eduit que l'ordre $4$ en $\hat{V}$, et lui seul, est susceptible de corriger le terme en $q^7$
de $\Gamma_\qq^{\rm Bel}$ de l'\'equation (\ref{eq:Bel_result}). Il fait l'objet d'une analyse d\'etaill\'ee dans la sous-section \ref{subsec:app_ordre4}.
On confirme aussi que l'ordre $2$ en $\hat{V}$ a une contribution dominante \`a $\Gamma_\qq$ en $q^5$. Ceci ne nous dispense pas d'une \'etude
rigoureuse de ses contributions sous-dominantes jusqu'à l'ordre $q^7$, effectu\'ee dans la sous-section \ref{subsec:app_ordreBel}.

\subsection{\'Etude à l'ordre 2 en $\hat{V}$}
\label{subsec:app_ordreBel}

Nous tronquons ici le hamiltonien effectif dans l'\'equation (\ref{eq:HeffdevpuisV}) \`a l'ordre 2 en $\hat{V}$ 
et voyons quelle valeur de $\Gamma_\qq$ en r\'esulte à l'ordre $q^7$.

\`A cet ordre en $\hat{V}$ correspondent en r\'ealit\'e
plusieurs diagrammes puisque $\hat{V}=\hat{H}_3+\hat{H}_4+\hat{H}_5+\ldots$, dont on peut estimer l'ordre dominant par
comptage de puissances. Si les deux facteurs $\hat{V}$ sont \'egaux \`a $\hat{H}_3$, il faut int\'egrer sur au moins un vecteur d'onde
ind\'ependant et l'ordre dominant est $q^5$ 
comme dans l'\'equation (\ref{eq:majaffinpair}). S'ils sont \'egaux \`a $\hat{H}_4$, il faut int\'egrer sur au moins deux vecteurs d'onde ind\'ependants
et l'ordre dominant est $q^9$ sans l'effet \g{petit d\'enominateur}, pas mieux avec l'effet \g{petit d\'enominateur}. Si l'un des facteurs $\hat{V}$
fait intervenir $\hat{H}_p$, $p\geq 5$, la contribution est encore plus n\'egligeable. On peut donc se limiter ici à $\hat{V}=\hat{H}_3$, et 
m\^eme à $\hat{V}=\hat{H}_3^{(\pm)}$ puisque les termes non r\'esonnants en $\hat{b}^\dagger\hat{b}^\dagger\hat{b}^\dagger$ et $\hat{b}\hat{b}\hat{b}$
de $\hat{H}_3$ font intervenir deux vecteurs ind\'ependants et contribuent en $O(q^8)$. 

\begin{figure}[t] 
\begin{center}
\includegraphics[width=0.15\textwidth,clip=]{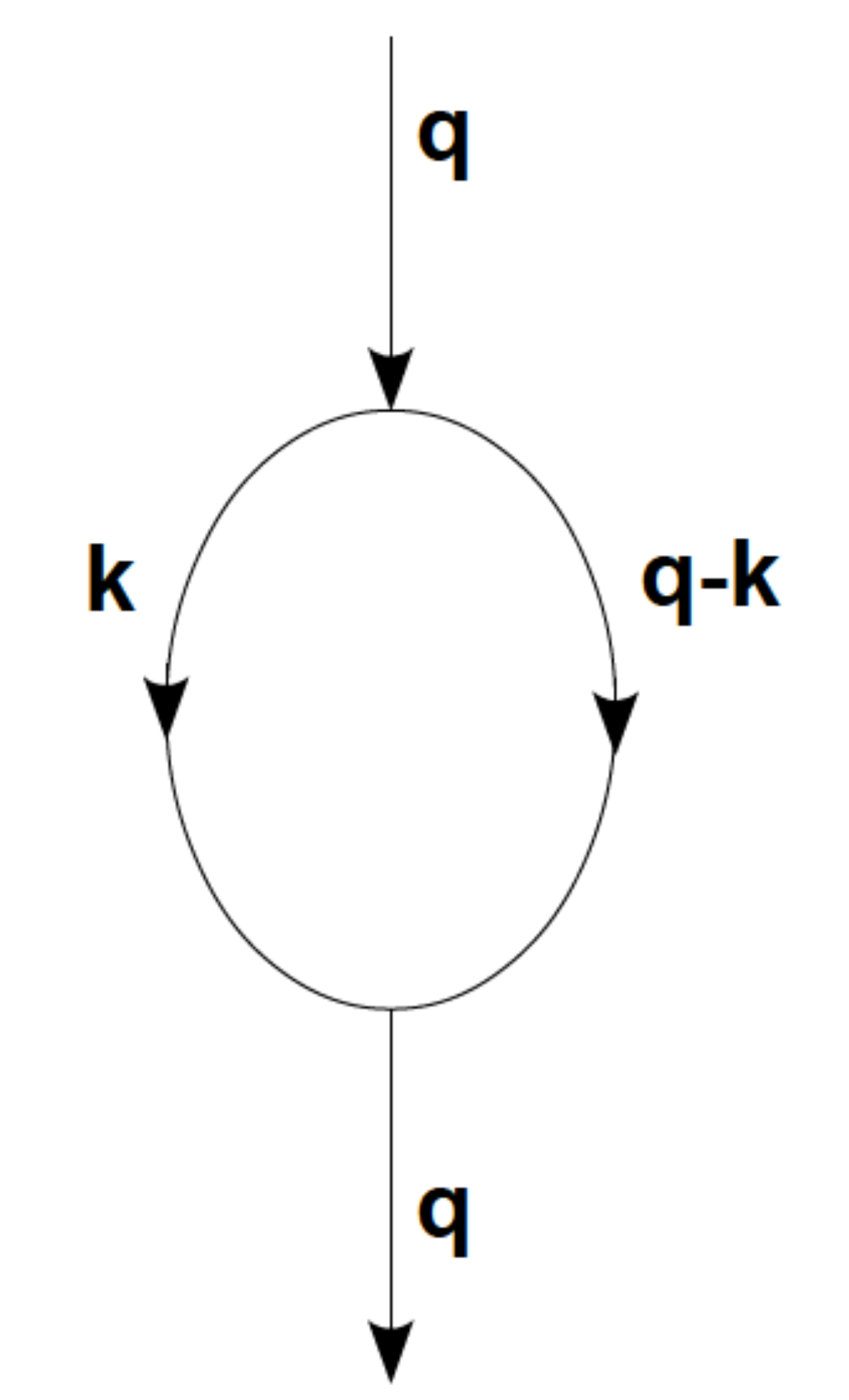}
\end{center}
\caption{\label{fig:diag2} Le diagramme d'ordre 2 en $\hat{V}$ contribuant au taux d'amortissement $\Gamma_\qq$ du phonon $\qq$ \`a l'ordre $q^7$. 
Il constitue le processus de Beliaev d'ordre dominant $q^5$.  Sur la figure, chaque vertex correspond à une action de $\hat{H}_3^{(\pm)}$.
}
\end{figure} 

Il reste ainsi un seul diagramme, celui de Beliaev, repr\'esent\'e sur la figure \ref{fig:diag2}. Introduisons la fonction \'energie
propre correspondante, dite de Beliaev, pour tout nombre complexe $z$ de $\im z >0$~:
\begin{multline}
\Delta \epsilon_\qq^{\rm Bel}(z)=\frac{1}{2} \sum_{\kk} \frac{|\langle\kk,\qq-\kk|\hat{H}_3^{(+)}|\qq\rangle|^2}
{z-(\epsilon_\kk+\epsilon_{\qq-\kk})}   \\
=\frac{2(mc^2)^2}{\rho} \int \frac{\dd^3k}{(2\pi)^3} 
\frac{[\mathcal{A}^{2\leftrightarrow 1}_{\rm hydro}(\kk,\qq-\kk;\qq)]^2}{z-(\epsilon_\kk+\epsilon_{\qq-\kk})}
\label{eq:DeQBelZ}
\end{multline}
où nous sommes pass\'es à la limite thermodynamique et avons utilis\'e la forme (\ref{eq:H3hydro}) du hamiltonien hydrodynamique.
\`A cet ordre d'approximation, le p\^ole $z_\qq$ associ\'e au phonon $\qq$ est solution de l'\'equation implicite 
\be
z_\qq^{\rm Bel} = \epsilon_\qq + \Delta \epsilon_\qq^{\rm Bel \downarrow }(z_\qq^{\rm Bel})
\label{eq:impli}
\ee
où la flèche vers le bas indique le prolongement analytique de la fonction \'energie propre du demi-plan sup\'erieur $\im z>0$
au demi-plan inf\'erieur $\im z < 0$ \`a travers la ligne de coupure situ\'ee sur l'axe r\'eel positif. S\'eparons $z^{\rm Bel}_\qq$ en partie r\'eelle
et partie imaginaire comme dans (\ref{eq:decompzq}) et d\'eveloppons le second membre en puissances de $\Gamma_\qq$~:
\be
z_\qq^{\rm Bel} = \epsilon_\qq + \Delta \epsilon_\qq^{\rm Bel\downarrow}(\epsilon_\qq) -\frac{\ii\hbar\Gamma_\qq}{2} 
\frac{\dd}{\dd z} \Delta \epsilon_\qq^{\rm Bel\downarrow}(\epsilon_\qq) + O({q}^9)
\ee
Il suffit ici de tronquer le d\'eveloppement de Taylor \`a l'ordre un~; en revanche, l'ordre z\'ero ne suffit pas
. \footnote{La contribution de Taylor d'ordre $n$ est $O(q^{2n+5})$ puisque la d\'eriv\'ee
d'ordre $n$ de la fonction \'energie propre, qui fait intervenir un d\'enominateur d'\'energie à la puissance  $n+1$, est d'ordre 
$q^8/q^{3(n+1)}$ compte tenu de l'effet \g{petit d\'enominateur}.} Les valeurs du prolongement analytique et de sa d\'eriv\'ee en $\epsilon_\qq$
s'obtiennent comme limites lorsque $\eta\to 0^+$ des fonctions non prolong\'ees en $z=\epsilon_\qq+\ii\eta$~:
\be
z_\qq^{\rm Bel} = \epsilon_\qq + \Delta \epsilon_\qq^{\rm Bel} (\epsilon_\qq+\ii\eta) -\frac{\ii\hbar\Gamma_\qq}{2} 
\frac{\dd}{\dd z} \Delta \epsilon_\qq^{\rm Bel} (\epsilon_\qq+\ii\eta) + O({q}^9)
\label{eq:taylore}
\ee
Le deuxième terme est le r\'esultat perturbatif ordinaire. Il conduit au taux d'amortissement $\Gamma_{\qq,{\rm pert}}^{\rm Bel}$ donn\'e
par la limite à temp\'erature nulle de l'\'equation (\ref{eq:GamqBel}), 
\be
\frac{\hbar \Gamma_{\qq,{\rm pert}}^{\rm Bel}}{2} = -\im \Delta \epsilon_\qq^{\rm Bel} (\epsilon_\qq+\ii\eta)
\ee
et sa valeur pour le gaz unitaire a \'et\'e calcul\'ee
jusqu'à l'ordre $q^7$ dans l'\'equation (\ref{eq:Bel_result_prov}).
Après prise de la partie imaginaire dans l'\'equation (\ref{eq:taylore}), nous trouvons cependant un facteur correctif non n\'egligeable
à cet ordre, premier effet d'une autocoh\'erence en $z_\qq$ dans l'\'equation implicite (\ref{eq:impli})~:
\be
\Gamma_\qq^{\rm Bel} = \Gamma_{\qq,{\rm pert}}^{\rm Bel} \left[1+\re \frac{\dd}{\dd z} \Delta \epsilon_\qq^{\rm Bel} (\epsilon_\qq+\ii\eta)
+O({q}^4)\right]
\label{eq:respersigmaprime}
\ee
avec
\begin{multline}
\frac{\dd}{\dd z} \Delta \epsilon_\qq^{\rm Bel} (\epsilon_\qq+\ii\eta)
= -\frac{2(mc^2)^2}{\rho} \\ \times \int \frac{\dd^3k}{(2\pi)^3} 
\frac{[\mathcal{A}^{2\leftrightarrow 1}_{\rm hydro}(\kk,\qq-\kk;\qq)]^2}{(\epsilon_\qq-\epsilon_\kk-\epsilon_{\qq-\kk}+\ii\eta)^2}
\end{multline}
Pour obtenir un \'equivalent de cette d\'eriv\'ee lorsque $\check{q}\equiv \hbar q/mc\to 0$, nous appliquons la m\^eme technique que dans la section \ref{subsec:ccadlk}.
Nous passons en coordonn\'ees sph\'eriques d'axe polaire la direction de $\qq$ et effectuons les changements d'\'echelle
$k=\bar{k}q$ et $\theta=\gamma^{1/2}\check{q}\check{\theta}$ sur le module et l'angle polaire du vecteur $\kk$, avec $\bar{k}<1$ pour b\'en\'eficier
de l'effet \g{petit d\'enominateur}. Dans l'int\'egrande
et donc dans l'amplitude de couplage (\ref{eq:Ahydro21}), nous passons à
la limite $\check{q}\to 0$ à $\bar{k}$ et $\check{\theta}$ fix\'es, avec les r\'esultats interm\'ediaires
\bea
|\qq-\kk|=q(1-\bar{k}) \left[1+\frac{\gamma\bar{k}\check{\theta}^2\check{q}^2}{2(1-\bar{k})^2} + O(\check{q}^4)\right] && \\
\label{eq:intermdeux}
\epsilon_\qq-\epsilon_\kk-\epsilon_{\qq-\kk} \sim \frac{\gamma\check{q}^3 mc^2}{8}\left[3\bar{k}(1-\bar{k})-\frac{4\bar{k}\check{\theta}^2}{1-\bar{k}}
\right] &&  \\
{[\mathcal{A}^{2\leftrightarrow 1}_{\rm hydro}(\kk,\qq-\kk;\qq)]}^2 \sim \frac{2}{9} \check{q}^3 \bar{k}(1-\bar{k}) &&
\eea
compte tenu du fait que $1+\Lambda_\mathrm{F}=8/9$ à la limite unitaire. L'int\'egration par rapport à $\check{\theta}$ sur $[0,+\infty[$
est imm\'ediate après le changement de variable $X=\check{\theta}^2$, et il reste
\be
\frac{\dd}{\dd z} \Delta \epsilon_\qq^{\rm Bel} (\epsilon_\qq+\ii\eta) \sim \frac{8 \check{q}^2}{9\pi^2\gamma}  \left(\frac{mc}{\hbar\rho^{1/3}}\right)^{3}
\int_0^{1} \dd\bar{k} \frac{\bar{k}^2(1-\bar{k})^2}{3\bar{k} (1-\bar{k}) +\ii\eta}
\label{eq:derivexpli}
\ee
Le th\'eorème de convergence domin\'ee nous autorise à faire tendre $\eta$ vers z\'ero sous le signe int\'egral pour obtenir
le r\'esultat (\ref{eq:Bel_result}), sachant que $\rho(\hbar/mc)^3=\sqrt{3}/(\pi^2\xi_B^{3/2})$.
{Notons en passant que le fait que la dérivée par rapport à $z$ (\ref{eq:derivexpli}) 
soit réelle à la limite $\eta\to 0^+$ signifie qu'un changement de la relation de dispersion réelle $q\mapsto\omega_\qq$ à l'ordre $q^5$ 
est sans effet sur le taux de décroissance $\Gamma_\qq$ à l'ordre $q^7$, ce que la méthode de calcul de la sous-section
\ref{subsec:tadpdlguat0} montrait déjà.}

\subsection{\'Etude à l'ordre 4 en $\hat{V}$}
\label{subsec:app_ordre4}

\begin{figure}[htb] 
\begin{center}
\includegraphics[width=0.4\textwidth,clip=]{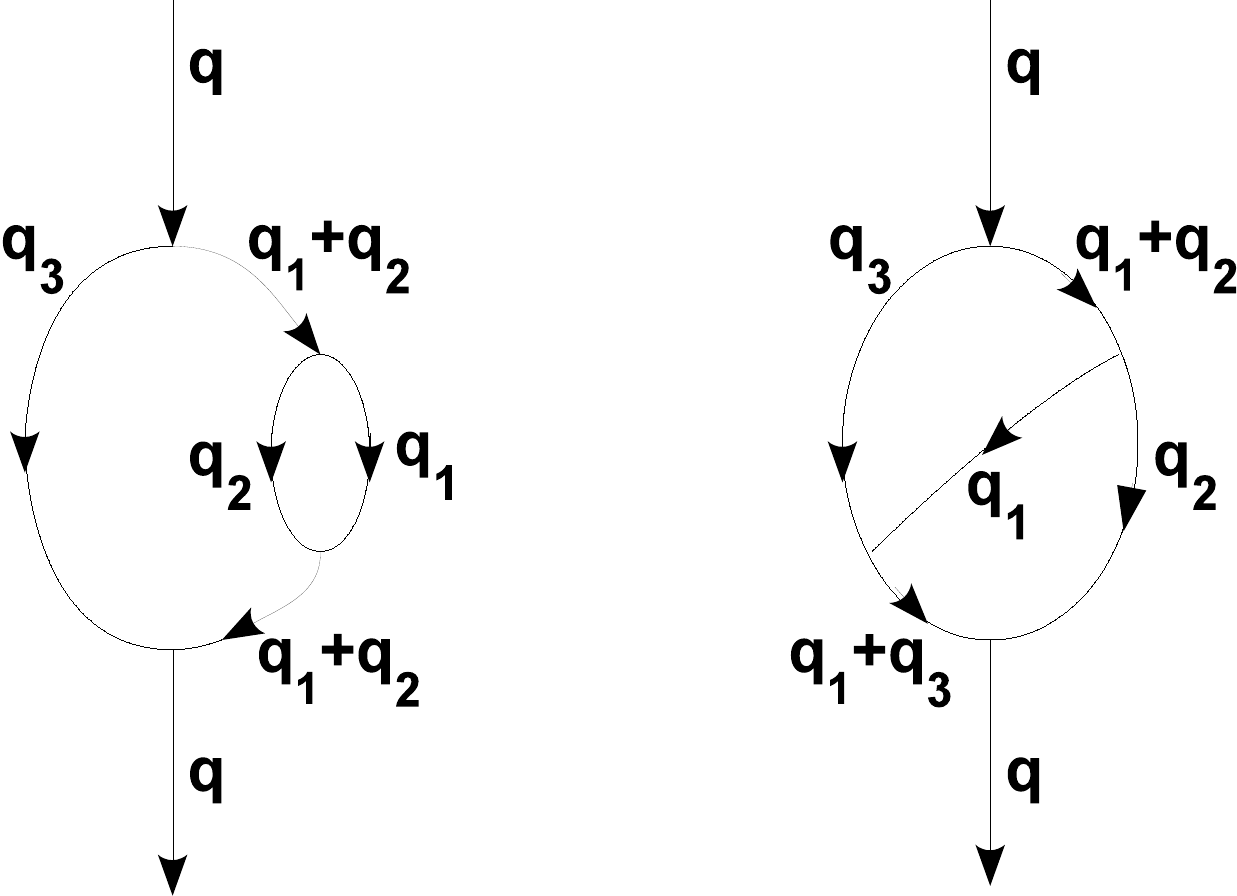}
\end{center}
\caption{\label{fig:diag4} Les deux types de diagrammes d'ordre 4 en $\hat{V}$ susceptibles de contribuer au taux d'amortissement
$\Gamma_\qq$ du phonon $\qq$ \`a l'ordre $q^7$ au travers de l'effet \g{petit d\'enominateur} en $q^3$. Celui de gauche (I)
est un processus de Beliaev
avec correction à une boucle (donc elle aussi beliaevienne) de la pulsation propre des phonons virtuels. Celui de droite (II)
est un processus de Beliaev
enrichi par une interaction entre les phonons virtuels, elle-m\^eme m\'edi\'ee par l'\'echange d'un phonon. Chaque vertex correspond
à une action de $\hat{H}_3^{(\pm)}$.}
\end{figure}

Nous avons vu dans la sous-section \ref{subsec:app_comptage} que l'ordre 4 en $\hat{V}$ est susceptible de contribuer \`a $\Gamma_\qq$
à l'ordre $q^7$. \'Ecrivons la correction correspondante $\delta\Gamma_\qq$ en la limitant tout de suite à son ordre dominant, 
c'est-\`a-dire en gardant $\hat{H}_3^{(-)}$
dans les deux premiers facteurs $\hat{V}$ et $\hat{H}_3^{(+)}$ dans les deux derniers, cet agencement \'etant impos\'e par la pr\'esence
des projecteurs $\hat{Q}$~:
\be
-\frac{\hbar\delta\Gamma_\qq}{2}=\im \langle\qq| \hat{W}^{(-)} \frac{\hat{Q}}{\epsilon_\qq+\ii\eta-\hat{H}_2} \hat{W}^{(+)} |\qq\rangle
\ee
o\`u l'on a introduit, pour $\varepsilon=\pm$, l'op\'erateur de couplage effectif $1\leftrightarrow 3$ au second ordre
\be
\hat{W}^{(\varepsilon)} \equiv \hat{H}_3^{(\varepsilon)} \frac{\hat{Q}}{\epsilon_\qq+\ii\eta-\hat{H}_2} \hat{H}_3^{(\varepsilon)}
\ee
Dans ces expressions, on peut utiliser pour $\hat{H}_3$ l'approximation (\ref{eq:H3hydro})
qu'en donne l'hydrodynamique quantique, mais
il faut inclure dans $\epsilon_\qq=\hbar\omega_\qq$ la correction cubique (\ref{eq:fleur}) au spectre lin\'eaire de l'hydrodynamique.
Comme les éléments de matrice de $\hat{H}_3$ sont r\'eels dans la base de Fock, les matrices qui y repr\'esentent $\hat{W}^{(\pm)}$
sont transpos\'ees l'une de l'autre. Apr\`es injection d'une relation de fermeture dans le sous-espace à trois phonons, on obtient
\begin{multline}
-\frac{\hbar\delta\Gamma_\qq}{2}=\im \frac{1}{3!} \sum_{\qq_1,\qq_2,\qq_3} 
\delta_{\qq_1+\qq_2+\qq_3,\qq} \\
\times \frac{(\langle\qq|\hat{W}^{(-)}|\qq_1,\qq_2,\qq_3\rangle)^2}
{\epsilon_\qq+\ii\eta-(\epsilon_{\qq_1}+\epsilon_{\qq_2}+\epsilon_{\qq_3})}
\label{eq:slfdus}
\end{multline}
avec, compte tenu de l'\'equation (\ref{eq:H3hydro}),
\begin{widetext}
\begin{multline}
\langle\qq|\hat{W}^{(-)}|\qq_1,\qq_2,\qq_3\rangle=\frac{4(mc^2)^2}{\rho L^3} 
\Big[\frac{\mathcal{A}_{\rm hydro}^{2\leftrightarrow 1}(\qq_1,\qq_2;\qq_1+\qq_2)\mathcal{A}_{\rm hydro}^{2\leftrightarrow 1}(\qq_1+\qq_2,\qq_3;\qq)}
{\epsilon_\qq+\ii\eta-(\epsilon_{\qq_1+\qq_2}+\epsilon_{\qq_3})} \\
+ \frac{\mathcal{A}_{\rm hydro}^{2\leftrightarrow 1}(\qq_2,\qq_3;\qq_2+\qq_3)\mathcal{A}_{\rm hydro}^{2\leftrightarrow 1}(\qq_2+\qq_3,\qq_1;\qq)}
{\epsilon_\qq+\ii\eta-(\epsilon_{\qq_2+\qq_3}+\epsilon_{\qq_1})}
+ \frac{\mathcal{A}_{\rm hydro}^{2\leftrightarrow 1}(\qq_1,\qq_3;\qq_1+\qq_3)\mathcal{A}_{\rm hydro}^{2\leftrightarrow 1}(\qq_1+\qq_3,\qq_2;\qq)}
{\epsilon_\qq+\ii\eta-(\epsilon_{\qq_1+\qq_3}+\epsilon_{\qq_2})}
\Big]
\end{multline}
La contribution du premier terme entre crochets à l'\'el\'ement de matrice, soit $P(\qq_1,\qq_2|\qq_3)$, 
correspond au processus de la figure \ref{fig:1d3}. Les deux autres contributions,
$P(\qq_2,\qq_3|\qq_1)$ et $P(\qq_1,\qq_3|\qq_2)$, s'en d\'eduisent par permutation circulaire. Comme le reste du sommande dans (\ref{eq:slfdus})
est invariant par permutation des trois vecteurs d'onde, on peut remplacer son num\'erateur $[P(\qq_1,\qq_2|\qq_3)+P(\qq_2,\qq_3|\qq_1)
+P(\qq_1,\qq_3|\qq_2)]^2$
par $3 [P(\qq_1,\qq_2|\qq_3)]^2+ 6 P(\qq_1,\qq_2|\qq_3) P(\qq_1,\qq_3|\qq_2)$, ce qui conduit à se ramener aux deux diagrammes
de la figure \ref{fig:diag4} et à l'expression int\'egrale
\begin{multline}
-\frac{\hbar\delta\Gamma_\qq}{2} =\lim_{\eta\to 0^+} \im \frac{8(mc^2)^4}{\rho^2}
\int\frac{\dd^3q_2\dd^3q_3}{(2\pi)^6}  \left\{
\frac{[\mathcal{A}_{\rm hydro}^{2\leftrightarrow 1}(\qq_1,\qq_2;\qq_1+\qq_2)\mathcal{A}_{\rm hydro}^{2\leftrightarrow 1}(\qq_1+\qq_2,\qq_3;\qq)]^2}
{[\epsilon_\qq+\ii\eta-(\epsilon_{\qq_1}+\epsilon_{\qq_2}+\epsilon_{\qq_3})] 
[\epsilon_\qq+\ii\eta-(\epsilon_{\qq_1+\qq_2}+\epsilon_{\qq_3})]^2} \right. \\
\left.
+\frac{2\mathcal{A}_{\rm hydro}^{2\leftrightarrow 1}(\qq_1,\qq_2;\qq_1+\qq_2)\mathcal{A}_{\rm hydro}^{2\leftrightarrow 1}(\qq_1+\qq_2,\qq_3;\qq)
\mathcal{A}_{\rm hydro}^{2\leftrightarrow 1}(\qq_1,\qq_3;\qq_1+\qq_3)\mathcal{A}_{\rm hydro}^{2\leftrightarrow 1}(\qq_1+\qq_3,\qq_2;\qq)}
{[\epsilon_\qq+\ii\eta-(\epsilon_{\qq_1}+\epsilon_{\qq_2}+\epsilon_{\qq_3})][\epsilon_\qq+\ii\eta-(\epsilon_{\qq_1+\qq_2}+\epsilon_{\qq_3})]
[\epsilon_\qq+\ii\eta-(\epsilon_{\qq_1+\qq_3}+\epsilon_{\qq_2})]}
\right\}
\end{multline}
où l'on a \'elimin\'e la sommation sur $\qq_1$ gr\^ace à la conservation de l'impulsion,
$\qq_1=\qq-(\qq_2+\qq_3)$, avant de passer à la limite thermodynamique. 
\end{widetext}

\begin{figure}[t] 
\hspace{3mm}
\begin{center}
\includegraphics[width=0.3\textwidth,clip=]{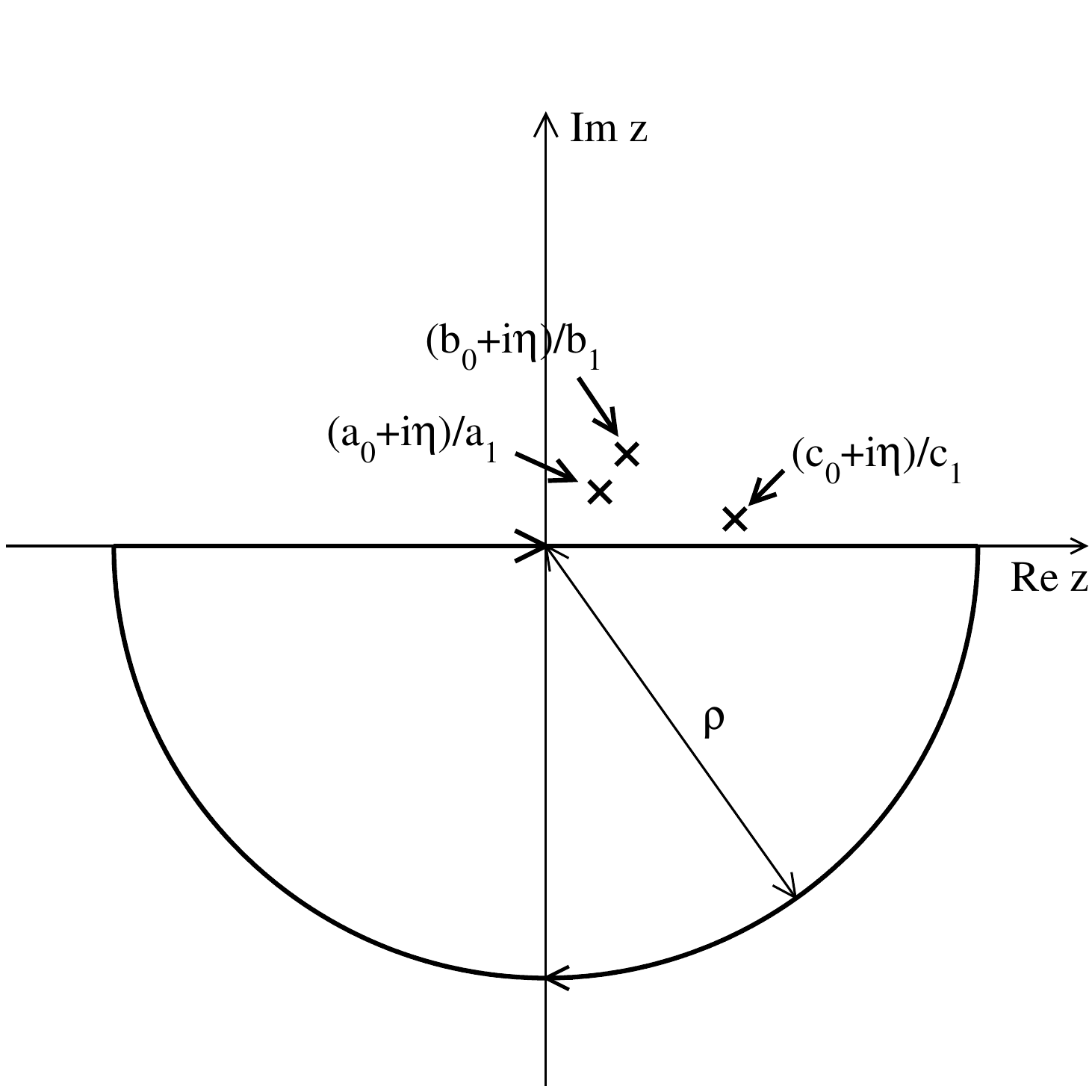}
\end{center}
\caption{Contour d'int\'egration dans le plan complexe \`a utiliser dans le th\'eor\`eme des r\'esidus pour \'etablir l'identit\'e (\ref{eq:resid}),
en trait gras. Les nombres r\'eels $a_0,a_1,b_0,b_1,c_0,c_1$ et $\eta$ sont strictement positifs.
Le rayon $\rho$ du demi-cercle tend vers l'infini.  Les croix
correspondent aux p\^oles de l'int\'egrande dans le cas o\`u ils sont tous simples, mais le r\'esultat vaut bien s\^ur dans le cas
g\'en\'eral.}
\label{fig:contour}
\end{figure}

En pratique, l'int\'egration sur $\qq_2$ et $\qq_3$ s'effectue
en coordonn\'ees sph\'eriques d'axe polaire $\qq$.
Pour \'evaluer la contribution à $\delta\Gamma_\qq$ des c\^ones d'\'emission de $\qq_2$ et $\qq_3$ 
d'ouverture angulaire $O(q)$ autour de $\qq$, nous redimensionnons les angles polaires $\theta_i$ comme suit~:
\be
\check{\theta}_i = \frac{\theta_i}{\gamma^{1/2} \check{q}}
\label{eq:thetaitildeq}
\ee
puis nous faisons tendre $\check{q}=\hbar q/mc$ vers $0$ \`a $\check{\theta}_i$ fix\'es. Le calcul est proche de celui de la section \ref{subsec:ccadlk}.
On utilise la repr\'esentation polaire (\ref{eq:reppolth})~:
\be
\check{\theta}_2 = R \cos\alpha  \qquad \qquad \check{\theta}_3 = R \sin\alpha
\ee
Les nombres d'onde $q_i$ doivent \^etre r\'e\'echelonn\'es par $q$, c'est-\`a-dire qu'on 
garde les  $\bar{q}_i \equiv \frac{q_i}{q}$
fix\'es dans le passage \`a la limite $\check{q}\to 0$, avec la contrainte issue de (\ref{eq:qq1q2q3}) et de la positivit\'e de $q_1$,
\be
\bar{q}_2+\bar{q}_3\leq 1
\ee
Donnons comme seul interm\'ediaire de calcul un d\'enominateur d'\'energie dans les \'el\'ements de matrice de $\hat{W}^{(-)}$,
\begin{multline}
\frac{\epsilon_\qq-(\epsilon_{\qq-\qq_3}+\epsilon_{\qq_3})}{mc^2}\!\underset{\check{q}\to 0}{=}\!\frac{\gamma \check{q}^3}{2} 
\left[\frac{3}{4}\bar{q}_3 (1-\bar{q}_3) -\frac{\bar{q}_3 R^2\sin^2\alpha}{1-\bar{q}_3}\right] \\ + O(\check{q}^5)
\end{multline}
et le d\'enominateur d'\'energie du sous-espace à 3 phonons,
\be\frac{\epsilon_\qq-(\epsilon_{\qq_1}+\epsilon_{\qq_2}+\epsilon_{\qq_3})}{mc^2} \underset{\check{q}\to 0}{=} \frac{\gamma \check{q}^3}{2}
(v-u R^2) +O(\check{q}^5)
\ee
dont les coefficients manifestement positifs (pour peu qu'on les \'ecrive convenablement) ressemblent beaucoup à ceux des
\'equations (\ref{eq:defualign}) et (\ref{eq:defvalign})~:
\begin{multline}
u\equiv\frac{\bar{q}_2\cos^2\alpha+\bar{q}_3\sin^2\alpha-\bar{q}_2\bar{q}_3(1-\cos\phi \sin 2\alpha)}{1-\bar{q}_2-\bar{q}_3} \\
= \frac{(\bar{q}_2\cos\alpha-\bar{q}_3\sin\alpha)^2+\bar{q}_2\bar{q}_3 (1+\cos\phi)\sin 2\alpha} {1-\bar{q}_2-\bar{q}_3} \\
+\bar{q}_2\cos^2\alpha+\bar{q}_3\sin^2\alpha 
\end{multline}
et
\be
v\equiv\frac{1}{4}[1-\bar{q}_2^3-\bar{q}_3^3-(1-\bar{q}_2-\bar{q}_3)^3]=\frac{3}{4}(1-\bar{q}_2)(1-\bar{q}_3)(\bar{q}_2+\bar{q}_3)
\ee
On obtient finalement
\begin{multline}
-\frac{\hbar \delta\Gamma_\qq}{2}\!\underset{\check{q}\to 0}{=}\! mc^2\frac{[3(1+\Lambda_{\rm F})]^4}{3\times 2\pi} 
\frac{\check{q}^7}{2^8\pi^4\gamma} \left(\frac{mc}{\hbar\rho^{1/3}}\right)^6 \\
\times (\im I^{(\mathrm{I})}+\im I^{(\mathrm{II})}) + o(\check{q}^7)
\end{multline}
avec le param\`etre $\Lambda_{\rm F}$ d\'efini à l'\'equation (\ref{eq:lambdaF}) 
{et les int\'egrales quintuples, issues respectivement
du diagramme de gauche (type I) et du diagramme de droite (type II) de la figure \ref{fig:diag4}, \`a prendre dans la limite $\eta\to 0^+$~:
\begin{widetext}
\begin{multline}
I^{(\mathrm{I})}=\lim_{\eta\to 0^+} \int_0^1 \dd\bar{q}_2 \int_0^{1-\bar{q}_2} \dd\bar{q}_3 \int_{-\pi}^{\pi} \dd\phi
\int_0^{\pi/2} \dd\alpha \int_0^{A^2/\check{q}^2}\frac{\dd X}{2} 
\frac{3X\sin\alpha\cos\alpha (1-\bar{q}_3)^2 (1-\bar{q}_2-\bar{q}_3) \bar{q}_2^3 \bar{q}_3^3}
{(v-uX+\ii\eta)[\frac{3}{4}\bar{q}_3(1-\bar{q}_3)-\frac{\bar{q}_3 X\sin^2\alpha}{1-\bar{q}_3}+\ii\eta]^2}
\label{eq:IexpI}
\end{multline}
\begin{multline}
I^{(\mathrm{II})}=\lim_{\eta\to 0^+} \int_0^1 \dd\bar{q}_2 \int_0^{1-\bar{q}_2} \dd\bar{q}_3 \int_{-\pi}^{\pi} \dd\phi \int_0^{\pi/2} \dd\alpha 
\\ \times \int_0^{+\infty}\frac{\dd X}{2}
\frac{6X\sin\alpha\cos\alpha (1-\bar{q}_2)(1-\bar{q}_3)(1-\bar{q}_2-\bar{q}_3) \bar{q}_2^3 \bar{q}_3^3}
{(v-uX+\ii\eta)[\frac{3}{4}\bar{q}_2(1-\bar{q}_2)-\frac{\bar{q}_2 X\cos^2\alpha}{1-\bar{q}_2}+\ii\eta]
[\frac{3}{4}\bar{q}_3(1-\bar{q}_3)-\frac{\bar{q}_3 X\sin^2\alpha}{1-\bar{q}_3}+\ii\eta]}
\label{eq:IexpII}
\end{multline}
\end{widetext}
Dans l'intégration sur le rayon $R$ nous avons effectué le changement de variable $X=R^2$. Ceci permet de voir facilement que l'intégrale sur
$R$, c'est-à-dire sur $X$, dans (\ref{eq:IexpII}) est de partie imaginaire nulle à la limite $\eta\to 0^+$, et donc que
\be
\im I^{(\mathrm{II})}=0
\label{eq:IexpIInul}
\ee
En effet, tous les facteurs au d\'enominateur de l'int\'egrande dans (\ref{eq:IexpII}) sont de la forme $\alpha-\beta X +\ii \eta$,
avec $\alpha>0$ et $\beta>0$. L'int\'egrale sur $X\in [0,+\infty[$ tombe donc sous le coup du th\'eorème élémentaire suivant~:

\noindent{\bf Th\'eor\`eme~:} {\sl On d\'efinit $a(X)=a_0-a_1X$, $b(X)=b_0-b_1X$ et $c(X)=c_0-c_1X$ pour tout $X\in\mathbb{R}$.
Si tous les coefficients $a_0, a_1, b_0, b_1, c_0, c_1$ sont $>0$ et
$\eta\in \mathbb{R}$, alors}
\begin{equation}
\lim_{\eta\to 0^+} \int_{\mathbb{R}^+} \dd X \im \frac{X}{[a(X)+\ii \eta] [b(X)+\ii \eta] [c(X)+\ii \eta]} = 0 \label{eq:theo}
\end{equation}

\noindent{\bf D\'emonstration~:}
On remarque d'abord que le th\'eorème serait \'evident si l'int\'egrale sur $X$ \'etait prise sur $\mathbb{R}^-$. En effet,
pour tout $\eta >0$ et pour tout $X<0$,
\begin{equation}
\left|\frac{X}{[a(X)+\ii \eta] [b(X)+\ii \eta] [c(X)+\ii \eta]}\right| < \frac{X}{a(X) b(X) c(X)}
\end{equation}
et le majorant est d'int\'egrale finie sur $\mathbb{R}^-$ si bien que l'on peut utiliser le th\'eor\`eme de convergence
domin\'ee pour \'echanger l'int\'egration sur $X$ et le passage \`a la limite $\eta\to 0^+$.
On peut donc, dans l'\'enonc\'e (\ref{eq:theo}), \'etendre le domaine d'int\'egration \`a $\mathbb{R}$ tout entier sans changer
le r\'esultat. Or nous disposons de l'identit\'e suivante pour tout $\eta>0$ non n\'ecessairement infinit\'esimal,
aussi bien en partie r\'eelle qu'en partie imaginaire~:
\begin{equation}
\int_{\mathbb{R}} \dd X \frac{X}{[a(X)+\ii \eta] [b(X)+\ii \eta] [c(X)+\ii \eta]} = 0 \label{eq:resid}
\end{equation}
Il suffit en effet d'utiliser le th\'eor\`eme des r\'esidus en refermant le contour d'int\'egration par le bas selon
un demi-cercle de rayon $\rho\to +\infty$
dans le demi-plan inf\'erieur~; la d\'ecroissance de l'int\'egrande en $1/X^2$ \`a grand $|X|$ le permet, qui conduit \`a une contribution
du demi-cercle en $O(1/\rho)$, tendant vers z\'ero. Comme les p\^oles de l'int\'egrande sont tous situ\'es dans le demi-plan sup\'erieur,
aucun n'est enserr\'e par le contour d'int\'egration, voir la figure \ref{fig:contour}, et il vient (\ref{eq:resid}) donc (\ref{eq:theo}). \footnote{On a supposé ici que la contribution dominante à $\delta\Gamma_\qq$ provient de la configuration \g{bicône} 
dans laquelle à la fois $\qq_2$ et $\qq_3$ sont dans le cône d'émission vers l'avant, c'est-\`a-dire d'ouverture angulaire $O(\check{q})$ autour de la direction de $\qq$.
On peut imaginer un sc\'enario plus subtil, dit \g{unic\^one}, dans lequel seul le vecteur $\qq_3$ se trouverait
dans ce c\^one, le vecteur $\qq_2$ faisant lui un angle $\approx \check{q}^0$ 
avec $\qq$. Dans ce cas, seul le d\'enominateur de $P(\qq_1,\qq_2|\qq_3)$ est $\approx \check{q}^3$, celui de $P(\qq_1,\qq_3|\qq_2)$
\'etant $\approx \check{q}$. Le terme crois\'e $P(\qq_1,\qq_3|\qq_2) P(\qq_1,\qq_2|\qq_3)$ est donc n\'egligeable par rapport
à $[P(\qq_1,\qq_2|\qq_3)]^2$. Par ailleurs, le d\'enominateur d'\'energie 
global $\epsilon_\qq+\ii\eta-(\epsilon_{\qq_1}+\epsilon_{\qq_2}+\epsilon_{\qq_3})$ de (\ref{eq:slfdus}) est lui aussi $\approx \check{q}$, ce
qui fait perdre un facteur $\check{q}^2$, mais ceci est exactement compens\'e par la perte au num\'erateur d'un facteur $\check{q}^2$
dans l'int\'egrale polaire $\int \dd \theta_2 \sin\theta_2$. L'ensemble semble donc devoir contribuer au m\^eme ordre $\check{q}^7$
que la configuration bic\^one. L'int\'egration sur l'angle polaire $\theta_3$ r\'e\'echelonn\'e comme dans (\ref{eq:thetaitildeq}), et avec
inclusion du jacobien tridimensionnel $\bar{q}_3^2$ de l'int\'egration sur $\bar{q}_3$,
conduit cependant à
\[
\int_0^{+\infty}\!\!\frac{\bar{q}_3^2 \check{\theta}_3\dd\check{\theta}_3}
{\left[\frac{3}{4} \bar{q}_3(1-\bar{q}_3) -\frac{\bar{q}_3\check{\theta}_3^2}{1-\bar{q}_3}+\ii\eta\right]^2}
=-\frac{\bar{q}_3(1-\bar{q}_3)/2}{\frac{3}{4} \bar{q}_3(1-\bar{q}_3)+\ii\eta} 
\]
de limite r\'eelle $-\frac{2}{3}$ lorsque $\eta\to 0^+$ m\^eme au sens des distributions.
Aussi la partie imaginaire n\'ecessaire à $\delta\Gamma_\qq$ ne pourrait-elle provenir que du d\'enominateur d'\'energie
global, qui sortirait un Dirac $\delta[\epsilon_\qq-(\epsilon_{\qq_1}+\epsilon_{\qq_2}+\epsilon_{\qq_3})]$~; mais, \`a l'ordre dominant
en $\check{q}$ où nous sommes d\'esormais, $\qq_3=\bar{q}_3 \qq$, avec $0<\bar{q}_3<1$, et la relation de dispersion est lin\'eaire,
si bien que l'argument du Dirac ne peut s'annuler que si $\qq_2$ et $\qq$ sont colin\'eaires et de m\^eme sens, en contradiction
avec l'hypothèse d'un $\qq_2$ en dehors du c\^one d'\'emission vers l'avant. La m\^eme argumentation vaut pour le taux d'amortissement
$\Gamma_\qq^{2\leftrightarrow 2}$ de la section \ref{subsec:ccadlk}, et justifie le fait que l'on n'y ait consid\'er\'e que la configuration
\g{bic\^one}.} \qed

On pourrait croire que le même raisonnement s'applique à la contribution (\ref{eq:IexpI}) et que $\im I^{(\mathrm{I})}=0$, auquel cas 
$\delta\Gamma_\qq=o(\check{q}^7)$ et il n'y aurait pas de correction \`a apporter au r\'esultat (\ref{eq:Bel_result}).
Ce n'est pas le cas, car nous sommes cette fois obligés de garder dans l'intégrale sur $X$ une valeur finie $A^2/\check{q}^2$ de la borne supérieure,
pour obtenir une valeur finie de $I^{(\mathrm{I})}$, ce qui bloque l'application du théorème. Ici, $A>0$ est une constante de coupure dont la valeur précise n'importe pas,
et la loi de puissance en $\check{q}^{-2}$ provient du fait que $\theta_i=O(1)$ donc $\check{\theta}_i=O(\check{q}^{-1})$ dans (\ref{eq:thetaitildeq}), 
et donc $X=R^2=\check{\theta}_2^2+\check{\theta}_3^2=O(\check{q}^{-2})$. En effet, si l'on remplace froidement $A^2/\check{q}^2$ par
$+\infty$ dans l'équation (\ref{eq:IexpI}), on trouve que l'intégrale sur $X$ diverge comme $\alpha^{-2}$ lorsque $\alpha$
tend vers zéro, ce qui conduit à une intégrale sur $\alpha$ logarithmiquement divergente en $\alpha=0$.
\footnote{Ce phénomène ne se produit pas dans la contribution
(\ref{eq:IexpII}) car ni le facteur contenant le terme $X\sin^2\alpha$, dangereux lorsque $\alpha\to 0$,
ni celui contenant le terme $X\cos^2\alpha$, dangereux lorsque $\alpha\to \pi/2$, ne sont élevés au carré.
Du coup, l'intégrale sur $X$ de $0$ à $+\infty$ diverge seulement comme $\ln \alpha$ ou $\ln(\frac{\pi}{2}-\alpha)$ lorsque $\alpha\to 0$
ou $\alpha\to \pi/2$.
Mathématiquement, en remplaçant dans l'équation (\ref{eq:IexpII}) l'intégrale $\int_0^{+\infty} \dd X f(X)$ par $-\int_{-\infty}^{0} \dd X f(X)$
en vertu de l'identité (\ref{eq:resid}), puis en utilisant le théorème de convergence dominée, nous pouvons justifier l'échange de la limite $\eta\to 0^+$
et de l'intégration sur $\bar{q}_2, \bar{q}_3, \phi$ et $\alpha$, et donc démontrer le résultat (\ref{eq:IexpIInul}).}
Pour avancer, nous coupons l'intervalle d'intégration sur $\alpha$ en deux, un sous-intervalle $[0,\nu]$ pour lequel la coupure de l'intégration
sur $X$ à $A^2/\check{q}^2$ est requise, et un sous-intervalle $[\nu,\pi/2]$ pour lequel l'intégration sur $X$ peut être étendue à $+\infty$~:
\begin{multline}
\int_0^{\pi/2} \dd \alpha \int_0^{A^2/\check{q}^2} \!\!\!\!\dd X f(\alpha, X) \to \int_0^{\nu} \dd \alpha \int_0^{A^2/\check{q}^2}\!\!\!\! \dd X f(\alpha, X)  \\
+ \int_{\nu}^{\pi/2} \dd \alpha  \int_0^{+\infty} \dd X f(\alpha, X) 
\end{multline}
La contribution du second bout à $\im I^{(\mathrm{I})}$ est nulle, en vertu du théorème. La contribution du premier bout ne dépend pas de la valeur de $\nu$.
Il suffit donc de la calculer explicitement en prenant $\nu$ suffisamment petit pour que l'on puisse approximer chaque contribution dépendant de $\alpha$
dans l'intégrande par son terme d'ordre dominant en $\alpha$, à savoir $\sin\alpha \simeq \alpha$, $\cos\alpha\simeq 1$,
$u\simeq u_0=\bar{q}_2 (1-\bar{q}_3)/(1-\bar{q}_2-\bar{q}_3)$. La dépendance en $\phi$ disparaît et l'intégration sur $\phi$ sort simplement un facteur $2\pi$.
On calcule aisément l'intégrale sur $\alpha$~:
\begin{multline}
\int_0^\nu \frac{\alpha \, \dd\alpha}{[\frac{3}{4}\bar{q}_3(1-\bar{q}_3)-\frac{\bar{q}_3 X\alpha^2}{1-\bar{q}_3}+\ii\eta]^2}= \\
\left[\frac{1-\bar{q}_3}{2\bar{q}_3X [\frac{3}{4}\bar{q}_3(1-\bar{q}_3)-\frac{\bar{q}_3 X\alpha^2}{1-\bar{q}_3}+\ii\eta]}\right]_{\alpha=0}^{\alpha=\nu}
\label{eq:apresintegalpha}
\end{multline}
On passe ensuite à l'intégration sur $X$. Le terme tout intégré $\alpha=\nu$ de l'équation (\ref{eq:apresintegalpha}), après multiplication par le facteur $X$
au numérateur de (\ref{eq:IexpI}) et division par le dénominateur $(v-u_0X+\ii\eta)$, donne un intégrande en $O(1/X^2)$~; on peut dans ce cas remplacer la borne
supérieure $A^2/\check{q}^2$ de l'intégration sur $X$ par $+\infty$ et utiliser la variante suivante du théorème pour montrer que sa contribution
à $\im I^{(\mathrm{I})}$ est exactement nulle~:
\be
\lim_{\eta\to 0^+} \int_{\mathbb{R}^+} \dd X \im \frac{1}{[a(X)+\ii \eta] [b(X)+\ii \eta]} = 0 \label{eq:vartheo}
\ee
Le terme tout intégré $\alpha=0$ de l'équation (\ref{eq:apresintegalpha}) conduit au contraire à une intégrale sur $X$ de partie imaginaire non nulle à la limite
$\eta\to 0^+$~:
\begin{multline}
\im \int_0^{A^2/\check{q}^2} \frac{(1-\bar{q}_3)\dd X}{2\bar{q}_3(v-u_0 X+\ii\eta)[\frac{3}{4}\bar{q}_3(1-\bar{q}_3)+\ii\eta]} =  \\
-\frac{1-\bar{q}_3}{2\bar{q}_3u_0} \im \frac{\left[\ln(v-u_0A^2\check{q}^{-2}+\ii\eta) - \ln(v+\ii\eta)\right]}{\frac{3}{4}\bar{q}_3(1-\bar{q}_3)+\ii\eta} \\
\underset{\eta\to 0^+}{\to} -\frac{\pi}{u_0} \frac{\Theta(u_0A^2\check{q}^{-2}-v)}{\frac{3}{2}\bar{q}_3^2}
\end{multline}
où $\Theta$ est la fonction de Heaviside et $\ln$ est la détermination habituelle du logarithme complexe, avec une ligne de coupure sur les réels négatifs.
Il reste donc
\begin{multline}
\im I^{(\mathrm{I})} = 2\pi^2 \int_0^1 \dd\bar{q}_2 \int_0^{1-\bar{q}_2} \!\!\!\!\dd\bar{q}_3 (1-\bar{q}_3) (1-\bar{q}_2-\bar{q}_3)^2 \bar{q}_2^2 \bar{q}_3 \\
\times \Theta(u_0A^2\check{q}^{-2}-v)  \underset{\check{q}\to 0}{\to}
\frac{\pi^2}{840}
\end{multline}
d'où la correction (\ref{eq:Gamma1d3}) à apporter au résultat (\ref{eq:Bel_result}).
}

Pour terminer, on peut se demander s'il existe une interpr\'etation physique aux diagrammes d'ordre $4$ en $\hat{V}$.
Nous sommes en mesure d'en proposer une pour celui de gauche (I) de la figure \ref{fig:diag4}. Ce diagramme prend en compte
l'effet d'une modification de la pulsation propre des phonons virtuels sur le d\'eplacement d'\'energie complexe $\Delta \epsilon_\qq^{\rm Bel}$
du phonon $\qq$ induit par le processus de Beliaev, cette modification \'etant elle-m\^eme imputable à cet ordre à un processus de Beliaev.
Plus quantitativement, introduisons la fonction \'energie propre de Beliaev (\ref{eq:DeQBelZ}),
en notant d\'esormais de fa\c{c}on plus rigoureuse l'\'energie propre non perturb\'ee des phonons de vecteur d'onde $\kk$
dans $\hat{H}_2$ sous la forme $\epsilon_\kk^{(2)}$, pour la distinguer de l'\'energie r\'eelle vraie $\epsilon_\qq$ 
(voir la note \thenoterenormalisation). 
Nous avons v\'erifi\'e alors que le changement d'\'energie complexe du phonon $\qq$ associ\'e au diagramme de gauche vaut exactement
\begin{multline}
\delta z_\qq^{\rm (I)}= \frac{1}{2} \sum_{\qq_3} \frac{(\langle \qq_3,\qq-\qq_3|\hat{H}_3|\qq\rangle)^2}{[\epsilon_\qq+\ii\eta
-(\epsilon_{\qq_3}^{(2)}+\epsilon_{\qq-\qq_3}^{(2)})]^2} \\
\times \left[\Delta \epsilon_{\qq_3}^{\rm Bel}(\epsilon_\qq+\ii\eta-\epsilon_{\qq-\qq_3}^{(2)}) 
+\Delta \epsilon_{\qq-\qq_3}^{\rm Bel}(\epsilon_\qq+\ii\eta-\epsilon_{\qq_3}^{(2)})\right]
\label{eq:reformul}
\end{multline}
On notera que l'argument $z$ de $\Delta \epsilon^{\rm Bel}_{\qq-\qq_3}$ vaut ici l'\'energie vraie $\epsilon_\qq$ diminu\'ee de l'\'energie non perturb\'ee
du phonon virtuel spectateur dans le diagramme (I), c'est-à-dire non partie prenante dans la boucle.
La somme de la correction (I) et du d\'eplacement de Beliaev perturbatif habituel du phonon $\qq$ peut donc s'interpr\'eter, \`a l'ordre 
$4$ en $\hat{H}_3$, comme un d\'eplacement de Beliaev pour une relation de dispersion complexe renormalis\'ee~:
\begin{widetext}
\be
\Delta \epsilon_\qq^{\rm Bel}(\epsilon_\qq+\ii\eta)+\delta z_\qq^{\rm (I)}\simeq \frac{1}{2} \sum_{\qq_3,\qq_4} 
\frac{(\langle\qq_3,\qq_4|\hat{H}_3|\qq\rangle)^2 \delta_{\qq_3+\qq_4,\qq}}
{\epsilon_\qq+\ii\eta-[\epsilon_{\qq_3}^{(2)}+\Delta\epsilon_{\qq_3}^{\rm Bel}(\epsilon_\qq+\ii\eta-\epsilon_{\qq_4}^{(2)})
+\epsilon_{\qq_4}^{(2)}+\Delta\epsilon_{\qq_4}^{\rm Bel}(\epsilon_\qq+\ii\eta-\epsilon_{\qq_3}^{(2)})]}
\label{eq:jolimaislong}
\ee
\end{widetext}
La formulation (\ref{eq:reformul}) donne d'ailleurs une autre m\'ethode de calcul que celle de l'\'equation (\ref{eq:IexpI}).
Avec les changements
d'\'echelle de la sous-section \ref{subsec:ccadlk} adapt\'es à l'existence des \g{petits d\'enominateurs}, et en utilisant (\ref{eq:intermdeux})
avec $(\qq,\kk)$ valant successivement $(\qq_3,\kk)$ et $(\qq,\qq_3)$, où $\kk$ est la variable d'int\'egration apparaissant
dans l'expression tir\'ee de (\ref{eq:DeQBelZ}) de $\Delta \epsilon_{\qq_3}^{\rm Bel}(\epsilon_\qq+\ii\eta-\epsilon_{\qq-\qq_3}^{(2)})$, nous trouvons
\begin{widetext}
\begin{multline}
\frac{\Delta \epsilon_{\qq_3}^{\rm Bel}(\epsilon_\qq+\ii\eta-\epsilon_{\qq-\qq_3}^{(2)})}{mc^2}=
-\frac{9(1+\Lambda_{\rm F})^2}{64\pi^2} \left(\frac{mc}{\hbar\rho^{1/3}}\right)^3
\bar{q}_3^5 \check{q}^5 \\ \times \int_0^1 \dd\bar{k}\, \bar{k}^2 (1-\bar{k})^2 
\left[\ii\pi +\ln \left(\frac{4\bar{q}_3\bar{k}A^2}{(1-\bar{k})\check{q}^2}\right)-\ln\left(\ii\eta+3\bar{q}_3(1-\bar{q}_3)+3\bar{q}_3^3\bar{k}(1-\bar{k})
-\frac{4\bar{q}_3\check{\theta_3}^2}{1-\bar{q}_3}\right)\right]
\label{eq:debelexpli}
\end{multline}
Ici $\bar{q}_3=q_3/q$, $\bar{k}=k/q_3$, $\check{\theta}_3=\theta_3/(\gamma^{1/2}\check{q})$, $\theta_3$ étant l'angle non orienté entre les vecteurs
$\qq$ et $\qq_3$. L'intégrale sur $\check{\theta}=\theta/(\gamma^{1/2}\check{q}_3)$, où $\theta$ est l'angle entre $\kk$ et $\qq$, a été effectuée explicitement sur
l'intervalle $[0,A/\check{q}_3]$ avec le même paramètre de coupure $A$ que dans l'équation (\ref{eq:IexpI}).
Apr\`es report de (\ref{eq:debelexpli}) dans l'\'equation (\ref{eq:reformul}), nous obtenons
\begin{multline}
\frac{\im \delta z^{\rm (I)}_\qq}{mc^2}=-\frac{9^2(1+\Lambda_{\rm F})^4}{64\pi^4\gamma} \left(\frac{mc}{\hbar\rho^{1/3}}\right)^6
\check{q}^7 \int_0^1 \dd\bar{q}_3\, \bar{q}_3^8 (1-\bar{q}_3) \int_0^1 \dd\bar{k}\, \bar{k}^2(1-\bar{k})^2  \\
\times \im \int_0^{+\infty} \dd X_3 \frac{\ii\pi-\ln\left(\ii\eta+3\bar{q}_3(1-\bar{q}_3)+3\bar{q}_3^3\bar{k}(1-\bar{k})
-\frac{4\bar{q}_3X_3}{1-\bar{q}_3}\right)}{\left(\ii\eta+3\bar{q}_3(1-\bar{q}_3)-\frac{4\bar{q}_3 X_3}{1-\bar{q}_3}\right)^2}
\end{multline}
Nous avons omis le terme logarithmique réel et indépendant de $\check{\theta}_3$ entre crochets dans l'équation (\ref{eq:debelexpli}) car sa contribution
au résultat final est clairement réelle dans la limite $\eta\to 0^+$. C'est ainsi que disparaît la dépendance en le paramètre de coupure $A$. 
L'intégrale sur {$X_3=\check{\theta}_3^2$} du bout avec logarithme se calcule par parties, bien entendu en dérivant le logarithme~; on trouve que sa contribution est nulle 
à la limite $\eta\to 0^+$ en utilisant l'identité (\ref{eq:vartheo}).
Finalement, seul le terme $\ii\pi$ importe, et il conduit exactement au résultat (\ref{eq:Gamma1d3}).
\end{widetext}


\providecommand*\hyphen{-}